\documentclass[12pt,lot, lof]{themes/puthesis}

\newcommand{\proquestmode}{}

\title{Early Detection of Research Trends}

\submitted{November 2018}  
\copyrightyear{2018}  
\author{Angelo Antonio Salatino}
\adviserfir{Prof. Enrico Motta}  
\advisersec{Dr. Francesco Osborne}
\department{Knowledge Media Institute}
\university{The Open University}
\degreetitle{Doctor of Philosophy}
\crest{\includegraphics[width=0.2\textwidth]{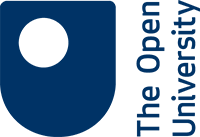}}
\collegeshield{\includegraphics[width=0.25\textwidth]{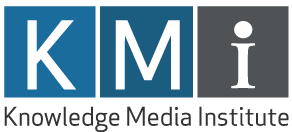}}

\usepackage{graphicx}

\usepackage{verbatim}

\usepackage{multirow}
\usepackage{longtable}

\usepackage{booktabs}

\usepackage{imakeidx}
\makeindex[columns=2, title=Index, 
           options= -s themes/example_style.ist]

\usepackage{csquotes}

\usepackage{color,soul}
\usepackage[bordercolor=white,backgroundcolor=gray!30,linecolor=black,colorinlistoftodos]{todonotes}

\usepackage{soul,siunitx,natbib}
\soulregister{\ref}{1}
\soulregister{\citep}{1}
\soulregister{\cite}{1}
\soulregister{\textit}{1}

\usepackage[inline]{enumitem}

\usepackage{epstopdf}

\usepackage[ruled,longend]{algorithm2e}

\usepackage{xcolor,colortbl}

\usepackage{subcaption}

\usepackage{amssymb}
\usepackage{amsmath}
\mathchardef\mhyphen="2D 

\usepackage{hyperref}

\usepackage{footmisc} 
\usepackage[colorinlistoftodos]{todonotes}\reversemarginpar

\usepackage{array}
\newcolumntype{L}[1]{>{\raggedright\let\newline\\\arraybackslash\hspace{0pt}}m{#1}}
\newcolumntype{C}[1]{>{\centering\let\newline\\\arraybackslash\hspace{0pt}}m{#1}}
\newcolumntype{R}[1]{>{\raggedleft\let\newline\\\arraybackslash\hspace{0pt}}m{#1}}

\usepackage{textcomp}

\usepackage{siunitx}

\usepackage[labelfont=bf]{caption}

\ifdefined\printmode

\usepackage{url}

\else

\ifdefined\proquestmode

\hypersetup{bookmarksnumbered}

\makeatletter
\hypersetup{pdftitle=\@title,pdfauthor=\@author}
\makeatother

\else

\hypersetup{colorlinks,bookmarksnumbered}

\makeatletter
\hypersetup{pdftitle=\@title,pdfauthor=\@author}
\makeatother

\fi 
\fi 

\ifodd 0
\renewcommand{\maketitlepage}{}
\renewcommand*{\makecopyrightpage}{}
\renewcommand*{\makeabstract}{}

\else

\abstract{
Being able to rapidly recognise new research trends is strategic for many stakeholders, including universities, institutional funding bodies, academic publishers and companies. The literature presents several approaches to identifying the emergence of new research topics, which rely on the assumption that the topic is already exhibiting a certain degree of popularity and consistently referred to by a community of researchers. However, detecting the emergence of a new research area at an embryonic stage, i.e., before the topic has been consistently labelled by a community of researchers and associated with a number of publications, is still an open challenge. In this dissertation, we begin to address this challenge by performing a study of the dynamics preceding the creation of new topics. This study indicates that the emergence of a new topic is anticipated by a significant increase in the pace of collaboration between relevant research areas, which can be seen as the ‘ancestors’ of the new topic. Based on this understanding, we developed Augur, a novel approach to effectively detect the emergence of new research topics. Augur analyses the diachronic relationships between research areas and is able to detect clusters of topics that exhibit dynamics correlated with the emergence of new research topics. Here we also present the Advanced Clique Percolation Method (ACPM), a new community detection algorithm developed specifically for supporting this task. Augur was evaluated on a gold standard of 1,408 debutant topics in the 2000-2011 timeframe and outperformed four alternative approaches in terms of both precision and recall.
}

\declarationofauthorship{

I, \MakeUppercase{\MyAuthor}, declare that this dissertation titled,`\MakeUppercase{\MyTitle}' and the work presented in it are my own. I confirm that:

\begin{itemize} 
\item[\tiny{$\blacksquare$}] This work was done wholly or mainly while in candidature for a research degree at this University.
 
\item[\tiny{$\blacksquare$}] Where any part of this thesis has previously been submitted for a degree or any other qualification at this University or any other institution, this has been clearly stated.
 
\item[\tiny{$\blacksquare$}] Where I have consulted the published work of others, this is always clearly attributed.
 
\item[\tiny{$\blacksquare$}] Where I have quoted from the work of others, the source is always given. With the exception of such quotations, this thesis is entirely my own work.
 
\item[\tiny{$\blacksquare$}] I have acknowledged all main sources of help.
 
\item[\tiny{$\blacksquare$}] Where the thesis is based on work done by myself jointly with others, I have made clear exactly what was done by others and what I have contributed myself.
\\
\end{itemize}

Signed:\\
\rule[1em]{25em}{0.5pt}  
 
Date:\\
\rule[1em]{25em}{0.5pt}  

}

\acknowledgements{
The PhD journey is a tough one. It is meant to last between 3 to 4 years in which you are involved in a wide range of activities, you face several challenges, and it requires an enormous amount of energies. It is not all doom and gloom, though. I have been fortunate enough to be constantly surrounded by amazing people, and I would like to use this space to express my gratitude to those who contributed in different ways to this journey.

First and foremost, I would like to thank Beppe. He is involved in this journey as much as I am. He helped me in applying to this studentship and provided useful insights while writing my research proposal. We know each other since 2007, we went to university together and we both started a PhD at the KMi. The ``barese couple'' I believe we were called at that time. Being with him here, made my move to the UK smoother. We were doing a lot of stuff together, except playing League of Legends. I grew a lot beside him. I owe him a debt of gratitude.

Special thanks go to my \textit{intellectual fathers} Enrico and Francesco, who gave me the opportunity to grow as a researcher under their mentorship. Their support and guidance have been crucial for the development of my work and skills. Enrico is an experienced supervisor. Every time I would face difficulties, he would jump on board and simplify my overcomplicated thoughts. Francesco is an engine of ideas. He is very knowledgeable, and he is able to break down problems and challenge ideas on the fly. It was a privilege being supervised by them.

I also would like to thank both my examiners, Kalina Bontcheva and Alun Preece, for making my defence enjoyable and fairly though, and for their valuable comments aiming at improving this dissertation. It was a very interesting discussion.

Throughout this journey, I also had the privilege to meet unique people, with whom I shared the pains and joys of this journey: Patrizia, Pinelopi, Loua, Tina. Always there providing advice, help and encouragement. To say thank you is not enough.

A great thank also goes to Alberto, Alessandro, Andrea, Thiviyan, Giorgio, Paco, Maria, Ilaria, Manu, Martin, Matteo, Simon, Marilena, and many others, who made my life in Milton Keynes always warm and enjoyable.

I extend my gratitude to all my colleagues in the Knowledge Media Institute (kmiers).
Their professional support, along with their unique positive spirit and attitude, make KMi one of the best and most unique places to work.

Special recognition goes to my family, which allowed me to follow my dreams and supported me in every possible mean. From them, I inherited two important values that allowed me also to succeed in this mission: patience and persistence.

I am also grateful to my friends in Sannicandro di Bari for making my homecoming as if I never left, and for tirelessly trying to convince me in moving back home.

Last but not least, I would like to thank Alessia for standing by me when the going was tough. 
Thank you all.
}

\dedication{
\newpage
To those who keep feeding my spirit of inquiry.
\newpage
}

\fi  

\begin{document}

\makefrontmatter


\part{Introduction and State of the Art}

\chapter{Introduction\label{ch:intro}}

The research environment changes and evolves rapidly: new research areas constantly emerge, whereas some others fade out. The ability to promptly recognise the emergence of new research topics is an important asset for anybody involved in the research environment, including journal editors, academic publishers, researchers, institutional funding bodies and other relevant stakeholders. Nowadays, as we are experiencing an exponential growth of research publications \citep{larsen2010}, keeping up with new trends is becoming progressively more challenging.


In the last two decades, very large repositories of scholarly data and other relevant sources have become available, opening the way to novel data-intensive approaches capable of detecting novel topics and their trends \citep{wu2016,he2009,duvvuru2012,bolelli2009}. 
However, a timely detection of research topics is still an open challenge.

In this work, we are going to face this challenge by creating a system able to anticipate the emergence of new research topics. Specifically, we define a research topic as a subject of study or issue that is of interest to the research community, and it is addressed in at least few research papers.
Subjects of study can encompass different research areas and application-dependent concepts. For instance, a paper introducing a new technology to classify data, is the event that triggers the emergence of new topic. Any paper discussing the implementation of such technology, its application, the release of a new efficient implementation, its evaluation and so on, are all part of the same research topic. 
We will provide further details about this definition in Section \mbox{\ref{sec:topicdefinition}}.

\section{Problem statement}\label{sec:problemstatement}
\begin{figure}[!h]
\begin{center}
\includegraphics[width=400px]{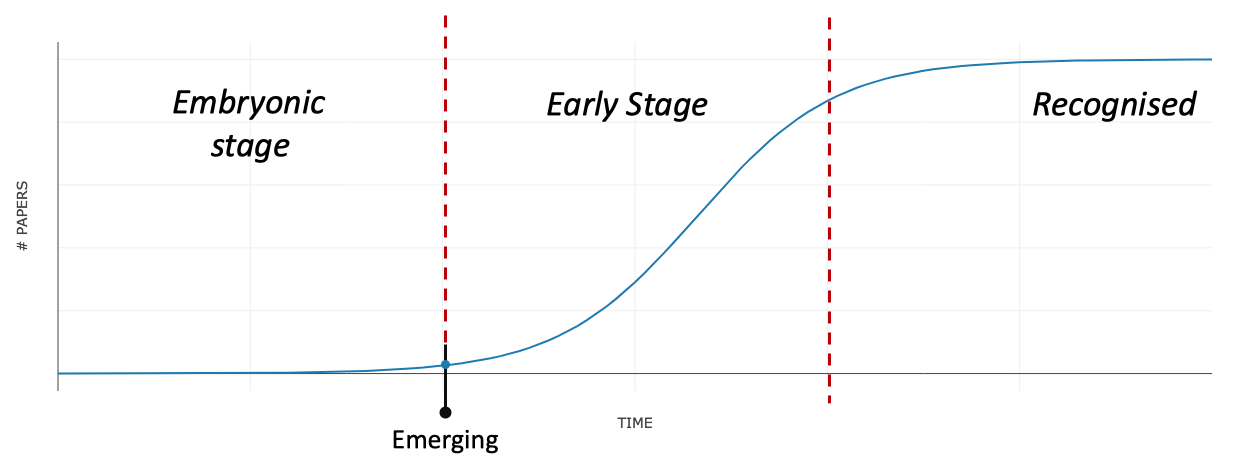}
\end{center}
\caption{Representation of a topic lifecycle in terms of published research papers.}
\label{fig:topic-timeline-pre}
\end{figure}

Looking deeply into the evolution of research topics we can observe that they go through a sequence of life stages. Specifically, we can distinguish three main stages: 
\begin{enumerate*}[label=(\roman*)]
\item embryonic stage,\index{embryonic stage}
\item early stage and\index{early stage}
\item recognised\index{recognised stage}
\end{enumerate*}, as showed in Fig. \ref{fig:topic-timeline-pre}.
A research topic, in its \textit{embryonic stage}, is still an idea or concept and it did not emerge, yet. Specifically, in this stage, a topic has not yet been explicitly labelled and recognised by a research community, but it is already taking shape, as evidenced by the fact that researchers from a variety of fields are forming new collaborations and producing new work, starting to define the challenges and the paradigms associated with the emerging new area.
A research topic in its \textit{early stage}, instead, has recently emerged and a relatively small group of researchers agree on certain theories which will allow the topic to thrive. As a result, there is a new label for it, and it is associated to a limited number of papers. Afterwards, when a research topic enters in its \textit{recognised stage}, it becomes mature and many researchers are actively producing and disseminating their results. The topic is then associated with a substantial number of papers.

As an example, for the \textit{Semantic Web}, we can observe that before its emergence in 2001, in its embryonic stage, it still was a concept in which communities of researchers from \textit{Artificial Intelligence}, \textit{World Wide Web}, \textit{Knowledge Representation}, and \textit{Knowledge-Based Systems} were joining their forces. After the 2001, the topic emerges and enters in its early stage, earning its own identity, in which an increasing number of practitioners started to work in it. Around the 2005, the \textit{Semantic Web} reaches its own maturity (recognised phase) with over 1500 papers published each year.


Current approaches for detecting research topics \citep{wu2016,he2009,duvvuru2012,bolelli2009}, focus on highlighting topics that are already associated with a number of publications and consistently referred to by a community of researchers. This limitation means that these solutions can only identify topics that already have an initial degree of consolidation, i.e., after a certain latency from their emergence. As a result, they can only provide limited value to stakeholders who wish to anticipate and promptly react to new developments in the research landscape. 

A strategy to address this problem would be to anticipate the emergence of new research topics by detecting them at their embryonic stage. This stage was initially observed by Thomas Kuhn\index{Kuhn Thomas}, who argued in his book \citep[page 86]{kuhn1970} that, topics might exist in the form of embryo. However, it has never been studied quantitatively, probably due to the complexity of the issue and the difficulty of formally defining the notion of research topic.


Nevertheless, in this dissertation we address this problem by introducing a novel framework, Augur, for identifying the appearance of new topics at an embryonic stage. Augur analyses networks of research topics, detects areas exhibiting a significant increase in the pace of collaboration, and produces clusters of topics correlated with the future emergence of new research areas. For instance, if available, before 2001, Augur would have allowed us to observe that previously less connected topics (e.g., \textit{Artificial Intelligence}, \textit{World Wide Web} and \textit{Knowledge-Based Systems}) were increasingly collaborating with each other\footnote{In this dissertation, we will use the expression ``collaboration between research area'' as a shortcut for ``collaboration between research communities associated with specific research areas''. The community of a research area is given by the authors who publish in the area in question.}. These dynamics indeed led to the emergence of a new research area, later labelled as \textit{Semantic Web} by Berners-Lee  \cite{berners2001}\footnote{It should be noted that this term was first introduced in the book Weaving the web \citep{berners1999}, but it was not really recognised by the scientific community until the Scientific American paper of 2001.}.

In the following sections, we discuss the motivation behind this dissertation, and detail our research questions, hypotheses, methodology and approach, and contributions.

\section{Motivation}
As already mentioned, understanding and reacting timely to new developments in the research landscape is critical for a variety of stakeholders. For instance, researchers need to stay up-to-date with new trends related to their topics and potentially interesting new research areas. Thanks to these insights, they can evolve their research agenda, ensuring they focus on ideas and concepts at the leading edge of the current landscape. 

Institutional funding bodies and companies also need to be aware of the latest research developments and promising trends. For instance, this knowledge can inform business decisions and suggest the selection of technologies on which to invest. 

Similarly, it is crucial for academic publishers and editors to know in advance new emerging topics with the aim of offering the most up to date and interesting content. For instance, a publisher can gain a competitive advantage by being the first to recognise the importance of a new trend, and thus publish a special issue or a journal about it. Indeed, financial support for this doctoral project comes from \textit{Springer Nature}, which is a global publishing company.

The undeniable potential of an approach for detecting novel research trends has recently attracted an increasing research interest to this area. However, as we will discuss in Chapter \ref{ch:pastwork}, current state-of-the-art solutions suffer from significant limitations when applied to the detection of trends at their early stage.

\section{Research questions}\label{sec:rqs}
The main research question investigated in this dissertation is:

\begin{center}
\textbf{Is it possible to detect a new research topic at the embryonic stage before it is consistently recognised by a research community (e.g., there is an established label for it)?}
\end{center}

We focused specifically on the creation of a novel approach that is able to detect the emergence of new research topics at their embryonic stage. Given the dimension of such problem, we articulated this main question in a set of related questions:

\begin{enumerate}[label={RQ\arabic*:},leftmargin=3cm]
\item Is it possible to precisely define the notion of established topic?
\item How early in the topic lifecycle is it possible to identify an emerging topic?
\item	What are the indicators that can be exploited to predict the emergence of new topics?
\item	Is it possible to develop an effective computational method that can support this prediction task?
\item	Are there commonalities between our approach to predicting the emergence of new topics and epistemological theories of research dynamics?
\item	What evaluation mechanisms are appropriate for this task? 
\end{enumerate}

In what follows, we will introduce and discuss these research questions. In the Conclusion (see Chapter \ref{ch:conclusion}), we will reprise each of them and summarise the answers according to the results showed in this dissertation. 

\subsection{RQ1: Is it possible to precisely define the notion of established topic?}\label{sec:rq1}
One of the recurring problems while reading the literature is the lack of widely-accepted definition for research topic. This issue is present also in \textit{Philosophy of Science}, as philosophers do not agree about a shared definition. This is mainly because research topics come in all shapes and sizes and therefore it is very hard to create a formal definition that would fit them all. 

This limitation has led many approaches that extract topics from collections of documents to propose their own definition, often tightly-coupled with the algorithmic method they employ. Indeed, some approaches use keywords as proxies for topics, others match terms to manually-curated taxonomies, and some others apply statistical techniques to associate topics to bags of words.

It is not clear when a topic can be considered established, but intuitively this assessment could derive from a number of indicators such as the number of papers addressing it, the number of researchers working on it, and the time elapsed since its debut. Acknowledging when and which topics can be considered established allows us to define a category of topics against which we can compare the emerging ones. Indeed, in Chapter \ref{ch:firststudy}, we compare emerging topics with established topics, and we suggest some indicators to determine when a research topic is definitely established.

\subsection{RQ2: How early in the topic lifecycle is it possible to identify an emerging topic?}\label{sec:rq2}



\begin{figure}[h]
\begin{center}
\includegraphics[width=400px]{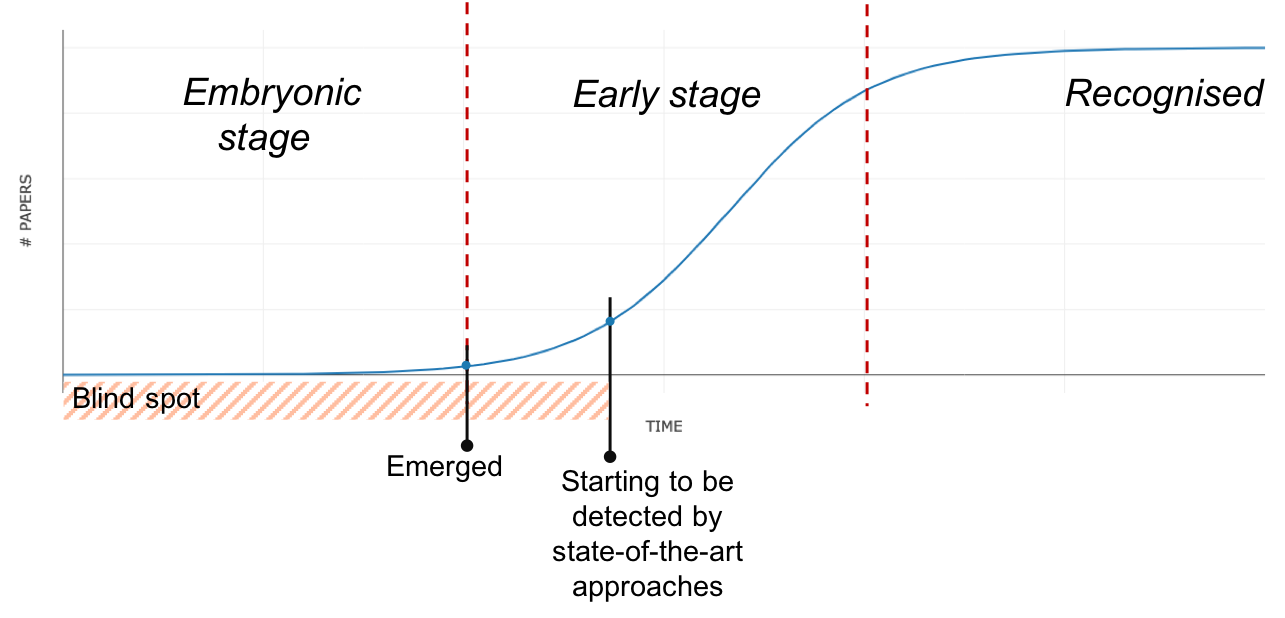}
\end{center}
\caption{Representation of a topic lifecycle, pointing out when it emerges, and it starts to be detected by state-of-the-art approaches.}
\label{fig:topic-timeline}
\end{figure}

As already mentioned in Section \ref{sec:problemstatement}, research topics go through different stages within their lifecycle.
The literature presents several approaches that are able to detect research topics, and as we will discuss in Chapter \ref{ch:pastwork}, they are able to perform their detection only later in their lifecycle (see Fig. \ref{fig:topic-timeline}). This is mainly because most of them perform a statistical analysis of the popularity of the labels associated with a topic across a certain time period (typically 2-4 years). 

It would be much more useful to detect topics in their very early stage and thus react before the currently available approaches, i.e., within the blind spot showed in Fig.~\ref{fig:topic-timeline}. Therefore, how early can we actually detect a new topic?

Our hypothesis is that it is possible to detect a topic at the embryonic stage. This endeavour will then seek to forecast new topics that are about to emerge rather than simply acknowledging that they emerged.

Therefore, the next question to answer is: how can we forecast the emergence of a new topic before it actually emerges?

In Section \ref{sec:topicdefinition}, we will analyse in more detail the different stages that a research topic goes through, while in Chapter \ref{ch:firststudy} and \ref{ch:secondstudy}, we will discuss how soon it is possible to forecast the emergence of a new topic.

\subsection{RQ3: What are the indicators that can be exploited to predict the emergence of new topics?}\label{sec:rq3}
A prediction system is a tool capable to make automatic and objective predictions about future or unknown events. These systems are usually built using statistical techniques such as predictive modelling, machine learning, and data mining to analyse historical data. This analysis consists of identifying patterns and latent relationships, given some indicators, that have predictive power.

In this doctoral work, this analysis is crucial to determine the main indicators that drive the emergence of a new research topic. We will then develop a prediction system with the ability to sense such patterns, with the aim of forecasting the emergence of new topics.

Therefore, what kind of indicators can be used to predict the emergence of new research topics? How can we analyse them? How we can be certain of their effectiveness? Will these indicators be general enough to consider the peculiarities of different fields (e.g., \textit{Computer Science}, \textit{Business}, \textit{Medicine} and so on)? Intuitively, there could be a wide array of relevant indicators that could anticipate the creation of a new research area. These may include a new collaboration between two or more research communities, the creation of interdisciplinary workshops, a rise in the number of experts working on a certain combination of topics, a significant change in the vocabulary associated with relevant topics, and so on.

In Chapter \ref{ch:firststudy}, we address this question by analysing some indicators associated with the emergence of novel research topics. In particular, we analyse whether the emergence of a novel research topic can be anticipated by a significant increase in the pace of collaboration and a change in the network density between its related research areas.

\subsection{RQ4: Is it possible to develop an effective computational method that can support this prediction task?}\label{sec:rq4}
We believe that it is possible to use the indicators discussed in the previous section for devising new methods to forecast research topics. For instance, knowing that the pace of collaboration between existent research areas is associated with the emergence of a new topic suggests that we can analyse these dynamics and predict the characteristic of the topics that may emerge in the following years. This hypothesis opens up several other research questions. What kind of technologies can be adopted for mining large datasets with this purpose? How can we represent yet unlabelled research topics? How can we produce a scalable method for detecting the relevant dynamics? 

\begin{figure}[!h]
\begin{center}
\includegraphics[width=200px]{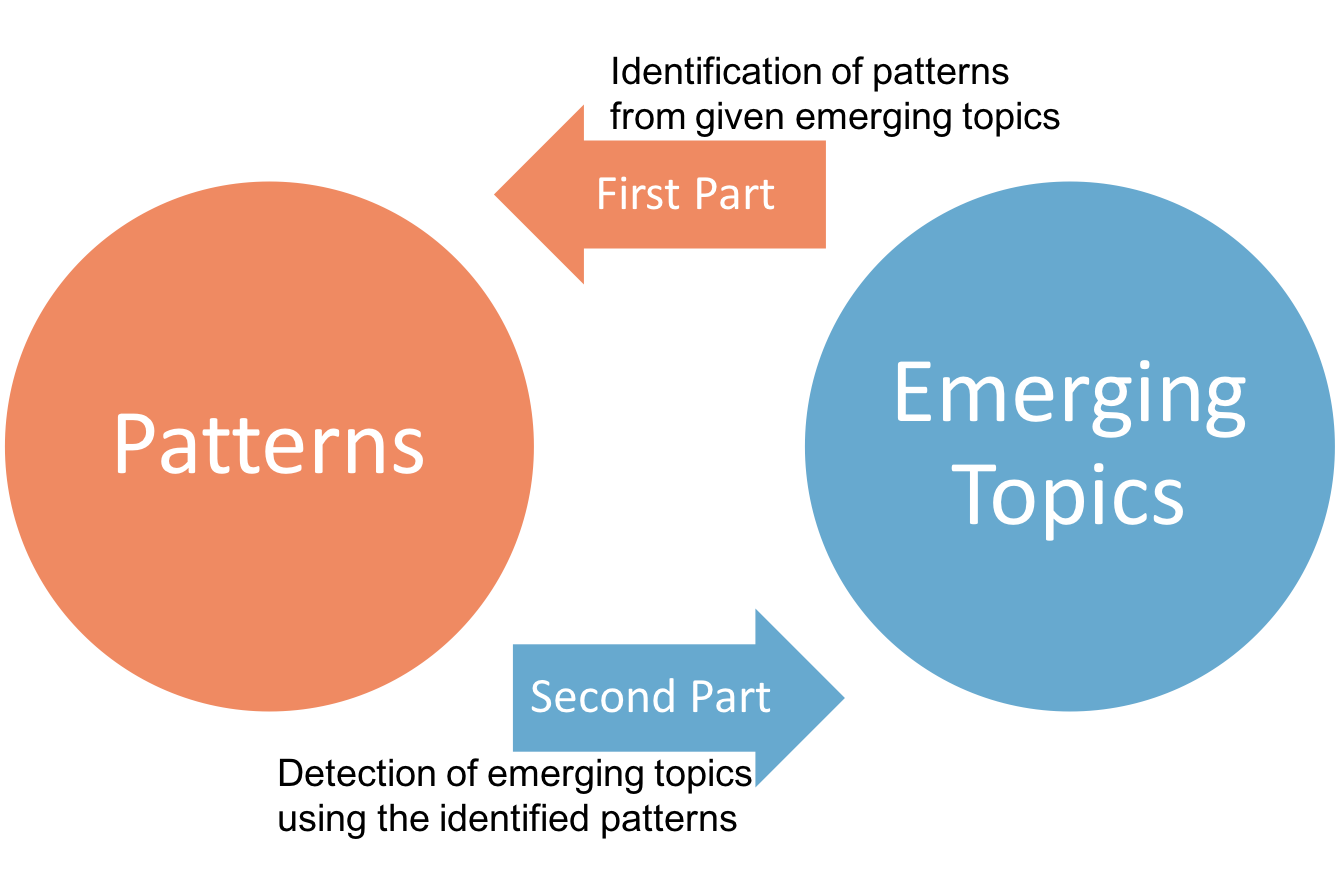}
\end{center}
\caption{Representation of the different stages of this doctoral work.}
\label{fig:stages-of-phd}
\end{figure}

To address these questions, we organised this doctoral work in two parts, as depicted by Fig. \ref{fig:stages-of-phd}. In the first part (Chapter \ref{ch:firststudy}), we perform a retrospective study for identifying indicators associated with the future emergence of research topics. In the second part (Chapter \ref{ch:secondstudy}), we exploit these indicators to produce a number of approaches for predicting the emergence of research topics and we evaluate them against a gold standard. 

We will address these questions in Chapter \ref{ch:firststudy} and \ref{ch:secondstudy}.

\subsection{RQ5: Are there commonalities between our approach to predicting the emergence of new topics and epistemological theories of research dynamics?}\label{sec:rq5}
In the literature, we can find several epistemological theories regarding the emergence of a new research area. One of the main contributions comes from \cite{kuhn1970}\index{Kuhn Thomas}, who theorised that science evolves through paradigm shifts. Research is pursued using a set of paradigms and, when these paradigms cannot cope with certain problems anymore, there is a paradigm shift that can lead to the emergence of a new scientific discipline. 
This shift is usually characterised by a number of activities, such as the exchange of theories and methods between scientific fields, the definition of new challenges and paradigms, the formation of new collaborations and so on. 

There are also other theories, which mention that different communities of researchers can be engaged in an exchange of tools, methodologies, and theories, leading then to the emergence of a new research area \citep{becher2001}.


In the following chapters, we will refer to some of these epistemological theories and we will highlight the similarities with our work.

\subsection{RQ6: What evaluation mechanisms are appropriate for this task? }\label{sec:rq6}
Evaluating our approach requires a formal framework for assessing the performance of a system predicting new research topics. Typically, a forecasting task is evaluated using historical data. In our case however, the research topics that we are trying to predict are not yet associated with a label. This makes the process more complex. What kind of information should be included in the gold standard? How can we compare the prediction of an emerging topic with the concrete dynamics of the research landscape? What are the metrics that we could use?

In Chapter \ref{ch:finalevaluation}, we will address these questions and discuss the evaluation of our approach.

\section{Research hypotheses}\label{sec:hypotheses}
In this section, we present three main hypotheses that are at the basis of this doctoral work.

\vspace{6mm} 
\textbf{Hypothesis 1:} \textit{Before being labelled and recognised by research communities, new topics go through an embryonic stage, in which researchers from different topics start to work on it.}

As we will see in Section \ref{sec:topiclifecycle}, a research topic goes through different stages during its lifecycle, such as the \textit{early stage} and the \textit{recognised phase}. Existing approaches are able to work in the last two stages, as they focus on topics that are already associated with a number of publications and for which the communities of researchers have already reached a consensus for a label. 

However, within the \textit{Philosophy of Science} there are some interesting theories about the emergence of new research topics, including the existence of an \textit{embryonic stage}. 

As already mentioned, at the embryonic stage a topic has not yet been explicitly labelled and recognised by a research community. However, the topic is already taking shape, as evidenced by the fact that researchers from a variety of fields are already working on it. 

Inspired by these theories, we hypothesise that such activity among researchers can be numerically measured and therefore it is possible to provide further evidence about its existence.

\vspace{6mm} 
\textbf{Hypothesis 2:} \textit{The emergence of a new research topic is anticipated by an increased rate of interaction of pre-existing topics, involved in developing this new area which is still in its embryonic stage.}

From a philosophical point of view, academic disciplines are specific branches of knowledge which together form the unity of knowledge that has been produced by the scientific endeavour. When two or more disciplines start to cooperate, they share their theories, concepts, methods and tools. The results of this cooperation may lead either to the creation of a new interdisciplinary research area or simply to a contribution in knowledge from one area to another. The basic hypothesis is that the creation of a topic is anticipated by a number of dynamics involving a variety of research entities, such as other topics, research communities, authors, venues and so on. Therefore, recognising these dynamics might enable a very early detection of emerging topics.

\vspace{6mm} 
\textbf{Hypothesis 3:} \textit{It is possible to create an automatic approach for detecting new emerging topics in their embryonic stage by analysing the dynamics of existing topics (i.e., observing their patterns of collaboration).}

\begin{figure}[h]
\includegraphics[width=\textwidth]{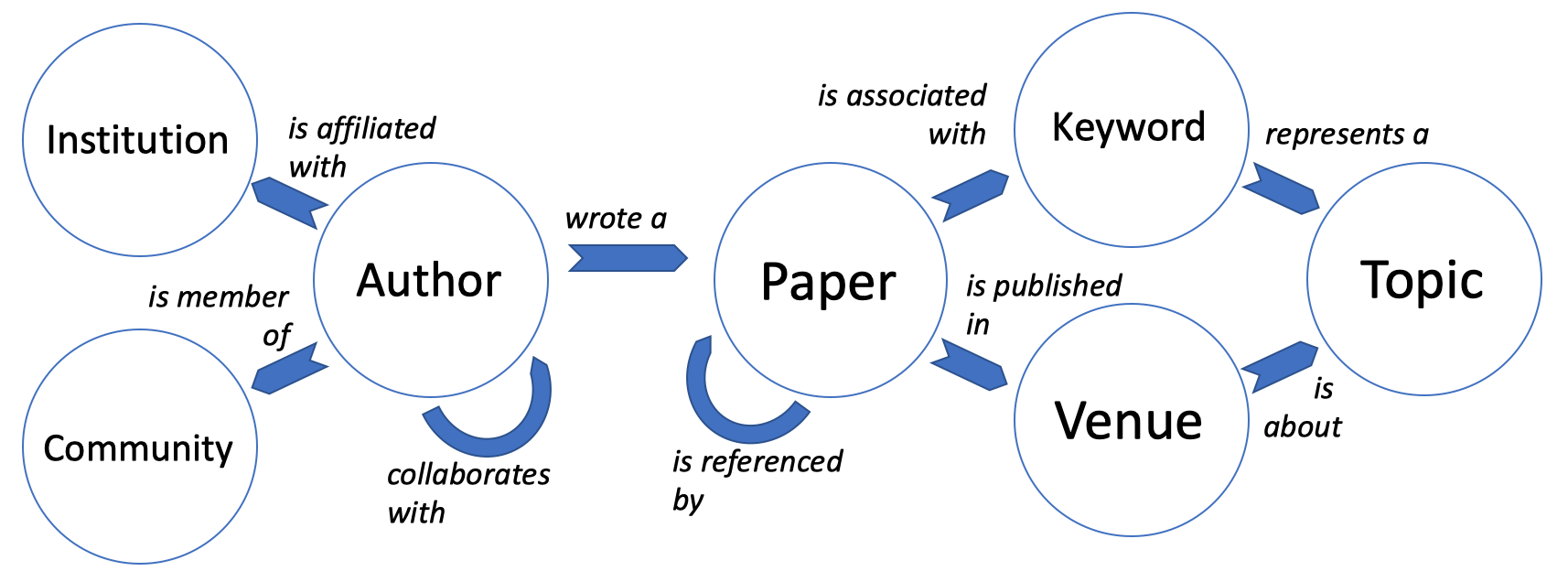}
\caption{Simplified model representing the scholarly meta-data and their relationships.}
\label{fig:scholarlydata}
\end{figure}

Scholarly data\index{scholarly data} can be used to analyse a large amount of research elements such as papers, authors, affiliations, venues, topics and communities \citep{osborne2014}. All these research elements are inherently interconnected by semantic relationships. For instance, Fig. \ref{fig:scholarlydata} shows some basic connections between the research elements according to our model. 

These connections can be used to derive a number of second order connections, e.g., authors are associated with a particular set of topics according to the papers they published. They also can be analysed in time to derive the dynamics that led to the emergence of a topic. 

Therefore, by exploiting the large variety of scholarly data, it should be possible to perform detection of topics at their embryonic stage.

\section{Research methodology}
For answering the research questions presented above, we use the \textit{experimental methodology}, which is widely adopted within \textit{Computer Science} for evaluating new solutions for given problems \citep{elio2011}. This experimental methodology consists of two different phases. Firstly, there is an exploratory phase in which the researchers investigate the different variables affecting the experimental group and measuring their effect. Then, researchers report the results of their evaluation with an analytical discussion stating what they had learned. 

This research effort started with an extensive analysis of the body of literature, which is reported in Chapter \ref{ch:pastwork}. We then formulated the research questions discussed in Section \ref{sec:rqs} and explored a number of ideas for addressing them. As a next step, we formulated our hypotheses regarding the existence of an embryonic stage for emerging topics.

\begin{figure}[h]
\includegraphics[width=\textwidth]{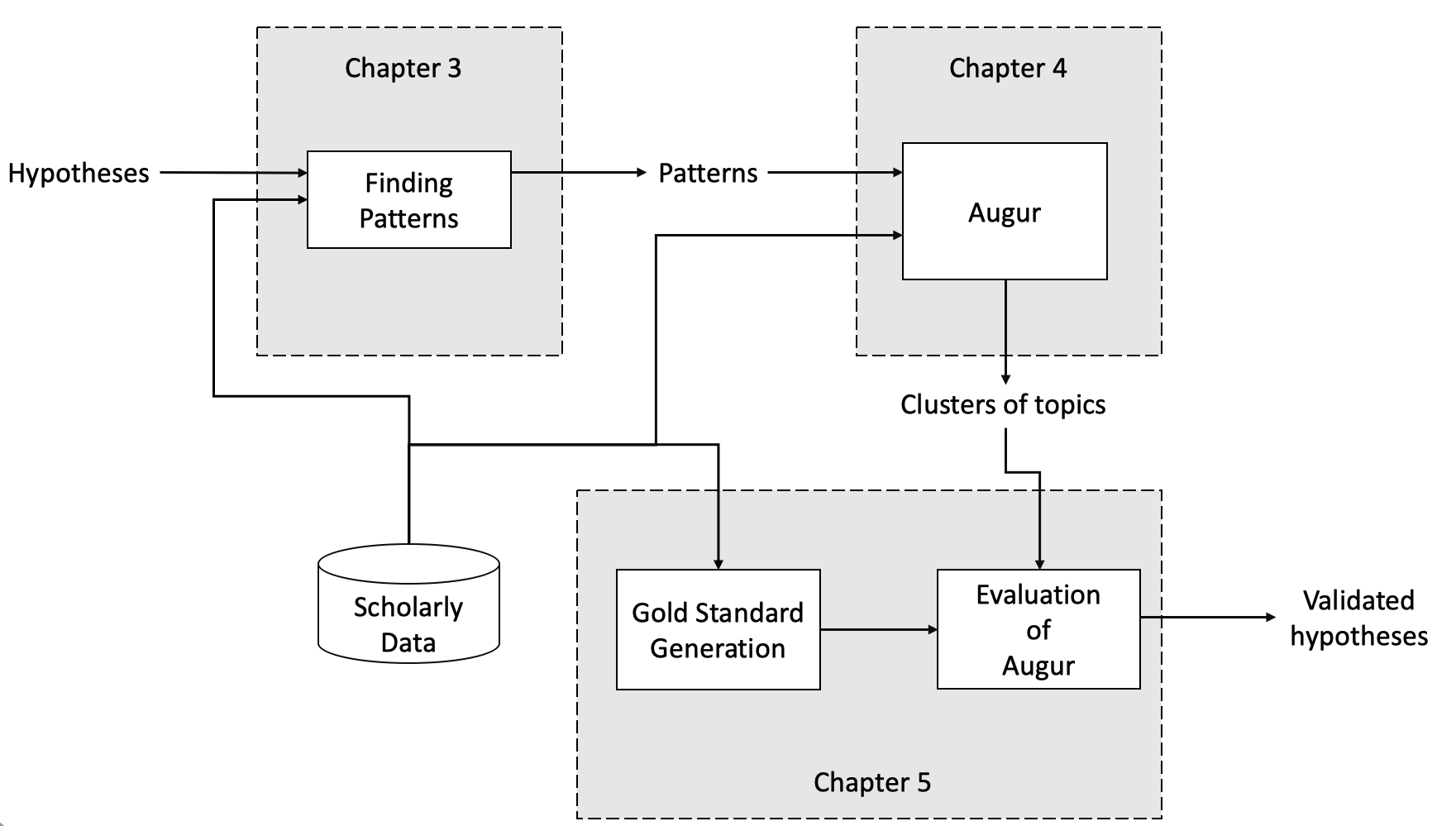}
\caption{Overview of Augur according to the dissertation narrative.}
\label{fig:approach-intro}
\end{figure}

Figure \ref{fig:approach-intro} summarises the process that we applied for going from these initial hypotheses to the creation and evaluation of Augur. We first conducted a preliminary study (described in Chapter \ref{ch:firststudy}) to investigate the hypothesis that the emergence of new topics can be anticipated by some indicators, such as an intensive collaboration among previously distant research communities. In this study, we generated a network of co-occurring topics from a sample of 3 million papers. We then compared the sections of the co-occurrence graphs where new topics are about to emerge with a control group of subgraphs associated with established topics. These graphs were analysed by using two novel approaches that integrate both statistics and semantics. The results provided evidence that the emergence of a novel research topic can be anticipated by a significant increase in the pace of collaboration between relevant research areas, which can be seen as the ``ancestors'' of the new topics. 

These dynamics and other insight yielded by the study were used to create Augur, a novel approach for predicting the emergence of research topics described in Chapter \ref{ch:secondstudy}. Finally, we evaluated Augur versus four alternative approaches on a dataset of 3 million papers (Chapter \ref{ch:finalevaluation}). 

\section{The Augur framework}
Augur is a novel framework that detects the emergence of new research areas by analysing topic networks and identifying clusters associated with an overall increase of the pace of collaboration.

Augur operates in three steps. Firstly, it creates an evolutionary network describing the collaboration of research topics in time. The evolutionary graph is a fully weighted graph in which nodes represent the topics, and arcs represent the collaboration between those two topics. Then node and arc weights represent how much topics and their inner collaboration are growing in time.

As a second step, Augur uses a novel clustering algorithm, the Advanced Clique Percolation Method (ACPM), for locating areas of the network that exhibit a significant increase in the pace of collaboration. ACPM crawls the evolutionary network and groups together nodes (i.e., topics) into sets of nodes (i.e., clusters), such that each cluster is densely connected internally (i.e., high pace of collaboration).

In the third and final step, Augur post-processes the results, by merging and filtering the resulting clusters. As some communities can overlap, Augur aggregates the most similar ones and for each cluster it returns the topics that have the most intense collaboration.

The output of the process is sets of already existent topics that are nurturing a new research area that should shortly emerge, associated with relevant authors and publications.

Augur was evaluated on the task of forecasting the emergence of 1,408 research topics in the 2000-2011 period and it outperformed four alternative approaches, successfully predicting many new research topics as soon as two years before they became explicitly recognised in the research community.

\section{Contributions and structure of this dissertation}
\subsection{Contributions}
The main contribution of this work concerns a new approach to the early detection of research topics. In particular, it presents the first approach able to detect research topics in the embryonic stage, that is, before they are labelled and recognised by the relevant research communities. This allows stakeholders such as academic editors, researchers, and institutional funding bodies, to react timely to changes in the research environment.

The contributions of the present dissertation can be summarised as follows: 

\begin{enumerate}
\item A comprehensive literature review that covers several aspects regarding the emergence of new research topics (Chapter \ref{ch:pastwork}).
\item New empirical evidence to fundamental theories in \textit{Philosophy of Science}, which are concerned with the evolution of scientific disciplines, supporting the existence of an embryonic stage for research topics (Chapter \ref{ch:firststudy}).
\item A new framework for the detection of new research topics at their embryonic stage (Chapter \ref{ch:secondstudy}).
\item Advanced Clique Percolation Method, a community detection algorithm developed for supporting this task (Chapter \ref{ch:secondstudy}).
\item A gold standard composed of 1408 debutant topics in the period 2000-2011, including their ancestors (Chapter {\ref{ch:finalevaluation}}).
\end{enumerate}

\subsection{Structure of the dissertation}
This dissertation is articulated in three parts, described as follows.

\subsubsection{Part I: Introduction and State of The Art}
Chapter \ref{ch:pastwork} provides the background for this thesis. We start by introducing the concept of research topic, discussing the different life stages it goes through. Then, we review the literature on the problem of the early detection of research trends through two different lenses. First, we take a technological perspective and discuss the approaches available in the state of the art for detecting topics and analyse their evolution in time. Then, we observe the literature with a philosophical lens and discuss the main contributions provided from historians of science and philosophers, such as Thomas Kuhn, Roy Anthony Becher and others. We provide also some details on community detection algorithm as approach we used to develop Augur.
Finally, we discuss the gap in the literature that we want to address.


\subsubsection{Part II: Augur}
This is the main part of the dissertation and includes two chapters which contain the details of the Augur framework. 
In Chapter \ref{ch:firststudy}, we describe our first study, which investigates some of the patterns that can lead to the emergence of new research topics. In Chapter \ref{ch:secondstudy}, we describe the Augur framework and the Advanced Clique Percolation Method, a novel community detection algorithm. 

\subsubsection{Part III: Evaluation and Conclusions}
The final part presents the evaluation of Augur and the discussion. In Chapter \ref{ch:finalevaluation}, we describe the gold standard and the evaluation process, and we report the result from the evaluation. Finally, Chapter \ref{ch:conclusion} closes this dissertation by providing an answer to the research questions, discussing the limitations of our solution, and outlining the future directions.

\section{Publications}
This dissertation is based on the following publication.

\vspace{6mm} 
\textbf{Chapter 1}
\begin{itemize}
\item \underline{Angelo A. Salatino}. ``Early Detection and Forecasting of Research Trends.'' ISWC-DC 2015, The ISWC 2015 Doctoral Consortium (2015): 49.
\end{itemize}

\vspace{6mm} 
\textbf{Chapter 3}
\begin{itemize}
\item \underline{Angelo A. Salatino}, Francesco Osborne, and Enrico Motta. ``How are topics born? Understanding the research dynamics preceding the emergence of new areas.'' PeerJ Computer Science 3 (2017): e119.
\item \underline{Angelo A. Salatino}, and Enrico Motta. ``Detection of Embryonic Research Topics by Analysing Semantic Topic Networks.'' International Workshop on Semantic, Analytics, Visualization. Springer, Cham, 2016. \textbf{(Best Paper Award)}
\end{itemize}

\vspace{6mm} 
\textbf{Chapter 4 and 5}
\begin{itemize}
\item \underline{Angelo A. Salatino}, Francesco Osborne, and Enrico Motta. ``AUGUR: Forecasting the Emergence of New Research Topics.'' In JCDL ’18: The 18th ACM/IEEE Joint Conference on Digital Libraries, June 3-7, 2018, Fort Worth, TX, USA
\end{itemize}

\vspace{6mm} 
Here follow other papers on the field of scholarly data analytics that I published during my PhD and that provided useful insights that fostered the creation of Augur:

\begin{itemize}
\item Francesco Osborne, \underline{Angelo A. Salatino}, Aliaksandr Birukou, and Enrico Motta. ``Automatic classification of springer nature proceedings with smart topic miner.'' In The 15th International Semantic Web Conference (ISWC 2016), 17-21 October 2016, Kobe, Japan, Springer International Publishing, 2016.
\item Amparo Elizabeth Cano-Basave, Francesco Osborne, and \underline{Angelo A. Salatino}. ``Ontology Forecasting in Scientific Literature: Semantic Concepts Prediction based on Innovation-Adoption Priors.'' In 20th International Conference on Knowledge Engineering and Knowledge Management (EKAW 2016), 19-23 November 2016, Bologna, Italy, Springer International Publishing, 2016.
\item Francesco Osborne, \underline{Angelo A. Salatino}, Aliaksandr Birukou, and Enrico Motta. ``Smart Topic Miner: Supporting Springer Nature Editors with Semantic Web Technologies.''  In (Poster and Demo session) @ The 15th International Semantic Web Conference (ISWC 2016), 17-21 October 2016, Kobe, Japan, 2016.
\item Francesco Osborne, Thiviyan Thanapalasingam, \underline{Angelo Salatino}, Aliaksandr Birukou, and Enrico Motta. ``Smart Book Recommender: A Semantic Recommendation Engine for Editorial Products.'' In (Poster and Demo session) @ The 16th International Semantic Web Conference (ISWC 2017), 21-25 October 2017, Vienna, Austria, 2017. 
\item Francesco Osborne, \underline{Angelo Salatino}, Aliaksandr Birukou, Thiviyan Thanapalasingam, and Enrico Motta. ``Supporting Springer Nature Editors by means of Semantic Technologies.''  In Industry Track @ The 16th International Semantic Web Conference (ISWC 2017), 21-25 October 2017, Vienna, Austria, 2017.
\item Andrea Mannocci, \underline{Angelo A. Salatino}, Francesco Osborne, and Enrico Motta. ``2100 AI: Reflections on the mechanisation of scientific discovery.'' In Workshop Re-Coding Black Mirror, at International Semantic Web Conference (ISWC) 2017 , 21-25 October 2017, Vienna, Austria, 2017.
\item \underline{Angelo A. Salatino}, Thiviyan Thanapalasingam, Andrea Mannocci, Francesco Osborne, Enrico Motta. ``The Computer Science Ontology: A Large-Scale Taxonomy of Research Areas.'' In The 17th International Semantic Web Conference (ISWC 2018), 8-12 October 2018, Monterey, California, USA, Springer International Publishing, 2018.
\item \underline{Angelo A. Salatino}, Thiviyan Thanapalasingam, Andrea Mannocci, Francesco Osborne, Enrico Motta. ``Classifying Research Papers with the Computer Science Ontology''. In (ISWC 2018 Posters \& Demonstrations and Industry Tracks) @ The 17th International Semantic Web Conference, 8-12 October 2018, Monterey, California, USA, 2018.
\end{itemize}

\chapter{Literature Review\label{ch:pastwork}}

Detecting topics and tracking their trends has attracted considerable attention in the last two decades. In the literature, we can find several approaches applied to different domains, such as news articles \citep{schultz1999,dai2010}, social networks \citep{cataldi2010,mathioudakis2010}, blogs \citep{gruhl2004,oka2006}, emails \citep{morinaga2004} and scientific literature \citep{bolelli2009,sun2016,osborne2014,erten2004,decker2007}.

In this chapter, we present the main approaches and theories available in the state of the art, with regard to the problem of the early detection of research trends. 


We start this chapter by showing how we defined a research topic within the context of our work, and the reasons why philosophers were not able to provide a widely accepted definition of it. In the same section, we also explain how we used the Computer Science Ontology to support our definition.

In Section \mbox{\ref{sec:topiclifecycle}} we describe the different stages of life in which a topic goes through.


Then, we review the literature through two different lenses. The first, in Section \mbox{\ref{sec:techback}}, is the technological lens in which we describe all the main approaches available to detect topics and keep track of their trends. We point out their main strengths, and also the limitations that we plan to address in this dissertation. The second lens, in Section \mbox{\ref{sec:philosophicalback}}, focuses on the philosophical perspective of the matter. We describe the main ideas from historians of science, such as Thomas Kuhn, which helps us in designing a new innovative approach that aims to fill the current gaps. 

We devote Section \mbox{\ref{sec:embryonicstage}} to describe the embryonic stage, a particular stage of the topic lifecycle that anticipates its emergence. In Section \mbox{\ref{sec:networkapproach}}, instead, we then briefly describe some community detection algorithms that can be used to identify topics in such stage.

 In conclusion, in Section \mbox{\ref{sec:limitations}} we highlight the main limitations and the gap in literature.

\section{Research topics}\label{sec:topicdefinition}\index{topic!definition}
Currently, the state of the art does not present any universally accepted definition of research topic. 

Indeed, as Becher and Trowler point out, ``the concept of academic discipline [and by extension research topic] is not altogether straightforward'' \citep[page 41]{becher2001}. On the same note, \cite{nordenstreng2007} mentions that ``the nature of the [academic] discipline often remains unclear, while its identity is typically determined by administrative convenience and market demand rather than analysis of its historical development and scholarly position within the system of arts and sciences.'' These claims suggest that disciplines can encompass complex combinations of bodies of knowledge, methods, norms and organisations \citep{sun2013}.

One might attempt to observe the characteristics of academic disciplines from different angles. For instance, from a philosophical point of view, academic disciplines are seen as branches of knowledge, which taken together, form the whole knowledge that has been created from the scientific endeavour. Academic disciplines have boundaries, within which we can find coherence in terms of theories, concepts, methodologies, and rules that allow us to test and validate our hypotheses. In general, the kind of questions the discipline tries to answer, the problems it tries to solve, the explanations it attempts to provide, as well as the kind of scientific language it uses, define the boundaries of an academic discipline. These boundaries make disciplines different from each other \citep[page 58]{becher2001}. 
From an anthropological perspective, academic disciplines consist of cohesive groups of scientists with a high degree of agreement about methods and contents. Scientific communities with their cultural practices shape their own academic discipline, as if they belong to different tribes \citep[page 23]{becher2001}. 

The lack of a widely accepted definition has led to the emergence of alternative ones.
For instance, \cite{jo2007} define topics as semantic units that can function as building blocks for discovering knowledge. For \cite{ding2011}, topics can function as categories of a set of documents and can have different facets, they can be domain-specific, application-dependent, and context-sensitive. 

Inspired by Kuhn's work, \cite{madlock2014} defines research topics as specific areas of study, methods or tools studied by different scientific communities. Indeed, we can see some similarities between this definition of topic and the definition of paradigm.

Many studies, such as \cite{upham2010}, \cite{chen2006}, \cite{morris2003}, and some others, consider research topics as research fronts. This concept of research front has been introduced by \cite{price1965}, who observed that among scholars, there is this tendency of citing the most recently published papers. Therefore, he sees those sets of papers that scientists actively cite as research fronts.

We can also find more practical definitions, such as the ones based on topic modelling, which consider topics as sets of words closely related to each other with a certain probability \citep{blei2003,blei2006,blei2012}. Other approaches are based on the simple adoption of keywords \citep{erten2004,decker2007}, or use conceptual labels \citep{afzal2016} to describe the different domains covered by a collection of documents. 
In Section \ref{sec:techback}, we will observe different approaches for characterising and detecting topics in a collection of documents. All those approaches differ from each other based on the kind of data they process (e.g., abstract, keywords, citations) to characterise the topic structure. 

The definition of research topic the we use for this dissertation stems from \cite{allan1998}.

In particular, we define research topic as a subject of study or issue that is of interest to the research community and it is addressed in at least few research papers. Subjects of study can encompass different research areas and application-dependent concepts. For instance, a paper introducing a new technology to classify data, is the event that triggers the emergence of new topic. Any paper discussing the implementation of such technology, its application, the release of a new efficient implementation, its evaluation and so on, are all part of the same research topic. Papers discussing another technology to perform classification are not likely to be on the same topic.

In practice, we define research topic as subject of studies whose label is associated to a related concept in an ontology of research areas, such as, \textit{Acoustics}\footnote{\url{https://physh.aps.org/concepts/40a5bd01-6544-4502-8321-458c33878bf3}} in the PhySH\footnote{\url{https://physh.aps.org/}} taxonomy, \textit{Hydrocodone}\footnote{\url{https://meshb.nlm.nih.gov/record/ui?ui=D006853}} in MeSH\footnote{\url{https://meshb.nlm.nih.gov/}}, or \textit{Web Ontology Language (OWL)}\footnote{\url{https://dl.acm.org/ccs/ccs.cfm?id=10003316&lid=0.10002951.10003260.10003309.10003315.10003316}} in ACM CCS\footnote{\url{https://www.acm.org/publications/class-2012}}.


For this doctoral work, the main data source is the Rexplore dataset (see Section \ref{sec:rexplore}), containing the co-occurrence networks between keywords, that are provided by authors when tagging their papers (see Section \ref{sec:cooccurrencenetwork}). However, since keywords appear in various forms, for example, singular, plural, acronym and so on, we use the Computer Science Ontology \citep{salatino2018b} produced by Klink-2 \citep{osborne2015}. In this way, we can abstract from the different keywords that may refer to the same research topic and also filter out all the keywords that do not represent topics.

In the following sections, we will provide further details regarding the Computer Science Ontology, that allowed us to define the concept of research topic. 

\subsection{The Computer Science Ontology}\label{sec:cso}\index{topic!ontology}
The Computer Science Ontology\index{Computer Science Ontology} (CSO) describes the relationships between research topics extracted from a corpus of 16 million publications, mainly from the field of \textit{Computer Science}. The version of CSO used in this studies includes more than \num{17000} concepts and over \num{70000} semantic relationship as well as 8 levels of granularity. Its main root is \textit{Computer Science}, however CSO includes also other topics as secondary roots, such as \textit{Semantics}, \textit{Linguistics}, \textit{Geometry} and so on. This ontology is produced using Klink-2 \citep{osborne2015} and it is regularly updated when the Rexplore dataset imports new scientific publications. Indeed, from the beginning of 2018, we released a new and more fine-grained version\footnote{Computer Science Ontology version 2.0 \url{http://w3id.org/cso/downloads}} of the ontology, including more than 26K concepts and over 226K semantic relationships. However, as these studies were already completed by the end of 2017, Augur currently does not take advantage of this new version of the ontology. 


The Klink-2 algorithm takes as input a set of scholarly keywords and their relationship, i.e., the number of times certain keywords co-occur in the different years. Klink-2 combines this information with knowledge from external sources (e.g., DBpedia, calls for papers, web pages) and with various semantic technologies and machine learning techniques, to automatically generating an ontology of research areas, according to three main semantic relations:

\begin{enumerate}[label=(\roman*)]
\item \textit{relatedEquivalent}\index{related equivalent}, indicates that there are two ways of referring to a particular research area and they can be treated as equivalent for the purpose of exploring research data, e.g., ``ontology matching'' and ``ontology mapping''; 
\item \textit{skos:broaderGeneric}\index{broader generic}, indicates that a topic is a sub-area of another one, e.g., \textit{Linked Data} is considered a sub-area of \textit{Semantic Web}; 
\item \textit{contributesTo}, indicates that the research output of one topic is relevant and therefore contributes to the research of another topic. For instance, research in \textit{Ontology Engineering} contributes to the Semantic Web, but arguably \textit{Ontology Engineering} is not a sub-area of the \textit{Semantic Web}, as there is plenty of research in \textit{Ontology Engineering} outside the context of \textit{Semantic Web} research.
\end{enumerate} 

In particular, for each pair of associated keywords (i.e., keywords appearing together in a certain number of papers), the algorithm aims to define their kind of relationship, based on how this relationship developed in time \citep{osborne2015}.

Among these three relationships, \textit{relatedEquivalent} and \textit{skos:broaderGeneric} are necessary to create the taxonomy of research topics and to handle different labels for the same research areas, while \textit{contributesTo} provides additional information that can be explored and used in other applications, such as analysing research data \citep{osborne2012}.

This ontology of research topics can be downloaded from the CSO Portal\footnote{CSO Portal \url{https://cso.kmi.open.ac.uk}}, which is a web application that enables users both to explore CSO and also provide granular feedbacks. For a given research topic, the portal displays all its relationships, allowing user to seamlessly browse the network of topics.

For this work, we rely on CSO because it allows to semantically enhance the co-occurrence network (see Section \ref{sec:cooccurrencenetwork}) by removing all keywords that do not refer to research areas and by aggregating keywords representing the same concept, i.e., keywords linked by a \textit{relatedEquivalent} relationship in the ontology. For instance, we aggregated keywords such as:
\begin{itemize}
\item ``ontology mapping'',  ``ontology matching'' and ``ontology alignment'', or 
\item ``knowledge-based systems'' and ``knowledge based systems'', or 
\item ``wireless ad hoc networks'', ``wireless ad hoc network'' and ``wireless ad-hoc networks'';
\end{itemize}
in single semantic research topics which will then group together all the publications associated with these keywords.

\section{Lifecycle of research topics}\label{sec:topiclifecycle}\index{topic!dynamics}
It is well known that science is a dynamic environment, and by constantly splitting and merging, communities of scholars shape its landscape \citep{di2017}. These social interactions among scientists guide the birth and evolution of research topics. 
 
\begin{figure}[!ht]
  \includegraphics[width=\linewidth]{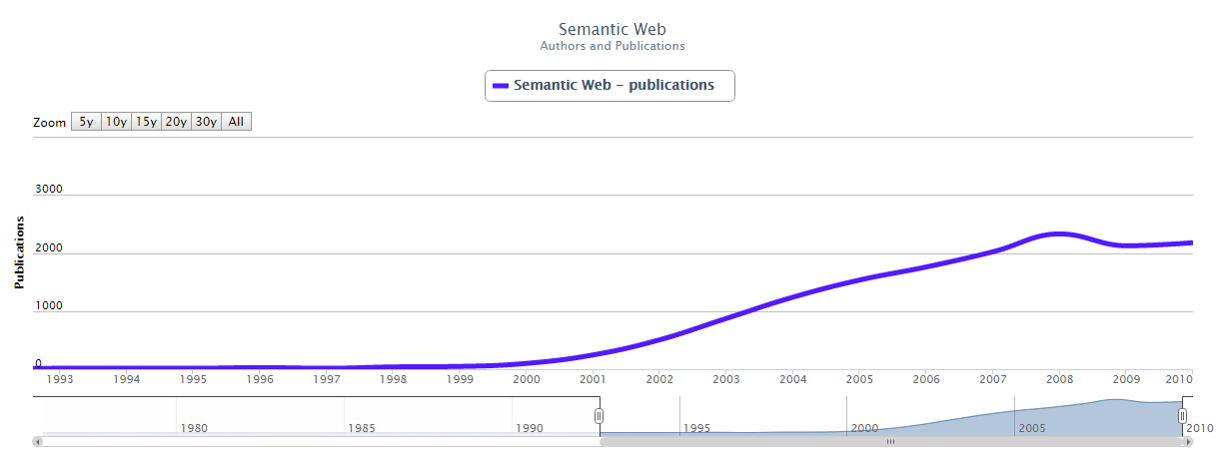}
  \caption{Evolution of the topic \textit{Semantic Web} in terms of number of publications per year. The chart has been produced using the Rexplore system.}
  \label{fig:semanticweblc}
\end{figure}

According to \cite{braun2000} the lifecycle of a research topic can be expressed as an S-curve, as depicted in Fig. \ref{fig:semanticweblc}, however there is no common agreement on the number of stages the topic goes through during its lifecycle. \cite{braun2000} characterise the lifecycle of a topic with four stages: birth, growth, maturity, and senility. Instead, \cite{tu2012} believe that a research topic evolves sequentially through five stages: initial stage, early stage, expansion stage, maturation stage, and decline stage. \cite{shneider2009} claims instead that each scientific discipline faces four stages, while match the activities pursued by scholars. At the first stage, scientists introduce a new language to describe the new subject matter. At the second stage, they develop the main techniques. The third stage is when the discipline is in its maturity, with scientists creating new insights, new answers and new questions. The fourth stage, instead, is more applicative. This last stage consists in deploying knowledge for practical purposes so that it can have an actual impact into the society.

In general, from the literature it is clear that when a new scientific area emerges, it goes through an initial phase, also called early stage\index{early stage}, in which a group of scientists agree on some basic tenets, build the conceptual framework and begin to establish a new scientific community. Afterwards, the research area enters into a recognised phase\index{recognised stage}, when it reaches its own maturity and a substantial number of authors become active in the area, producing and disseminating results. In Fig \ref{fig:topic-timeline-pre}, at page \pageref{fig:topic-timeline-pre}, we can observe how these stages are located within a timeline. In Fig. \ref{fig:semanticweblc}, instead, we can see an example of a specific case regarding the \textit{Semantic Web} showing the amount of scientific papers published per year, in this area. Through this graph it is possible to observe that it emerges around the year 2001, in agreement with the Tim Berners Lee's seminal paper \citep{berners2001}, therefore it directly moves to the early stage when several authors start to define scientific field until it gets mature few years after.

As we will see throughout this chapter, all the approaches available in literature can identify new research topics only once they have emerged and they have been explicitly labelled and recognised by researchers. Therefore, the problem of detecting research topics, before they emerge, still needs to be addressed.
For this reason, we will rely on the hypothesised embryonic stage, as initially conceived by Kuhn \citep{kuhn1970}, to analyse and forecast the emergence of new research topics.




\section{Technological background}\label{sec:techback}
The detection of research trends involves two main tasks: 
\begin{enumerate*}[label=(\roman*)]
\item identifying topics within a collection of documents, and 
\item determining their development in time (i.e., tracking their trends).
\end{enumerate*}
In literature, we can find studies focusing only on detecting topics, and some other studies aiming to detect trends, which include technologies for both detecting topics and analysing their evolution. 
Since, there are different studies tackling different elements of the problem, we believe we can provide a better literature review by separately looking at the different technologies, that aim to resolve the tasks of topic detection and tracking.
To this end, when discussing such approaches, rather than reviewing the whole studies, we will focus on the specific technologies adopted for performing each of these tasks, independently.


Therefore, we organise this section in two parts. Section \ref{sec:topicdet}, considers all the main approaches for detecting topics within corpora, while Section \ref{sec:topictrack} will describe the main techniques for tracking the evolution of topics. 

\subsection{Topic detection in the scientific literature}\label{sec:topicdet}
The aim of topic detection\index{topic detection} is to break the whole collection of documents and to arrange them into groups of topics. Different approaches will produce different results when run on the same corpus 
also because the concept of topic does not have a widely accepted definition. Indeed, the definitions adopted by different approaches usually depend on the scope of the application. As a matter of fact, sometimes topics are algorithmically defined, which means that they are defined only implicitly as the output of an algorithm.

One of the very first attempts dates back to 1983, when in a seminal paper, \cite{callon1983} introduced the method of co-word analysis\index{co-word analysis} to study the content of documents. Their main assumption is that ``key words\footnote{The key words are different form the keywords that authors use nowadays to tag their papers. In the context of \cite{callon1983} work, a key word or macro-term is a signal word that acts as agent of contextualisation.}'' (or macro-terms) within papers constitute an adequate description of their content. In this way, the key words used for describing the content of a publication can be used as the basic building blocks to characterise the structure of science. In particular, this approach extracts those key words and measures their co-occurrence within the documents. Then, it creates a network, in which the nodes are words and the links represent whether their co-occurrence value exceeds a certain threshold. The role of each macro-term of an article, is to help in placing that article in the corresponding area of the network. The approach, then, organises this network of key words into group of words according to the frequency with which they appear together. The resulting clusters will represent the themes or concepts in the scientific literature. While Callon and his colleagues focused only on few case studies, when the computational power of machines increased, many other research groups extended this work and focused on larger case studies, such as analysing the interactions between basic and technological research \citep{callon1991}, mapping knowledge evolution in the field of \textit{Scientometrics} \citep{courtial1994}, and studying a number of fields, such as \textit{Software Engineering} \citep{coulter1998}, \textit{Information Research} \citep{ding2001}, \textit{Ethics} and \textit{Dementia} \citep{baldwin2003}, \textit{Geographic Information System} \citep{tian2008}, and others. However, the main drawback is that such approach relies on the appearance of specific words and it does not consider the context, therefore it cannot handle issues of synonymy and polysemy.

Another study that laid the groundwork for this field was the \textit{Topic Detection and Tracking} (TDT)\index{topic detection and tracking (TDT)} program developed by the Defense Advanced Research Projects Agency (DARPA) \citep{allan1998}. Although the domain of interest for this program was on broadcasted news, it articulated the problem and provided interesting insights on how to automatically extract topics and organise a stream of documents. To detect topics, they perform cluster analysis so that items are grouped by their relatedness within bins that represent topics. Allan and his colleagues tested different solutions for clustering \citep{allan1998}. One approach is based on \cite{eichmann1999} and uses \textit{term frequency inverse document frequency} index (tf-idf)\index{tf-idf} to select the most significant words and then it computes the cosine similarity between all pairwise news to create clusters. Another approach uses the Kullback-Leibler distance as metric to group together news on similar topic \citep{allan1998}.

One of the main contributions of this work was to formalise the problem of Topic Detection and Tracking, which helped this field to flourish. Indeed, in the literature we can find many approaches based on one of the tasks formalised by \cite{allan1998}. 

Allan and his colleagues organised the problem around five research tasks: \textit{Topic Tracking}, \textit{Story Link Detection}, \textit{Clustering} (or \textit{Topic Detection}), \textit{First Story Detection}, and \textit{Story Segmentation}.
In particular, the \textit{clustering} task helps in finding the thematic structure within the corpus and it mainly consists of technologies to summarise the content of documents and group them based on the topics they discuss.

All the available approaches in literature can be classified based on the kind of information they use. For collections of scientific documents, the main information exploited by the approaches for detecting topics are titles, abstracts, keywords, full content, and citations. Some approaches rely on just one set of information (e.g., keywords), while some others can combine more than one set of information (e.g., the combination of abstract and keywords), with the aim of improving clustering performance. However, it is also worth to point out that sometimes the corpus itself can introduce some constraints over the designed approach. For instance, some approaches cannot take advantage of the full content of papers simply because this information may not be available.

In what follow, we discuss the main approaches for topic detection according to the kind of information they adopt, such as:
\begin{enumerate*}[label=(\roman*)]
\item abstract or full content,
\item citations,
\item keywords and taxonomies, and
\item hybrid solutions, which combine more than one entity.
\end{enumerate*}

\subsubsection{Abstract or full content}

Approaches relying on either or both abstracts and full content\footnote{Approaches using the full content take advantage of the full text of a document.} of documents, usually aim to synthesise the content of scientific papers first. This process of dimensionality reduction is accomplished by extracting some meaningful words, so that now papers are treated as ``bag of words''\index{bag of words}, with the assumption that these words can reflect its content. 

One of the first approaches that helped to extract meaningful words from the content of documents is the already mentioned tf-idf. This technique computes an index for each word, reflecting how important they are and then only those with higher value are retained \citep{salton1988}. Then, text documents are clustered on the basis of shared words, using cosine similarity \citep{schultz1999}.
\cite{roche2010} used a similar approach on the field of \textit{Molecular Biology}. However, while the tf-idf can identify sets of words that are discriminative for documents, it does not capture position in text, semantics and co-occurrences in different documents. Besides, the dimensionality reduction is very little and it does not provide enough information about the inter- or intra-document statistical structure.

Hence, the next generation of approaches for topic detection adopted topic modelling. One of the most influential approaches is the Latent Dirichlet Analysis\index{Latent Dirichlet Analysis (LDA)} developed by \cite{blei2003}. LDA is a three-level hierarchical Bayesian model to retrieve latent patterns in texts. The basic idea is that each document is modelled as a mixture of topics, where a topic is a multinomial distribution over words characterised as a discrete probability distribution defining the likelihood that each word will appear in a given topic. LDA aims to discover the hidden\textendash or latent\textendash structure which is words per topic and topic per document. In order to do so, it computes the conditional distribution of the hidden variables (topics) given the observed variables (words) \citep{blei2012}. 

A major difference between tf-idf and LDA is that the former works at lexical level, for example a document discussing networks might be considered dissimilar from a document about graphs when adopting tf-idf. On the other hand, since networks and graphs frequently co-occur in articles, they are likely to be associated with the same topic, so the LDA is likely to produce a similar representation for these articles.

Other approaches that fall within the same category of topic modelling\index{topic models} are the Latent Semantic Analysis (LSA), the Probabilistic Latent Semantic Analysis (pLSA) and the Correlated Topic Model (CTM).

The Latent Semantic Analysis was actually one of the first attempts to cope with the shortcomings of tf-idf\index{tf-idf}, as it can achieve significant reduction in dimensionality, in large collections \citep{deerwester1990}. In particular, LSA counts the frequencies of each word per documents and then it uses the singular value decomposition (SVD) technique to reduce the number of rows while preserving the similarity structure among columns. In this way, LSA is able to group together the documents, which use similar words, as well as the meaningful words, shared by a similar set of documents. In addition, it can capture some basic linguistic aspects such as synonymy and polysemy. Such patterns are used to detect the latent components within the collection of documents.

An alternative approach to the LSA is the Probabilistic Latent Semantic Analysis (pLSA) developed by \cite{hofmann1999}. The pLSA models words of documents as samples from a mixture model, then by means of the Expectation-Maximization algorithm, it performs the extraction of topics that are represented as multinomial random variables, thereby components of the mixture. In developing the LDA, Blei and his colleagues, aimed to fix some weaknesses introduced by the pLSA, such as \begin{enumerate*}[label=(\roman*)]
\item it learns the topic mixtures only for documents seen in the training phase, so basically it is not able to assign probability to previously unseen documents; and 
\item the number of parameters increase linearly with the size of the corpus, indicating that the model suffers from overfitting problems \citep{blei2003}\end{enumerate*}. 

The Correlated Topic Model \citep{blei2006} uses the logistic normal distribution instead of the Dirichlet, to solve the fact that LDA fails to model the correlation between topics. This approach fits the process of extracting topics from scientific text corpora since it is natural to expect that subset of the underlying latent topics will be highly correlated. For instance, a scientific paper about \textit{Genetics} is likely to be also about \textit{Health} and \textit{Disease}. 

\cite{lee2010} and \cite{alghamdi2015} analysed in more detail the main differences among these four methods for topic modelling. In Table \ref{tab:topicmodels}, are reported the main characteristics and limitation summarised from \cite{lee2010}. As we can see, each method has its strengths and weaknesses, and the selection of one method over another mainly depends on the kind of application a researcher has in mind \citep{lee2010}. 
However, LDA is the most frequently used approach for modelling topics. Indeed, since its introduction, LDA has been extended and adapted in several applications. Some approaches include temporal text mining, author-topic analysis, supervised topic models, latent Dirichlet co-clustering and LDA based bio-informatics \citep{wang2006,shen2008}. Other extensions of LDA include the hierarchical LDA where topics are grouped together in a hierarchy \citep{griffiths2004} and the relational topic model which is a combination of topic model and network model for collections of linked documents \citep{chang2010}. 

\begin{table}[ht]
\centering
\caption{Characteristics and limitations of the four methods for topic modelling. This table is an extract of \cite{lee2010}.}
\label{tab:topicmodels}
\begin{tabular}{|L{3cm}|L{5.5cm}|L{5.5cm}|}
\hline
\textbf{Method}                               & \textbf{Characteristics}                                                                                                                   & \textbf{Limitations}                                                     \\ \hline
Latent Semantic Analysis (LSA)                & i)        Can get from the topic if there are any synonym words.                                                                           & i)        Hard to determine the number of topics.                        \\
                                              & ii)       Reduces dimensionality of tf-idf using the svd method.                                                                           & ii)       Hard to interpret loading values with probability meaning.     \\
                                              & iii)     Not robust statistical background.                                                                                                & iii)     Difficult to label a topic with the words within the topic      \\ \hline
Probabilistic Latent Semantic Analysis (PLSA) & i)        It can generate each word from a single topic; even though various words in one document may be generated from different topics. & i)        At the level of documents, PLSA cannot do probabilistic model. \\
                                              & ii)       Partially handles polysemy.                                                                                                      &                                                                          \\ \hline
Latent Dirichelet Allocation (LDA)            & i)        Need to pre-process documents to remove stop-words, first.                                                                       & i)        Cannot model the relations among topics.                       \\
                                              & ii)       Can handle long documents                                                                                                        &                                                                          \\ \hline
Correlated Topic Model (CTM)                  & i)        Uses the logistic normal distribution to create relations among topics.                                                          & i)        Computationally expensive.                                     \\
                                              & ii)       Allows the occurrences of words in other topics and topic graphs                                                                 & ii)       Contains general words within topics.                          \\ \hline
\end{tabular}
\end{table}

However, one of the main issues with the LDA, and actually recurring in all the approaches for topic modelling, is that topics are bereft of labels. These approaches represent a topic with a set of words that are also representative for the topic itself, however it is very difficult to select one of them as label, at least automatically. As an illustration, in Table \ref{tab:ldaexample}, there is an example of word distribution, which we manually associated to the semantic concept of \textit{Video-stream} \citep{cano2016}.

\begin{table}[ht]
\centering
\caption{Bag of words associated with the concept of Video Streaming, extracted using LDA. This example is extracted from \cite{cano2016}.}
\label{tab:ldaexample}
\begin{tabular}{|l|}
\hline
\begin{tabular}[c]{@{}l@{}}video, stream, network, rate, system, application, adapt, bandwidth, packet, internet\end{tabular} \\ \hline
\end{tabular}
\end{table}

Another approach that uses the full text to determine the structure of topics within the collection of documents is \cite{morinaga2004}. This approach uses the Finite Mixture Model to detect topics and in addition by analysing the changes in the extracted components (clusters of words), it can track the emergence of new topics. This is done by using an improved version of Kleinberg's approach \citep{kleinberg2003} which will be described in the next section about topic evolution. In their work, Morinaga and his colleagues represent a topic as a single mixture component and then they use the finite mixture to model the topic structure. Then, by dynamically learning the finite mixture model and tracking the changes of main components they can recognise a change in the topic structure and therefore identify topics trends. However, this approach was evaluated on an email corpus and therefore is not clear how will it would perform on scientific literature.

\cite{chavalarias2013} developed a method to automatically reconstruct representations of the evolution of science. They analysed a collection of \num{200000} articles from the Thomson Web Of Science corpus related to research in Embryology. With the CorText Manager tool they extracted a list of \num{2000} n-grams representing the most salient terms. Then they derived a co-occurrence matrix upon which, with the clique detection approach, they performed clustering analysis to discover patterns in the evolution of science. Another network-based approach is from \cite{sayyadi2013}. Sayyadi and his colleague developed an approach that extracts all meaningful words from the content of papers. Next, they represent the collection of documents as keywords co-occurrence network. Then, upon this network they run a community detection algorithm to locate clusters of keywords that can represent topics.

\subsubsection{Citations}

Today, we can find many approach using citation networks\index{citation networks}\index{networks!citation}, and most of them are based on the principle of clustering scientific documents by means of the co-citation analysis, introduced by \cite{small1973}. The use of citations for detecting topics has been explored in many different ways and some approaches combine citations with other entities such as keywords and abstracts.

One of the first approaches using citation for detecting topics was described by \cite{small1977}. Small identified ``hot fields'' by clustering highly co-cited documents each year and linking them through time. This analysis has been performed more recently in \cite{small1999} and \cite{boyack2014a} arguing that even if the thresholds and normalizations have changed along time, the basic process is still valuable. 

Similar studies come from \cite{upham2010} and \cite{small2014}. The first study used the ISI corpus (which now is Web of Science\footnote{Web of Science \url{http://wokinfo.com/}}) to identify 20 emerging topics within the years 1999-2004, representing co-citation clusters. The study of \cite{small2014}, instead, performs co-citation analysis over Scopus data, allowing them to perform a more global analysis, rather than focusing on a confined area.

\cite{chen2006} designed a tool called CiteSpace which combines co-citation analysis and burst detection \citep{kleinberg2003} to identify new emerging trends. In particular, they perform a progressive network analysis which synthesises slices of networks and focuses on critical nodes which play a crucial role in the evolution of network over time. CiteSpace has been employed to characterise new emerging topics within the field of \textit{Mass-extinction} and \textit{Terrorism} \citep{chen2006}, \textit{String Theory}, \textit{Gene Targeting} and \textit{Peptic Ulcer} \citep{chen2009} and \textit{Regenerative Medicine} \citep{chen2012}. According to their results, two of the main factors related to the emergence of new topics are the burst in citation and the betweenness centrality of their papers. This suggests that such emerging topics or clusters were the result of collaborations between two or more existing topics.

Another study is from \cite{jo2007}, who developed an approach to detect topics combining distributions of terms (i.e., n-grams) with the distribution of the citation graph related to publications containing that term. Their work is based on the intuition that if a term is relevant to a particular topic, documents containing that term will have a stronger connection than randomly selected ones. In their first approximation, that term would also represent the label for that topic, but they extended their algorithm so that it can consider topics as a relation of multiple terms. 

The main drawback with these approaches based on co-citation analysis is that they are able to assign each document to only one topic. A document is rarely monothematic. Therefore, these approaches can create clusters of documents that are not coherent.

\subsubsection{Keywords and taxonomies}

Another category of approaches for detecting topics in a collection of documents use keywords. Keywords\index{keywords} (also called \textit{subject terms}), are terms carefully selected by scholars to identify the research focus of their manuscript \citep{abrahamson1996,mccloskey1998}. They represent the key concepts and the main ideas of papers, as well as reflecting their novelty. Such representation, is also beneficial for search engines for returning relevant papers to a given query. 

In the literature there are many approaches aiming to detect topics which rely on keywords. 
For instance, \cite{duvvuru2012,duvvuru2013} analysed keywords and the networks of their co-occurrence in order to monitor how the link weights develop in time. In this way, they aim to detect research trends as well as emerging research areas. However, we can argue that using keywords as a proxy for topics brings several drawbacks. Indeed, \cite{osborne2015} point out that keywords tend to be noisy and do not always represent research topics, such as the keyword ``case study''. Moreover, they also cannot handle synonymy and polysemy, since many keywords can refer to the same topic or a single keyword can represent more topics. \cite{yi2012} developed an approach that tackles some of the issues of \cite{duvvuru2012}. Before creating the keyword network\index{keyword networks}\index{networks!keyword}, the authors designed a procedure to clean the keyword set. For instance, when multiple keywords are considered similar, they are transformed into a single form, such as ``agent'' and ``agents'' into ``agent''. In addition, if keywords consist of multiple independent keywords, they are split, like ``efficiency and effectiveness'' into ``efficiency'' and ``effectiveness''. However, few details are provided regarding the implementation of this procedure. 

\cite{decker2007} developed an approach that generates paper-topic relationships by exploiting keywords and words extracted from the abstracts in order to analyse the trends of topics on different timescales. In particular, this approach matches a collection of scientific manuscripts with a manually curated taxonomy of topics\index{taxonomy of topics}. This taxonomy has been manually crafted using topics appearing in proceedings, and then complemented with keywords and terms from the abstracts \citep{cameron2010}. Then, by analysing the changes in the number of publications associated with topics, they can detect research trends. Similarly, \cite{erten2004} adopted the ACM Computing taxonomy for analysing the evolution of topic graphs to monitor research trends. Since, these last two approaches rely on taxonomies of topics, they can be considered as an improvement with respect to the approach of \cite{duvvuru2012}, which simply uses keywords. However, the main issue is that both taxonomies are manually curated. Indeed, focusing on the field of \textit{Computer Science}, which we know it is a constantly expanding field, manual taxonomies tend to become quickly outdated. Not to mention that the last update of the ACM Computing taxonomy occurred in 2012\footnote{The 2012 ACM Computing Classification System \url{http://www.acm.org/publications/class-2012}} (6 years before, at the time of writing), replacing its previous version released in 1998\footnote{ACM Computing Classification System ToC \url{http://www.acm.org/about-acm/class}}. In this way, new emerging topics do not get the chance to be included in the taxonomy, preventing the process of clustering to be accurate enough.

\cite{herrera2010} offered an approach for categorising physics literature described through the Physics and Astronomy Classification Scheme (PACS)\footnote{PACS - Physics and Astronomy Classification Scheme \url{https://publishing.aip.org/publishing/pacs}} codes. In particular, they create a scientific concept network in which each node represents a PACS code, which refers to a specific topic in Physics. Two nodes are then linked if the two corresponing codes co-occur together in at least one paper. In order to study the evolution of different fields, the authors used the Clique Percolation Method to identify communities in the PACS network and linked together similar communities along time using the method proposed by \cite{palla2007b}. The PACS scheme, which has now been replaced by the Physics Subject Headings (PhySH)\footnote{PhySH - Physics Subject Headings \url{https://physh.aps.org/about}}, was maintained by the American Institute of Physics (AIP) and we can argue that it had the same issues of the ACM Computing taxonomy. On the other hand, PhySH has the advantage of being crowdsourced with the support of authors, reviewers, editors, and organisers of scientific conferences, so that it is constantly updated with new terms. A similar analysis in the field of \textit{Medicine} has been performed by \cite{ohniwa2010}, using the Medical Subject Heading (MeSH)\footnote{MeSH - Medical Subject Headings \url{https://www.nlm.nih.gov/mesh/}}. The MeSH taxonomy is maintained by the National Library of Medicine of the United States and has similar characteristics to PhySH, since it is constantly updated collecting new terms as they appear in the scientific literature or in emerging areas of research.

Another approach falling into the same category is Klink-2 of \cite{osborne2015}, which is an algorithm to automatically generate an ontology of research areas in \textit{Computing}, the Computer Science Ontology\footnote{Computer Science Ontology \url{http://cso.kmi.open.ac.uk}}\index{Computer Science Ontology}. Klink-2 is able to detect relationships and then build a semantic network of research areas, from keywords associated with a collection of documents, exploiting semantic technologies, machine learning and knowledge from external sources (e.g., DBpedia, calls for papers). Similarly to the approaches of \cite{decker2007,duvvuru2012,erten2004}, Klink-2 also relies on keywords, but it adds a conceptual layer aiming to resolve some of the aforementioned drawbacks removing the keywords that do not represent a topic and introducing three relationships between keywords: \textit{skos:broaderGeneric}, \textit{relatedEquivalent} and \textit{contributesTo}. This automatically generated ontology presents two main advantages compared to the manually crafted ones, e.g., the ACM Computing taxonomy. Firstly, the ontology is automatically generated and therefore can be easily updated to include new emerging topics by re-running Klink-2 over an updated collection of documents. The second advantage is that Klink-2 can create relations between a wider set of terms, compared to the other taxonomies. This semantic characterisation of research topics has been beneficial in many applications, such as \textit{Rexplore}, a platform for exploring and making sense of scholarly data \citep{osborne2013}, \textit{Smart Topic Miner} for supporting academic editors in classifying scholarly publications \citep{osborne2016}, \textit{Smart Book Recommender} to support the Springer Nature editorial team in promoting their publications \citep{osborne2017} and the detection of topic-based research communities \citep{osborne2014}.

In brief, the state of the art presents several approaches for detecting research topics using keywords. Some approaches directly rely on keywords, but as we have already observed they tend to suffer from polysemy, synonymy and a number of other issues. Instead, other approaches only indirectly rely on keywords, using taxonomies of topics, which produce more accurate results. However, those approaches to topic detection that rely on automatically generated taxonomies tend of course to be more comprehensive and up to date.

\subsubsection{Hybrid approaches}

So far, we have analysed approaches that aim to find topics using some important entities which can be extracted from scientific papers, such as the abstract or the full content, the citations and the keywords. In this section we describe the hybrid approaches that use a different kind of metadata such as titles, venues (i.e., journals, conferences), and authors. While some approaches might exclusively use one of these entities, some others use a combination of them.

Authors are the main agents within the scientific endeavour. 
They perform scientific research, report and publish results. The information about them can be beneficial and provide interesting insights when investigating the topical structure of a corpus. In particular, author networks, in which links represent co-authorship relation, are used by a variety of approaches to analyse the research landscape \citep{zhou2006}. 
Other approaches describe authors using the probabilistic topics model described in the previous section. For example, \cite{rosen2004} proposed the author-topic model (ATM) which improves LDA \citep{blei2003} by including authorship information. Starting from the LDA assumption, in which every document is a distribution of topics, their approach further extends the concept of topics as distributions over both words and authors. 
Their model returns the set of topics that appear in the corpus and identifies which topics are relevant to which authors. 
The same research team aimed to improve the results of ATM by proposing the Probabilistic Author-Topic model \citep{steyvers2004}, which models documents as a composition of multiple topics, topics as probability distribution over words, and authors as probability distribution over topics. 
Instead of using the LDA, they extend the Probabilistic Latent Semantic Analysis introduced by \cite{hofmann1999}. 
Nevertheless, the probabilistic model is quite simple and disregards some aspects such as topic correlation and author interaction. \cite{bolelli2009} further extended the ATM introducing the Segmented Author-Topic model (S-ATM), a generative model that uses temporal ordering of documents in order to identify topic evolution and then exploits citations to evaluate the weights for the main terms in documents. 

Conferences and journals can also provide significant insights in the contest of studying the evolution of scientific fields and identifying topics. Indeed, publication venues can influence the shape of a research topic over time. 
In the state of the art, several approaches exploit publication venues and they can be classified as approaches extending topics models and approaches based on network analysis. For instance, \cite{tang2008} presented a unified topic model for simultaneously modelling the topical distribution of papers, authors, and conferences called Author-Conference-Topic (ACT) model. This model further extends the Author-Topic model of \cite{rosen2004}, including conference and journal information. In particular, they present three different implementations of the model, which differ from each other in the way they model the association between authors, distribution of topics and conference information. 
Yan and his colleagues \citep{yan2012}, adopted the ACT model to demonstrate that communities of scholars and research topics are intertwined and co-evolve, instead of being two disconnected entities. However, \cite{wang2012} identify a limitation of the Author-Conference-Topic model, pointing out that this model extracts topics for the corpus of documents and then it maps them to the research areas available within the ``call for papers'' of conferences. 
This operation is not always possible because the latent topics extracted with the LDA may not match the conference topics. Therefore, Wang and his colleagues presented a further extension, the Author Conference Topic Connection (ACTC) model, which aims to improve this aspect by adding subject information of the conference and the latent mapping information between subjects and topics.

Examples of network-based approaches using information from venues are \cite{sun2016}, \cite{boyack2005} and \cite{pham2011}. \cite{sun2016} used a similar approach to \cite{herrera2010} and \cite{duvvuru2012}, mentioned above, as they created a network of entities in order to identify scientific fields and analyse their evolution. However, the approach of Sun and her colleagues differs from the latter ideas since they measure the similarity between conferences using author co-occurrences. In particular, for each year, they create a network in which each node represents an active conference in that year and the link between them is determined based on the amount of authors that publish in both conferences. Once the network is created, they identify scientific fields by detecting communities within the network using the Louvain method \citep{blondel2008}. 

Another approach, that has the potential to be extended and used to find topics is from \cite{boyack2005}. In this approach, the authors built a map representing the structure of all of science using journal citation data. They designed eight similarity measures to group together journals based on how they cite each other. Although this approach serves for visualisation purposes, the authors show that each cluster of journals is related to a particular research field. Indeed, with the same strategy, \cite{pham2011} analysed the development of Computer Science disciplines by visualising the knowledge network at different time points. Their approach combines data on venues and citations coming from DBLP and CiteSeerX, respectively. Initially, they created a venue knowledge network in which both journals and conferences are linked to each other based on their citations. Then, they identify sub-areas of Computer Science by clustering the network and tracing their development. Such journal maps, have been investigated by other research teams, such as \cite{leydesdorff2012} and \cite{leydesdorff2013}. In these studies, the authors aggregated data from the Journal Citation Reports (JCR) and Social Sciences Citation Index (SSCI) to create global maps of science based on the journal-journal citation relations. In addition, we can find other approaches which share plenty of similarities with the ones just mentioned \citep{borner2010,van2010,boyack2008}. 

\cite{fried2014} proposed an approach that creates maps of Computer Science (MoCS) by taking advantage of titles of papers from the DBLP database. They firstly extract prominent words and phrases and then calculate their similarity based on co-occurrence in titles. Next, these words are clustered based on their similarity values, providing a visual representation of topic space. The main reason why authors are using only titles in their analysis is mainly due to the corpus, as the available information in DBLP is limited to titles, authors and venues. However, there are some drawbacks as the title of a paper may not necessarily reflect all the topics that are discussed within the paper.

As we can see, the state of the art proposes plenty of approaches to identify topics within a collection of documents. The design and the adoption of a certain approach can be influenced by the comprehensiveness of the corpus and whether certain entities are available. To this end, we organised this review by grouping together approaches that use similar set of entities, and towards the end we also presented approaches that combine more than one entity.
Before analysing the limitations of the state of the art, we devote the next section to review the approaches that aim to track the evolution of topics, with particular attention on how they aim to detect emerging topics.


\subsection{Topic evolution}\label{sec:topictrack}

The task of topic detection deals with the static component of a topic, i.e., its identification within a collection of documents. Instead, topic evolution\index{topic evolution} involves the dynamic component of topics, i.e., their development over time. 

In particular, \cite{allan1998} considered two tasks: First Story Detection and Tracking. 

The First Story Detection (FSD) task aims to recognise new emerging topics that had not been discussed earlier. Specifically, this task monitors the stream of incoming documents and identifies the presence of a new topic. For instance, a good FSD system should detect the first articles reporting about the \textit{Semantic Web} around 2001 or \textit{Deep Learning} and \textit{Cloud Computing} in the middle of the previous decade (2000-2009). In this case, however, this task is only able to detect topics once they are already emerged, rather than anticipating, i.e., forecasting, their emergence. Considering such limitation, we chose to rely on a network-based approach instead, which we will discuss in Section {\ref{sec:networkapproach}}.

Alternatively, the task of Tracking identifies new articles that discuss previously known topics. The system performing such task, is aware of the topics present in a collection of documents and, when analysing the stream of incoming articles, it should be able to sort them properly. This means that for each new document, it will extract its topics and classify them with the documents already in the collection, which discuss the same topics. At this point, the system can perform an analysis (i.e., using statistical methods) to observe the state and understand the development of each topic. Some technologies to help in achieving this task were quite mature already twenty years ago. Indeed, the state of the art could already provide approaches to extract topics from documents, such as PLSA \citep{hofmann1999} and some similarity measures to aggregate documents according to their topics.

Basically, these two tasks allow users to keep track and analyse the dynamicity of topics, from the moment they emerge onwards. However, even if they are two different tasks, some approaches combine them together so that, when organising the set of incoming documents, they are able to analyse the state of each topic as well as identifying new topics. Indeed, as we will describe later, \cite{di2017} developed an approach for observing the development of topics within a corpus in different time windows. With their approach, they could also observe the emergence of new topics that had no antecedents in previous time windows.

In general, in the literature we can find several approaches that aim to track the development of topics as well as their emergence. Some of them use custom metrics relying on the number of documents associated to the topic \citep{kleinberg2003,ho2014} or number of authors \citep{guo2011}, while some other approaches perform more complex analysis \citep{bettencourt2009,jo2007}. A second category of approaches use citation analysis which focus on determining the citation patterns between documents \citep{shibata2008,jo2007}, others might use co-word analysis which, as already seen, studies the co-occurrence of words within documents \citep{furukawa2015,morinaga2004,di2017}, as well as hybrid studies \citep{bettencourt2009}. Finally, a third kind of approaches exploit overlay mapping techniques to build maps of science and rely on human experts to assess emerging topics \citep{leydesdorff2013,boyack2005}. 

Although, many of these approaches are capable of both tracking the development of topics over time and acknowledge their emergence, for the purposes of this study, we will review them according to their ability to detect new emerging topics.

In what follow, we discuss the main approaches according to the technique they use:
\begin{enumerate*}[label=(\roman*)]
\item burst detection,
\item shift in citation patterns,
\item connectedness in co-authorship network,
\item co-word analysis, and
\item overlay mapping.
\end{enumerate*}

\subsubsection{Burst detection}

According to \cite{small2014}, two of the main features associated with the emergence are novelty (or newness) and growth. A method for identifying emerging topics, is the burst detection\index{burst detection} algorithm of \cite{kleinberg2003}, which detects the rapid growth of term usage. This approach has inspired a good number of methods for detecting research trends \citep{mane2004,jo2007,decker2007,ke2004,boyack2004,he2011}. Some of these approaches have already been mentioned in the previous section.
The burst detection algorithm has also been incorporated in larger tool sets such as Citespace II \citep{chen2006}, the Network Workbench \citep{borner2010b} and Sci2\footnote{Science of Science (Sci2) tool \url{http://sci2.wiki.cns.iu.edu/}}.

The overall idea behind Kleinberg's approach is to analyse a stream of documents and find features whose behaviour is ``bursty'', or in other words they occur with high intensity over a limited period of time. The approach uses a probabilistic automaton having different states corresponding to the frequencies of each individual word. There are as many automata as the number of words, and they change state when the frequency of their associated word changes significantly: at the beginning or end of the period of burst. This approach can be used to identify topics and concepts that came into adoption and were actively discussed for a period of time. However, it performs burst analysis for every word (including stop words), therefore it must be included in a pipeline that first pre-processes the documents \citep{kleinberg2003}.

Because identifying and tracking emerging trends can be very difficult when relying only on raw text, the task is usually accomplished taking into account more information, such as citation networks. In Section \ref{sec:topicdet}, we mentioned the approach of \cite{jo2007} and CiteSpace \citep{chen2006} which combine distributions of terms and citation networks to detect topics. Their primary goal, however, is to facilitate the analysis of emerging trends, which is inspired by the burst detection approach of \cite{kleinberg2003}. Similarly, \cite{jo2007}, developed an approach that combined distributions of terms (i.e., n-grams) with the distribution of the citation graph related to publications containing that term. In particular, the authors assume that if a term is relevant for a topic, documents containing that term will have a stronger connection than randomly selected ones. Then, their algorithm aimed at locating the set of terms having citation patterns that exhibit synergy. Their results showed that the algorithm is effective and it is also able to detect new emerging topics. However, their approach introduces a time lag, since it takes time for the citation network of a term to become tightly connected.


\subsubsection{Shift in citation patterns}

In the state of the art, we can find several approaches that identify the emergence of a new area, by analysing the shifts in citations patterns over time \citep{morris2003,morris2005,takeda2009,shibata2008,small2009,aastrom2007}. These approaches are based on the intuition that the fact that two previously unconnected, or loosely connected, areas are getting close to each other, may indicate the emergence of a new area that can extend previous research. \cite{takeda2009} use the citation network and by means of a modularity-based community detection algorithm, they detect part of the networks representing topics where they then manually identify emerging trends. However, such manual analysis could be accomplished because the authors were focusing on the relatively small area of \textit{Optics}, whilst we can argue that such approach might not be scalable when broadening the domain. \cite{morris2003} use co-citation networks\index{citation networks}\index{networks!citation} to cluster documents that are bibliographically coupled \citep{kessler1962}, or in other words they share a certain amount of cited papers. However, this approach suffers from the same issues mentioned for \cite{takeda2009}. In addition, these approaches do not provide any statistical measure to assess the emergence of a new topic, as their evaluation is performed by human expert.
 
On the other hand, \cite{shibata2008} introduced topological measures to detect the emergence of new topics without the support of experts. In this study, the authors performed clustering over the citation network and labelled each cluster with its most representative word. To detect the emerging topics, the approach tracks the age of the cluster. An interesting claim the authors make, which is actually in line with our view, is that this approach suffers of time lag. In fact, this issue is quite common for each approach based on citation and co-citation networks. Sometimes, newly published papers might need 2 years before being cited and therefore they tend to be underrepresented in such networks. Indeed, \cite{shibata2008} advise that to detect the emergence of new topics, these approaches should be complemented with information coming from other entities such as venues.

A slightly different approach, that uses citations but does not perform a topological analysis over the network, comes from Clarivate Analytics\footnote{\url{https://clarivate.com/}}. Each year, from 2013 onwards, Clarivate Analytics publishes a report called \textit{Research Fronts}\index{research fronts} revealing a selection of key research fronts (i.e., hot and new emerging ones). According to ``Research Fronts 2017''\footnote{Research Fronts 2017 \url{https://web.archive.org/web/20180515085946/https://clarivate.com.cn/research_fronts_2017/2017_research_front_en.pdf} (web archived)} which identifies 100 hot research fronts and 43 emerging ones, ``a research front consists of a core of highly cited papers along with the citing papers that have frequently co-cited the core.''

To select hot research fronts, they group the whole set of research fronts (9,690) within 10 macro-areas. Then, for each of these 10 groups they select the top 10 research fronts based on the average year of their highly cited papers (core papers). The results are then presented with the identified core papers, affiliated countries, and institutions.

For selecting the emerging research fronts, instead, they select specialties whose core papers have been published in the previous two years (for this report, from 2015 onwards). 

These emerging research fronts are then analysed and interpreted by human experts to understand the ongoing trends and assess their significance. 

In this approach we can see two main issues. As with the other approaches based on citations analysis, this approach suffers of a time lag between the actual emergence of a research topic (i.e., emerging research fronts) and the actual time it is identified. Even though this time lag can reach at most two years, in the worst-case scenario we can acknowledge the emergence of a new topic two years later. The second issue concerns the potentially low recall of the approach. Even if the report does not provide any statistical values about precision and recall in identifying hot and emerging research fronts, based on the methodology, it is easy to recognise such issue. Indeed, many novel topics might not be identified because their core papers did not reach a high enough number of citations in the previous two years.


\subsubsection{Co-authorship networks}

Authors and co-authorship networks\index{co-authorshiop networks}\index{networks!co-authorship} have also been considered to understand how the growth in the number of authors could impact on the emergence of new topics. \cite{guo2011} proposed a model that looked at three different indicators of emergence. They used frequency of keywords, growing number of authors, and interdisciplinarity of cited references. The last indicator is computed using the year-average Rao-Stirling diversity index \citep{rao1982,stirling2007}. Their results show that the emergence of new topics is consistent with the pattern in which bursts of keywords are followed by the rapid growth in the number of authors and by an increase in the interdisciplinarity of cited references.

\cite{bettencourt2009} also examined co-authorship networks to identify patterns associated with the emergence of new research fields. In particular, they observed three main patterns, such as: 
\begin{enumerate*}[label=(\roman*)]
\item the average number of edges per node increases, meaning that the network around such nodes is becoming denser, 
\item the average path length between two nodes remains stable or decreases, if the diameter of the network is changing, and 
\item there is an increase in the number of edges in the largest component.
\end{enumerate*}
These patterns can all be seen as indicators of an increased connectedness within the co-authorship network. Therefore the emergence of a new community of scholars is considered as an indicator of emerging topics.

\subsubsection{Co-word approaches}

Among the approaches using co-word analysis\index{co-word analysis} the most representative is from \cite{furukawa2015}. The authors proposed a method which analyses the development of conference networks to indicate the emergence of topics. In particular, using co-word analysis, they firstly created a progressive conference network, in which nodes represent conferences, and links represents their similarity in terms of keywords extracted from the published papers. Then, as indicator for emerging topics, they consider conferences that are becoming similar and thus they are collapsing over each other. 

\cite{di2017} designed an approach for observing how topics evolve over time. After splitting the collection of documents according to different time windows, the approach selects two consecutive slices of the corpus, it extracts topics with LDA, and it analyses how these topics changed from one time window to the other. The main assumption is that when comparing the topics generated in two adjacent time windows, it is possible to observe how topics evolve as well as capture their birth and death. 

\cite{morinaga2004} developed an algorithm to detect the emergence of a new topic, employing a probabilistic model called Finite Mixture Model. With this algorithm, the authors dynamically learned the structure of topics from the documents in each year. Then, they analysed the changes in the extracted components to track emerging topics. However, their analysis has not been tested on a collection of scientific documents, therefore it is not clear how it would perform when attempting to forecast the emergence of new research topics.

\subsubsection{Overlay mapping}

One of the first approaches based on the overlay mapping\index{overlay mapping} is from \cite{boyack2005}, who mapped the ``backbone of science''. In their approach, they firstly split the corpus along time, and for each time window (i.e., year), they collected clusters of words and associated them to research fields. Then, by observing the clusters in two consecutive years they were able to both match similar topics along time and also identify new clusters related to new topics. In the same way, \cite{leydesdorff2013} generated overlay maps to support policy makers in locating bodies of research that cross their traditional disciplinary boundaries. These approaches based on overlay mapping can be interesting as they enable users to visually analyse areas with a rapid increase in the number of papers, within the global context of science. However, they are able to provide only a coarse-grained perspective and miss weak interactions between research areas. Indeed, \cite{rafols2010} advise to use them in combination with other maps providing more detailed perspectives.

In brief, the state of the art presents several approaches for detecting the emergence of research trends and, like the approaches for topic detection, they can be classified according to the kind of analysis and kind of entities they rely on. Although each approach has its own limitations, these can be generalised with respect to the state of the art as a whole. Approaches based on citation analysis introduce a time lag, which does not allow users to recognise new emerging topics promptly. Approaches using co-word analysis focus on already recognised topics, associated either with a label or, in the case of topics model, with a bag of words. The overlay maps provide a broad perspective of science, however they lose important details that could help in detecting new emerging topic. In addition, all these approaches are able to identify a new topic only after it has emerged. 
To the best of our knowledge, the state of the art does not offer any approach able to forecast the appearance of new research topics years before they actually emerge.

\section{Philosophical background}\label{sec:philosophicalback}
The problem of detecting the emergence of new research topics can be framed also within the literature in \textit{Philosophy of Science}. Philosophy of Science, is a branch of \textit{Philosophy}, aiming to answer fundamental questions around science and its endeavour, and it deals with the understanding of the evolution of scientific paradigms within a discipline. 

One of the main contributions to this field comes from Thomas Kuhn\index{Kuhn Thomas} and his landmark book, \textit{The Structure of Scientific Revolution}. In this book, Kuhn introduces the concept of paradigm, to express the idea that disciplines are organised around a certain way of thinking, a set of assumptions, legitimate theories or a framework able to explain empirical phenomena in that discipline or field. Indeed, he defines it as ``a universally recognized scientific achievements that, for a time, provide model problems and solutions for a community of practitioners'' \citep[page~viii]{kuhn1970}.

Kuhn also describes the dynamics involving the influence of a new scientific discipline as a \textit{paradigm shift}, in which a revolution in ideas, knowledge and research project can lead to ``the successive transition from one paradigm to another via revolution'' \citep[page~12]{kuhn1970}. This kind of phenomena occurs when a paradigm cannot cope with anomalies, which lead to a crisis that will persist until a new outcome redirects research through a new paradigm. Kuhn's intuition was that science is an alternating of progress and changes. Hence he designed a model, in particular a cycle model, which explains how science evolves. This model consists of three main stages: pre-paradigm, normal science and revolution.

In the pre-paradigm phase, there are several incompatible and incomplete theories and there is no consensus on them. If a group of scientists concur with some theories, as a new scientific community they build a conceptual framework and ultimately, they disseminate their methods, terminologies and also the kind of experiments that will contribute to the progress in knowledge.

In the stage of normal science, science progresses within the existing paradigm accumulating knowledge. The research is firmly based upon one or more past scientific achievements, which are acknowledged from a particular scientific community in order to provide the basis for further developments.

In the revolution phase, the boundaries of the fields are crossed, and some unanswerable questions are discovered. This triggers a crisis derived from the fact that an old paradigm cannot explain some important observations and then the model is no longer capable to solve current problems. The revolution starts when a new paradigm challenges the previous one to encompass explanations and resolve some outstanding and generally recognised problems. The new paradigm settles in when it has a number of influential supporters and thus a new cycle begins all over again.

As evidence of his claims, Kuhn analysed several historical examples of discovery, especially from the history of \textit{Physics} with the shift from Newtonian mechanics to Einstein's special theory of relativity and \textit{Cosmology} from the Ptolemaic system to the Copernican heliocentrism.

Kuhn's work had profound influence on the \textit{History}, \textit{Philosophy} and \textit{Sociology of Science} although it attracted plenty of criticism. However, for what concerns this doctoral work we can take a valuable insight from his work. A scientific field is composed by a set of paradigms, each of them addressesing a different issue. Researchers pursue their research within the defined set of paradigms. When the set of existing paradigms cannot cope with the current challenges, they fall into a crisis, which leads new paradigms to replace the old ones. The emergence of such new paradigms can foster the emergence of a new field. All these dynamics leave trails that can be analysed both on a qualitative level, as done by historians of science (i.e., Kuhn), and on a quantitative level as already performed by \cite{yi2012,upham2010,di2017,chen2017} and so on. As an example, we can think of an exchange of theories between two not-so-distant scientific fields, which help to create a brand-new paradigm capable of solving recent challenges. 

Another contribution regarding the evolution of science comes from Becher and Trowler and their book \textit{Academic Tribes and Territories}. In this book, the authors analysed the linkages between academic knowledge (i.e., \textit{territories}) and those who inhabit it (\textit{academic tribes}). In particular, when focusing on the different cultures of academics, they argue that the scientific area specifies the characteristics and the structures of the knowledge domain, within which a group of academics place their attitudes, values, activities and cognitive styles \citep{becher2001}. This line of thought is very similar to Kuhn's idea regarding the connection between paradigms and scientific communities, stating that ``a paradigm is what the members of a scientific community share, and, conversely, a scientific community consists of men who share a paradigm'' \citep[page~176]{kuhn1970}. 

Becher and Trowler concluded that those characteristics of scientific community are quite similar to the social aspects of a tribe, and thus they coined the term ``academic tribes''. In addition, citing the work of \cite{clark1962}, they claim that as research fields are proceeding towards specialised environments, scientists across different disciplines have fewer and fewer things in common, in terms of education, experiences and daily problems to cope with. As a result, they become less compatible and less prone to cooperate with each other. However, as the authors also point out, the existing barriers between communities of scholars are not so high, therefore friendly relationships can be established for mutual benefit \citep{becher2001}. Surowiecki in \textit{The Wisdom of Crowds} has a similar view, ``collaboration allows scientists to incorporate many different kinds of knowledge \dots{•} makes it easier for scientists to work on interdisciplinary problems'' \citep[page~161]{surowiecki2005}.

Collaboration puts together the efforts of different disciplines and guarantees a diversity in perspectives. This can definitely have a role in reshaping the set of theories and methodologies and therefore leading to the emergence of new paradigms, as well as new research areas.

Here, we want to emphasise that there is a strong connection between groups of academics and their respective fields. In particular, the activities, attributes, and cognitive styles of the former are related to the structures and characteristics of the latter.
It is important also to point out that the dynamics involving the different research entities can potentially lead to a new paradigm. 
Since such dynamics are discoverable and can be analysed, they could theoretically be exploited to forecast a scientific revolution or the emergence of a new topic, when it is still an embryo.
In the next section, we will discuss more in detail the embryonic stage as part of the topic lifecycle.

\section{Embryonic stage}\label{sec:embryonicstage}
Discussing the literature from a philosophical angle helped us to open up new lines of thought in which we could develop interesting hypotheses. In particular, we observed that the social activities of communities of scholars can suggest that the development of new topics is encouraged by the cross-fertilisation of established research areas, and recognised that multidisciplinary approaches foster new developments and innovative thinking. \cite{sun2013}, \cite{osborne2014} and \cite{price1986} provided empirical evidence to these theories by analysing the social dynamics of researchers and their effects on the formation and lifecycle of research communities and topics.

All these theories support the idea of another phase, which we name \textit{embryonic stage}\index{embryonic stage}, occurring before the the \textit{early stage}, as showed in Fig \ref{fig:topic-timeline-pre} at page \pageref{fig:topic-timeline-pre}.
In this phase, a topic has not yet been explicitly labelled and recognised by a research community, but it is already taking shape, as evidenced by the fact that researchers from a variety of fields are forming new collaborations and producing new work, starting to define the challenges and the paradigms associated with the emerging new area. Kuhn also pointed out that ``often a new paradigm emerges, at least in embryo, before a crisis has developed far or been explicitly recognized'' \citep[page 86]{kuhn1970}.

In fact, it can be argued that a number of topics start to exist in an embryonic way, often as a combination of other topics, before being officially identified and then named by researchers. For example, the \textit{Semantic Web} emerged as a common area for researchers working on \textit{Artificial Intelligence}, \textit{World Wide Web} and \textit{Knowledge-Based Systems}, before being acknowledged and labelled in the 2001 paper by \cite{berners2001}. According to Fig. \ref{fig:semanticweblc} at page \pageref{fig:semanticweblc}, we can locate the embryonic stage of the \textit{Semantic Web} in the years before 2001 when different communities of scholars were joining forces to create this new interdisciplinary research topic. 

The existence of the embryonic stage for research topics is crucial because it would allow us to anticipate their emergence. Indeed, in the first study presented in Chapter \mbox{\ref{ch:firststudy}, we empirically confirmed that the existence of this stage can be numerically measured.}





\section{Network-based perspective}\label{sec:networkapproach}

To tackle the problem of forecasting new research trends, we developed a network-based approach that looks for topics in the embryonic stage by analysing the patterns of other existing topics. Specifically, we relied on clustering algorithms, that scan the co-occurring networks of topics and identify clusters (i.e., areas of topics) having an intense collaboration.


In \textit{Network Science}, these clusters or areas are also called communities and the action of clustering is often referred as community detection \citep{fortunato2010}. In this section, we will provide further details to understand the concept of community and how the underlying community structure of a network can be detected. Then we will introduce the approach that Augur uses to identify communities within the evolutionary networks.

\subsection{Community detection}\index{community detection}
Most real-world complex networks are usually organised in compartments or, in other words, areas in which nodes are more connected between themselves than to the rest of the network \citep{lancichinetti2009}. These groups of nodes are usually called communities, clusters of cohesive groups or modules that can have their own role and/or function. Community detection can help to explain the organisation of complex networks and their function. 

The idea of communities and in particular social communities is intuitively clear. People tend to form groups with their co-workers, family, and friends. And these groupings have been widely studied for a long time in \textit{Sociology} \citep{coleman1964, freeman2004, moody2003} and \textit{Social Anthropology }\citep{kottak1991, scupin1998}. Communities have also been analysed in networks that emerged from fields like \textit{Computer Science} \citep{flake2002,dourisboure2007}, \textit{Engineering} \citep{kahng2011,guimera2005b}, \textit{Natural Science} \citep{guimera2005,kepes2007,zhang2005}, \textit{Economics} \citep{burt2000,cowan2004}, \textit{Politics} \citep{fowler2006,fowler2008,zhang2008,waugh2009,porter2007} and so on. 

To analyse such networks and detect communities, the state of the art proposes different methods, that are mostly based on tools and techniques imported from other disciplines, like \textit{Computer Science}, \textit{Physics}, \textit{Biology}, \textit{Applied Math} and \textit{Social Science}. However, it is always hard to determine which algorithm performs best for a given case, since the concept of community has not been rigorously defined \citep{fortunato2010}. Indeed, many algorithms differ from each other because they implement their own definition of community. Their general guideline is, as already mentioned, that a community is a connected subnetwork whose nodes have more links to other nodes in the same community than to the nodes that belong to other communities \citep{radicchi2004}. 

However, even if we cannot know a priori which algorithm to use, the state of the art proposes the notion of  \textit{modularity} as a way to assess the goodness of a given partition \citep{newman2006mod}. Modularity is a quality function that measures, for each partition, the fraction of links that fall within them versus the fraction one would expect if links were placed randomly. Therefore, high values of modularity suggest that most of the links fall into the partitions rather than in random distributions. Hence these partitions represent potential communities. This indicator has been widely adopted, to the extent that many algorithms for community detection aim to maximise this function, e.g. Edge Betweenness \citep{newman2004}, Fast Greedy \citep{clauset2004}, Leading Eigenvector \citep{newman2006}, Spin Glass \citep{reichardt2006}, Louvain \citep{blondel2008} and many other versions \citep{radicchi2004,blondel2008,brandes2008}. Other methods available in the state of the art that might not directly use modularity are based on spectral analysis \citep{donetti2004,capocci2005}, random walks \citep{pons2005}, spin-spin interactions \citep{blatt1996,reichardt2004}, label propagation \citep{raghavan2007} and many other techniques \citep{arenas2006,boccaletti2007,long2007,papadopoulos2009}.

Another feature to take into account when detecting communities within networks is the kind of partition. In \textit{Network Science}, a partition is defined as the division of the network in an arbitrary number of clusters, such that each node belongs to only one clusters. This is called \textit{crisp} partitioning\index{crisp partitioning}. However, a node is rarely confined into a single community. For instance, in social networks an individual can belong to different communities consisting of family, friends, co-workers, old schoolmates and so on. Other individuals from the same communities can in turn belong to other communities, making the network a space of intertwined overlapping communities. This partitioning is instead called \textit{fuzzy}\index{fuzzy partitioning}.

Overlapping communities\index{overlapping communities} are not limited to social networks. As \cite{osborne2013} pointed out, in a scientific collaboration network, researchers can belong to different communities of scholars because they might be involved in multidisciplinary collaborations. Also, in biological networks we can observe that different clusters of diseases overlap because the same genes are often involved in different disorders \citep{barabasi2011}. We can argue that also in topic networks, and by extension, evolutionary network, is possible to observe a similar phenomenon given that a topic can be involved in the process of fostering the emergence of more than one new area. 

The state of the art proposes different algorithms that can detect overlapping communities. The most popular is the Clique Percolation Method proposed by \cite{palla2005}, which has been extended so that it can work for weighted networks \citep{farkas2007} and directed networks \citep{palla2007}. Other algorithms are the Link Clustering \citep{ahn2010, evans2009}, the Fuzzy C-Means \citep{bezdek1981,zhang2007}, and others \citep{baumes2005,nepusz2008}.

As part of assessing the validity of our approach (see evaluation in, Chapter \ref{ch:finalevaluation}), we compared our results against some approaches, such as Fast Greedy \citep{clauset2004}, Clique Percolation Method \citep{palla2005}, Fuzzy C-Means \citep{bezdek1981}, and Leiden Eigenvector \citep{newman2006}, which failed to provide a good characterisation of the community structure within the evolutionary networks. However, those experiments allowed us to learn some interesting peculiarities of the evolutionary networks. This understanding allowed us to design the Advanced Clique Percolation Method, which extends the Clique Percolation Method \citep{palla2005}, and works much better on this task.

\section{Limitations of existing methods}\label{sec:limitations}
In Section \ref{sec:techback}, we presented an overview of the main approaches and directions for detecting topics and tracking their evolution. As discussed, a good number of methods deal with the detection of new emerging topics, but they can be applied only on already emerged topics, associated either with a label or a bag of words. Indeed, they can identify topics that \textit{recently} emerged, and exhibit dynamics related to their adoption by a research community, e.g., a good number of publications containing a keyword and a new pattern in the citation graph.
On the other hand, the state of the art is still missing a comprehensive solution for detecting topics that are \textit{about to} or \textit{will} emerge. 

In Section \ref{sec:philosophicalback}, we observed the literature with another lens, discussing the evolution of science from a philosophical perspective. There is a significant body of knowledge claiming that the emergence of new scientific fields is tied to the dynamics of group of scholars that exchange theories and methodologies. This form of collaboration can be analysed and it has great potential for supporting the detection of new emerging topics.

To sum up, on the one hand we have the currently available approaches, which are unable to identify new topics before their actual emergence, but on the other hand we a have solid philosophical background suggesting that is possible to spot clear signs of evolutionary stages that can lead to the emergence of a new topic. 

In the following chapter, we will elaborate on how to close the gap and specifically design an approach that is able to take into account the aforementioned philosophical insights for detecting new topics at an embryonic stage.

{\let\newpage\relax
\part {AUGUR}}
In ancient Rome, a religious official who observed natural signs, especially the behaviour of birds, interpreting these as an indication of divine approval or disapproval of a proposed action.

Today, an automatic system that observes natural signs, especially the behaviour of topics, interpreting these as an indication of the emergence of new areas.

\chapter{First study: analysis of dynamics\label{ch:firststudy}}

\section{Introduction}

The goal of our research is to detect the emergence of new research topics. In the previous chapter, we contextualised the topic and we analysed its anatomy (Section \mbox{\ref{sec:topicdefinition}}). Then, we observed the different life stages it goes through. 
We found also a number of interesting theories in \mbox{\textit{Philosophy of Science}} \citep{herrera2010,sun2013,nowotny2013,kuhn1963,becher2001}, supporting the idea that the life of a research topic starts with an\mbox{ \textit{embryonic stage}}. Specifically, in this stage, a topic has not yet been explicitly labelled and recognised by a research community, but it is already taking shape, as evidenced by the fact that researchers from a variety of fields are forming new collaborations and producing new work, starting to define the challenges and the paradigms associated with the emerging new area. For this reason, we hypothesise that topics can actually be detected in such phase.

Hence, our research question is: ``\textbf{How we can detect new emerging topics at the embryonic stage?}''

In Section \ref{sec:embryonicstage}, we mentioned that the development of new topics is usually fostered by the cross-fertilisation of established research areas. Therefore, to try to answer the question above, we can analyse the \textit{dynamics} of the ``procreators'' or ``ancestors''\index{ancestors} of new emerging topics. In particular, we can analyse the interactions between topics and those between entities linked to these topics, such as publications, authors, venues. For example, the sudden appearance of some publications concerning a combination of previously uncorrelated topics may suggest that some pioneer researchers are investigating new possibilities and maybe shaping a new emerging area.

In this chapter, we present a study that aims to uncover key elements associated with the research dynamics preceding the creation of novel topics. In particular, we compare the pace of collaboration and the network density within sections of the co-occurrence graph associated to new emerging topics (treatment set) and established topics (control set). Our results show that the pace of collaboration and the density measured in the sections of the network that will give rise to a new topic are significantly higher than those in the control set. 
Furthermore, these findings support our hypothesis that the existence of an embryonic phase can be numerically measured and also yield new empirical evidence consistent with the already available theories in \textit{Philosophy of Science}. 

The results, and in general the overall study, presented in this chapter have led to two articles:
\begin{itemize}
\item \citep{salatino2016} at SAVE-SD workshop co-located with The Web Conference (once World Wide Web conference) and
\item \citep{salatino2017} published by PeerJ Computer Science.
\end{itemize}
The raw data, which include the list of both emerging and established topics as well as their procreators is available at \url{https://osf.io/bd8ex/}.

The remainder of the chapter is organised as follows. In Section \ref{sec:dataset} we describe the data we will use to carry out our experiments. In Section \ref{sec:testing-topics}, we describe the group of topics upon which we performed our analysis. In Section \ref{sec:pocnd}, we describe the dynamics that we aim to investigate. Then, we describe the experimental approach consisting of two main steps:
\begin{enumerate}[label={Step \arabic*.},leftmargin=3cm]
\item \textbf{Selection phase}, described in Section \ref{sec:selection-phase-first}, in which we select the emerging topics and their counterpart as control set. Then for each topic we select and extract the portion of network related to them, prior to their emergence.
\item \textbf{Analysis phase}, described in Section \ref{sec:analysis-phase-first}, integrates statistics and semantics, and aims to analyse the differences within the networks extracted for the two groups of topics. To measure the two dynamics this phase uses two methods:
\begin{enumerate}[label*={ \alph*.},leftmargin=1.8cm]
  \item Cliques for measuring the \textbf{pace of collaboration}, described in Section \ref{sec:analysis-phase-poc} \index{pace of collaboration}
  \item Triad-census for measuring the \textbf{network density}, which is described in Section \ref{sec:analysis-phase-triad} \index{network density}
\end{enumerate}
\end{enumerate}
Finally, we evaluate and discuss our results, we highlight some limitations and we summarise the main conclusions.

\section{Dataset}\label{sec:dataset}\index{Rexplore!dataset}

In this section, we briefly describe the Rexplore dataset \mbox{\citep{osborne2013}} which includes different metadata mainly in the field of \mbox{\textit{Computer Science}}. Then, we will describe some network datasets we will use to carry out our experiments.

\subsection{Rexplore}\label{sec:rexplore}
\begin{figure}[h]
\includegraphics[width=\textwidth]{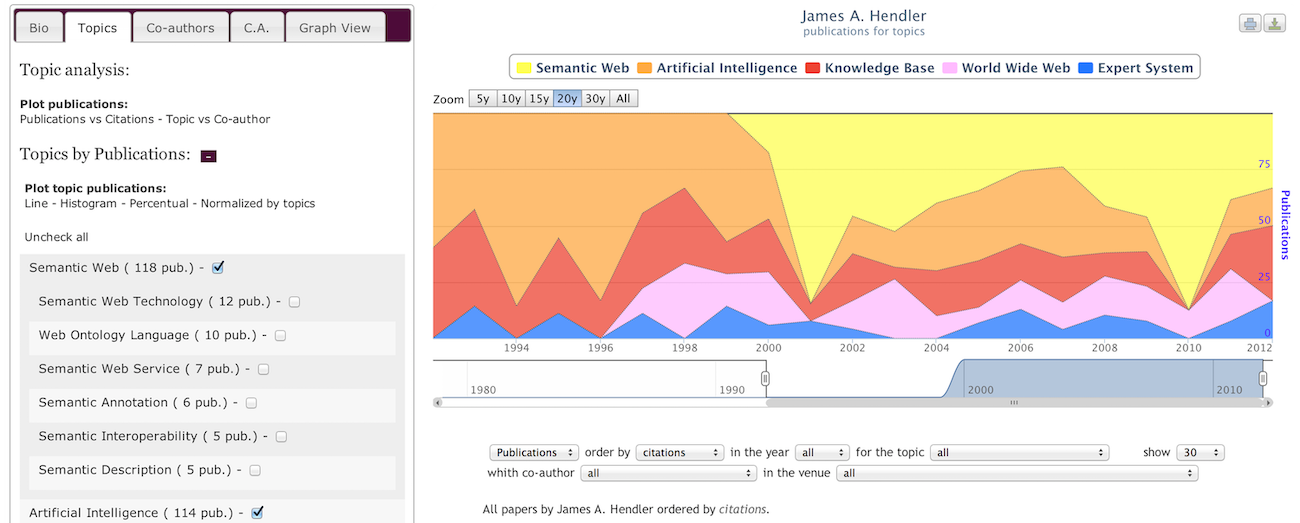}
\caption{Screen capture of Rexplore while looking at the research trajectory of the renowned \textit{Artificial Intelligence} researcher James A. Hendler.}
\label{fig:rexplore3}
\end{figure}

Rexplore\index{Rexplore}\footnote{Link to Rexplore: \url{https://technologies.kmi.open.ac.uk/rexplore/}.} is a web application that combines large-scale data mining, data visualisation and semantic technologies to explore and make sense of the research environment.
In particular, Rexplore offers an advanced graphical interface, including a variety of innovative and fine grained visualisations, which support users in: 
\begin{enumerate*}[label=(\roman*)]
 \item performing advanced search using a variety of parameters, on authors, papers, research areas and research group;
 \item investigating how research communities develop in time, whether they merge, split, grow, contract and so on; 
 \item exploring the research trajectory of researchers, in terms of their main areas of interest and contributions in time, as shown in Fig. \ref{fig:rexplore3};
 \item identifying a variety of interesting relations between researchers, e.g., recognising authors who share similar research trajectories;
 \item detecting important trends in research, such as, significant migrations of researchers from one area to another,
 \item classifying  book collections, authors, conferences and other research entities according to the associated research topics.
\end{enumerate*}

Rexplore builds on a very large collection of data that integrates scholarly data from major commercial publishers, including \textit{Scopus}, \textit{Springer} and \textit{Microsoft Academic Search}.

At the moment, the Rexplore corpus includes metadata on 16 million papers and 11 million authors, mostly covering the field of \textit{Computer Science}, and partially other fields like \textit{Biomedics}. This metadata describes different attributes of research papers, e.g., titles, abstract, keywords, number of citations, references, venues and many others, as well as information about from authors, such as name, organisation, number of citations, h-index and so on. 

Rexplore relies also on the Computer Science Ontology, to represent the different topics the scientific papers are associated with. 

In the following sections, we will describe in more detail some segments of the Rexplore dataset that we have used in our experiments. In particular, we will describe the co-occurence network and the semantic-enhanced topic network.

\subsection{Co-occurrence network}\label{sec:cooccurrencenetwork}\index{co-occurrence network}\index{networks!co-occurence}\index{networks!keyword}\index{keyword networks}
The co-occurrence network, or simply \textit{keyword network}, is a graph\footnote{There is a subtle distinction between graph\index{graph} and network. \textit{Network} is generally a concept, like network of web documents, while a \textit{graph} is the mathematical representation of a network. Throughout this dissertation we will use them interchangeably.} describing the interaction between keywords.
Specifically, the co-occurrence network connects the different keywords, used by authors to tag their papers, and this connection is determined by the amount of scientific publications in which they appear together (i.e., co-occur) in a particular year. 
The Rexplore dataset contains a co-occurrence network per each year $t$, and naturally the amount of co-occurrences between a couple of keyword can change over time. Therefore, it is possible to perform a diachronic analysis to understand whether the interaction between particular keywords is increasing or decreasing.
In more detail, the co-occurrence network is a fully weighted graph, as described in Eq. \ref{eq:cooccurrencegraph}.

\begin{equation}
G_{yea{r_t}}\, = \,(V_{yea{r_t}},\,E_{yea{r_t}},p_{yea{r_t}},w_{yea{r_t}}),\,\,\,\,p\,:\,\,V \to \mathbb{R},\,\,w\,:\,\,E \to \mathbb{R}
\label{eq:cooccurrencegraph}
\end{equation}

In particular, $V$ is the set of keywords while $E$ is the set of links representing the co-occurrences between keywords. The node weights in $p$ represent the number of publications in which the keyword appears in year $t$, while the link weights in $w$ is equal to the number of publications in which two keywords co-occur together in the same year $t$. The union of all these co-occurrence graph over time represents the whole collection co-occurrence networks, as showed in Eq. \ref{eq:cooccurrencegraphall}.
\begin{equation}
{co \mhyphen occurence\,graphs} = \bigcup\limits_{years} {{{\rm{G}}_{{\rm{year}}}}} 
\label{eq:cooccurrencegraphall}
\end{equation}

\begin{figure}[!ht]
\includegraphics[width=\textwidth]{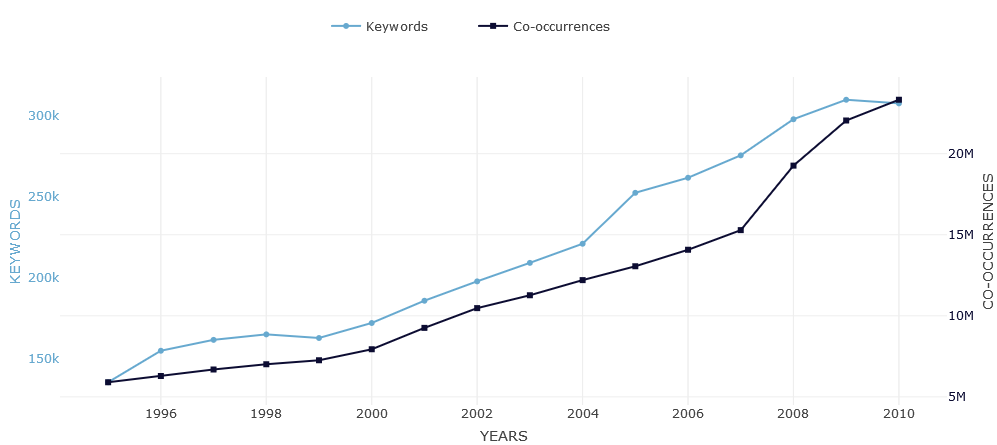}
\caption{Size of 16 co-occurrence networks from 1995 to 2010. The number of keywords correspond to the number of nodes in $V$ and the number of co-occurrences correspond to the number of links in $E$.}
\label{fig:cooccs}
\end{figure}

In Fig. \ref{fig:cooccs} we reported the number of unique keywords (light blue line), for each year from 1995 to 2010, which are also the nodes of the co-occurrence network. In addition, we reported the number of links (black line) between these keywords which correspond to the amount of interactions in terms of co-occurrences between keywords.

\subsection{Semantic-Enhanced Topic Network}\label{sec:topicnetwork}\index{topic network} \index{networks!topic}\index{networks!semantic-enhanced topic}
The \textit{Semantic-Enhanced Topic Network}, also referred to simply as \textit{topic network}\index{topic network}, is a network that represents the interactions between research topics. This network is the result of a semantic enhancement of the co-occurrence network using the Computer Science Ontology\index{Computer Science Ontology}. 

CSO allowed us to filter out all keywords that do not refer to research areas, i.e., ``case study'', ``error'', ``methodology'', and so forth. Furthermore, using the \textit{relatedEquivalent} relationship from the ontology, we were able to aggregate different keywords representing the same concept. For example, it has been possible to aggregate keywords like ``semantic web'', ``semantic web technology'' and ``semantic web technologies'' in one single semantic topic. It also allowed us to associate a keyword to a related acronym, e.g., ``resource description framework'' and ``rdf'', or even more complex associations, e.g., ``fuzzy c mean'', ``fuzzy c-means (fcm)'', ``fuzzy c-means'', ``fuzzy c means clustering'', ``fuzzy c-means clustering'', ``fuzzy c-means algorithms'' and ``fuzzy c-means algorithm''.

More formally, like the co-occurrence network, the topic network is also a fully weighted graph, as described in Eq. \ref{eq:cooccurrencegraph}. As before, the set of nodes $V$ represents topics while $E$ is the set of links representing the topic co-occurrences. The node weight in $p$ represents the number of publications in which the topic appears in the year $t$, while the link weight in $w$ is equal to the number of publications in which two topics co-occur together in the same year $t$.

\begin{figure}[!ht]
\includegraphics[width=\textwidth]{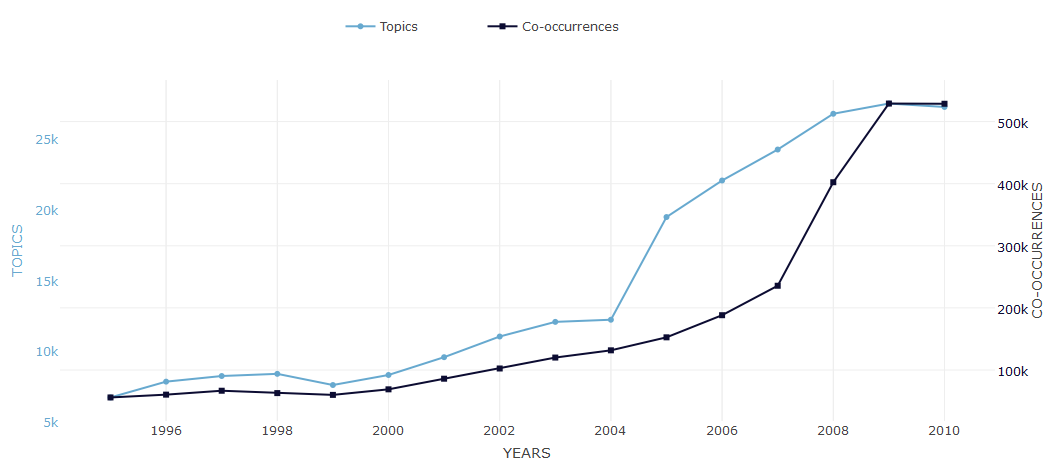}
\caption{Size of 16 topic networks from 1995 to 2010. The number of topics correspond to the number of nodes in $V$ and the number of co-occurrences correspond to the number of links in $E$.}
\label{fig:topicnet}
\end{figure}

In Fig. \ref{fig:topicnet} we show the number of unique topics (light blue line), for each year from the 1995 to 2010, which are also the nodes of the topic network. In addition, we also show the number of links (black line) between these topics, which correspond to the amount of collaborations between these topics.

\section{Testing topics}\label{sec:testing-topics}
With this study, we want to understand whether the emergence of a new research topic can be encouraged by new intense collaborations among its related topics, presumably its ``ancestors''\index{ancestors}. To this end, we randomly selected a set of 75 new topics in the decade 2000-2010, so that we can aim to select their procreators and understand their dynamics, in the years prior to their emergence. This group of topics is also referred as the \textit{treatment set}. A topic emerges in the year, also called \textit{year of debut}\index{year of debut}, in which its label first appears in a research paper. Although the process of selecting emerging topics was completely random, we aimed at having a balanced number of topics over the years. In this way, we avoided the emergence of a bias due to the time in which they were analysed.

\begin{figure}[ht]
  \includegraphics[width=\linewidth]{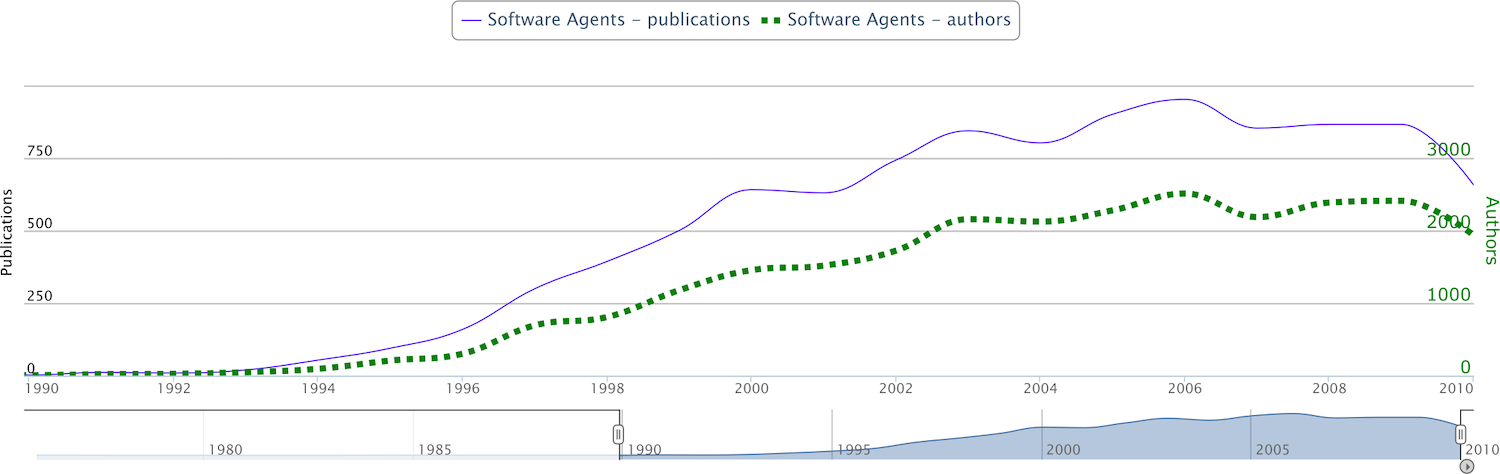}
  \caption{Evolution of the topic \textit{Software Agents} in terms of number of authors and number of publications per year. The chart has been produced using the Rexplore system.}
  \label{fig:softwareagents}
\end{figure}

For the sake of comparison, as a control set, we selected a set of 100 established topics (also referred to as the \textit{non-debutant group}) and assigned them a random \textit{year of analysis}\index{year of analysis} within the decade 2000-2010. A topic is considered established if: i) it debuted before 2000, so that it can be in its recognised phase at the time when we analyse the emerging topics, ii) it appears in the CSO Ontology, iii) it is associated each year with a substantial and consistent number of publications. As an instance, Fig. \ref{fig:softwareagents} shows the evolution over time of the well-established topic \textit{Software Agents}, in terms of number of active authors and publications. 

The figure shows that the topic made its debut in 1993 and in the year 2000 reached a rate of over 500 publications per year with more than 1500 authors working on it. It can thus be considered established in the context of our study. 

Table \ref{tab:different-years} compares topics in both treatment and control sets, showing when they emerged and in which years we performed their analysis.

\begin{table}[ht]
\centering
\caption{Comparing year of debut and year of analysis for both treatment and control sets.}
\label{tab:different-years}
{\renewcommand{\arraystretch}{1.2}%
\begin{tabular}{lll}
\hline
\textbf{Testing topics}                  & \textbf{Emerging}    & \textbf{Year of analysis} \\ \hline
Debutant topics          &  $2000 \mhyphen2010$           & \textit{year of debut} $(2000 \mhyphen10)$ \\
Non-debutant topics          & before 2000           & randomly assigned in decade $2000 \mhyphen10$  \\ \hline
\end{tabular}}
\end{table}

Both treatment and control sets are crucial for this analysis. In particular, investigating the dynamics within the sections of networks associated with the two sets and comparing their results allow us to understand whether the investigated dynamics are exclusive for the emerging topics or they appear in both sets. In this way, we can provide further evidence to the existence of the embryonic stage.

\section{Pace of collaboration and network density}\label{sec:pocnd}
For this study, we hypothesise that the existence of an embryonic phase can be measured and that it is possible to detect new topics within this stage by analysing the dynamics of existent topics which presumably are also their ancestors. In particular, the dynamics we aim to investigate are the pace of collaboration and network density.

As already mentioned in the previous chapter, collaboration between scientists of different disciplines is a key driving force for the scientific enterprise. Indeed, collaboration makes it easier to work on interdisciplinary problems, allows researchers to incorporate different kinds of knowledge and guarantees diversity in perspective. To this end, we consider collaboration as the central core of our investigation and we analyse both increase in the pace of collaboration between topics and also the creation of brand-new collaborations over time.


The analysis of the pace of collaboration between two or more topics aims to observe whether the topics are strengthening their ties. In particular, if the percentage of papers shared by these topics is increasing it means that these topics are increasing their collaboration and they are getting closer to each other. In addition, this growing closeness between topics can eventually lead to a new interdisciplinary field. For instance, it is well-acknowledged that the communities interested in \textit{Artificial Intelligence}, \textit{World Wide Web} and \textit{Knowledge Based Systems} started to collaborate on novel ideas, giving rise to a novel research area later labelled \textit{Semantic Web} by \cite{berners2001}. In Section \ref{sec:analysis-phase-poc}, we will observe in more detail how we compute the pace of collaboration between the ancestors of a debutant topic.
 
On the other hand, to observe whether topic networks are expanding in time and therefore whether within the network some topics are introducing new collaborations, we use the notion of network density. This is important because the fact that a number of authors from previously unrelated research communities or topics are starting to collaborate together may also suggest the emergence of a new interdisciplinary research area. In Section \ref{sec:analysis-phase-triad}, we will show a method for evaluating the creation of new collaborations between topics.

\section{Selection phase}\label{sec:selection-phase-first}

The selection phase is the first step of the study and, as already mentioned, it aims to select and extract portions of the topic networks that are related to topics of both treatment and control set, in a few years prior to the year of analysis. Afterwards, the analysis phase (in Section \ref{sec:analysis-phase-first}) will analyse the differences within the networks belonging to the two sets, suggesting the dynamics that may be involved in the process of creation of a new topic.
 
In our previous study \citep{salatino2016}, we firstly analysed two well-known topics, \textit{Semantic Web} (debuting in 2001) and \textit{Cloud Computing} (debuting in 2006), so that we could tune our approach, and then we tested 50 topics debuting in the period between 2000 and 2010 and 50 non-debutant topics. Here we report a follow-up study, in which this set of topics has been extended to 75 topics debuting in the same period and 100 non-debutant topics \citep{salatino2017}. In particular, from 50 debutant topics, in the treatment set, we iteratively added new topics until we reached 75 topics and data saturation \citep{fusch2015}, to the extent that the results of the analysis did not significantly vary with the inclusion of new data points. On the other hand, for the control set, we randomly assigned another year of analysis to each topic, with the consequence of doubling the amount of non-debutant topics (from 50 to 100).


After selecting the testing topics (topics in both groups), we can proceed with selecting and extracting the portion of network related to them. 
We assume that a new emerging topic will continue to collaborate with the topics that contributed to its creation for a certain time after its debut. Indeed, as part of the nurturing phase, its procreator will still collaborate with the emerging topic until it reaches its maturity. This assumption was also discussed and tested in a previous study \citep{osborne2012}, where it was used to find historical subsumption links between research areas. Being able to select the procreators of a given topic allows us to analyse the dynamics involved in their interactions before the new topic emerged.

\begin{figure}[ht]
  \includegraphics[width=\linewidth]{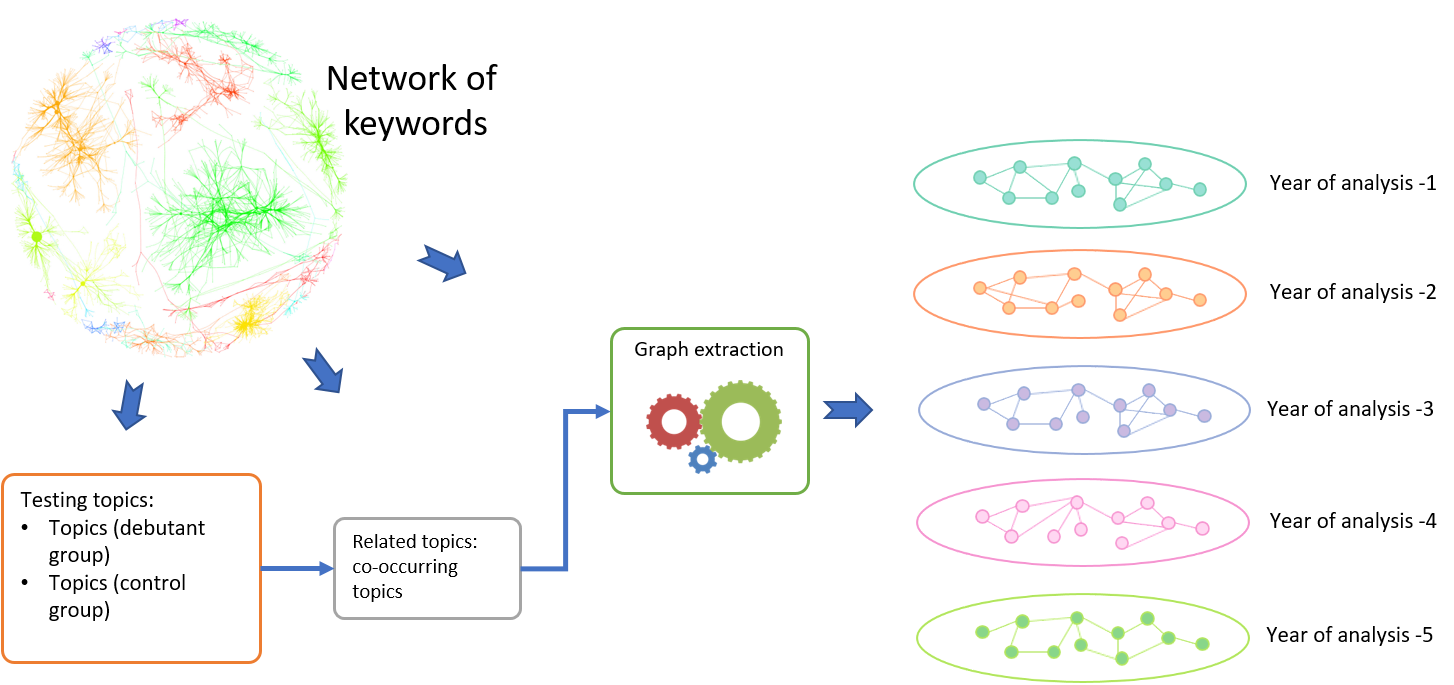}
  \caption{Workflow representing all the steps for the selection phase.}
  \label{fig:selectionphase}
\end{figure}

To this end, for each debutant topic we firstly selected its 20 most co-occurring topics from the year of debut to 2014 and then we extracted the portion of topic networks containing such topics in the five years preceding the year of debut. For instance, if a topic \textit{A} made its debut in 2003, we analyse the portion of network containing its most co-occurring topics in the 1998-2002 timeframe. However, since we want to analyse how the dimension of these subgraphs can influence the results, we also tested solutions with 40 and 60 most co-occurring topics. We repeated the same procedure on the topics in the control set, using the year of analysis we randomly assigned (within the decade 2000-2010), rather than their actual year of debut. The workflow representing this approach is showed in Fig. \ref{fig:selectionphase}, while its pseudocode is available in Code \ref{alg:selection-phase}.

In brief, the selection phase associates to each topic in both treatment and control sets a graph ${G^{topic}}$, shown in Eq. \ref{eq:selectionGraph}.

\begin{equation}
\label{eq:selectionGraph}
{G^{topic}} = \,G_{year - 5}^{topic}\, \cup \,G_{year - 4}^{topic}\, \cup \,G_{year - 3}^{topic}\, \cup \,G_{year - 2}^{topic}\, \cup \,G_{year - 1}^{topic}
\end{equation}
This graph ${G^{topic}}$ corresponds to a section of collaboration networks related to a debutant topic in the five years prior to its emergence (or year of analysis for non-debutant topics). In particular, each year   corresponds to the induced subgraph of the topic networks (see Section \ref{sec:topicnetwork}) containing its \textit{n} most co-occurring topics (or procreators $V_{year - i}^{topic}$), the links between them $E_{year - i}^{topic}$, and also both node and link weights ($P_{year - i}^{topic}$ and $W_{year - i}^{topic}$, as reported in Eq. \ref{eq:indecedSelectionGraph}).

\begin{equation}
\label{eq:indecedSelectionGraph}
G_{year - i}^{topic} = {\mkern 1mu} (V_{year - i}^{topic},{\mkern 1mu} {\mkern 1mu} E_{year - i}^{topic}{\mkern 1mu} ,\,{\mkern 1mu} P_{year - i}^{topic},\,{\mkern 1mu} W_{year - i}^{topic}{\mkern 1mu} )
\end{equation}

The networks associated to the debutant topics included 1,357 unique topics, while the ones associated to the control set included 1,060 topics. 
\newline
\newline
\begin{algorithm}[H]
\label{alg:selection-phase}
\setstretch{1}
\SetKwInOut{Input}{Input}\SetKwInOut{Output}{Output}

\Input{List of Testing Topics $db.topics$ and $ndb.topics$}
\Output{Networks containing their ancestors $db.networks$ and $ndb.networks$}
\SetKwProg{Fn}{def function}{}{end definition}\SetKwFunction{Fextract}{ExtractNetwork}%
\BlankLine
\Fn(\tcc*[h]{returns networks}){\Fextract{topic, year}}{

	related.topics$\leftarrow$ SelectCoOccurringTopics(topic, from=year, to=2017)\;
%
\For{$yr\leftarrow (year-5)$ \KwTo $year$}{
	ancestors.network[yr]$\leftarrow$ InducedSubgraph(related.topics,year)\;
}
return(ancestors.network)\;
}
\BlankLine
\BlankLine
\tcp*[h]{Algorithm starts here}\\
\ForEach{topic in db.topics}{
	year$\leftarrow$ GetYearOfSoftDebut(topic)\;
	db.networks[topic]$\leftarrow$ ExtractNetwork(topic,year)\;
}

\ForEach{topic in ndb.topics}{
	year$\leftarrow$ AssignRandomYear(from=2000, to=2010)\;
	ndb.networks[topic]$\leftarrow$ ExtractNetwork(topic,year)\;
}

\caption{Extracting the portion of topic networks associated to topics in both treatment and control set.}
\end{algorithm}
\vspace*{0.7cm}

\section{Analysis phase}\label{sec:analysis-phase-first}
In this second phase, we assess the dynamics in the graphs associated to both the two sets with two approaches: cliques-based and triad-based. The first approach, described in Section \ref{sec:analysis-phase-poc}, measures the pace of collaboration and in particular it transforms the graphs in 3-cliques, then it associates to each of them a measure reflecting the strength of collaboration between relevant topics, and finally it analyses how these collaborations change over time. The second approach, described in Section \ref{sec:analysis-phase-triad}, evaluates the differences between the graphs associated to the two sets by measuring the changes in the graph density using the triad census technique \citep{davis1967}. We devote the following subsections to describe the two mentioned approaches, while their results will be evaluated in Section \ref {evaluation-first-study}.

\subsection{Pace of collaboration through cliques} \label{sec:analysis-phase-poc} 
\begin{figure}[ht]
\centering
  \includegraphics[width=150px]{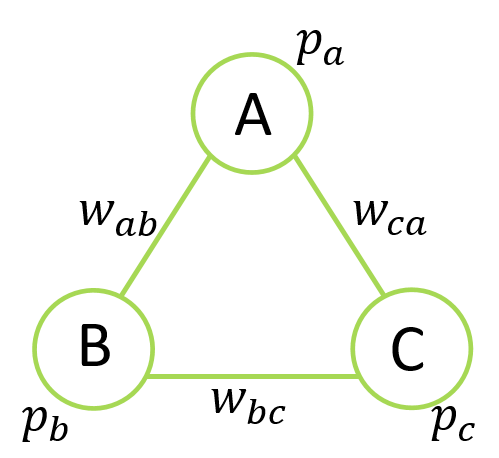}
  \caption{An instance of a 3-clique containing node $\{A,B,C\}$ with node weights $\{p_a,p_b,p_c\}$ and link weights $\{w_{ab},w_{bc},w_{ca}\}$.}
  \label{fig:clique}
\end{figure}

\index{pace of collaboration}
In this first approach, we measure the pace of collaboration of the graphs associated to the topics, in both treatment and control sets, by analysing the diachronic activity of triangles of their collaborating topics. To this end, we firstly extract all 3-cliques from the five sub-graphs associated to each topic under analysis. A 3-clique\index{clique}, as shown in Fig. \ref{fig:clique}, is a complete sub-graph of order three in which all nodes are connected to one another. A 3-clique plays an important role in many applications \citep{eppstein2010} since it can model small groups of entities close to each other \citep{luce1949}.
 
\begin{figure}[ht]
  \includegraphics[width=\linewidth]{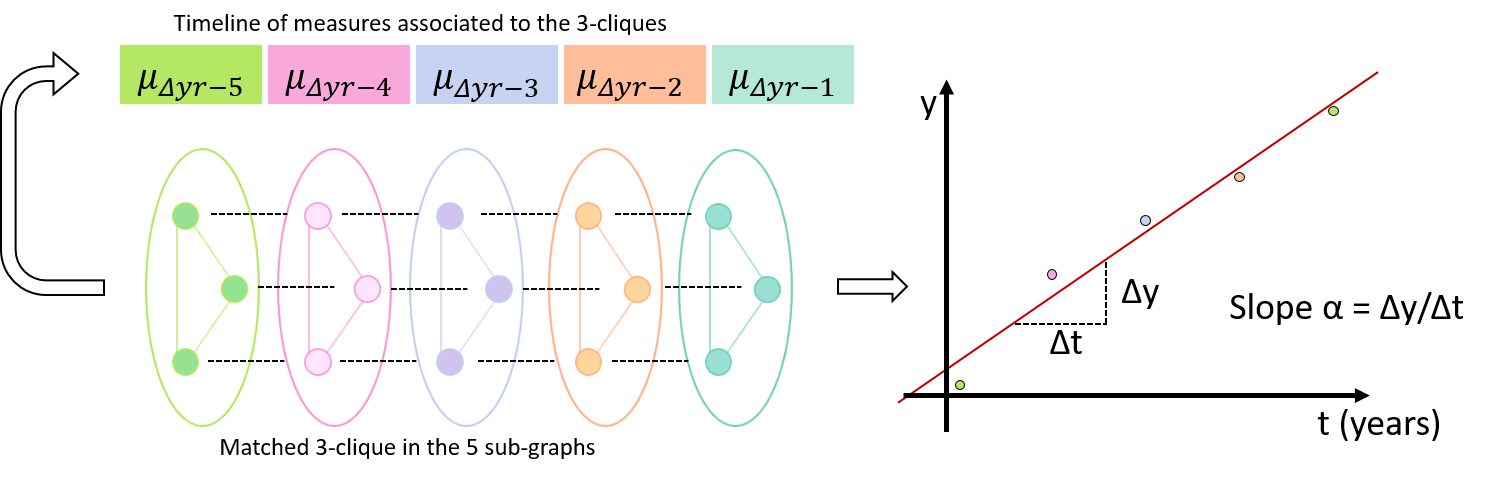}
  \caption{Main steps of the analysis phase.}
  \label{fig:analysisphaseclique}
\end{figure}

To study the dynamics preceding the debut of each topic, we analyse the evolution of the same 3-clique in subsequent years. Figure \ref{fig:analysisphaseclique} summarises the process. Considering a 3-clique having nodes $\{A,B,C\}$, we quantify its \textit{collaboration index} $\mu_\Delta$ in a year by taking into account both node weights $\{p_a,p_b,p_c\}$ and link weights $\{w_{ab},w_{bc},w_{ca}\}$.

\begin{equation}
\begin{array}{l}
{\mu _{A - B}} = \,mean\left( {P(A|B),P(B|A)} \right)\\
{\mu _{B - C}} = \,mean\left( {P(B|C),P(C,B)} \right)\\
{\mu _{C - A}} = \,mean\left( {P(C|A),P(A|C)} \right)\\
{\mu _\Delta } = \,mean\left( {{\mu _{A - B}},{\mu _{B - C}},{\mu _{C - A}}} \right)
\end{array}
\label{eq:poc}
\end{equation}

The index $\mu_\Delta$ is computed by aggregating the three coefficients $\mu_{(A-B)}$, $\mu_{(B-C)}$ and $\mu_{(C-A)}$ as illustrated by Eq. \ref{eq:poc}. The \textit{strength of collaboration} $\mu_{(x-y)}$  between two nodes of the topic network, \textit{x} and \textit{y}, is computed as the mean of the conditional probabilities $P(y|x)$ and $P(x|y)$, where $P(y|x)$ is the probability that a publication associated with a topic \textit{x} will be associated also with a topic \textit{y} in a certain year. The advantage of using conditional probabilities instead of the number of co-occurrences is that the value $\mu_{(x-y)}$  is normalised to the number of publications associated to each topic. Finally, $\mu_\Delta$ is computed as the mean of the strengths of collaboration of the three links in a 3-clique. In our previous experiment \citep{salatino2016}, we found that the harmonic mean provides better results than the arithmetic mean.
The evolution of the 3-clique collaboration pace can be represented as a timeline of values in which each year is associated with its average strength of collaboration, as in Eq. \ref{eq:timelineofpoc}. We assess the increase of the collaboration pace in the period under analysis by computing the slope of the linear regression of these values. 
\begin{equation}
\mu _{\Delta {\rm{time}}}^{clique} = [{\mu _{\Delta {\rm{yr}} - {\rm{5}}}},{\mu _{\Delta {\rm{yr}} - {\rm{4}}}},{\mu _{\Delta {\rm{yr}} - {\rm{3}}}},{\mu _{\Delta {\rm{yr}} - {\rm{2}}}},{\mu _{\Delta {\rm{yr}} - {\rm{1}}}}]
\label{eq:timelineofpoc}
\end{equation}
Initially, we tried to determine the increase in the collaboration pace exhibited by a clique by simply taking the difference between the first and last values of the timeline $({\mu _{\Delta {\rm{yr}} - {\rm{5}}}}-{\mu _{\Delta {\rm{yr}} - {\rm{1}}}})$. However, as reported in \citep{salatino2016} this method ignores the other values in the timeline and can thus neglect important information. For this reason, we applied instead the linear interpolation method on the five measures using the least-squares approximation to determine the linear regression of the time series $f(x)=\alpha x+\beta$. The slope $\alpha$, showed in Eq. \ref{eq:pocdef}, is then used to assess the increase of collaboration in a clique. In particular, $\overline {\mu _{\Delta {\rm{yr}}}}$ represents the mean value of the five collaboration indices within $\mu _{\Delta {\rm{time}}}^{clique}$, $\overline {year}$ is the mean value of all the years the extracted cliques refer to $\{ year_{t},year_{t - 1},year_{t - 2},year_{t - 3},year_{t - 4}\}$, and $year_{t - i}$ is the instance value from that set.

When $\alpha$ is positive the degree of collaboration between the topics in the clique is increasing over time, while, when it is negative, the number and intensity of collaborations are decreasing. 

\begin{equation}
\alpha\, = \,\frac{{\sum\limits_{i = 0}^4 {(yea{r_{t - i}}\, - \,\overline {year} )(\,{\mu _{\Delta {\rm{yr}} - {\rm{i}}}}\, - \,\overline {\mu _{\Delta {\rm{yr}}}} )} }}{{\sum\limits_{i = 0}^4 {{{(yea{r_{t - i}}\, - \,\overline {year} )}^2}} }}
\label{eq:pocdef}
\end{equation}

Finally, the overall pace of collaboration\index{pace of collaboration} associated to each sub-graph (and therefore each testing topic) is measured by computing the average of all slopes associated with the timelines of 3-cliques. 

In brief, for each testing topic we select a subgraph of related topics in the five years preceding the year of debut (or \textit{analysis} for topics in the control set). We then extract the 3-cliques and associate each of them with a vector representing the evolution of their pace of collaboration. The trend of each clique is computed as the angular coefficient of the linear regression of these values. Finally, the increase in the pace of collaboration of a subgraph is obtained by averaging these values.

\subsection{Density of network using triad census} \label{sec:analysis-phase-triad} 
\begin{figure}[ht]
\centering
  \includegraphics[width=350px]{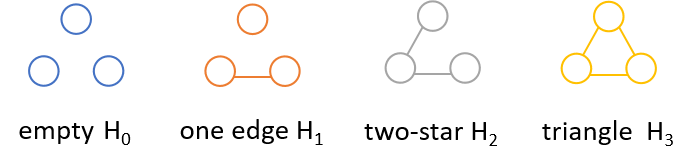}
  \caption{The four isomorphism classes of triad. The triad census consists in counting the frequencies of \textit{$H_i$} of the input graph.}
  \label{fig:triadcensus}
\end{figure}

\index{network density}
The second approach employs the \textit{triad census}\index{triad census} \citep{davis1967} to measure the change of topology and the increasing density of the subgraphs during the five year period. The triad census of an undirected graph, also referred as \textit{global 3-profiles}, is a four-dimensional vector representing the frequencies of the four isomorphism classes of triad, as shown in Fig. \ref{fig:triadcensus}. 
The triad census summarises structural information in networks and is useful to analyse the structural properties of social networks. It has been applied to several scenarios, such as identifying spam \citep{kamaliha2008,o2012}, comparing networks \citep{prvzulj2007}, and analysing social networks \citep{faust2010,ugander2013}.

In this study, we use triad census \index{triad census} to describe all the sub-graphs $G_{year - i}^{topic}$ associated to a input topic in terms of frequencies of \textit{Hi} (see Fig. \ref{fig:triadcensus} and then evaluate how the frequencies of \textit{empties} (\textit{H0}), \textit{one edges} (\textit{H1}), \textit{two-stars} (\textit{H2}) and \textit{triangles} (\textit{H3}) change in time. Figure \ref{fig:triadcensus} illustrates the four classes of triads for an undirected graph in the case of topic networks. An increase in the numbers of triangles suggests the appearance of new collaboration clusters among previously distant topics.
Differently from the previous approach, the triad census treats the topic networks as binary graphs. It does not consider the weight of the links, but only their existence. Hence, it is useful to assess how including links with different strengths affects the analysis. To this end, we performed three experiments in which we considered only links associated with more than 3, 10 and 20 topic co-occurrences.

We initially perform the triad census over the five graphs associated to each input topic. For example, Table \ref{tab:artbeecolonies} shows the results of the triad census over the five sub-graphs associated with the debutant topic \textit{Artificial Bee Colonies}. 

\begin{table}[ht]
\centering
\caption{Frequencies of $H_i$ obtained performing triad census on the debutant topic \textit{Artificial Bee Colonies}}
\label{tab:artbeecolonies}
{\renewcommand{\arraystretch}{1.2}%
\begin{tabular}{l|l|l|l|l}
\textbf{Graph}  & $H_0$  & $H_1$  & $H_2$  & $H_3$   \\ \midrule
$G_{(year-5)}^{topic}$ & 446 & 790 & 807 & 882  \\
$G_{(year-4)}^{topic}$ & 443 & 854 & 915 & 1064 \\
$G_{(year-3)}^{topic}$ & 125 & 486 & 967 & 1698 \\
$G_{(year-2)}^{topic}$ & 100 & 410 & 908 & 1858 \\
$G_{(year-1)}^{topic}$ & 68  & 486 & 849 & 2251 \\ \bottomrule
\end{tabular}}
\end{table}

\begin{figure}[ht]
\centering
  \includegraphics[width=300px]{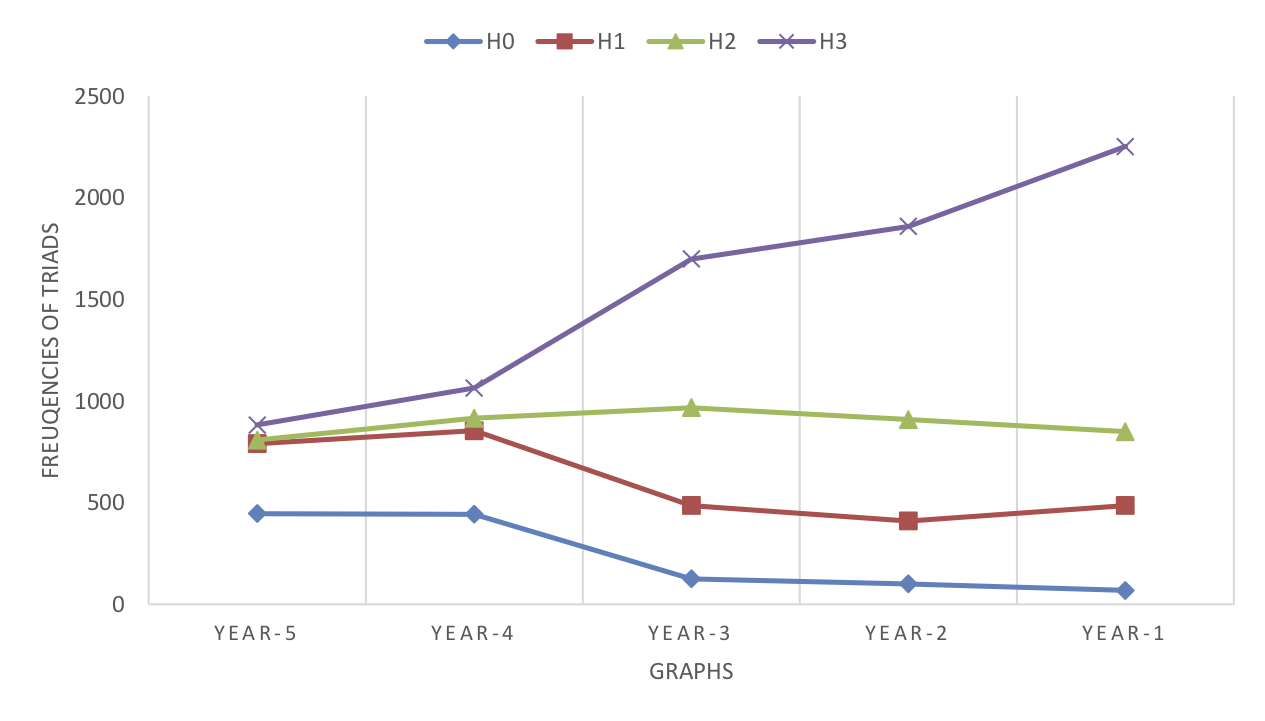}
  \caption{Development in time of the frequencies of $H_i$ in the network related to the emergence of \textit{Artificial Bee Colonies}.}
  \label{fig:beecolonies}
\end{figure}

Next, we check whether the co-occurrence graph was becoming denser by analysing the change of frequencies associated with $H_i$ (see Fig. \ref{fig:beecolonies}). We first calculate the percentage growth of each \textit{$H_i$} (Eq. \ref{eq:growth}) and then compute their weighted summation (Eq. \ref{eq:growthindex}), here labelled \textit{growth index}. We empirically tested other solutions for aggregating the various contributions (e.g., considering only \textit{$H_3$}, summing the values, weighing the sum in a variety of ways) and found that the use of a growth index provides the best discrimination between the two classes of graphs.

\begin{equation}
\label{eq:growth}
\% Growth{H_i} = \frac{{(H_i^{Yr - 1} - H_i^{Yr - 5})*100}}{{H_i^{Yr - 5}}}
\end{equation}
\begin{equation}
\label{eq:growthindex}
GrowthIndex_{topic} = \sum\limits_{i = 0}^3 {i \cdot \% Growth{H_i}}
\end{equation}

Although, we can argue that the number of triangles (\textit{$H_3$}) can by itself be a fair indicator of the density, the \textit{growth index}, showed in Eq. \ref{eq:growthindex}, takes into account all the isomorphism classes. Indeed, previous studies by \citep{faust2010,holland1976} have shown that all four classes of triads are useful for computing useful properties of the network, including transitivity, intransitivity and density.  Considering only $H_3$ might fail to detect some subtler cases, characterised for example by a contemporary increase of $H_2$ and decrease of $H_1$ and $H_0$.

In brief, this approach receives the same input as the previous. For each of the five subgraphs associated to a topic, we perform the triad census obtaining the different frequencies \textit{$H_i$} in different years. We then analyse them diachronically to quantify the increase in density.

\section{Evaluation}\label{evaluation-first-study}
In a recent paper \citep{salatino2016}, we reported the results of a preliminary evaluation that allowed us to fine tune the final approach for measuring the pace of collaboration. In particular, we focused on two well-known topics such as \textit{Semantic Web} and \textit{Cloud Computing} and the results of such approach allowed us to select the best methods for determining the strength of collaboration in Eq. \ref{eq:poc} (by means of harmonic mean) and the pace of collaboration in Eq. \ref{eq:timelineofpoc} (using the least-squared method). 

In this section, instead, we want to focus on the results obtained with the two previously discussed approaches using 75 debutant topics in the treatment set and 100 non-debutant topics in the control set.


\subsection{Measuring the pace of collaboration}\label{clique-based}
\begin{figure}[ht]
\centering
  \includegraphics[width=400px]{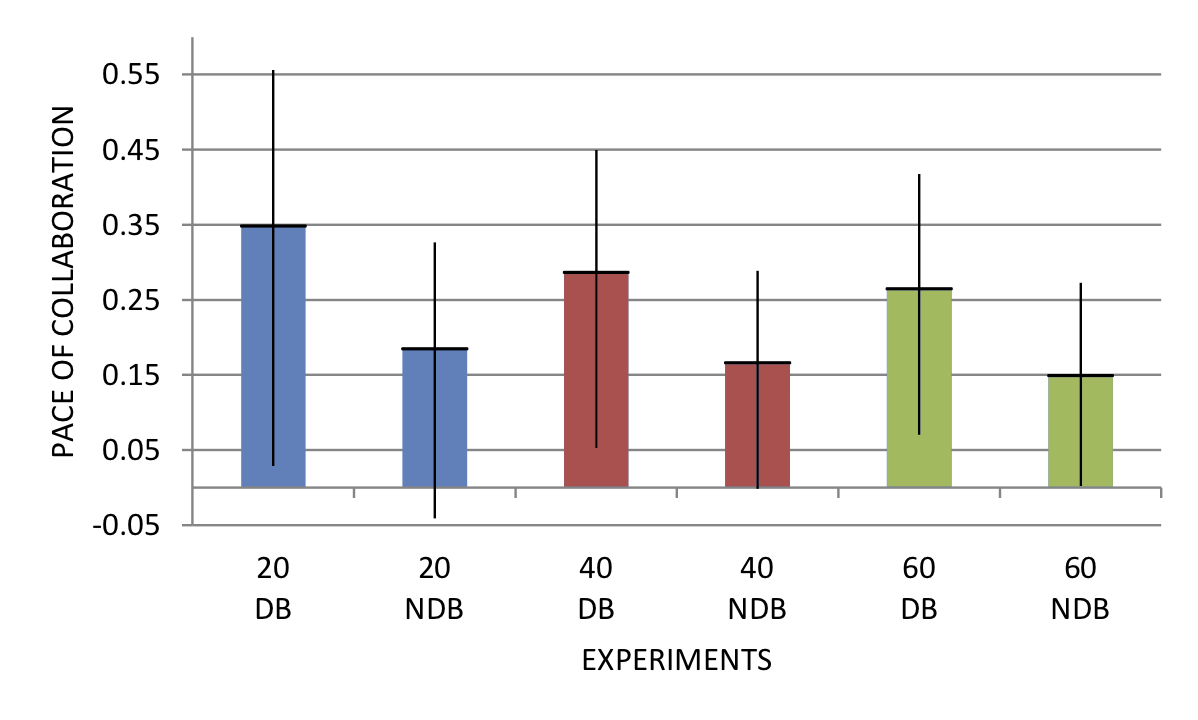}
  \caption{Average collaboration pace of the sub-graphs associated to the treatment (DB) and control set (NDB), when selecting the 20, 40 and 60 most co-occurring topics. The thin vertical lines represent the ranges of values.}
  \label{fig:clique1}
\end{figure}

\index{pace of collaboration}
We applied the cliques-based method, described in Section \ref{sec:analysis-phase-poc}, on the subgraphs associated to topics in the treatment and control sets. Figure \ref{fig:clique1} reports the results obtained by using subgraphs composed by the 20, 40 and 60 topics with the highest co-occurrence. Each bar shows the mean value of the average pace of collaboration for the debutant (DB) and non-debutant (NDB) topics, while the thin vertical lines represent their range of values. The results support the initial hypothesis: the pace of collaboration of the cliques within the portion of network associated with the emergence of new topics is positive and higher than the ones of the control set. Interestingly, the pace of collaboration of the control set is also slightly positive. Further analysis revealed that this behaviour is probably caused by the fact that the topic network becomes denser and noisier in time. 

\begin{figure}[ht]
\begin{subfigure}{.5\linewidth}
\centering
\includegraphics[width=220px]{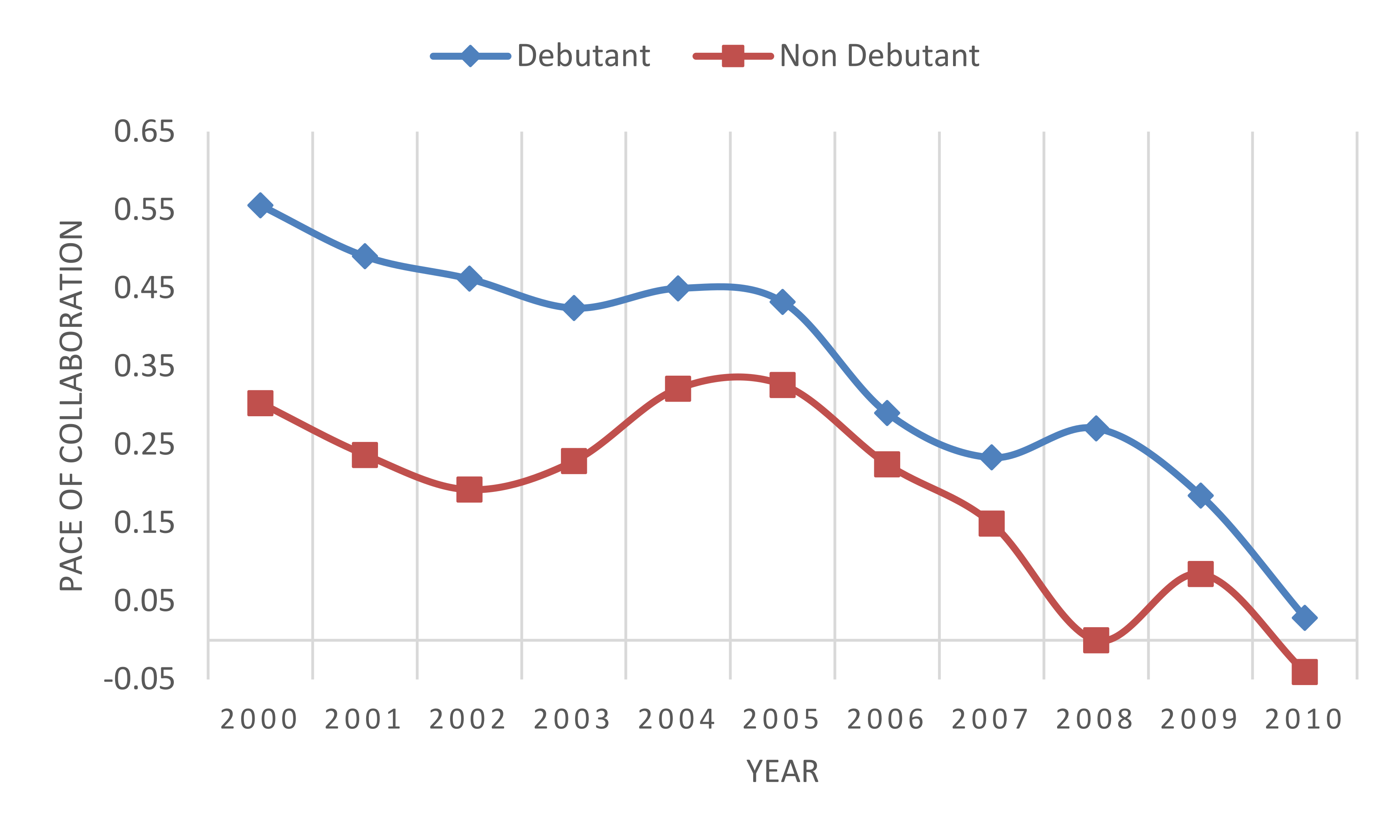}
\caption{}
\label{fig:clique-res-a}
\end{subfigure}%
\begin{subfigure}{.5\linewidth}
\centering
\includegraphics[width=220px]{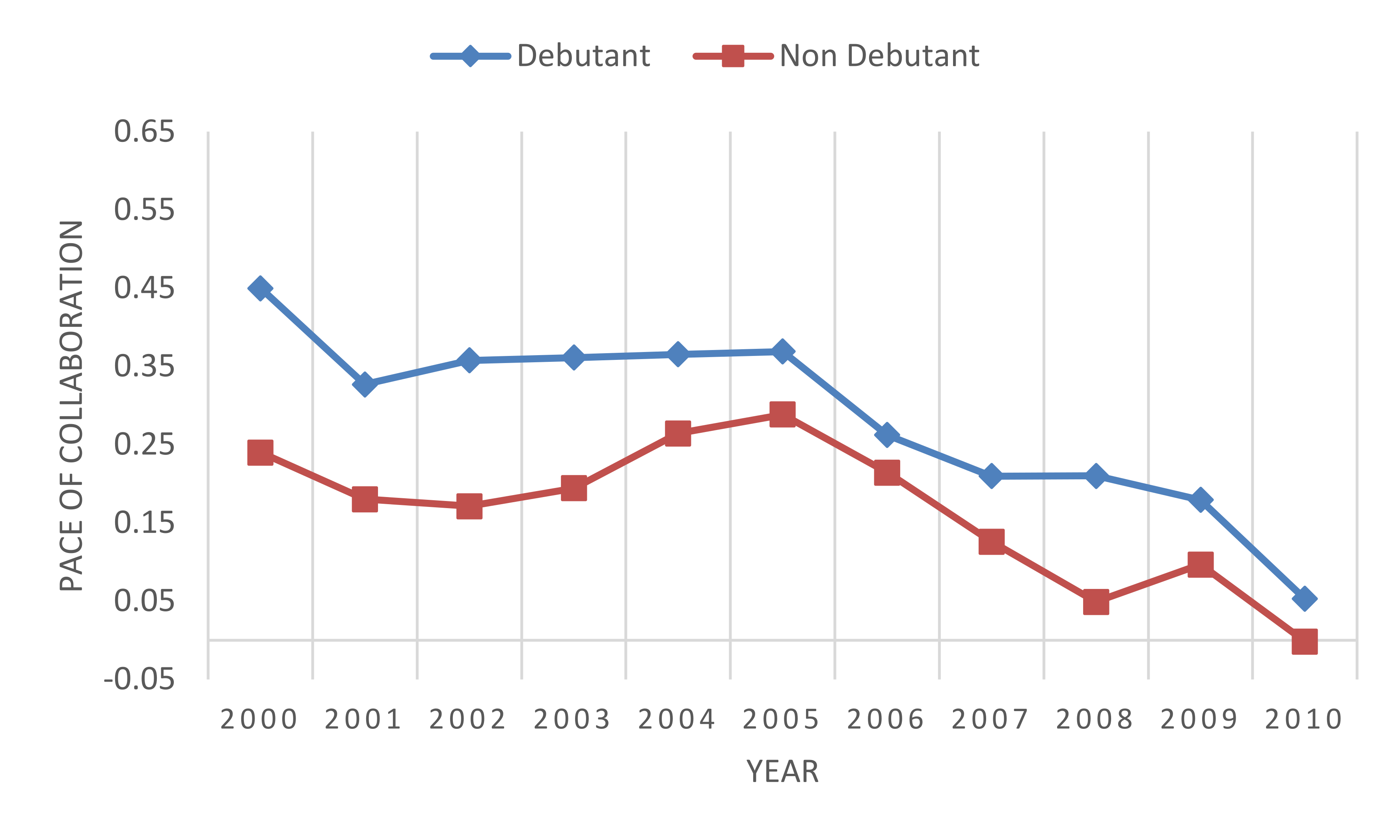}
\caption{}
\label{fig:clique-res-b}
\end{subfigure}\\[1ex]
\centering
\begin{subfigure}{.5\linewidth}
\centering
\includegraphics[width=220px]{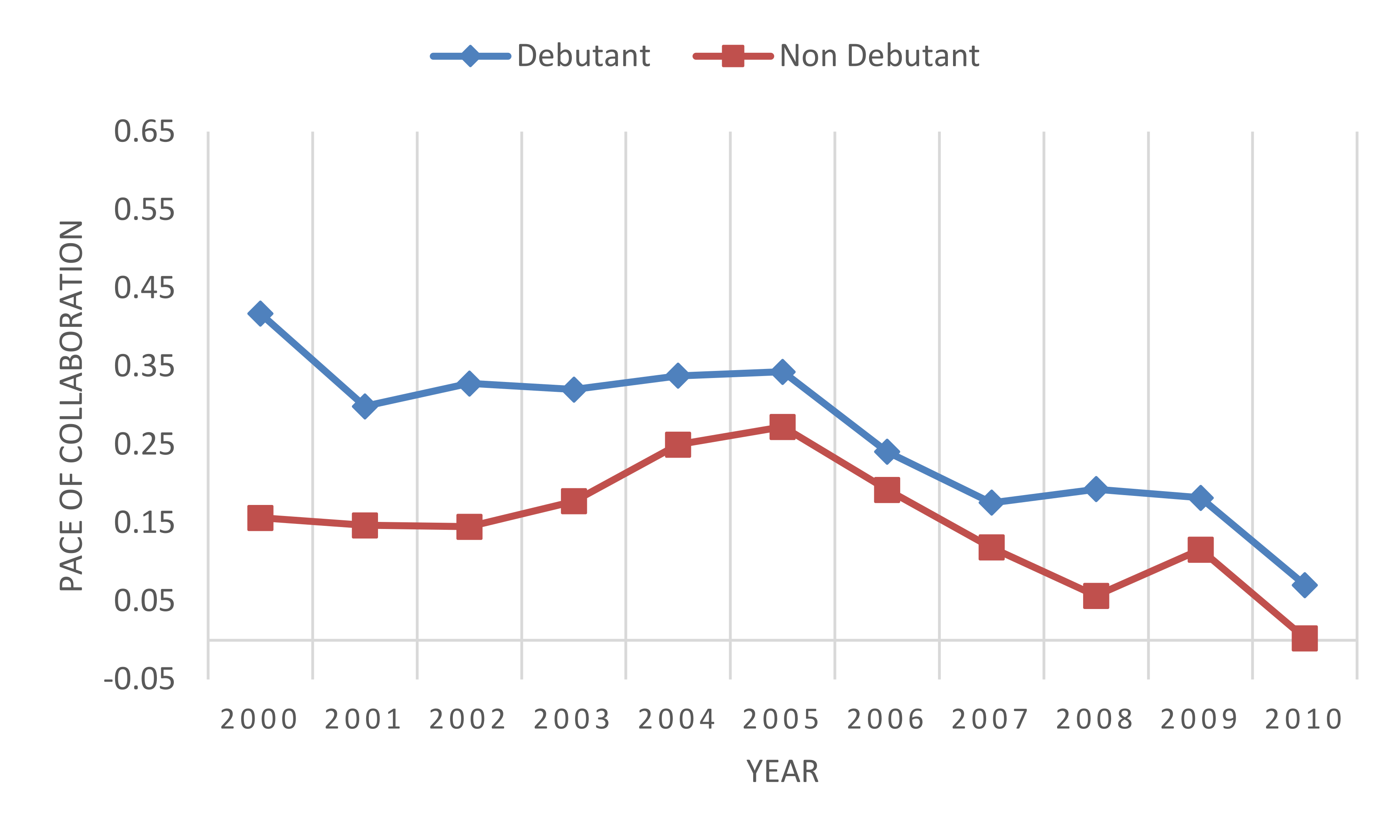}
\caption{}cliques
\label{fig:clique-res-c}
\end{subfigure}
\caption{Average collaboration pace per year of the sub-graphs related to input topics in both debutant and control sets considering their 20 (a), 40 (b) and 60 (c) most co-occurring topics. The year refers to the year of analysis of each topic.}
\label{fig:clique-res}
\end{figure}

Since the pace of collaboration shows significant changes within the period considered, we studied its behaviour across the 2000-2010 interval. Figure \ref{fig:clique-res-a}, \ref{fig:clique-res-b} and \ref{fig:clique-res-c}, show the yearly average of collaboration pace when considering the 20, 40 and 60 most co-occurring topics. In all cases, the collaboration pace for the debutant topics is higher than the one for the control set. We can also notice that in the last five years the overall pace of collaboration suffered a fall for both debutant and non-debutant topics. As outlined above, this may also be due to the fact that the topic network became denser and noisier in the last years of the analysed period. Moreover, the most recent debutant topics often have an underdeveloped network of co-occurrences, which may result in a suboptimal selection of the group of topics to be analysed in the previous years. Therefore, simply selecting the 20 most co-occurring topics may not allow us to highlight the real dynamics preceding the topic creation. 

With the null hypothesis: \enquote{\textit{The differences in the pace of collaboration between the debutant topics and topics in the control set result purely from chance}}, we ran Student\textquotesingle s t-test on the sample of data provided by this approach, to verify whether the two sets belong to different populations. 
The test yielded $p < 0.0001$ which allowed us to reject the null hypothesis\footnote{We consider $p < 0.0001$ as a conventional statistical representation to indicate an extremely high statistical significance ( $>500$ times stronger than the conventional 0.05 threshold for claiming significance). It includes all mathematical outcomes below 0.0001, which are essentially equivalent in assessing excellent significance.}. 
In particular, assuming that there are no differences between the two distributions, then there is less than 0.01 probability that a difference like this (at least as extreme as this) will be observed. 

The results of the Student\textquotesingle s t-test also suggest that the experiment involving the 60 most co-occurring topics, represented in Fig. \ref{fig:clique-res-c}, provides a better discrimination of debutant topics from non-debutant ones. For the sake of completeness, in Table \ref{tab:clique-p-values} we report the p-values yielded by each experiment.

As an example, Table \ref{tab:cliques-poc} compares the collaboration pace of 23 debutant topics with the collaboration pace of the control set in the same year. We can see how the appearance of a good number of well-known topics, which emerged in the last decade, was anticipated by the dynamics of the topic network.

\begin{table}[ht]
\centering
\caption{P-values obtained performing the Student\textquotesingle s t-test over the distributions of both debutant and control sets considering their 20, 40 and 60 most co-occurring topics. The best result is in bold.}
\label{tab:clique-p-values}
{\renewcommand{\arraystretch}{1.2}%
\begin{tabular}{lll}
\hline
\textbf{Experiment}                  & \textbf{p-value}    & \textbf{Associated chart} \\ \hline
20 most co-occurring topics          & $4.22\cdot10^{-2}$           & Fig. \ref{fig:clique-res-a} \\
40 most co-occurring topics          & $6.84\cdot10^{-2}$           & Fig. \ref{fig:clique-res-b} \\
\textbf{60 most co-occurring topics} & \textbf{$4.64\cdot10^{-45}$} & Fig. \ref{fig:clique-res-c} \\ \hline
\end{tabular}}
\end{table}

In conclusion, the results confirm that the portions of the topic network in which a novel topic will eventually appear exhibit a measurable fingerprint, in terms of increased collaboration pace, well before the topic is recognised and labelled by researchers. 

\begin{table}[ht]
\centering
\caption{Collaboration pace of the sub-graphs associated to selected debutant topics versus the average collaboration pace of the control set in the same year of debut.}
\label{tab:cliques-poc}
\begin{tabular}{lC{3cm}C{3cm}}
\hline
\textbf{Topic (year of debut)}          & \textbf{Collaboration Pace} & \textbf{Standard Collaboration pace} \\ \hline
Service discovery (2000)                & 0.455                       & 0.156                                \\
Ontology engineering (2000)             & 0.435                       & 0.156                                \\
Ontology alignment (2005)               & 0.386                       & 0.273                                \\
Service-oriented architecture (2003)    & 0.360                       & 0.177                                \\
Smart power grids (2005)                & 0.358                       & 0.273                                \\
Sentiment analysis (2005)               & 0.349                       & 0.273                                \\
Semantic web services (2003)            & 0.349                       & 0.177                                \\
Linked data (2004)                      & 0.348                       & 0.250                                \\
Semantic web technology (2001)          & 0.343                       & 0.147                                \\
Vehicular ad hoc networks (2004)        & 0.342                       & 0.250                                \\
Mobile ad-hoc networks (2001)           & 0.342                       & 0.147                                \\
P2p network (2002)                      & 0.340                       & 0.145                                \\
Location based services (2001)          & 0.331                       & 0.147                                \\
Service oriented computing (2003)       & 0.331                       & 0.177                                \\
Ambient intelligence (2002)             & 0.289                       & 0.145                                \\
Social tagging (2006)                   & 0.263                       & 0.192                                \\
Community detection (2006)              & 0.243                       & 0.192                                \\
Cloud computing (2006)                  & 0.241                       & 0.192                                \\
User-generated content (2006)           & 0.240                       & 0.192                                \\
Information retrieval technology (2008) & 0.231                       & 0.057                                \\
Web 2.0 (2006)                          & 0.224                       & 0.192                                \\
Ambient assisted living (2006)          & 0.224                       & 0.192                                \\
Internet of things (2009)               & 0.221                       & 0.116                                \\ \hline
\end{tabular}
\end{table}

\subsection{Measuring the network density}\label{triad-based}

We applied the triad-based methods, described in Section \ref{sec:analysis-phase-triad} on the subgraphs composed by the 60 most co-occurring topics, since this configuration provided the best outcomes in previous tests. We performed multiple tests by filtering links associated with less than 3, 10 and 20 co-occurrences, to understand how collaboration strength influences the outcome.

\begin{figure}[ht]
\begin{subfigure}{.5\linewidth}
\centering
\includegraphics[width=220px]{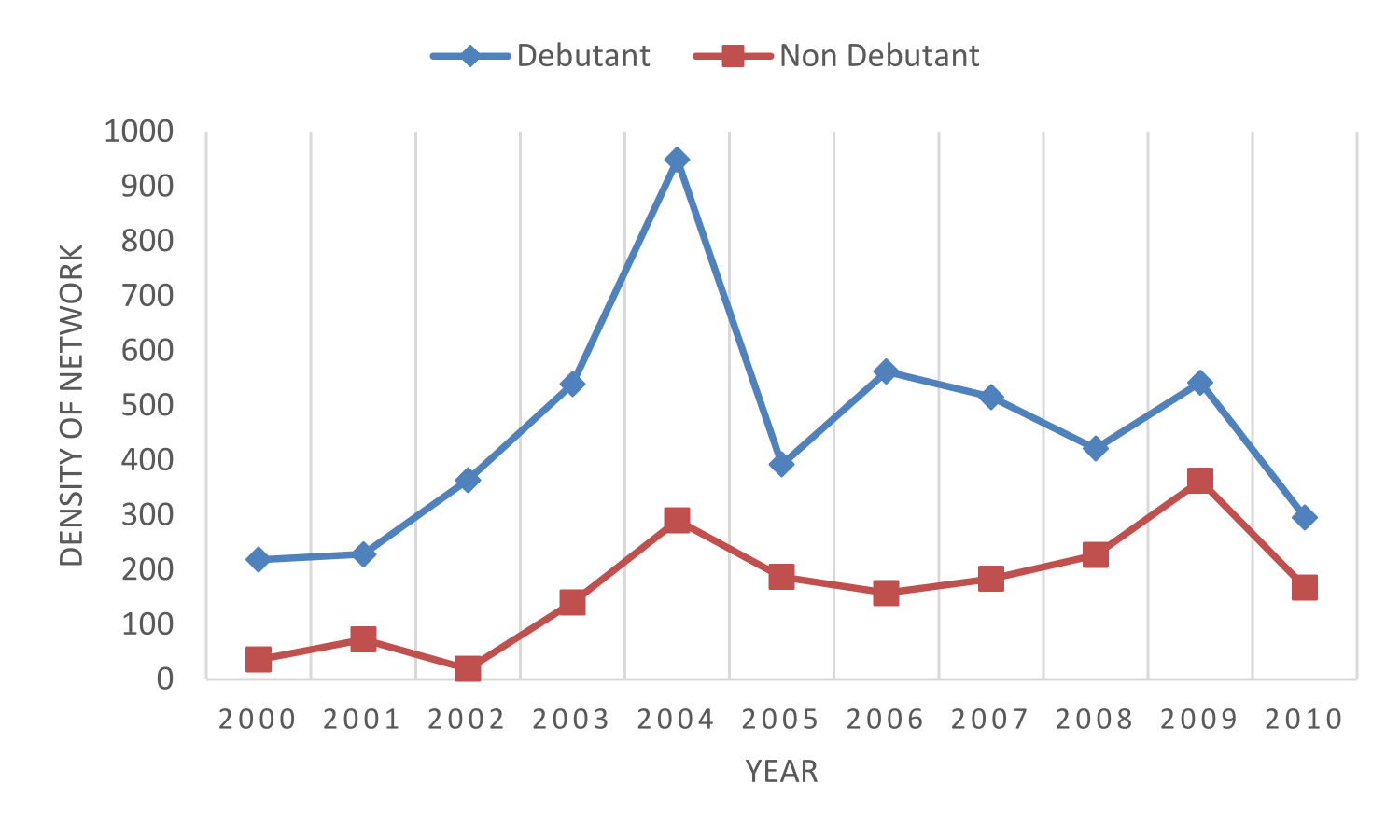}
\caption{}
\label{fig:triad-res-a}
\end{subfigure}%
\begin{subfigure}{.5\linewidth}
\centering
\includegraphics[width=220px]{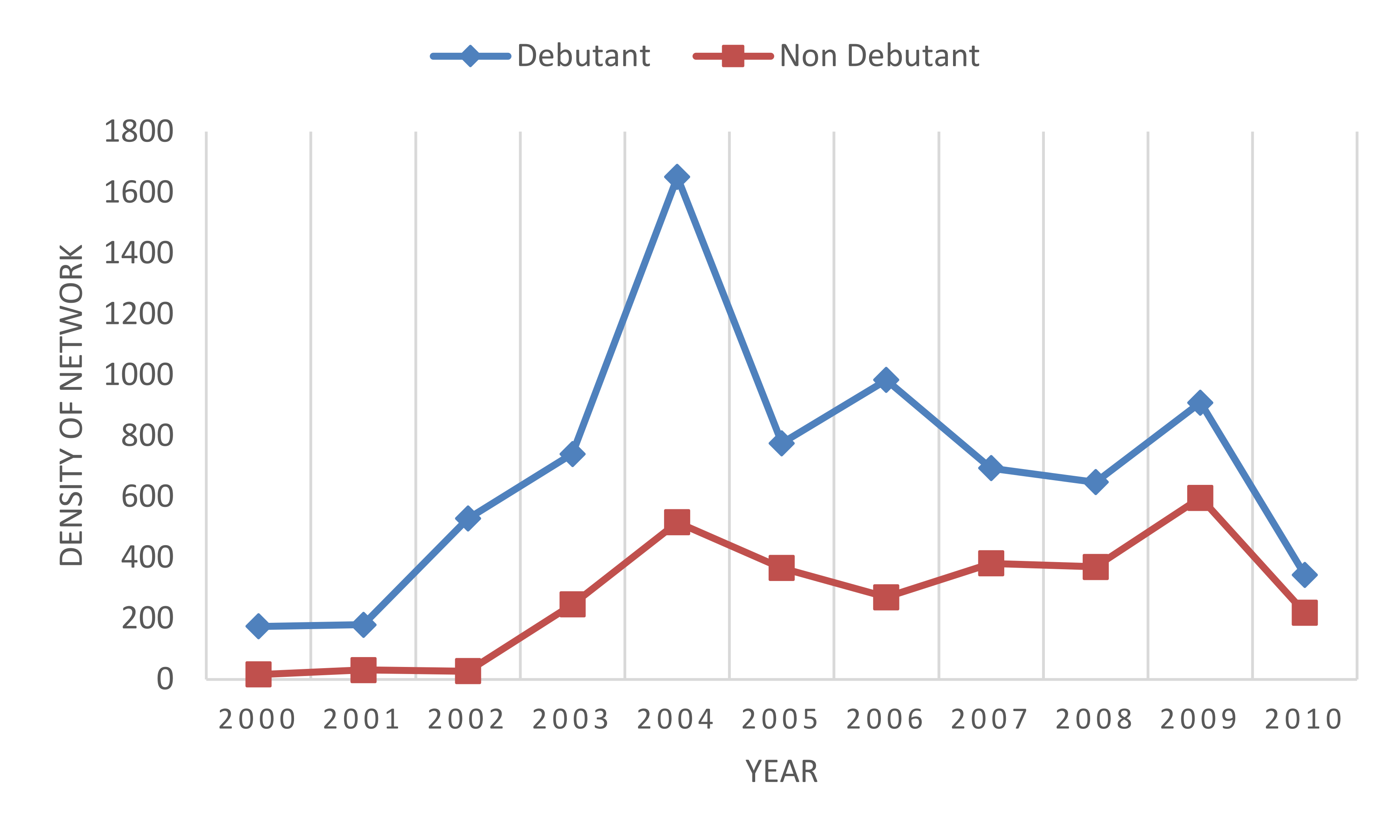}
\caption{}
\label{fig:triad-res-b}
\end{subfigure}\\[1ex]
\centering
\begin{subfigure}{.5\linewidth}
\centering
\includegraphics[width=220px]{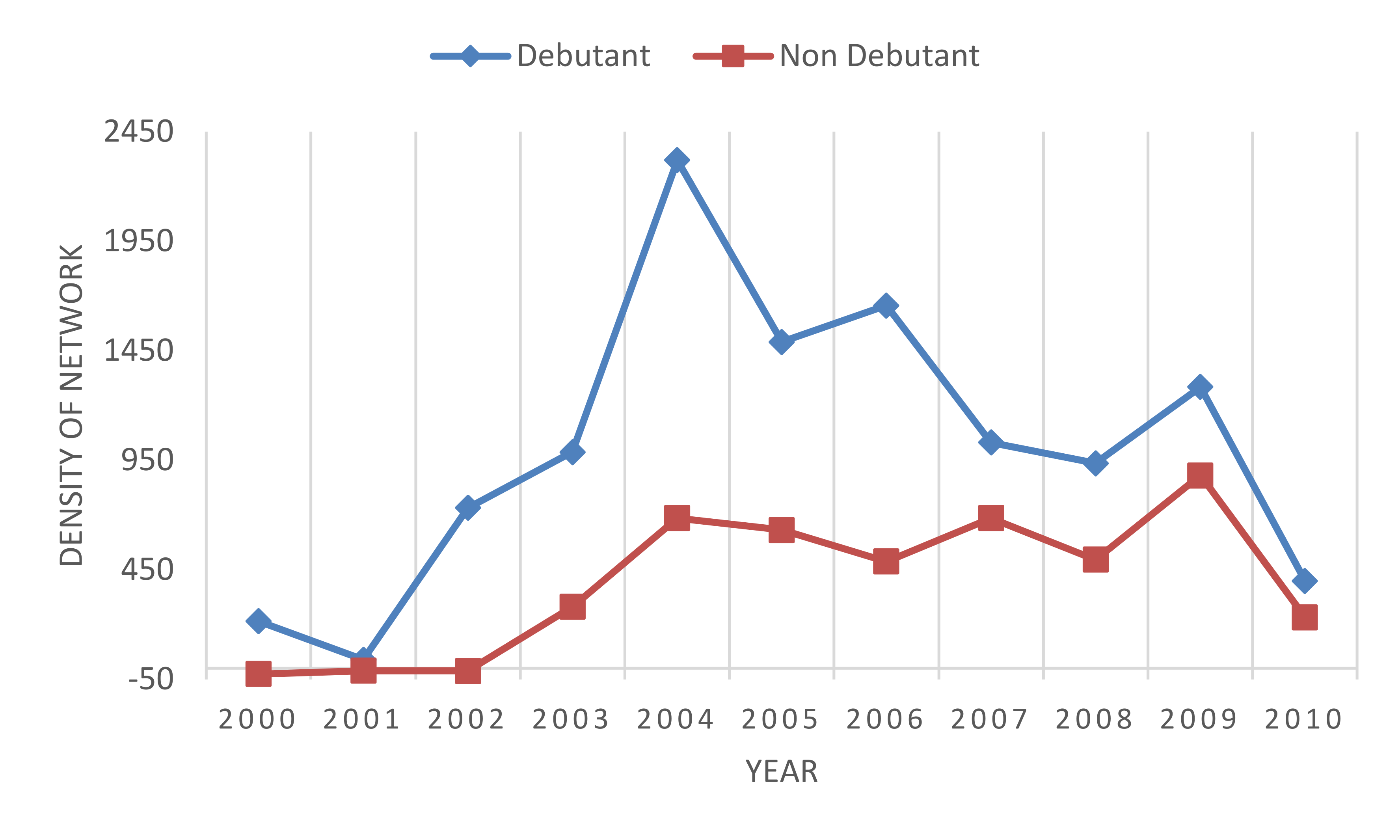}
\caption{}
\label{fig:triad-res-c}
\end{subfigure}
\caption{Average growth index per year of the sub-graphs related to the topics in both debutant and non-debutant groups considering their 60 most co-occurring topics and filtering links having with less than 3 (a), 10 (b) and 20 (c) publications.}
\label{fig:triad-res}
\end{figure}

Figure \ref{fig:triad-res-a} reports the average value of the growth indexes when discarding links with less than 3 co-occurrences. The approach allows us to discriminate well the portion of networks related to debutant topics from the ones related to the control set. In particular, the density of network associated with the debutant topics is always higher than its counterpart. Figure \ref{fig:triad-res-b} and \ref{fig:triad-res-c} report the results obtained by removing links with less than 10 and 20 co-occurrences. 

With the null hypothesis ``\textit{The differences in growth index between the debutant topics and topics in the control set result purely from chance}", we ran Student\textquotesingle s t-test over the two distributions of growth indexes, for all three experiments. 
For each experiment, the test yielded $p < 0.0001$ allowing us to reject the null hypothesis. Assuming that there are no differences between the growth-indexes of both sets of topics, then there is less than a 0.01 probability that a difference like this (at least as extreme as this) will be observed.
More details about the computed p-values per each experiment performed in this triad-based study can be found in Table \ref{tab:triad-p-values}.
 
Similarly to the previous experiment, we used the results of the Student\textquotesingle s t-test to understand among the three tests which one could provide better discrimination between the two classes of topics. In particular, the results in Table \ref{tab:triad-p-values} suggest that the experiment in which we discard links with less than three co-occurrences provides a better discrimination of debutant topics from non-debutant ones. This suggests that considering weak connections is more beneficial for discriminating the two groups. Regarding the peak in the year 2004, we performed a qualitative investigation, which allowed us to determine that its presence is mostly caused by the debut of a number of topics associated with particularly strong underlying dynamics, such as \textit{Linked Data}, \textit{Pairing-based Cryptography}, \textit{Microgrid} and \textit{Privacy Preservation}.


\begin{table}[ht]
\centering
\caption{P-values obtained performing the Student's t-test over the distributions of both debutant and control sets considering their 60 most co-occurring topics filtering links having with less than 3, 10 and 20 publications. The best result is bolded.}
\label{tab:triad-p-values}
{\renewcommand{\arraystretch}{1.2}%
\begin{tabular}{lll}
\hline
\textbf{Experiment}                        & \textbf{p-value}             & \textbf{Associated chart }      \\ \hline
\textbf{less than 3 publications} & \textbf{$6.43\cdot10^{-16}$} & Fig. \ref{fig:triad-res-a} \\
less than 10 publications         & $1.69\cdot10^{-11}$          & Fig. \ref{fig:triad-res-b}          \\
less than 20 publications         & $3.52\cdot10^{-10}$          & Fig. \ref{fig:triad-res-c}         \\ \hline
\end{tabular}}
\end{table}

\begin{table}[!b]
\centering
\caption{Growth indexes of sub-graphs associated to selected debutant topics versus the average growth index of the control set in the same year of debut (Standard Growth Index).}
\label{tab:triads-ghs}
\begin{tabular}{lC{3cm}C{3cm}}
\hline
\textbf{Topic (year of debut)  }                 & \textbf{Growth Index} & \textbf{Standard Growth Index }\\ \hline
Service discovery (2000)                & 290.29       & 35.97                 \\
Ontology engineering (2000)             & 207.22       & 35.97                 \\
Ontology alignment (2005)               & 399.60       & 186.89                \\
Service-oriented architecture (2003)    & 628.07       & 140.17                \\
Smart power grids (2005)                & 637.53       & 186.89                \\
Sentiment analysis (2005)               & 354.10       & 186.89                \\
Semantic web services (2003)            & 439.85       & 140.17                \\
Linked data (2004)                      & 590.81       & 289.94                \\
Semantic web technology (2001)          & 465.53       & 72.71                 \\
Vehicular ad hoc networks (2004)        & 859.44       & 289.94                \\
Mobile ad-hoc networks (2001)           & 87.31        & 72.71                 \\
P2p network (2002)                      & 305.28       & 18.92                 \\
Location based services (2001)          & 595.90       & 72.71                 \\
Service oriented computing (2003)       & 422.92       & 140.17                \\
Ambient intelligence (2002)             & 308.34       & 18.92                 \\
Social tagging (2006)                   & 429.77       & 157.69                \\
Community detection (2006)              & 583.21       & 157.69                \\
Cloud computing (2006)                  & 695.79       & 157.69                \\
User-generated content (2006)           & 485.89       & 157.69                \\
Information retrieval technology (2008) & 552.14       & 227.02                \\
Web 2.0 (2006)                          & 387.42       & 157.69                \\
Ambient assisted living (2006)          & 940.79       & 157.69                \\
Internet of things (2009)               & 580.33       & 167.86                \\ \hline
\end{tabular}
\end{table}

Table \ref{tab:triads-ghs} shows a selection of debutant topics and their growth indexes compared with the growth index of the control set in the same year. If we compare this table to Table \ref{tab:cliques-poc}, we can see that the two methods used in this study reflect the same dynamics.

Hence, the results from this second experiment confirm our initial hypothesis too. In addition, per Table \ref{tab:triad-p-values}, the results from the t-test also suggest that the first experiment, which ignores the links associated with less than 3 publications, better discriminates the two populations.

\section{Examples}
The results of this first study also allowed us to get interesting insights on the creation of some well-known research topics. In particular, analysing the most promising cliques extracted within the selected networks related to the debutant topics we can have a glimpse of the main areas that most certainly contributed to their emergence. In the following, we will focus on \textit{Semantic Web}, \textit{Deep Learning} and \textit{Cloud Computing} as examples.

\subsection{Semantic Web}
Within the network associated to the \textit{Semantic Web}, we can locate the cliques which exhibited the steepest slope, which are listed in Table \ref{tab:semweb}.
We can see that \textit{Semantic Web} was anticipated in the 1996-2001 timeframe by a significant increase in collaboration of the \textit{World Wide Web} area with topics such as \textit{Information Retrieval}, \textit{Artificial Intelligence}, and \textit{Knowledge Based Systems}. This is consistent with the initial vision of the \textit{Semantic Web}, defined in the 2001 by the seminal work of Tim Berners-Lee \citep{berners2001}. However, there are other topics that surely have played a role in this dynamic, such as \textit{Knowledge Representation} and \textit{Search Engines}.

\begin{table}[!ht]
\centering
\caption{Ranking of the cliques with highest slope value for the \textit{Semantic Web}.}
\label{tab:semweb}
{\renewcommand{\arraystretch}{1.2}%
\begin{tabular}{@{}llll@{}}
\toprule
\textbf{Topic 1}        & \textbf{Topic 2 }                & \textbf{Topic 3}                  & \textbf{Slope} \\ \midrule
world wide web & information retrieval & search engines & 2.529 \\
world wide web & user interfaces & artificial intelligence & 1.125 \\
world wide web & knowledge based systems & knowledge representation & 0.982 \\
world wide web & artificial intelligence & knowledge representation & 0.974 \\
world wide web & information retrieval & metadata & 0.897 \\
world wide web & user interfaces & knowledge representation & 0.885 \\
world wide web & knowledge based systems & artificial intelligence & 0.84 \\
world wide web & information retrieval & websites & 0.834 \\
websites & metadata & information services & 0.822 \\
user interfaces & search engines & websites & 0.812 \\ \bottomrule
\end{tabular}}
\end{table}

\begin{figure}[!ht]
\centering
  \includegraphics[width=400px]{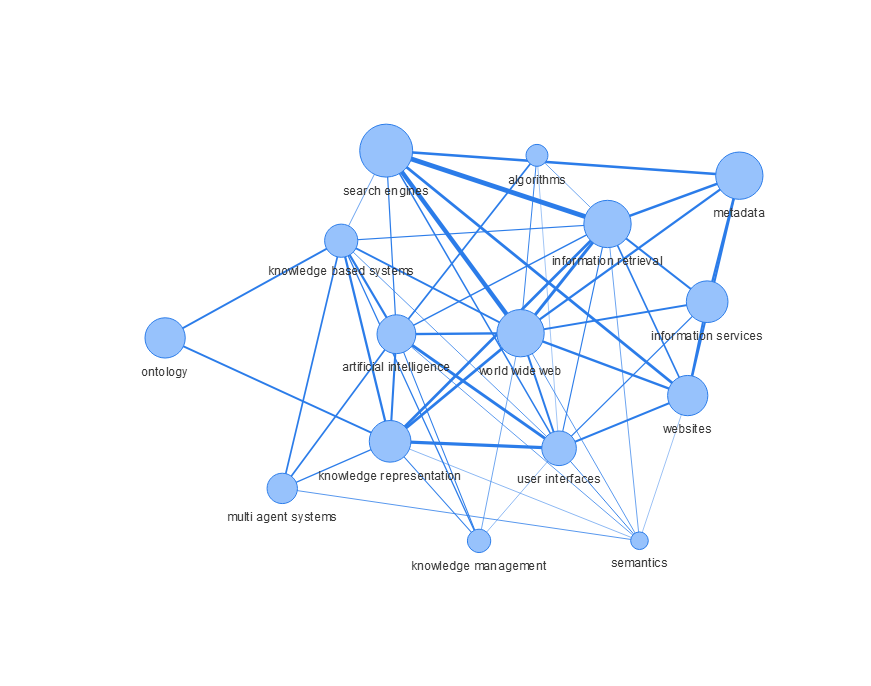}
  \caption{Network suggesting the emergence of the \textit{Semantic Web} in the years prior its emerging. The thickness of an edge reflects the pace of collaboration between the nodes it connects. The size of a node reflects the centrality of the node in this small-world network.}
  \label{fig:semantic-web}
\end{figure}

In Fig. \ref{fig:semantic-web} there is a snapshot of the network that shows the pace of collaboration between the active research areas that contributed to the emergence of the \textit{Semantic Web}.

\subsection{Deep Learning}
\begin{figure}[!htpb]
\centering
  \includegraphics[width=300px]{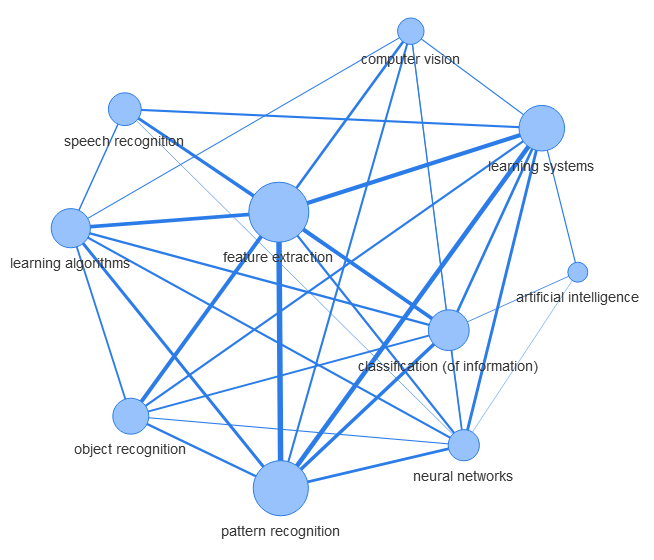}
  \caption{Network envisioning the emergence of the \textit{Deep Learning}. The thickness of an edge reflects the pace of collaboration between the nodes it connects. The size of a node reflects the centrality of the node in this small-world network.}
  \label{fig:deep-learning}
\end{figure}

In Fig. \ref{fig:deep-learning} there is a snapshot of the network that shows the pace of collaboration between the active research areas that contributed to the emergence of \textit{Deep Learning}. In Table \ref{tab:deep-learning} we show the most active collaborations that eventually led to the emergence of this research area. In particular, we can see that the most active areas include \textit{Neural Networks}, \textit{Classification}, \textit{Pattern Recognition}, \textit{Object Recognition} and so on.

\begin{table}[!htpb]
\centering
\caption{Ranking of the cliques with highest slope value for the \textit{Deep Learning}.}
\label{tab:deep-learning}
{\renewcommand{\arraystretch}{1.2}%
\begin{tabular}{ L{4cm} L{4cm} L{4cm} L{1.5cm}}
\toprule
\textbf{Topic 1}        & \textbf{Topic 2 }                & \textbf{Topic 3}                  & \textbf{Slope} \\ \midrule
classification & pattern recognition             & feature extraction              & 1.156 \\
learning systems                & classification & pattern recognition             & 1.043 \\
classification & object recognition              & feature extraction              & 1.038 \\
learning systems                & classification  & feature extraction              & 0.986 \\
learning algorithms             & classification  & feature extraction              & 0.953 \\
learning algorithms             & classification & pattern recognition             & 0.920 \\
learning algorithms             & object recognition              & feature extraction              & 0.879 \\
neural networks                 & learning algorithms             & classification  & 0.843 \\
neural networks                 & learning systems                & classification & 0.819 \\ 
\bottomrule
\end{tabular}}
\end{table}

\newpage
\subsection{Cloud Computing}
\begin{figure}[!htpb]
\centering
  \includegraphics[width=400px]{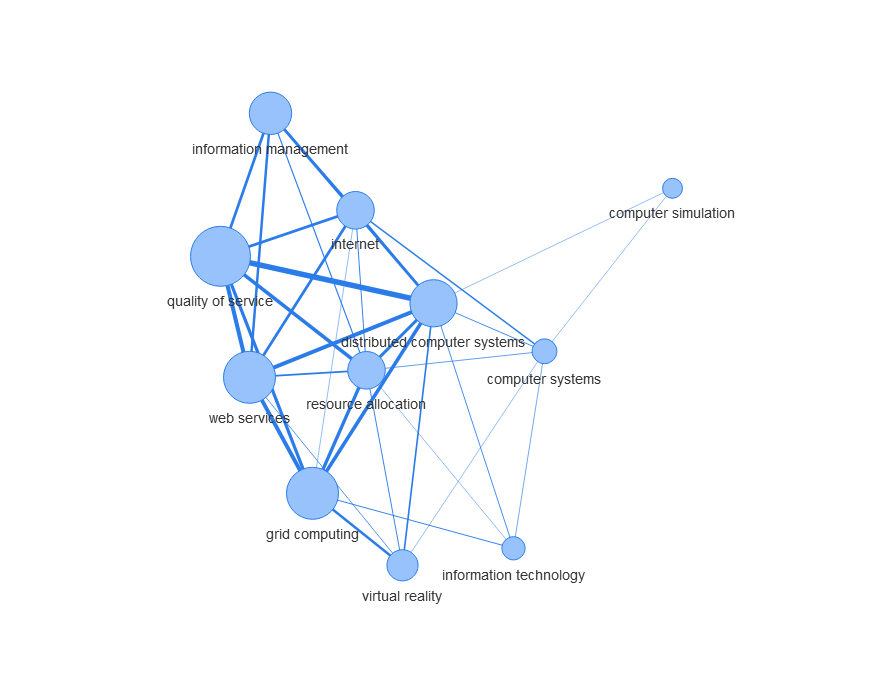}
  \caption{Network envisioning the emergence of the \textit{Cloud Computing}. The thickness of an edge reflects the pace of collaboration between the nodes it connects. The size of a node reflects the centrality of the node in this small-world network.}
  \label{fig:cloud-computing}
\end{figure}

\textit{Cloud Computing}, as we can see from Table \ref{tab:cloudcomp}, was anticipated by an increase in the collaboration between topics such as \textit{Grid Computing}, \textit{Web Services}, \textit{Distributed Computer Systems}, \textit{Information Management}, \textit{Quality of Service} and \textit{Internet}. 
Figure \ref{fig:cloud-computing} shows the network that represents the main interactions between the topics that fostered the emergence of \textit{Cloud Computing}.

\begin{table}[!htpb]
\centering
\caption{Ranking of the cliques with highest slope value for the \textit{Cloud Computing}.}
\label{tab:cloudcomp}
{\renewcommand{\arraystretch}{1.2}%
\begin{tabular}{llll}
\toprule
\textbf{Topic 1}        & \textbf{Topic 2 }                & \textbf{Topic 3}                  & \textbf{Slope} \\ \midrule
grid computing               & distributed computer systems & web services                 & 1.252 \\
quality of service           & distributed computer systems & web services                 & 1.245 \\
grid computing               & distributed computer systems & resource allocation          & 1.131 \\
internet                     & quality of service           & web services                 & 1.087 \\
grid computing               & distributed computer systems & quality of service           & 1.036 \\
grid computing               & quality of service           & resource allocation          & 1.007 \\
web services                 & distributed computer systems & information management & 0.991 \\
internet                     & distributed computer systems & web services                 & 0.886 \\
grid computing               & quality of service           & web services                 & 0.879 \\
quality of service           & resource allocation          & web services                 & 0.812 \\
quality of service           & information management       & web services                 & 0.805 \\

 \bottomrule
\end{tabular}}
\end{table}

\newpage
\section{Discussion and limitations}\label{sec:discussion-first-study}
In this study, we analysed the topic networks with the aim of experimentally confirming our hypothesis that the emergence of new research areas is anticipated by an increased interaction of pre-existing topics. We examined the pace of collaboration (via the cliques-based method described in Section \ref{sec:analysis-phase-poc}) and the change in topology (via the triads-based method described in Section \ref{sec:analysis-phase-triad}) in portions of the network related to debutant topics, showing that it is possible to effectively discriminate areas of the topic graph associated with the future emergence of new topics. The first approach showed that the subgraphs, associated with the emergence of a new topic, exhibit a significantly higher pace of collaboration than the control set of subgraphs associated with established topics. Similarly, the second approach showed that the graphs associated with a new topic display a significantly higher increase in their density than the control set. We can thus confirm that these two dynamics are good indicators for detecting embryonic topics.
 
Interestingly, the ability of these two approaches in discriminating the debutant graph from the control set varies with the time interval considered. Looking at the best results obtained for both experiments, it appears that the cliques-based approach (Fig. \ref{fig:clique-res-c}) works better in the first years of the decade (2000-2004), while the triads-based method (Fig. \ref{fig:triad-res-a}), performs better in the central period (2004-2007). As a possible future work, we can try to better understand whether these behaviours are associated with specific characteristics of the network. In addition, we also intend to explore other dynamics correlated with the emergence of new research topics. Indeed, we believe that pace of collaboration and network density are just two of a variety of dynamics that could be exploited to understand the patterns that precede the emergence of research topics. These dynamics may involve also other dimensions such as authors (e.g., a rise in the number of experts working on a certain combination of topics), venues (e.g., the creation of interdisciplinary workshops), citations and other scholarly entities, including new collaborations between two or more research communities \citep{osborne2014} or a significant change in the vocabulary associated with relevant topics \citep{cano2016}, and so on.

Ultimately, the results of these two experiments allow us to effectively discriminate specific sections of the topic networks and suggest that a significant increase in the rate of collaboration between existing topics provides a strong indicator for predicting the emergence of new research areas. Indeed, these dynamics could be used as the starting point for developing new automatic methods, which could detect the emergence of a new research topic well before this becomes explicitly recognised and established. However, while these results are satisfactory, our analysis presents some limitations, which we shall address in the next study. 

In particular, we identified the relevant subgraph during the selection phase simply by selecting the \textit{n} most co-occurrent topics of the topic under analysis. This solution allows us to compare graphs of the same dimension, however it introduces three issues.

First of all, it assumes that all topics derive from the same number of research areas, which is an obvious simplification. Emerging topics may have a different nature, based on their origin, development patterns, interactions of pioneer researchers, and so on. Therefore, each of them will be linked to a different number of established research areas. A manual analysis of the data suggests that using a constant number of co-occurring topics is one of the reasons why the overall pace of collaboration and growth index associated with the emergent topics are not much higher than the ones of the control set. When selecting too many co-occurring topics, we may include less significant research areas or, alternatively, research areas that started to collaborate with the topic in question only after its emergence. Conversely, when selecting too few topics, the resulting graph may exclude some important ones.

A second limitation is in the selection of the procreators. In this study, for each topic, we selected the most co-occurring topics, from its year of debut to 2014. We can argue that with such approach, we can fetch topics that not necessarily are considered its procreators. Indeed, as we already mentioned, each new emerging topic can have a different development pattern. And, if the debutant topic becomes mature enough to form new intense collaborations with other topics, those topics might also be part of the most co-occurring ones and be wrongly classified as ancestors. In addition, if a debutant topic helps to foster new areas, these new areas can also belong to the most co-occurring topics, since it is likely that they will have an intense collaboration, as part of the nurturing phase. 
To this end, considering the ancestors as only the topics which collaborated the most, can definitely be a limitation. 
A possible solution is to classify the relationship between the emerging topic and each most co-occurring topic by observing their pattern of collaboration, and then select only the topics that have a very closed relationship with the emerging topic in its first period of life. 

Lastly, the selection phase performed in our study cannot be reused in a system capable of automatically detecting embryonic topics, since it requires knowledge of the set of topics with which the embryonic topic will co-occur in the future. However, this could be fixed by developing techniques that are able to select promising subgraphs according to their collaboration pace and density. The application of this technique in a realistic setting would however require a scalable method for identifying promising topic graphs. Indeed, the approach developed in our second study (see next chapter) generates an evolutionary graph in which: i) links are weighted according to the pace of collaboration between the two relevant topics and ii) community detection algorithms are applied to select portions of the network characterised by an intense collaboration between topics. We expect that this solution will be able to detect at a very early stage that ``something'' new is emerging in a certain portion of the topic graph, even if it may not be able to accurately define the topic itself. It would thus allow relevant stakeholders to react very quickly to developments in the research landscape.

\section{Conclusions}
In this study, we hypothesised that research topics go through an embryonic stage, in which, while they are not yet consistently labelled or associated with a considerable number of publications, they can nonetheless be detected by analysing the dynamics between already existing topics. 

To confirm this hypothesis, we performed an experiment on 75 debutant topics in \textit{Computer Science}, which led to the analysis of a topic network including about 2000 topics extracted from a sample of 3 million papers in the 2000-2010 time interval. 

The results confirm that the creation of novel topics is anticipated by a significant increase in the pace of collaboration and density of the portions of the networks in which they will appear. These findings provide contributions of potential value to research in \textit{Philosophy of Science}. 
Firstly, they support our hypothesis that the embryonic stage is part of the topic lifecycle and it can be measured numerically.
Secondly, they bring new empirical evidence to fundamental theories in \textit{Philosophy of Science}, which are concerned with the evolution of scientific disciplines, e.g., \cite{herrera2010}, \cite{kuhn1970}, \cite{nowotny2013}, and \cite{sun2013}. Lastly, they highlight that new topics tend to be born in an environment in which previously less interconnected research areas start to cross-fertilise and generate new ideas. This suggests that interdisciplinarity is one of the most significant forces that drives innovation forward, allowing researchers to integrate a diversity of expertise and perspectives, and yielding new solutions and new scientific visions. Hence, the results of our analysis could be used to support policies that promote interdisciplinary research.
 
The next step is to exploit the dynamics discussed in this study to create a fully automatic approach for detecting embryonic topics. This approach will crawl the different topic networks and reveal areas within them exhibiting such dynamics. The aim is to produce a robust method to be used by researchers and companies alike for gaining a better understanding of where research is heading. In the next chapter, we present our second study that develops such an automatic approach, showing how we propose to solve the scalability issue and how we locate the dynamics within the networks.

\chapter{Second study: early detection of topics}\label{ch:secondstudy}
\section{Introduction}
In the previous chapter, we described the first study aiming to discover some interesting dynamics between established research topics that could be used to forecast the emergence of a new research area. In addition, this study gave us new insights on how science evolves which are consistent with Kuhn's vision of the scientific revolution. We explored these findings to design a second study aiming to effectively detect the emergence of new research topics.

\begin{figure}[ht]
\centering
  \includegraphics[width=400px]{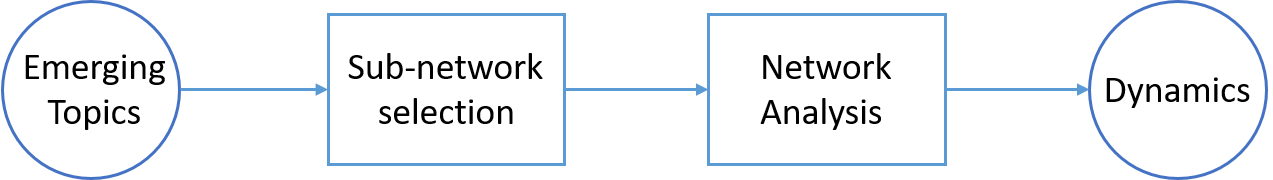}
  \caption{Workflow for the first study in Chapter \ref{ch:firststudy} (forward) and for this study (backward).}
  \label{fig:idea}
\end{figure}

For the first study, we took advantage of some prior knowledge: we knew which research areas emerged and when (year of debut). We then selected some of their co-occurring topics, which could be considered as their ancestors, and performed an analysis to detect the most promising dynamics, i.e., pace of collaboration and network density prior to their emergence, which anticipated the emergence of new research areas. A representation of this pipeline can be found in Fig. \ref{fig:idea}.

The second study, instead, walks this flow backwards. Knowing that certain dynamics exist and can be measured, we aim to effectively detect the emergence of new research areas. In particular, the resulting algorithm crawls the different topic networks and locates clusters (i.e., groups of topics) where there is an overall increase of the pace of collaboration. 


\begin{figure}[ht]
\centering
  \includegraphics[width=400px]{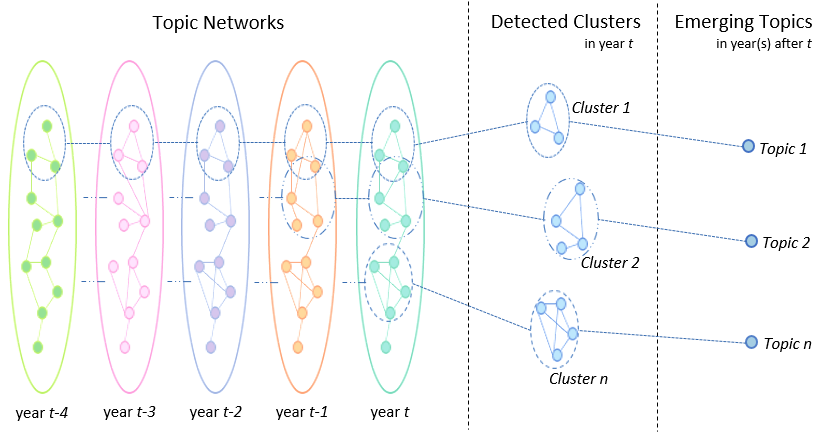}
  \caption{Representation of both input and output of the second study. On the left there are the topic networks that will be processed in search of interesting sub-networks, while on the right there are the matched debutant topics. The dashed circles within the topic networks represent clusters of interest that are analysed over time and can potentially lead to the emergence of new research topics.}
  \label{fig:showconcept}
\end{figure}

However, this approach brings new challenges, such as ensuring its effective computability. In the first study, given a debutant topic, the analysis was located within small portions of the network containing its related topics, in the five years prior its emergence. Similarly, in this second study, we assume that the dynamics involving the development of a new research topic can be observed in the five years prior to its emergence. To this end, the designed algorithm will have to process five complete topic networks at once, each one representing one year. For instance, to discover the emerging topics in 2003, the algorithm will investigate dynamics that developed in the lustrum 1998-2002, as represented in Fig. \ref{fig:showconcept}. However, this poses several challenges because even the processing of a singles topic network, considering its dimension in terms of number of nodes and connections between nodes (see Section \ref{sec:topicnetwork}), is computationally expensive \citep{lancichinetti2009}. Therefore, processing five topics networks at once can easily lead to scalability issues. 
In addition, as \cite{fortunato2010} points out, the analysis of dynamic networks (i.e., topic networks), in which nodes and links appear and disappear over time, is hard and in its infancy.

\begin{figure}[ht]
\centering
  \includegraphics[width=400px]{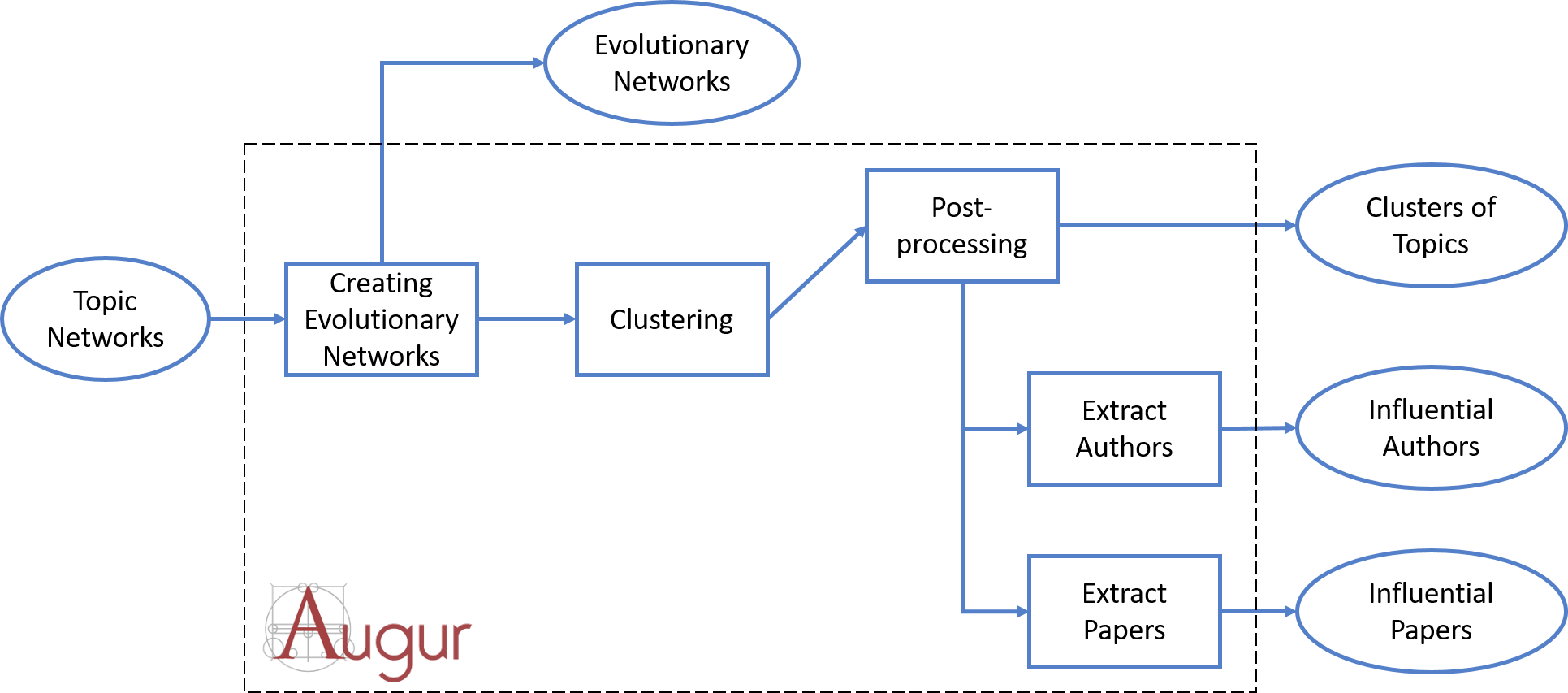}
  \caption{Workflow of Augur.}
  \label{fig:workflow-augur}
\end{figure}

In this chapter, as part of our second study, we propose a way to effectively predict research trends. This approach is called \textit{Augur}, whose workflow is depicted in Fig. \ref{fig:workflow-augur} and consists of three main stages: 

\begin{enumerate}[label={Stage \arabic*.},leftmargin=3cm]
\item \textbf{Creating Evolutionary Networks}. Augur takes in input the topic networks and converts them into evolutionary networks, which are static representations of the dynamicity of the topics networks that compress multiple topic networks into one (see Section \ref{sec:evolutionary-network});
\item \textbf{Clustering}. This involves locating within such evolutionary networks, areas or groups of topics that exhibit an increase of the pace of collaboration (see Section \ref{sec:clustering-acpm});
\item \textbf{Post-Processing}. This stage further enhances the outcome with information regarding active authors and relevant papers. This information can be useful for the user to make sense of the results (see Section \ref{sec:post-processing-acpm}).
\end{enumerate}


In brief, the aim of Augur is to return a set of clusters for a given year that could eventually lead to the emergence of new research areas in the following years. In addition, each cluster will be associated with additional information, such as active authors and relevant papers. This study alongside its results, which will be described and evaluated in the next chapter, have been presented at the 18th ACM/IEEE Joint Conference on Digital Libraries \citep{salatino2018}.

\section{Evolutionary networks}\label{sec:evolutionary-network}\index{evolutionary networks}\index{networks!evolutionary}
The main objective of this study is to provide an algorithm that explores sequential topic networks and provides as output groups of topics exhibiting an increase in the pace of collaboration. A na\"ive solution would start with identifying some interesting clusters (or subnetworks) from the topic network in $year_{t-4}$. This can be simply done either extracting all the 3-cliques, as already performed in the first study, or employing some clustering algorithm that can aggregate a larger set of topics based on their topology. Once all the clusters from that topic network ($year_{t-4}$) have been registered as initial set, the algorithm would proceed iteratively in the following networks ($year_{t-3}$, $year_{t-2}$ and so on). This could be accomplished by matching clusters in different years and modifying the initial set of clusters depending on whether the topics contained in each cluster are increasing their collaboration or not. This approach is depicted in Fig. \ref{fig:showconcept}, where clusters are identified by blue dashed circles and matched with straight dashed lines along time. However, this approach poses several challenges: 
\begin{enumerate*}[label=\roman*)]
\item as already mentioned, processing five topic networks in one iteration is computationally expensive; \item since topics are extremely dynamic and can develop new relationships in time, it is difficult to match the same cluster in two consecutive years --- indeed also \cite{fortunato2010} argues that techniques for such analysis are still premature; and \item the algorithm would need to take into account the possibility that a set of topics start their collaboration in the middle of the time-frame ($year_{t-4}$, $year_{t}$), let's say from the topic network $year_{t-2}$. 
\end{enumerate*}
In this section, we propose the notion of evolutionary network as a solution that can cope with all these challenges.

An evolutionary network is a network that gathers information from different topic networks and uses the pace of collaboration to represent the gradual development of topics and their relationships over time. Positioning the cursor over a particular year, the evolutionary network will provide a view over a set of topics and their interaction in recent years. In the previous study, we observed that five years can be enough to observe the dynamics preceding the emergence of a new area. In this study we make a similar assumption and let the evolutionary network contemplate a similar timespan. For instance, the evolutionary network of the year 2000 will contain a snapshot of the interactions between topics in the years (1996, 2000), while the evolutionary network of 2001 will show the interaction of topics from 1997 to 2001, and so on. In brief, an evolutionary network will provide information on topics and how their collaboration evolved in the last five years and this helps to cope with the aforementioned challenges. In particular, considering that one evolutionary network can represent the interactions of topics in a period of five years, the algorithm can work on one network at once rather than five. Indeed, the aim is to locate clusters of topics within the evolutionary network, where there is an overall increase in the pace of collaboration, rather than in the five topic networks. This approach allows us also to address the second challenge, since the clusters returned from the evolutionary graph will already represent the areas in which there is an increase of the pace of collaboration. Finally, since the evolutionary network takes into account the pace in which the relationships between topics evolved, topics that started a collaboration later in the timespan will also be part of the evolutionary network, so that the algorithm can process also their interaction. Here follows the formal representation of the evolutionary graph and how it is mined from the five topic networks that it includes.

As any other network, an evolutionary network can also be mathematically represented by a graph as showed by Eq. \ref{eq:evolutionarygraph}.

\begin{equation}
G_{yea{r_t}}^{evol}\, = \,(V_{yea{r_t}}^{evol},\,E_{yea{r_t}}^{evol},p_{yea{r_t}}^{evol},w_{yea{r_t}}^{evol}),\,\,\,\,p\,:\,\,V \to \mathbb{R},\,\,w\,:\,\,E \to \mathbb{R}
\label{eq:evolutionarygraph}
\end{equation}

The evolutionary graph $G_{yea{r_t}}^{evol}$ is then a fully weighted graph composed by $V_{yea{r_t}}^{evol}$ which is the set of vertices, $E_{yea{r_t}}^{evol}$ the set of edges that connect those vertices, while $p_{yea{r_t}}^{evol}$ and $w_{yea{r_t}}^{evol}$ are the weights respectively for the sets of vertices and edges. However, since the evolutionary network is generated taking into account five topic networks, there is a function that maps these topic networks (also represented by graphs, $G_{yea{r_t}}^{topic}$, as defined in Section \ref{sec:topicnetwork}) into an evolutionary network, as showed in Eq. \ref{eq:evolutionarynetwork2}.

\begin{equation}
\begin{gathered}
  G_{yea{r_t}}^{evol}\, = \,\,f(G_{yea{r_t}}^{topic},\,G_{yea{r_{t - 1}}}^{topic},\,G_{yea{r_{t - 2}}}^{topic},\,G_{yea{r_{t - 3}}}^{topic},\,G_{yea{r_{t - 4}}}^{topic}) \hfill \\
  V_{yea{r_t}}^{evol}\, = \,unique(V_{yea{r_t}}^{topic}\, \cup \,V_{yea{r_{t - 1}}}^{topic}\, \cup \,V_{yea{r_{t - 2}}}^{topic}\, \cup \,V_{yea{r_{t - 3}}}^{topic}\, \cup \,V_{yea{r_{t - 4}}}^{topic}) \hfill \\
  E_{yea{r_t}}^{evol}\, = \,unique(E_{yea{r_t}}^{topic}\, \cup \,E_{yea{r_{t - 1}}}^{topic}\, \cup \,E_{yea{r_{t - 2}}}^{topic}\, \cup \,E_{yea{r_{t - 3}}}^{topic}\, \cup E_{yea{r_{t - 4}}}^{topic}) \hfill \\ 
\end{gathered} 
\label{eq:evolutionarynetwork2}
\end{equation}

\begin{figure}[ht]
\centering
  \includegraphics[width=400px]{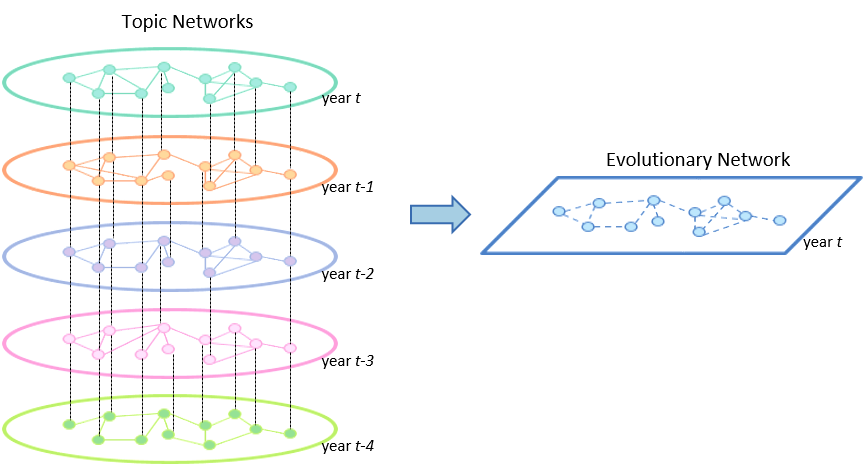}
  \caption{Representation of the evolutionary network.}
  \label{fig:evolutionarynetwork}
\end{figure}

This mapping between the topic networks and the evolutionary networks is also depicted in Fig. \ref{fig:evolutionarynetwork}.

To ensure that all the topics and their interactions are included in the evolutionary network, both the set of vertex $V_{yea{r_t}}^{evol}$ and the set of edges $E_{yea{r_t}}^{evol}$ need to include all nodes and edges of the five topic networks. To this end, the algorithm needs to select the unique set of different topics $V_{yea{r_t}}^{topic}$ within the topic networks, as well as the unique set of all edges $E_{yea{r_t}}^{topic}$ within the different topic networks. These sets will also include nodes that perhaps appeared only in one of the five topic networks and edges that connected two nodes only once out of five. Once the set $V_{yea{r_t}}^{evol}$ and $E_{yea{r_t}}^{evol}$ have been populated, we can proceed in computing the sets $p_{yea{r_t}}^{evol}$ and $w_{yea{r_t}}^{evol}$ which report the different values of the pace of collaboration.

In the first study we defined the pace of collaboration as the rate at which the number of publications two topics share changes in time (see Section \ref{sec:analysis-phase-poc}). More specifically, the pace of collaboration is the slope $\alpha$ of the line $f(x)=\alpha x+\beta$ that best fits the different collaboration indices (${\mu _\Delta }$) along time. This line is computed using the linear least-squares regression. 

To compute the link weights of the evolutionary network we use a similar approach. For each edge, we firstly compute the \textit{strength of collaboration} of all links, according to Eq. \ref{eq:strengthofedges}, and then we compute the \textit{pace of collaboration}, as showed in Eq. \ref{eq:evolutionarynetworkedges}.

In particular, given a link connecting node $u$ and node $v$, with Eq. \ref{eq:strengthofedges} we compute the strength of their collaboration ($\hat w_{u,v}^{topic}$) by firstly normalising the weight of their link ($w_{u,v}^{topic}$) against the number of publications of both nodes ($p_u^{topic}$ and $p_v^{topic}$), and then we compute the harmonic mean of these two values. A similar approach has been presented by Eq. \ref{eq:poc} at page \pageref{eq:poc}.

\begin{equation}
\hat w_{u,v}^{topic}\, = \,HarmonicMean(\frac{{w_{u,v}^{topic}}}{{p_v^{topic}}},\frac{{w_{u,v}^{topic}}}{{p_u^{topic}}})
\label{eq:strengthofedges}
\end{equation}

Next, with Eq. \ref{eq:evolutionarynetworkedges}, we compute the pace of collaboration ($w_{u,{v_{year}}}^{evol}$) as the slope of the line that fits the five measures (strength of collaboration) obtained from the same link in the five topics networks.

\begin{equation}
w_{u,{v_{year}}}^{evol}\, = \,\frac{{\sum\limits_{i = 0}^4 {(yea{r_{t - i}}\, - \,\overline {year} )(\,\hat w_{u,{v_{year - i}}}^{topic}\, - \,\overline {\hat w_{u,v}^{topic}} )} }}{{\sum\limits_{i = 0}^4 {{{(yea{r_{t - i}}\, - \,\overline {year} )}^2}} }}
\label{eq:evolutionarynetworkedges}
\end{equation}

In Eq. \ref{eq:evolutionarynetworkedges}, $\overline {\hat w_{u,v}^{topic}}$ represents the mean value of the five weights, $\overline {year}$ is the mean value of all the years the topics networks refer to $\{ year_{t},year_{t - 1},year_{t - 2},year_{t - 3},year_{t - 4}\}$, and $year_{t - i}$ is the instance value from that set.

In case the instance of an edge is missing in one of the topic networks, perhaps because the two topics did not have any co-occurrence in that year, the strength of collaboration $\hat w_{u,{v_{year - i}}}^{topic}$ for that year will be zero. 

For the sake of completeness, the evolutionary network includes the set of node weights $p_{yea{r_t}}^{evol}$. These weights report the pace of growth of each node computed according to Eq. \ref{eq:evolutionarynetworknodes}. Also in this case we use the least-squared method. In particular, the weight of a given \textit{k-th} vertex ($p_{{k_{year}}}^{evol}$) is the slope of the line that best fits the weights of the same \textit{k-th} vertex in the different topic networks ($p_{{k_{year - i}}}^{topic}$).

\begin{equation}
p_{{k_{year}}}^{evol}\, = \,\frac{{\sum\limits_{i = 0}^4 {(yea{r_{t - i}}\, - \,\overline {year} )(\,p_{{k_{year - i}}}^{topic}\, - \,\overline {p_k^{topic}} )} }}{{\sum\limits_{i = 0}^4 {{{(yea{r_{t - i}}\, - \,\overline {year} )}^2}} }}
\label{eq:evolutionarynetworknodes}
\end{equation}

In such equation, $p_{{k_{year - i}}}^{topic}$ is the number of publications the k-th topic received in $year_{t - i}$, while $\overline {p_k^{topic}}$ is the mean value of all the weights of the same node in the different topic networks or the average value of publications received in that period of five years. 
In case the instance of a vertex is missing in a topic network ($year_{t - i}$), implying that for such topic there are zero publications in that year, the weight $p_{{k_{year - i}}}^{topic}$ is zero. 

In brief, the evolutionary network $G_{yea{r_t}}^{evol}$ can be seen as a static representation or a snapshot of all the dynamics between topics in a period of five years. The network contains all the topics within the topic networks and reports whether or not these topics had a collaboration in this period. In addition, the node weight represents the increase in number of publications on a topic and the link weight the pace of collaboration between two topics.

\section{Cluster of established topics}\label{sec:clustering-acpm}
The creation of evolutionary networks is the first phase of Augur. With it, we move from the domain of collaboration between topics to the domain of dynamics, so that we can focus our attention on detecting areas where there is this intense collaboration between topics.
In this step, Augur clusters nodes that exhibit a strong collaboration pace.

In Section \mbox{\ref{sec:networkapproach}} we described the concept of community and how the underlying community structure of a network can be detected. To this end, we performed some experiments with several community detection algorithms available in literature (see evaluation in, Chapter \mbox{\ref{ch:finalevaluation}}), such as Fast Greedy, Clique Percolation Method and others. However, these algorithms failed to provide a good characterisation of the community structure within the evolutionary networks. Nonetheless, they allowed us to learn some interesting peculiarities of such networks. Indeed, this understanding allowed us to design the Advanced Clique Percolation Method, which extends the Clique Percolation Method \mbox{\citep{palla2005}}, and works much better on this task. Before proceeding in its description (Section \mbox{\ref{sec:acpm}}), we will briefly describe the standard Clique Percolation Method.

\subsection{Clique percolation method}\label{sec:cpm}\index{clique percolation method}

The Clique Percolation Method (CPM) is a popular approach for analysing the community structure of networks and belongs to the family of fuzzy community detection algorithms. This approach has been enormously successful in many applications in many domains, such as the examination of social networks \citep{palla2007,gonzalez2007}, the analysis of cancer-related proteins in protein interaction networks \citep{jonsson2006a,jonsson2006b}, document clustering \citep{gao2006} and so on. 

CPM differs from other methods of clustering particularly on its definition of community. Indeed, the community (or \textit{component}, \textit{k-clique community} or \textit{k-clique percolation cluster}) is defined as the union of all k-cliques that are connected and can be reached from each other across a set of adjacent \textit{k-cliques} \citep{everett1998,batagelj2003,derenyi2005}. \index{k-clique}\index{clique}
A \textit{k-clique} is a complete (i.e., fully connected\footnote{k-cliques have been part also of the first study, in Chapter \ref{ch:firststudy}. Indeed, an example of 3-clique is in Fig. \ref{fig:clique} at page \pageref{fig:clique}.}) subgraph of \textit{k} vertices and the adjacency between two k-cliques is determined on whether they share \textit{k-1} nodes (hence, if they differ for only one node). The overall idea of community for the CPM is that it consists of several complete subgraphs (\textit{k-cliques}) that tend to share many of their nodes.
In particular, the Clique Percolation Method consists of two main steps:
\begin{enumerate}[label={Step \arabic*.},leftmargin=3cm]\label{item:cpm}
\item Locating \textit{k-cliques} within the network
\item Creating the clique graph (\textit{k-clique adjacency graph}), in which the vertices represents the \textit{k-cliques} of the original graph, and there is an edge between two vertices when the two \textit{k-cliques} are adjacent
\end{enumerate}

\begin{figure}[ht]
\centering
  \includegraphics[width=250px]{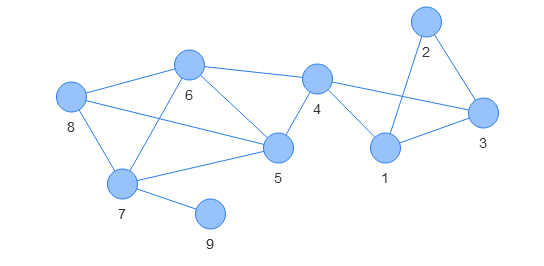}
  \caption{Example of network. Our goal is to identify all cliques and create the clique graph.}
  \label{fig:testgraph}
\end{figure}

\begin{figure}[!ht]
\centering
  \includegraphics[width=250px]{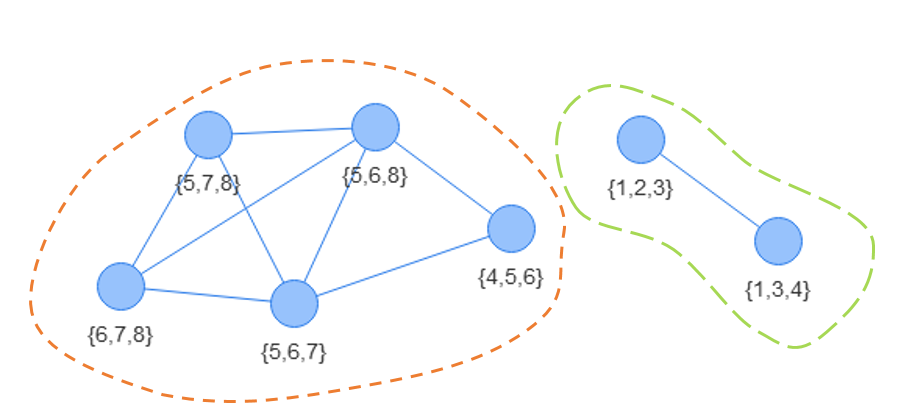}
  \caption{Clique graph or \textit{k-clique adjacency graph}. The two connected components of this network map into the communities of the original network.}
  \label{fig:cliquegraphs}
\end{figure}

\begin{figure}[!ht]
\centering
  \includegraphics[width=250px]{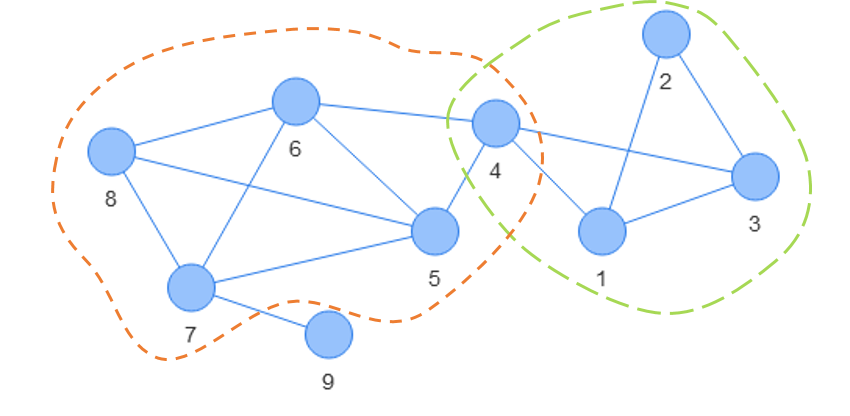}
  \caption{The two communities shown in Fig. \ref{fig:cliquegraphs} are now mapped on the original network inFig. \ref{fig:testgraph}.}
  \label{fig:communitygraph}
\end{figure}

Communities are then defined as the connected components of such graph. For instance, given the network depicted in Fig. \ref{fig:testgraph}, with $k = 3$ the algorithm will locate the 3-cliques $\{1,2,3\}$, $\{1,3,4\}$, $\{4,5,6\}$, $\{5,6,7\}$, $\{5,6,8\}$, $\{5,7,8\}$ and $\{6,7,8\}$. Then the algorithm creates the clique graph, using the clique adjacency matrix (Table \ref{tab:adjacencymatrix}), in which the cliques are connected between themselves when they share two (\textit{k-1}) nodes \citep{everett1998}. The resulting graph of \textit{k-clique adjacency graph} in Fig. \ref{fig:cliquegraphs}, contains two connected components orange and green, which are the two final communities of the network, as reported in Fig. \ref{fig:communitygraph}. In Code \ref{alg:cpm} there is the pseudocode for the Clique Percolation Method.

\begin{table}[ht]
\centering
\caption{The overlap matrix. This matrix is viewed as an adjacency matrix whose nodes are cliques of the original network. The elements of the matrix indicate whether pairs of cliques are adjacent (share 2 nodes) or not in the graph.}
\label{tab:adjacencymatrix}
{\renewcommand{\arraystretch}{1.5}%
\begin{tabular}{c|c|c|c|c|c|c|c|}
\cline{2-8}
                                         & \textbf{\{1,2,3\}} & \textbf{\{1,3,4\}} & \textbf{\{4,5,6\}} & \textbf{\{5,6,7\}} & \textbf{\{5,6,8\}} & \textbf{\{5,7,8\}} & \textbf{\{6,7,8\}} \\ \hline
\multicolumn{1}{|l|}{\textbf{\{1,2,3\}}} &                    & 1                  &                    &                    &                    &                    &                    \\ \hline
\multicolumn{1}{|l|}{\textbf{\{1,3,4\}}} & 1                  &                    &                    &                    &                    &                    &                    \\ \hline
\multicolumn{1}{|l|}{\textbf{\{4,5,6\}}} &                    &                    &                    & 1                  & 1                  &                    &                    \\ \hline
\multicolumn{1}{|l|}{\textbf{\{5,6,7\}}} &                    &                    & 1                  &                    & 1                  & 1                  & 1                  \\ \hline
\multicolumn{1}{|l|}{\textbf{\{5,6,8\}}} &                    &                    & 1                  & 1                  &                    & 1                  & 1                  \\ \hline
\multicolumn{1}{|l|}{\textbf{\{5,7,8\}}} &                    &                    &                    & 1                  & 1                  &                    & 1                  \\ \hline
\multicolumn{1}{|l|}{\textbf{\{6,7,8\}}} &                    &                    &                    & 1                  & 1                  & 1                  &                    \\ \hline
\end{tabular}}
\end{table}

\begin{algorithm}[H]
\label{alg:cpm}
\setstretch{1}
\SetKwInOut{Input}{Input}\SetKwInOut{Output}{Output}

\Input{Evolutionary Network $network$}
\Output{Communities $comms$}
\BlankLine

cliques$\leftarrow$ FindCliques(network, k=3)\;
adj.mat$\leftarrow$ matrix(length(cliques),init=0) \tcc*[r]{adjacency matrix}
\For{$i\leftarrow 1$ \KwTo $length(cliques)$}{
\For{$j\leftarrow 1$ \KwTo $length(cliques)$}{
	\If(\tcp*[f]{k-1}){SharedNodes(cliques[i],cliques[j]) == 2}{
	adj.mat[i,j]$\leftarrow$ 1 \tcc*[r]{adjacent cliques}
	}
}
}

clique.graph$\leftarrow$ GraphFromAdjacencyMatrix(adj.mat)\;
components$\leftarrow$ FindConnectedComponents(clique.graph)\;
comms$\leftarrow$ ConvertComponents(components)\tcc*[r]{converts components of cliques in components of topics: communities}

return(comms)\;

\caption{Clique Percolation Method.}
\end{algorithm}

\vspace{1cm}
As we will see in the evaluation chapter (Chapter \ref{ch:finalevaluation}) the Clique Percolation Method is not able to characterise the community structure of the evolutionary networks. In particular, there are two main limitations:
\begin{enumerate}[label={Limitation \arabic*.},leftmargin=3cm]
\item This method does not take into account the link weights, treating them as either they exist or not;
\item Some areas of the evolutionary networks appear to be particularly dense and the algorithm is unable to split such areas in further communities.
\end{enumerate}
The first one can be considered as a major weakness, as each link in the evolutionary network represents a different pace of collaboration between the two topics it connects. In the next section we will describe the Advanced Clique Percolation Method, which we developed to overcome these limitations.


\subsection{Advanced Clique Percolation Method}\label{sec:acpm}\index{clique percolation method!advanced}\index{advanced clique percolation method}
The \textit{Advanced Clique Percolation Method} (ACPM) is a solution that aims to tackle the limitations of the algorithm described in the previous section. In the first instance, ACPM takes into account the link weights in a refined way, so that also the process of creating the clique graph and locating communities considers such weights. Secondly, this new algorithm redefines the concept of community, to handle the presence of very large communities within the network.

The Advanced Clique Percolation Method consists of four steps:
\begin{enumerate}[label={Step \arabic*.},leftmargin=3cm]
\item Locating \textit{k-cliques} within the network
\item Measuring pace of collaboration per \textit{k-cliques} and filtering noise
\item Creating the \textit{k-clique adjacency graph}
\item Locating local maxima and extracting neighbourhoods
\end{enumerate}
In the following subsections, we will describe ACPM step-by-step and we will show how we integrated the link weights in Step 2 and how we redefined the concept of community in Step 4. In Code \ref{alg:acpm} is reported the pseudocode for the Advanced Clique Percolation Method.

\subsubsection{Step 1: Locating k-cliques within the network}

The first step is similar to the original method as the approach locates all the 3-cliques\index{cliques} within the evolutionary network. At this stage, the algorithm explores the topology of the network and gathers together nodes that belong to triangles. In Code \ref{alg:acpm} this step is accomplished with the first line of code.

\subsubsection{Step 2: Measuring pace of collaboration per k-cliques and filtering noise}

The standard clique percolation method works on binary networks \index{binary network}\index{networks!binary}(i.e., undirected networks with links without weights) and an arbitrary network can always be converted into a binary network, simply by inducing the graph that contains only those links whose weight is higher than the weight threshold \textit{w}. Indeed, the standard method uses the link weights only to filter them according to their value, so that it can create the binary network upon which locating the 3-cliques. However, we can argue that filtering the network using a static threshold is not an optimal solution. Considering that we want to promote 3-cliques that show intense activity of collaboration (high pace of collaboration), a link having weight below \textit{w} can be part of them and thus be still meaningful.

The same research team that developed the Clique Percolation Method, proposed a new approach that shifts from filtering links to filtering \textit{k-cliques} based on their intensity \citep{farkas2007}. After finding cliques, the approach computes the intensity of a clique as the geometric mean of the weights of its links. Then the community or weighted k-clique percolation cluster is considered as the set of adjacent k-cliques having an intensity greater of an intensity threshold \textit{I} \citep{onnela2005}. In this way, a k-clique containing weak links (low weights) can be still part of the percolation cluster as long as it contains edges with large weights that help it to exceed the threshold \textit{I}.

The Advanced Clique Percolation Method will then use a similar approach and will compute the intensity of 3-cliques after propagating all the links of the evolutionary network. However, rather than computing the geometric mean as done in \citep{onnela2005}, the algorithm will compute the harmonic mean as already done for the first study in Chapter \ref{ch:firststudy}, so that the intensity matches with the pace of collaboration of the 3-clique. 

To this end, for each retrieved clique, the algorithm computes its intensity or pace of collaboration\index{pace of collaboration} as the harmonic mean of the link weights ($W_{ab}$, $W_{bc}$, $W_{ca}$, please refer to Fig. \ref{fig:clique} at page \pageref{fig:clique}), as showed in Eq. \ref{eq:paceofcollclique}. 

\begin{equation}
PaceOfCollaboration(clique)\, = \,\frac{3}{{{W_{ab}}^{ - 1} + {W_{bc}}^{ - 1} + {W_{ca}}^{ - 1}}}
\label{eq:paceofcollclique}
\end{equation}
Next, the algorithm filters some k-cliques that have very low value of pace of collaboration.
In Code \ref{alg:acpm} this step is accomplished from line 2 to 6.

\subsubsection{Step 3: Creating the k-clique adjacency graph}
ACPM creates the k-clique adjacency graph $G = {\text{ }}(V,E,W)$. $V$ is the set of vertices representing the identified 3-cliques from the original graph, $E$ is the set of links connecting adjacent k-cliques that share \textit{k-1} vertices, and $W$ is the set containing the node weights (i.e, pace of collaboration of each clique computed using Eq. \ref{eq:paceofcollclique}). This step is very similar to Step 2 of the standard clique percolation method (see page \pageref{item:cpm}), with the only difference that ACPM propagates the link weights throughout the different steps.
In Code \ref{alg:acpm} this step is accomplished from line 7 to 19.

\subsubsection{Step 4: Locating local maxima and extracting neighbourhoods}

In this final step, the algorithm identifies the communities within the evolutionary network. The standard clique percolation method defines communities as connected components of the \textit{k-clique adjacency graph}, and as we will see in the evaluation chapter (see Chapter \ref{ch:finalevaluation}), it returns very large communities. With ACPM we want to address this limitation by redefining the concept of community.

When using the standard method, in the literature we can find some solutions that can help to cope with the presence of very large communities, such as those from \cite{palla2005} and \cite{farkas2007}. However they appear to be unsuitable in the context of this work.

In particular, \cite{palla2005} suggest to monitor how communities change, by trying different values of link weights \textit{w} and tuning the value \textit{k} for the dimension of cliques. Such analysis produces a similar effect to changing the resolution in a microscope. Indeed, increasing the threshold \textit{w} leads the communities to shrink and fall apart, as fewer cliques will be formed. Whereas, increasing the dimension of cliques \textit{k} makes the communities smaller, more cohesive and more fragmented. However, changing the values of \textit{w} and \textit{k} is not an optimal solution for two reasons.

Firstly because, as already mentioned in Step 2, we want to promote 3-cliques that, while having some links with weight below \textit{w}, still show a high pace of collaboration. Therefore we cannot increase too much the value of \textit{w}. 

Secondly, if we increase the value of \textit{k}, communities become more granular, and this prevents smaller cliques from joining a community. By definition, a component (or \textit{k-clique community}) contains all adjacent k-cliques and does not include cliques with smaller values of \textit{k}. For instance, in the network in Fig. \ref{fig:testgraph}, the node 9 belongs to the 2-clique $\{7,9\}$ but does not belong to any community simply because it is not part of any 3-cliques. In this way, increasing \textit{k} to 4 would mean not including all 3-cliques that do not belong to 4-cliques. Since we acknowledged from the first study (see Chapter \ref{ch:firststudy}) that all 3-cliques are important for the analysis of the dynamics of each triad of topics, we cannot raise the number of \textit{k} to 4. 

In Step 2, we computed the clique intensity (with Eq. \ref{eq:paceofcollclique}). In the case of very large weighted communities, \cite{farkas2007} suggest to increase the threshold \textit{I} and filter more 3-cliques based on their intensity. Indeed, a low intensity threshold allows a large number of k-cliques to participate in clusters, while increasing this threshold prevents the emergence of large modules that could mask the details of smaller communities. However, changing the values of \textit{I} is not an optimal solution for the following reason.

\begin{figure}[ht]
\begin{subfigure}{.33\linewidth}
\centering
\includegraphics[width=150px]{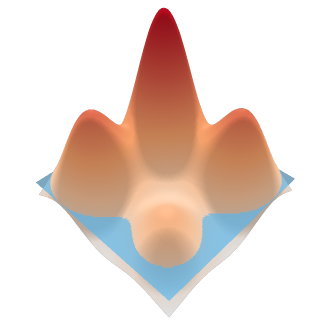}
\caption{}
\label{fig:island1}
\end{subfigure}%
\begin{subfigure}{.33\linewidth}
\centering
\includegraphics[width=150px]{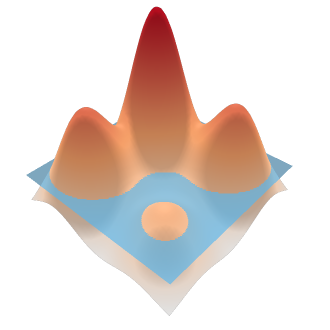}
\caption{}
\label{fig:island2}
\end{subfigure}
\begin{subfigure}{.33\linewidth}
\centering
\includegraphics[width=150px]{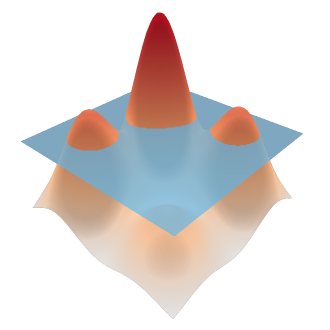}
\caption{}
\label{fig:island3}
\end{subfigure}
\caption{Approximative representation of the k-clique adjacency graph as island with complex terrain. The sea level represents the filter. With low tide, in (a) the returned network is almost similar to the original one. Increasing the level of the water, as in (b) the network splits in two components. With high tide, as in (c), the network further splits in more components but some of them might be covered.}
\label{fig:island}
\end{figure}

\index{k-clique adjacency graph}\index{clique graph}
Differently from the standard method, the \textit{k-clique adjacency graph} (or clique graph) $G$ in ACPM is a weighted graph. Indeed, it contains the node weights representing the different values of pace of collaboration of 3-cliques.
Therefore, we can imagine this clique graph as a large island with a complex terrain, as showed in Fig. \ref{fig:island1}. The height of each point on the terrain is defined as the value of their intensity (or pace of collaboration). We can also imagine that the intensity threshold \textit{I} is the level of water around this island and the action of raising this threshold has the effect of leaving more portion of the landscape underwater. Indeed, as the water raises, more valleys get flooded and the main island splits into smaller islands, as showed in Fig. \ref{fig:island2}. These new islands (or peaks) are groups of ``closely related'' vertices (k-cliques) that locally dominate according to the values of their pace of collaboration \citep{batagelj2006}. However, it is also possible to raise the water high enough to further split the islands but at the same time we can submerge important pieces of landscape or peaks, as showed in Fig. \ref{fig:island3}. We can argue that for the effective detection of new emerging topics, every peak can be relevant. 

In brief, we can use this threshold to filter out some background noise, such as all the k-cliques that have very low pace of collaboration and that can slow the computation of the algorithm. However, we cannot increase the threshold too much, as some important areas of the clique graph, which are relevant for the effective detection of new emerging areas, can fall below the threshold and be filtered out. 

So far, the state of the art has not provided any effective solution to further split these very large weighted communities. Hence, we need to investigate new solutions, such as redefining the concept of community.

\begin{figure}[ht]
\centering
  \includegraphics[width=180px]{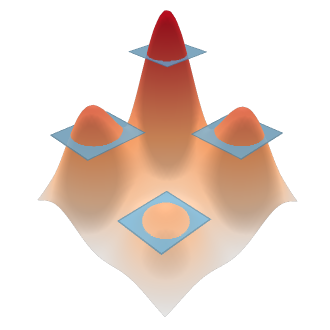}
  \caption{Graphical representation of the k-clique adjacency graph. Rather than filtering with the sea level, we locate local maxima (peaks).}
  \label{fig:localmaxima}
\end{figure}

We are interested in communities as areas of the evolutionary network in which there is an intense collaboration between topics that eventually could lead to the emergence of new topics. As already mentioned, these peaks represent the areas of the network where a set of closely related topics (by means of k-cliques) are exhibiting intense collaboration. We can certainly focus just on the peaks of the network. Therefore, once the algorithm creates the \textit{k-clique adjacency graph} we can proceed in locating all the peaks and selecting the topics with the most active collaboration, as showed in Fig. \ref{fig:localmaxima}.

In brief, with this fourth step, ACPM aims to identify the communities as peaks of the network. This consists of two main steps: 
\begin{itemize}
\item locating the local maxima within the adjacency graph, which in turn consists of: 
\begin{itemize}
\item converting the clique graph into an overlap matrix $M$,
\item locating local maxima in the matrix $M$.
\end{itemize}
\item selecting their neighbourhoods which contain all the closely related topics that have an intense collaboration.
\end{itemize}

Firstly, the algorithm converts the clique graph into an overlap matrix $M$, according to Eq. \ref{eq:cliqueadjacencymatrix}.

\begin{equation}
{m_{i,j}} = \left\{ \begin{gathered}
  {W_j}\,\,\,\,\,\,\,\,if\,(i,j)\, \in \,E \hfill \\
  {W_i}\,\,\,\,\,\,\,\,\,if\,i = j \hfill \\
  0\,\,\,\,\,\,\,\,\,\,\,if\,(i,j)\, \notin \,E \hfill \\ 
\end{gathered}  \right.\,\,\,\,\,\,\,\,with\,\,{m_{i,j}} \in {\mathbb{R}^{\left| V \right| \times \left| V \right|}}
\label{eq:cliqueadjacencymatrix}
\end{equation}

This overlapping matrix $M$ is similar to the adjacency matrix but instead of reporting ones\footnote{The adjacency matrix of a graph is a matrix with rows and columns labelled by graph vertices, with a 1 or 0 in position ($V_i,V_j$) respectively if $V_i$ and $V_j$ are adjacent or not.} wherever there is a link between the \textit{i-th} and \textit{j-th} cliques, it reports the pace of collaboration of the \textit{j-th} clique ($W_{j}$). In addition, in the main diagonal, when $j=i$, the matrix reports the pace of collaboration of \textit{i-th} clique ($W_{i}$). To this end, each row \textit{i} of the matrix $M$, reports at different positions the pace of collaboration of all the nodes to which the \textit{i-th} clique is connected, plus its pace of collaboration at position \textit{i}. In this way, we can analyse from the perspective of each clique (row) whether the clique itself is a maximum. 

Secondly, once the matrix M is created, the algorithm locates the maximum value in each row and its position. If the maximum is positioned in the main diagonal (\textit{i-th} row, \textit{i-th} column) then the \textit{i-th} clique is a local maximum. This is because, since each row represent the connection that a given clique has, if it exhibits the highest value it means that among the nodes it is connected, it is the local maximum.

As the algorithm knows which cliques are considered local maxima, it can proceed with the extraction of their neighbourhoods. To extract the neighbourhood of a given local maximum we can select their ego network\index{ego networks}\index{networks!ego}. The ego network is a network consisting of a central node (\textit{ego})\index{ego node}, the nodes it is directly connected (\textit{alters})\index{alter nodes}, and the ties between them \citep{freeman1982}. In the context of this work, an ego network consists of the induced subgraph containing a given local maximum clique and the cliques that are directly connected to it. The size of the neighbourhood or ego network is given by the ``order''. When the order is 0, the ego network includes only the ego node. When the order is 1, the ego network includes the ego node plus its immediate neighbours. ACPM extracts the ego network with order 2, meaning that it will contain the ego node, the immediate neighbours and the neighbours of neighbours, and all the links between these cliques. This is because using order 1 might not allow us to observe the phenomenon in its completeness, since we are extracting the local maximum clique and its immediate alters. Using order 2, instead, slightly broadens this space that might contain more nodes and can allow us to locate the topics that are exhibiting an increase of the pace of collaboration. 

As output, ACPM returns the ego networks of cliques converting them into clusters of topics.
In Code \ref{alg:acpm} this step is accomplished from line 19 to 31.

\subsubsection{Difference between CPM and ACPM}

In brief, the most important differences between CPM and ACPM is that the newly designed approach gives more value to the link weights, rather than simply filtering the links based on a threshold. In addition, the whole concept of community is redefined. Instead of considering communities as connected components of the k-clique adjacency graph that in many occasions can appear very large, it considers communities as the peaks of the network characterised as composition of local areas exhibiting an intense collaboration between topics.

In the next section, we describe the final result of Augur, i.e. the actual result that the user receive as outcome. In particular, we describe how we post-processed the communities to retrieve the final set of clusters with their topics. In addition, to help the user to understand the results, the algorithm will provide, for each cluster, some relevant authors and papers.
\newline
\newline
\begin{algorithm}[H]
\label{alg:acpm}
\LinesNumbered 
\setstretch{1}
\SetKwInOut{Input}{Input}\SetKwInOut{Output}{Output}

\Input{Evolutionary Network $network$, Threshold $threshold$}
\Output{Communities $comms$}
\BlankLine

cliques$\leftarrow$ FindCliques(network, k=3)\;
\For{$i\leftarrow 1$ \KwTo $length(cliques)$}{
poc.cliques[i]$\leftarrow$ PaceOfCollaboration(cliques[i])
}

cliques$\leftarrow$     cliques[which(poc.cliques $>$ threshold)]\;
poc.cliques$\leftarrow$ poc.cliques[which(poc.cliques $>$ threshold)]\;
\BlankLine
m$\leftarrow$ matrix(length(cliques), init=0) \tcp*[f]{overlap mat. initialized with 0}

adj.mat$\leftarrow$ matrix(length(cliques), init=0) \tcp*[f]{adjacency matrix}

\For{$i\leftarrow 1$ \KwTo $length(cliques)$}{
\For{$j\leftarrow 1$ \KwTo $length(cliques)$}{
	\uIf{(i $\neq$ j) $\&$ (SharedNodes(cliques[i],cliques[j]) $==$ 2)}{
	adj.mat[i,j]$\leftarrow$ 1 \tcc*[r]{adjacent cliques}
	m[i,j]$\leftarrow$ poc.cliques[j] \tcp*[f]{according to Eq. \ref{eq:cliqueadjacencymatrix}}
	}
	\ElseIf{j $==$ i}{
	m[i,j]$\leftarrow$ poc.cliques[i]\;
	}
}
}
clique.graph$\leftarrow$ GraphFromAdjacencyMatrix(adj.mat)\;
pos.max$\leftarrow$ MaxPosition(m, mod="rowwise")\tcp*[f]{find position of max vals}
\BlankLine
k$\leftarrow$ 1\;
\For{$i\leftarrow 1$ \KwTo $length(pos.max)$}{
\If{pos.max[i] == i}{
	local.maxima[k]$\leftarrow$ i\;
	k$\leftarrow$ k+1\;
}
}
\BlankLine
\For{$i\leftarrow 1$ \KwTo $length(local.maxima)$}{
components[i] $\leftarrow$ EgoNetwork(clique.graph, local.maxima, order = 2)\;
}
\BlankLine
comms$\leftarrow$ ConvertComponents(components)\tcc*[r]{converts components of cliques in components of topics: communities}

return(comms)\;
\caption{Advanced Clique Percolation Method.}
\end{algorithm}
\vspace*{0.7cm}

\section{From cluster of topics to embryonic topics}\label{sec:post-processing-acpm}
In the previous sections, we devoted our attention to the main phases of Augur, which are the creation of the evolutionary networks and the clustering of these networks, using the Advanced Clique Percolation Method. However, the results are still far from being conclusive. Indeed, the selection of ego networks from local maximum as community, can draw in links that are not necessarily high in pace of collaboration. To this end it is important to further post-process the results of the previous step.

In addition, to support the user in making sense of such results, Augur also provides a set of active papers and relevant authors associated to each community. 

\subsection{Cluster aggregation and main topics}
Given an evolutionary network, the Advanced Clique Percolation Method returns several communities whose topics are involved in an intense collaboration. These communities have been extracted locating the local maxima within the clique graph and then returning their ego network. As already mentioned, the dimension of the ego network strongly depends on the order value, which has been set to 2. However, in this case, the returned communities may contain links of the evolutionary network that not necessarily have high pace of collaboration. Indeed, regardless of its weight, if a link is placed nearby a local maximum it will become part of the community.
Furthermore, in very dense zones of the evolutionary network, retrieving an ego networks of order 2 can lead to communities of topics that go from 50 to 200, which is arguably a very large dimension for a community. Therefore, Augur needs to further process these results by removing less significant nodes.

To this end, we ranked the links within each community according to their pace of collaboration and we observed that their values follow a long-tailed distribution. Observing these distributions across all the retrieved clusters, we found that, on average, the pace of collaboration within a cluster significantly drops after 15 links.

Therefore, for each community, Augur ranks the links by their weights in descending order and selects the first 15, which potentially embody the most active collaborations between topics. In this way the topics of the final clusters will be the topics that are connected by the selected 15 links.
However, since ACPM returns overlapping communities, it can happen that two (or more) clusters share a subset of their topics simply because their mutual topics had the most active collaborations in both communities. Therefore, before returning the result, Augur checks all the clusters removing duplicates and merging clusters that have Jaccard similarity greater than 0.7. An empirical analysis suggested that using 0.7, the two compared sets can be considered almost similar.


In Code \ref{alg:filteringcommunities} there is the pseudocode for filtering communities.
\newline
\newline
\begin{algorithm}[H]
\label{alg:filteringcommunities}
\setstretch{1}
\SetKwInOut{Input}{Input}\SetKwInOut{Output}{Output}
\SetKw{KwBy}{by}
\Input{List of returned communities $comms$}
\Output{Clusters of topics $clusters$}
\BlankLine
k $\leftarrow$ 1\;
\ForEach{community in comms}{
links $\leftarrow$ GetLinks(community)\;
links $\leftarrow$ sort(links, descendig = true)\;
links $\leftarrow$ links[1:15]\tcp*[f]{select the first 15 links}
\BlankLine
clusters[k][topics] $\leftarrow$ TopicsInLinks(links)\; 
clusters[k][links] $\leftarrow$ links\;
k $\leftarrow$ k + 1\;
}

\BlankLine
clusters $\leftarrow$ RemoveDuplicates(clusters)\;
clusters $\leftarrow$ MergeClusters(clusters, sim=0.7)\tcp*[f]{with similarity above 0.7}
\BlankLine
\tcc{The result contains information about the most important topics and links that connect these topics}
return(clusters)

\caption{Algorithm for post-processing the clusters of topics.}
\end{algorithm}
\vspace*{0.7cm}


\subsection{Active authors}\label{sec:influential-authors2}
To help the user in making sense of the returned cluster of topics, Augur provides a set of relevant authors. Augur performs an analysis on the five years prior to the detection of the cluster and identifies the authors that are actively publishing in as many topics as possible within the cluster.
For instance, if ACPM is identifying the community structure within the evolutionary network of 2001, the analysed topics networks will range from 1997 to 2001. The most active authors for such communities will be selected in the same period.

Given a cluster in $year_{t}$, its most active authors are evaluated based on:
\begin{itemize}
\item \textbf{The number of topics and links within the community he or she is actively publishing from $year_{t-4}$ to $year_{t}$.} From Rexplore, it is possible to select the list of authors that were publishing in a topic in a particular time range ($year_{t-4}$, $year_{t}$). Since an author can belong to different topics, for each author we counted the number of topics he or she was actively publishing in that period. Moreover, we also counted the number of links within the community he or she was part of. For instance, if in that period, an author was active in five research areas of the community, the algorithm will consider him or her as belonging to five nodes and also count the number of links that connect these nodes in the community. This is because an author can be considered more central and active if he or she belongs to as many links as possible within the community (especially those with the highest pace of collaboration).
\item \textbf{Cumulative weight according to the link weights of the evolutionary network.} Within the community, different links can have different weights based on the pace of collaboration of the two topics they connect. We can argue that authors belonging to the topics with highest pace of collaboration can be the most active. Therefore, for each author, we selected the links connecting the topics, in which he or she published, and we summed up all the link weights.
\item \textbf{Number of publications until $year_{t}$.} We selected for each author the number of papers he or she had published until $year_{t}$. This number has been normalised with the number of topics he or she was publishing.
\item \textbf{Topics of the community he or she is publishing from $year_{t-4}$ to $year_{t}$.} Information about the topics within the community the author is publishing in the time range ($year_{t-4}$, $year_{t}$).
\end{itemize}
The final output consists of a ranking of the authors based on the number of associated topics (first parameter), then according to the cumulative weight (second parameter). The other two parameters can support the user in identifying other important authors more comprehensively. Code \ref{alg:selectinfluentialauthors} shows the pseudocode for extracting the active authors.
\newline
\newline

\begin{algorithm}[H]
\label{alg:selectinfluentialauthors}
\setstretch{1}
\SetKwInOut{Input}{Input}\SetKwInOut{Output}{Output}
\SetKw{KwBy}{by}
\Input{List of clusters $clusters$, years in which the clusters have been found $years$}
\Output{List of active authors $inf.authors$}
\BlankLine
\For{$i\leftarrow 1$ \KwTo $length(clusters)$}{
topics $\leftarrow$ clusters[i][topics]\tcp{see Code \ref{alg:filteringcommunities}}
links$\leftarrow$ clusters[i][links]\;
year$\leftarrow$ years[i] \tcp{year when the cluster was discovered}
\ForEach{topic in topics}{
authors[topic] $\leftarrow$ RetrieveAuthors(topic,from=(year-5), to=year)\;
}
\BlankLine
inf.authors[names]$\leftarrow$ unique(authors)\;
inf.authors[num.topics]$\leftarrow$ CountTopics(inf.authors[names],topics)\;
inf.authors[num.links] $\leftarrow$ CountLinks(inf.authors[names],links)\;
inf.authors[cum.weight]$\leftarrow$ SumWeights(inf.authors[num.links], links)\;
inf.authors[num.publications]$\leftarrow$ RetrievePublications(inf.authors[names],from=(year-5), to=year)\;
inf.authors[topics]$\leftarrow$ GetTopics(inf.authors[names],topics)\;
\BlankLine
\BlankLine

inf.authors$\leftarrow$ sort(inf.authors, by=(num.topics, num.weight))
}
return(inf.authors)
\caption{Algorithm for selecting the most active authors.}
\end{algorithm}
\vspace*{0.7cm}

\subsection{Relevant papers}\label{sec:influential-papers2}
Alongside the most active authors, for a given cluster, Augur provides also a set of relevant papers. These papers are important research outputs that are laying the basis of the new potential area. As per the active authors, also the relevant papers are analysed in the same period of analysis of the evolutionary network that generated such cluster. 

In particular, for a set of topics within the cluster at $year_{t}$, the relevant papers are evaluated according to:
\begin{itemize}
\item \textbf{The number of topics in which they appear.} From Rexplore, it is possible to select all the papers published within a topic in a time period ($year_{t-4}$, $year_{t}$). Since a paper can describe content belonging to different topics, for each paper we counted the number of topics it is associated with. We then retained the papers that appear at least in two topics.
\item \textbf{Cumulative weight according to the link weights of the evolutionary network.} As in the case of active authors, we can argue that the papers belonging to the topics with the highest pace of collaboration can be the most relevant. Therefore, for each paper, we observed the links connecting the topics associated to the paper, and we summed up all their weights.
\item \textbf{Weight based on the authors that wrote the paper.} This value considers the number of citation the authors that wrote the paper received at the time the paper was written. This value has then been normalised by the number of its authors. The idea is that a particular paper written by a senior researcher, can push the boundaries towards a new interdisciplinary field.
\item \textbf{Number of citations.} Rexplore reports the number of citations a paper received over time. To this end, we summed the number of citations each paper received from the year it has been published (within the range ($year_{t-4}$, $year_{t}$)) until $year_{t}$.
\end{itemize}
The final output consists of a ranking of relevant publications based on the number of topics they appear (first parameter), then according to the cumulative weight (second parameter). The other parameters can support the user in identifying other important papers more comprehensively. Code \ref{alg:selectinfluentialpapers} shows the pseudocode for selecting the relevant papers.
\newline
\newline
\begin{algorithm}[H]
\label{alg:selectinfluentialpapers}
\setstretch{1}
\SetKwInOut{Input}{Input}\SetKwInOut{Output}{Output}
\SetKw{KwBy}{by}
\Input{List of clusters $clusters$, years in which the clusters have been found $years$}
\Output{List of relevant papers $inf.papers$}
\BlankLine
\For{$i\leftarrow 1$ \KwTo $length(clusters)$}{
topics $\leftarrow$ clusters[i][topics]\tcp{see Code \ref{alg:filteringcommunities}}
links$\leftarrow$ clusters[i][links]\;
year$\leftarrow$ years[i] \tcp{year when the cluster was discovered}
\ForEach{topic in topics}{
papers[topic] $\leftarrow$ RetrievePapers(topic,from=(year-5), to=year)\;
}
\BlankLine
inf.papers[title]$\leftarrow$ unique(papers)\;
inf.papers[num.topics]$\leftarrow$ CountTopics(inf.papers[title],topics)\;
inf.papers[num.links] $\leftarrow$ CountLinks(inf.papers[title],links)\;
inf.papers[cum.weight]$\leftarrow$ SumWeights(inf.papers[num.links], links)\;
inf.papers[authors]$\leftarrow$ GetInfoAboutAuthors(inf.papers[title])\;
inf.papers[num.citation]$\leftarrow$ RetrieveCitations(inf.papers[title], until=year)\;

\BlankLine
\BlankLine

inf.papers$\leftarrow$ sort(inf.papers, by=(num.topics, num.weight))
}
return(inf.papers)
\caption{Algorithm for selecting the most relevant papers.}
\end{algorithm}
\vspace*{0.7cm}

\section{Conclusions: from the approach to the evaluation}
In this chapter we introduced Augur, a new framework for the early detection of topics, and we discussed its main components: the evolutionary network and the Advanced Clique Percolation Method (see Fig. {\ref{fig:workflow-augur}}).
In particular, thanks to the insights gained from the first study in Chapter {\ref{ch:firststudy}}, we designed the evolutionary network as a way to represent our input data. This strategy allowed us to address the scalability issue associated with the simultaneous computation of several large topic networks. 

In addition, we designed the Advanced Clique Percolation Method, a novel community detection algorithm that could exploit the peculiarities of the evolutionary network and could also cope with the presence of very large communities.


The next chapter is devoted to the evaluation of Augur, which has a twofold purpose: 
\begin{enumerate*}[label=\roman*)]
\item assess its validity, and 
\item evaluate the effectiveness of the Advanced Clique Percolation Method against other community detection algorithms, such as Fast Greedy \citep{clauset2004}, Leading Eigenvector \citep{newman2006}, Fuzzy C-Means \citep{bezdek1981} and the standard Clique Percolation Method \citep{palla2005}.
\end{enumerate*}

\part{Evaluation and Conclusion}
\chapter{Evaluation of Augur\label{ch:finalevaluation}}
\section{Introduction}\label{sec:final-evaluation-intro}

In this chapter, we evaluate \textit{Augur} and other alternative approaches on a gold standard of historical data. In particular, we want to measure to what extent the clusters produced by our approach are effective indicators of the emergence of new topics.
As we will see, the approach that provides better results, in terms of precision and recall, is the Advanced Clique Percolation Method.

More specifically, in the sections that follow: there will be a description of how the evaluation was designed (Section \ref{sec:from-clusters-2-eval}); and a description of the gold standard (Section \ref{sec:gold-standard}); in particular how it was compared with the returned clusters and the results obtained from the different clustering algorithms with respect to the adopted measures (Section \ref{sec:labelling-the-clusters}). We will then see how these results have been improved by means of appropriate filters (Section \ref{sec:semanticfilters}) and semantically enhanced (Section \ref{sec:semanticenhancement}). Finally, we will discuss the obtained results in more detail (Section \ref{sec:discussion}).

The results presented in this chapter have been published at the 18th ACM/IEEE Joint Conference on Digital Libraries \citep{salatino2018}.

The data collected during the evaluation and the gold standard are available at \url{http://doi.org/10.21954/ou.rd.8281010}

\section{From the clusters to the evaluation}\label{sec:from-clusters-2-eval}



We evaluated Augur by selecting topic networks from the year 1995 to 2009, which yielded 11 evolutionary networks. Table \ref{tab:evaluation-schema} reports the complete mapping between topic networks and evolutionary networks. For instance, the evolutionary network of the year 2005 takes into account the topic networks in the years 2001-2005.

\begin{table}[ht]
\centering
\caption{Mapping between \textit{topics networks} and \textit{evolutionary graphs}.}
\label{tab:evaluation-schema}
\begin{tabular}{C{4cm} | C{3cm}}
\hline
\textbf{Topic Networks of the Year} & \textbf{Evolutionary Graphs of the Year} \\ \hline
1995-1999                  & 1999                            \\
1996-2000                  & 2000                            \\
1997-2001                  & 2001                            \\
1998-2002                  & 2002                            \\
1999-2003                  & 2003                            \\
2000-2004                  & 2004                            \\
2001-2005                  & 2005                            \\
2002-2006                  & 2006                            \\
2003-2007                  & 2007                            \\
2004-2008                  & 2008                            \\
2005-2009                  & 2009                            \\ \hline
\end{tabular}
\end{table}

Given the evolutionary networks, we run different community detection algorithms, including the Advanced Clique Percolation Method, to produce the set of clusters that will be then evaluated against the gold standard.
Specifically, the evaluation aims to answer the question: ``\textbf{How many clusters returned each year by the algorithm effectively indicate a new emerging area?}''

To answer this question, it is important to understand how many of the returned clusters, for a given algorithm, map to the ancestors of emerging topics. 

In the first study (see Section \ref{ch:firststudy}), we analysed the emergence of a debutant topic using the network of its \textit{n} most co-occurring topics in the five years prior its emergence. For example, for a topic emerging in 2006, we would analyse the networks from year 2001 to 2005. In the context of this study, instead, we need to reverse this concept and, for instance, if we are analysing topic networks from 2001 to 2005 (evolutionary network of 2005), we can expect that those areas exhibiting such dynamics will lead to the emergence of a new topic in the following year, or at most two years. Indeed, for this evaluation, we will compare the clusters obtained in a given year with the debutant topics of the following two years. To this end, for the gold standard, we included all the topics that emerged in the period 2000-2011. Table \ref{tab:match-evg-deb} reports the mappings between the clusters and the debutant topics. As an example, the clusters obtained from the evolutionary network of the 2005, will be compared with the topics that emerged in the years 2006 and 2007. 

\begin{table}[ht]
\centering
\caption{Mapping the \textit{evolutionary graphs} and their clusters with the debutant topics in the subsequent years. For instance, the clusters obtained from the evolutionary network of the 2005 will be compared with the topics emerged in the years 2006 and 2007.}
\label{tab:match-evg-deb}
\begin{tabular}{ C{5cm} | C{5cm}}
\hline
\textbf{Evolutionary Graph and Clusters of the Year} & \textbf{Debutant Topics of the Year} \\ \hline
1999                        & 2000-2001          \\ 
2000                        & 2001-2002          \\ 
2001                        & 2002-2003          \\ 
2002                        & 2003-2004          \\ 
2003                        & 2004-2005          \\ 
2004                        & 2005-2006          \\ 
2005                        & 2006-2007          \\ 
2006                        & 2007-2008          \\ 
2007                        & 2008-2009          \\ 
2008                        & 2009-2010          \\ 
2009                        & 2010-2011          \\ \hline
\end{tabular}
\end{table}

In general, the evaluation consists of finding the match between clusters and debutant topics. This process can also be seen as ``labelling'' the clusters, in which the cluster receives a label with the name of the topic that is potentially nurturing.

However, before proceeding with the evaluation, it is important to produce the gold standard, containing the topics that emerged in the year 2000-2011, as reported in Table \ref{tab:match-evg-deb}. 

In the next section, we will show how we created the gold standard so that we could evaluate both Augur and the different clustering algorithms.

\section{Gold Standard}\label{sec:gold-standard}
Very often an evaluation is carried out to compare the results of a given algorithm against a set of results determined a priori to be correct, also known as gold standard. In the context of this work, the results that represent the truth are the debutant topics that emerged from 2000 to 2011. 

This information is already available within the Rexplore database since for each topic, there is information about the year of debut. However, only knowing the emerging topics and their year of debut might not be enough for evaluating the clusters. A domain expert can analyse the interaction of some topics in a certain year and suggest, based on his or her judgement, that such interaction might lead to the emergence of one of the topics beheld by the gold standard. Therefore, the evaluation would need several domain experts with different areas of expertise. This process cannot be achieved in an automatic way, unless we enrich the debutant topics with additional information. 


To automatise the evaluation we need to associate to each debutant topic a set of ancestors, which will be matched against the retrieved clusters.

In brief, the gold standard designed for this evaluation consists of gathering together the list of emerging topics with their year of debut, as we will see in Section \ref{sec:debutant-topics}. Then in Section \ref{sec:extraction-related-topics}, we will describe how we selected the ancestors for each debutant topic.

\subsection{Debutant topics}\label{sec:debutant-topics}

The Rexplore corpus is very diverse. It integrates entities of different dimensions such as topics, authors, papers and much more, through different semantic relationships. With this wide variety of interconnections, for a given a topic, one can easily retrieve the number of published papers per year, the number of publishing authors, the year of debut and so on. In particular, Rexplore computes the year of debut for a topic as the year in which the label of that topic has been used for the first time as keyword in a paper. For example, the label of the topic \textit{Cloud Computing}, according to the corpus in the database, made its first appearance in the year 2006. 

Since the scope of this doctoral work is confined within the \textit{Computer Science}, to build the gold standard, we initially filtered the set of topics covered by Rexplore.

\begin{figure}[ht]
\centering
  \includegraphics[width=\linewidth]{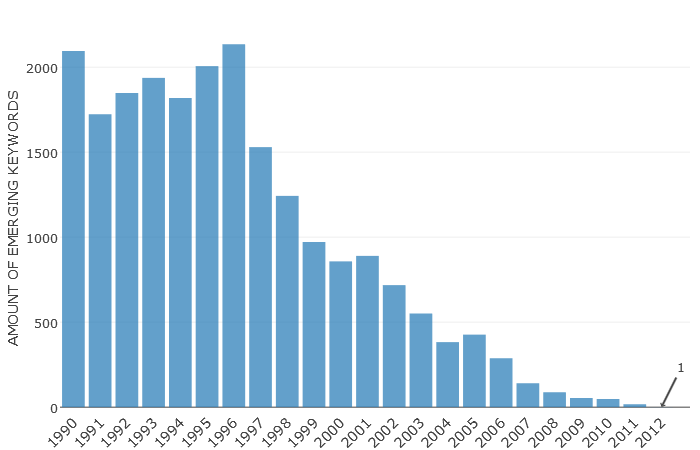}
  \caption{Histogram reporting the number of keywords that had their first appearance (hard debut) in the years between 1990 and 2012.}
  \label{fig:emerging-keywords}
\end{figure}

Rexplore contains 21772 keywords that made their debut between the years 1990 and 2012, and they cover areas within fields like \textit{Chemistry}, \textit{Genetics}, \textit{Biology} and so on. All these keywords are scattered along time according to the histogram in Fig. \ref{fig:emerging-keywords}. To filter this large and diverse set of keywords we used the Computer Science Ontology which allowed us to remove both the keywords that do not represent topics and those that do not belong to the \textit{Computer Science} field.

After filtering, we then retrieved all the emerging topics that have their year of debut into the range 2000-2011, according to Table \ref{tab:match-evg-deb}. 
To minimise the noise that can be caused by the spurious occurrence of the label associated with a topic in the literature, we use the 
concept of soft debut, in which a topic makes its debut only when it has reached at least 5 publications. As an example, the topic Deep Learning makes its first appearance in 2003, and reached five publications in 2005.
In this way, we can be more certain that a new label recognised by multiple authors  was introduced. We thus computed the soft debut of each topic. After analysing the difference between the original debut (herein labelled hard debut) and the soft debut, it appears that most of the topics have their soft debut at most two years after their hard debut.

\begin{figure}[ht]
\centering
  \includegraphics[width=\linewidth]{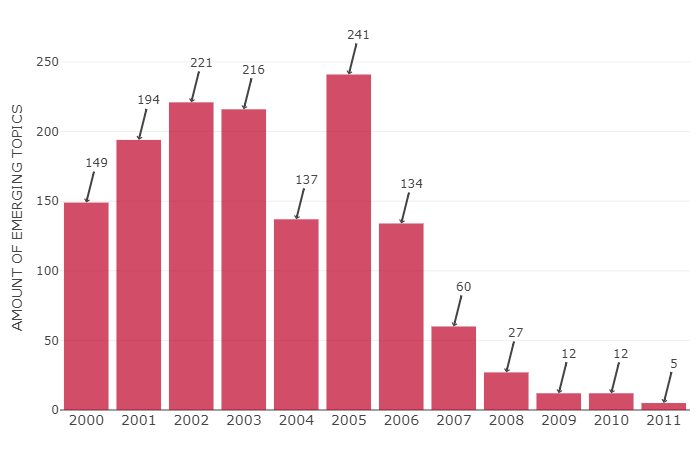}
  \caption{Histogram reporting the number of topics that had their soft debut in the years between 2000 and 2011.}
  \label{fig:emerging-topics}
\end{figure}

In Fig. \ref{fig:emerging-topics}, it is reported a histogram of all the 1,408 topics emerged within the field of \textit{Computer Science} from 2000 to 2011.

Unfortunately, the number of debutant topics drastically decreases in the second part of the analysed period. This is probably due to some missing data in our dataset. We still included the years after 2006 in the analysis for the sake of completeness, however this issue prevents us from trusting the results of the evaluation for years after 2006. As future work, we plan to analyse other scholarly datasets to provide a gold standard that will cover also more recent years.

Knowing all the debutant topics under the field of \textit{Computer Science} that emerged in the period between 2000 and 2011, we can proceed in enhancing them with their ancestors as described in the next section.

\subsection{Extraction of related topics}\label{sec:extraction-related-topics}

The extraction of related topics is inspired by the selection phase designed in the first study, and described in Section \ref{sec:selection-phase-first} at page \pageref{sec:selection-phase-first}. For each debutant topic, the algorithm aims to select a set of topics, considered its ancestors, that could potentially have helped it to arise as a result of a multidisciplinary interaction.
 
In the first study, the main assumption was that a new topic will continue to collaborate with its ancestors for a certain time after its emergence, and the algorithm locates these ancestors by selecting the \textit{n} most co-occurring topics. However, in the closing notes of Chapter \ref{ch:firststudy}, we pointed out some drawbacks of this approach. We can argue that for a given topic, selecting the most co-occurring topics, from the year of debut until now, can lead to topics that are not necessarily its ancestors. 

Indeed, if the debutant topic becomes mature enough to form new intense collaborations with other topics, those topics might also be part of the most co-occurring ones and be wrongly classified as ancestors. In addition, if the debutant topic helps to foster new areas, those new areas can also belong to the most co-occurring topics since it is likely that they will have an intense collaboration, as part of the nurturing phase.
 



To this end, when generating our gold standard, we consider as ancestors\index{ancestors} the top 25 topics, which mostly collaborated with the debutant topic during its initial five years of activity, and were already existing when the debutant topic emerged.


In practice, given a debutant topic, the algorithm firstly selects all the co-occurring topics in its first five years of activity, which are also older than the debutant topic.  
Then, for each co-occurring topic, we observe its \textit{intensity of collaboration}\index{index of collaboration} by computing the distance between the amount of papers the debutant topics received in the first five years and the amount of papers in which the debutant topics and the related topic appear together. 
In this way, the algorithm can locate those topics that mostly collaborated with the debutant topic in its first five years of life.

In detail, considering ${\vec P_{dt}}\,$ (paper vector of the debutant topic) the non-zero vector containing the amount of papers published in the emerging topic in the first five years of activity, and ${\vec C_{dt,rt}}\,$ (collaboration vector of the debutant topic and its related topic) the vector containing the amount of papers the two topics appear together in the same five years, the intensity of collaboration $IoC({\vec P_{dt}},{\vec C_{dt,rt}})$ between a debutant topic and a related topic is the Euclidean distance between these two vectors, as defined in Eq. \ref{eq:class-of-collaboration}.
\begin{equation} \label{eq:class-of-collaboration}
\begin{array}{l}
{{\vec P}_{dt}}\, = \,({p_{year}},{p_{year + 1}},{p_{year + 2}},{p_{year + 3}},{p_{year + 4}})\\
{{\vec C}_{dt,rt}}\, = \,({c_{year}},{c_{year + 1}},{c_{year + 2}},{c_{year + 3}},{c_{year + 4}})\\
IoC({{\vec P}_{dt}},{{\vec C}_{dt,rt}}) = \sqrt {\sum\limits_{k = 0}^4 {{{({p_{year + k}} - {c_{year + k}})}^2}} } 
\end{array}
\end{equation}

If the distance between the collaboration vector and the paper vector of the debutant topics is close to zero, it means that the debutant topic and the related topic had a very intense relationship in the first five years of life of the debutant topic and that the related topic had a major role in shaping the debutant topic. Therefore, we can consider this related topic as ancestor. 

By contrast, if the distance between the collaboration vector and the paper vector is near to one, it means that the debutant topic in its first five years of life had a very mild relationship with the related topic and the latter had a minor role in shaping the debutant topic. 

Accordingly, after performing all the intensities of collaboration between the debutant topic and all the retrieved co-occurring topics, the algorithm ranks them in ascending order based on the Euclidean distance. For the evaluation, we selected the first 25 meaningful topics that were influential for the debutant topic. In Table {\ref{tab:sample-ancestors}} we report a sample of ancestors extracted for the topic \textit{Semantic Search} debuting in 2003. Code \ref{alg:extracting-topics} instead reports the pseudocode that describes this process.

The result of this process is a gold standard that contains topics in the \textit{Computer Science} between 2000-2011 and their ancestors.
\newline
\newline
\begin{algorithm}[H]
\label{alg:extracting-topics}
\setstretch{1}
\SetKwInOut{Input}{Input}\SetKwInOut{Output}{Output}

\Input{List of Debutant Topics $deb.topics$}
\Output{List of ancestors $ancestors$}
\BlankLine
\ForEach{topic in deb.topics}{
	year$\leftarrow$ GetYearOfSoftDebut(topic)\;
	related.topics$\leftarrow$ SelectCoOccurringTopics(topic, from=year, to=year+5)\;
	p.dt$\leftarrow$ NumberOfPapers(topic, from=year, to=year+5)\;
	\ForEach{rt in related.topic}{
	c.dt.rt$\leftarrow$ NumCoOccurringPapers(topic, rt, from=year, to=year+5)\;
	ioc[rt]$\leftarrow$ sqrt(sum((p.dt $-$ c.dt.rt)$ ^ 2$))\tcc*[r]{Euclidean distance}
}
temp.ancestors$\leftarrow$ sort(related.topic, values=ioc, increasing=TRUE)\;
ancestors[topic]$\leftarrow$temp.ancestors[1:25]\tcc*[r]{keep only the best 25}
}
return(ancestors)\;
\caption{Extracting the ancestors of debutant topics.}
\end{algorithm}
\vspace*{0.7cm}


\begin{table}[ht]
\centering
\caption{Example of ancestors extracted form the \textit{Semantic Search} topic debuting in 2003.}
\label{tab:sample-ancestors}
\begin{tabular}{L{3.9cm}|l}
\hline
\textbf{Debutant Topic} & Semantic Search                                                                                                                                                                                                                                                                                                     \\ \hline
\textbf{Soft debut}     & 2003                                                                                                                                                                                                                                                                                                                \\ \hline
\textbf{some ancestors (alphabetical order)} & \begin{tabular}[c]{@{}l@{}}data mining\\ database systems\\ information retrieval\\ internet\\ knowledge based systems\\ metadata\\ multimedia systems\\ natural language processing systems\\ ontology\\ query processing\\ search engines\\ semantic web\\ semantics\\ web services\\ world wide web\end{tabular} \\ \hline
\end{tabular}
\end{table}

\section{Labelling the clusters}\label{sec:labelling-the-clusters}
As already mentioned, Augur returns a set of clusters that each year are extracted from the evolutionary networks. Each cluster contains a ``team of topics'' that should lead to the emergence of a new area in the near future.


The aim of this evaluation is to compare and match the returned clusters in each year with the debutant topics in the next two years, as reported in Table \ref{tab:match-evg-deb}. However, as previously mentioned, comparing the emerging topics with the clusters, using only their label, is not feasible. 
For this reason, the debutant topics have been supplied with information regarding their ancestors, so that the algorithm can match the topics in each cluster against the ancestors of each debutant topic.

This is consistent with the aim of the evaluation, because the topics within the clusters 
will eventually lead to the emergence of a new topic, while the ancestors of a debutant topic, by definition, are the topics that bred this new topic and helped it to arise. Therefore, if the topics in the cluster are analogous to the ancestors of a debutant topics, we can assume that the cluster is correctly forecasting the debutant topic.

The algorithm that runs the evaluation, henceforth also called \textit{evaluator}, compares the clusters and the debutant topics in an \textit{all-vs-all} fashion. The algorithm computes the Jaccard Index between the \textit{i-th} cluster ${C_i}$ and the ancestors of the \textit{k-th} debutant topic  ${D_k}$, showed in Table \ref{tab:cluster-ancestors-example}, using Eq. \ref{eq:jaccard1}. However, considering only the syntactical representation of topics is restrictive. Indeed, a topic can have more than one label referring to it. Therefore, if the cluster ${C_i}$ and the debutant topic ${D_k}$ contain the same topic but with different labels, the equation will fail to match the topic.

\begin{table}[!htbp]
\centering
\caption{Example of cluster ${C_i}$ returned by Augur and ancestors ${D_k}$ of the topic \textit{Resource Description Framework} available in the gold standard.}
\label{tab:cluster-ancestors-example}
\begin{tabular}{|p{2.5cm}|p{11.5cm}|}
\hline
\textbf{Cluster                                                                                                                                                                                                                                                                                                                                                                      ${C_i}$}  &  query languages, information retrieval, artificial intelligence, world wide web, management information systems, relational database systems, computational geometry, image analysis, data reduction, knowledge acquisition, three dimensional computer graphics, membership functions, graph theory, theorem proving, vectors \\ \hline
\textbf{Ancestors of debutant topic ${D_k}$} &   query languages, information retrieval, artificial intelligence, world wide web, management information systems, natural language processing systems, multi agent systems, expert systems, xml, search engines, semantics, metadata, intelligent agents, interoperability, html, algorithms, data communication systems, security of data, schemas, linguistics, resource allocation, constraint theory, software agents, network protocols, websites  \\ \hline
\end{tabular}
\end{table}

To tackle this problem, we semantically enhanced the equation by employing the Computer Science Ontology. Particularly, Eq. \ref{eq:jaccard1} includes also the topics which have a \textit{same-as} relationship with the topics appearing in the cluster.

\begin{equation} \label{eq:jaccard1}
\begin{array}{l}
{C_i} \to i \mhyphen th\,Cluster\\
{D_k} \to ancestors\,of\,k \mhyphen th\,Debutant\,topic\\
S{A_i} \to \,same \mhyphen as\,\,of\,i \mhyphen th\,Cluster\\
J\left( {{C_i},{D_k},S{A_i}} \right)\, = \,\cfrac{{\left| {({C_i} \cup S{A_i})\,\, \cap \,\,{D_k}} \right|}}{{\left| {{C_i} \cup {D_k}} \right|}}
\end{array}
\end{equation}

We enhanced the clusters, so that even if a topic in ${C_i}$ does not have a syntactic match with an ancestor of a debutant topic in ${D_k}$, its \textit{related-equivalent} in $S{A_i}$ might have it. This works also in the opposite direction, if we had enhanced ${D_k}$, so that even if one ancestor was not matching a topic in a cluster ${C_i}$, one of its \textit{related-equivalent} could have helped to match. However, we did not enhance both ${C_i}$ and ${D_k}$ simultaneously, to avoid that the same topic could match multiple times.
In addition, we did not include $S{A_i}$ in the denominator, because although the $S{A_i}$ topics have been merged with the ${C_i}$, they actually play the role of substitutes for the actual topics in the cluster. 

The similarity index $J\left( {{C_i},{D_k},S{A_i}} \right)\,$ between the clusters and the debutant topics falls in the range 0 to 1. If similarity is very near 0 the two sets share only few topics, while for values of 0.1 to 0.4 there is already a good number of matching topics. When the similarity is near 1, that the two sets are almost identical.  We consider a positive match between a debutant topic and a cluster only when their similarity is above a threshold \textit{t}, as stated as follow:

\[\left\{ \begin{array}{l}
match:\,\,\,\,\,\,\,\,\,\,J\left( {{C_i},{D_k},S{A_i}} \right) \ge \,t\\
mismatch:\,\,\,J\left( {{C_i},{D_k},S{A_i}} \right)\, < \,t\,\,\,\,\,\,\,
\end{array} \right.\]

The similarity threshold \textit{t} cannot be defined a priori and indeed its detection will be part of the evaluation. The algorithm that runs the comparisons between the set of clusters $C$ and the set of ancestors of debutant topics $D$, will track all the different similarities into a similarity matrix (${\mathbb{R}^{\left| C \right| \times \left| D \right|}}$). In this way, we can compute the performance for varying values of similarity. The algorithm will iteratively set a threshold and then compute the values of precision and recall based on the matched clusters and the matched debutant topics. Eventually, we would like to earn some insights on the similarity threshold that guarantees the significance of the match and that keeps high the performance of Augur. 
Code \ref{alg:labellingclusters}, describes the general workflow for the process of evaluation.

In the next section, we will define the values of precision and recall employed to assess the performance of Augur. Then we will compare five alternative clustering algorithms:
\begin{enumerate*}[label=(\roman*)]
\item Fast Greedy Algorithm,
\item Leading Eigenvector Algorithm,
\item Fuzzy C-Means,
\item Clique Percolation Method, and
\item Advanced Clique Percolation Method.
\end{enumerate*}
\pagebreak

\begin{algorithm}[!htbp]
\setstretch{1}
\SetKwInOut{Input}{Input}\SetKwInOut{Output}{Output}
\SetKw{KwBy}{by}
\Input{List of Debutant Topics $deb.topics$, list of Clusters $set.clusters$, list of years $years$}
\Output{$precision$ and $recall$ along time}
\BlankLine
\ForEach{year of years}{
	clusters$\leftarrow$ GetClustersOfYear(set.clusters, year)\;
	topics$\leftarrow$ GetDebutantTopics(deb.topics, year+1, year+2)\tcc*[r]{in the two following years}
	\ForEach{cl in clusters}{
		same.as$\leftarrow$ GetSameAsOfCluster(cl)\;
		\ForEach{tp of topics}{
		ancestors$\leftarrow$ GetAncestorsOfTopic(tp)\;
		sim.matrix[cl,tp]$\leftarrow$ $\cfrac{|(cl \cup same.as) \cap ancestors|}{|cl \cup ancestors|}$\tcc*[r]{Jaccard Index in Eq. \ref{eq:jaccard1}}
		}
	}
	\For{$i\leftarrow 0$ \KwTo $1$  \KwBy $0.001$}{
		temp.mat$\leftarrow$ BinarizeSimilarityMatrix(threshold=i)\;
		temp.precision[i]$\leftarrow$ GetMatchedClusters(temp.mat)/length(clusters) \;
		temp.recall[i]$\leftarrow$ GetMatchedTopics(temp.mat)/length(topics) \;
		
	}
	precision[year]$\leftarrow$temp.precision\;
	recall[year]$\leftarrow$temp.recall\;
}
\tcc{Now precision and recall, each year, are arrays containing values of precision and recall varying with the different similarity values.}
\caption{Process of evaluation.}
\label{alg:labellingclusters}
\end{algorithm}

\subsection{Measures of performance}\label{sec:performance}

Precision and Recall are often used to evaluate the performance in tasks such as information retrieval and pattern recognition. To evaluate the clusters returned by Augur, we will also consider precision and recall. In this section, we will provide a definition of these measures taking into account the number of returned clusters, the amount of debutant topics and the number of matches between these two sets.

If we assume that there is a \textit{one-to-one} relationship between clusters and ancestors of debutant topics, so that every retrieved cluster has only one corresponding emerging topic and vice versa, we can define precision and recall using the confusion matrix in Table \ref{tab:confusion-matrix}.

\begin{table}[ht]
\centering
\caption{Confusion matrix for a binary classifier.\index{confusion matrix}}
\label{tab:confusion-matrix}
\renewcommand{\arraystretch}{2}
\begin{tabular}{L{2cm}L{2cm}|C{3cm}|C{3cm}|}
\cline{3-4}
                                                                                                                          &                                        & \multicolumn{2}{c|}{\cellcolor[HTML]{1A9641}Clusters}                     \\ \cline{3-4} 
                                                                                                                          &                                        & \cellcolor[HTML]{A6D96A}Retrieved & \cellcolor[HTML]{A6D96A}Not Retrieved \\ \hline
\multicolumn{1}{|c|}{\cellcolor[HTML]{E66101}}                                                                            & \cellcolor[HTML]{FDB863}Identified     & True Positive                     & False Negative                        \\ \cline{2-4} 
\multicolumn{1}{|c|}{\multirow{-2}{*}{\cellcolor[HTML]{E66101}\begin{tabular}[c]{@{}c@{}}Debutant\\ topics\end{tabular}}} & \cellcolor[HTML]{FDB863}Not Identified & False Positive                    & True Negative                         \\ \hline
\end{tabular}
\end{table}

Comparing the set of clusters against the set of debutant topics leads to four different cases:
\begin{itemize}
\item \textit{true positives} represent the number of clusters that have been matched with the debutant topics;
\item \textit{false positives} represent the retrieved clusters that have not been matched with any topic;
\item \textit{false negatives} are the debutant topics that do not have any matched cluster;
\item although not quantifiable, the \textit{true negatives} represent the amount of not found cluster that effectively do not correspond to any debutant topic.
\end{itemize}

The values of precision\index{precision} and recall\index{recall} are respectively described in Eq. \ref{eq:precision} and Eq. \ref{eq:recall}.

\begin{multline}\label{eq:precision}
  Precision\, = \,\frac{{True\,Positive}}{{True\,Positive\, + False\,Positive\,}}\, = \, \hfill \\
   = \,\frac{{\left| {\left\{ {retrieved\,clusters} \right\} \cap \left\{ {debutant\,topics} \right\}} \right|}}{{\left| {\left\{ {retrieved\,clusters} \right\}} \right|}}\, = \,\frac{{\# \,matchings}}{{\# \,retrieved\,clusters}} \hfill \\ 
\end{multline} 

\begin{multline}\label{eq:recall}
  Recall\, = \,\frac{{True\,Positive}}{{True\,Positive\, + False\,Negative\,}}\, = \, \hfill \\
   = \,\frac{{\left| {\left\{ {retrieved\,clusters} \right\} \cap \left\{ {debutant\,topics} \right\}} \right|}}{{\left| {\left\{ {debutant\,topics} \right\}} \right|}}\, = \,\frac{{\# \,matchings}}{{\# \,identified\,topics}} \hfill \\ 
\end{multline} 

However, the \textit{one-to-one} relationship between clusters and ancestors of debutant topics is a stretched assumption. Indeed, we can argue that a cluster can foster the emergence of more than one new topic and similarly two or more different clusters can share the same subset of ancestors
that match the same debutant topic. Therefore, precision and recall, respectively defined in Eq. \ref{eq:precision} and Eq. \ref{eq:recall} are not suitable to assess the validity of the approach.

\begin{figure}[ht]
\centering
  \includegraphics[width=\linewidth]{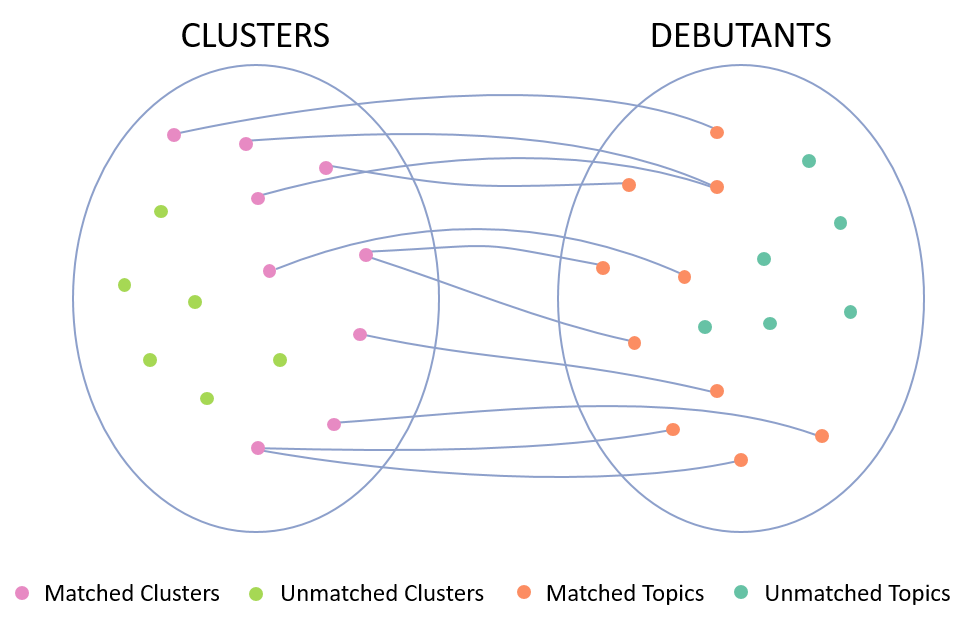}
  \caption{Graphical explanation of the \textit{many-to-many} relationship between clusters and debutant topics. Since the number of matched cluster can be different from the matched debutant topics, the value of precision and recall need to take into account this difference.}
  \label{fig:matching-clusters-debutants}
\end{figure}

Since there is a \textit{many-to-many} relationship between clusters and topics, as represented in Fig. \ref{fig:matching-clusters-debutants}, to evaluate this system it is important to observe this relationship from different angles (matched clusters, and matched debutant topics) and focus on the following two specific questions:
\begin{enumerate}
\item \textbf{How good is our system in identifying portion of topic networks (clusters) that eventually will lead to the emergence of new topics? (perspective of the cluster)}
\item \textbf{How good is our system in identifying debutant topics that have been matched with clusters? (perspective of the debutant topics)}
\end{enumerate}

Those two questions shape the definition of true positive mentioned above giving it a dual definition. Taking into account the first question it is possible to define the true positives as the clusters that have been matched with debutant topics. Considering the second question, instead, the true positives are the number of debutant topics that have been matched with the retrieved clusters. In a \textit{one-to-one} relationship, these two quantities can have the same value, while in a \textit{many-to-many} this is not the case.

Therefore, we define \textbf{precision} as the fraction of clusters that have been successfully matched with debutant topics (also known as \textit{correctness}), and \textbf{recall} as the fraction of topics that have been successfully matched by clusters (also known  as \textit{completeness}), as defined in Eq. \ref{eq:precision2} and Eq. \ref{eq:recall2}.

\begin{equation}\label{eq:precision2}
Precision\, = \,\frac{{\left| {\left\{ {retrieved\,clusters} \right\} \ltimes \left\{ {debutant\,topics} \right\}} \right|}}{{\left| {\left\{ {retrieved\,clusters} \right\}} \right|}}\, = \,\frac{{\# \,matched\,clusters}}{{\# \,retrieved\,clusters}}
\end{equation}

\begin{equation}\label{eq:recall2}
Recall\, = \,\frac{{\left| {\left\{ {retrieved\,clusters} \right\} \rtimes \left\{ {debutant\,topics} \right\}} \right|}}{{\left| {\left\{ {debutant\,topics} \right\}} \right|}}\, = \,\frac{{\# \,matched\,topics}}{{\# \,identified\,topics}}
\end{equation}

The symbol $\ltimes$ is called \textit{left semijoin}\index{left semijoin}\index{semijoin} and the result of $\left\{ {retrieved\,clusters} \right\} \ltimes \left\{ {debutant\,topics} \right\}$ is the set of the clusters for which there is a match with the debutant topics. Similarly, the sign $\rtimes$ is called \textit{right semijoin}\index{right semijoin} and the result of $\left\{ {retrieved\,clusters} \right\} \rtimes \left\{ {debutant\,topics} \right\}$ is the set of debutant topics that received a match from the clusters.

Since, the gold standard and the tools to evaluate are now both available, we can proceed in assessing the validity of Augur and of some community detection algorithms.

\subsection{Description of the main community detection algorithms}
In the previous chapter, we introduced the concept of community detection as a way to perform clustering over networks. In addition, we talked about some algorithms that are available in the literature, which can be classified according to two main families: 
\begin{itemize}
\item a \textit{crisp} community detection algorithm produces mutually exclusive clusters;
\item a \textit{fuzzy} algorithm provides overlapping communities so that a node can belong to more than one cluster.
\end{itemize}

As already anticipated, crisp algorithms are not suitable for the purpose of this work as it is expected that a topic can lead to the emergence of more than one other topic. However, for the sake of completeness, we tested them and indeed their results allowed us to gain a better understanding of evolutionary networks.
In particular, for the family of crisp community detection algorithms, we used Fast Greedy \citep{clauset2004} and Leading Eigenvector \citep{newman2006}. Instead, as fuzzy community detection algorithms, we evaluated Fuzzy C-Means \citep{bezdek1981}, the Clique Percolation Method \citep{palla2005} and the Advanced Clique Percolation Method proposed in this thesis.

In the following sections, we describe the results we obtained from all the tested algorithms. In Table \ref{tab:listofalgorithms} we report all the employed algorithms with information about their implementation and the section describing their results.

\begin{table}[!ht]
\centering
\caption{List of the community detection algorithms used for the evaluation.}
\label{tab:listofalgorithms}
{\renewcommand{\arraystretch}{1.6}%
\begin{tabular}{L{3.9cm} p{3.9cm} L{3.3cm} p{2.5cm}}
\hline\textbf{Algorithm}               & \textbf{Refercence}  & \textbf{Implementation used} & \textbf{Results}                \\ \hline
Fast Greedy                      & \cite{clauset2004} & R: iGraph           & Section \ref{sec:result-fg}   \\ 
Leading Eigenvector             & \cite{newman2006}  & R: iGraph           & Section \ref{sec:result-le}   \\ 
Fuzzy C-Means                    & \cite{bezdek1981}  & R: e1071            & Section \ref{sec:result-fcm}  \\ 
Clique Percolation Method        & \cite{palla2005}   & R: developed from scratch    & Section \ref{sec:result-cpm}  \\ 
Advan. Clique Percolation Method & \cite{salatino2018} and this work          & R: developed from scratch  & Section \ref{sec:result-acpm} \\ \hline
\end{tabular}}
\end{table}

\subsection{Results with Fast Greedy Algorithm}\label{sec:result-fg}
\begin{figure}[ht]
\centering
  \includegraphics[width=\linewidth]{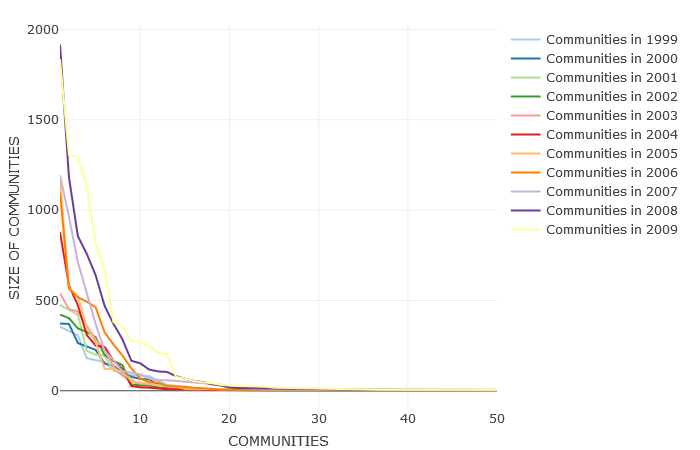}
  \caption{Distribution of the community sizes obtained from the Fast Greedy algorithm in the different years.}
  \label{fig:results-clusters-fg}
\end{figure}

We adopted the implementation of the Fast Greedy algorithm available within the iGraph\footnote{\url{http://igraph.org/r/doc/cluster_fast_greedy.html}} package for R, which is the most comprehensive package for \textit{Network Science}. The algorithm takes as input the 11 evolutionary networks, according to Table \ref{tab:match-evg-deb} at page \pageref{tab:match-evg-deb}, and returns several communities along a time dimension. In Table \ref{tab:fastgreedydata}, are reported the number of communities returned for each year and we can observe that, as a consequence of the constant widening of evolutionary networks over time, the algorithm returns an almost ever increasing number of communities. Another interesting phenomenon is that the sizes of the communities returned each year by the algorithm have a long-tailed distribution, as showed in Fig. \ref{fig:results-clusters-fg}. In particular, there is a small number of communities with a substantial number of topics and a large number of communities with just a few topics. In Table \ref{tab:fastgreedydata}, we also reported the dimension of the first ten communities to highlight this phenomenon along time.

\begin{table}[ht]
\centering
\caption{Number of communities and the size of the first ten communities using the Fast Greedy algorithm over the different evolutionary networks}
\label{tab:fastgreedydata}
\begin{tabular}{l|c|l}
\hline
\textbf{Year} & \textbf{\# communities} & \textbf{Dimension of the first 10 communities}                 \\ \hline
1999 & 50		      & 355, 329, 306, 180, 170, 159, 116, 87, 86, 84         \\
2000 & 52             & 372, 370, 264, 246, 226, 150, 138, 100, 77, 66        \\
2001 & 38             & 475, 447, 421, 221, 199, 196, 110, 107, 65, 29        \\
2002 & 51             & 422, 402, 345, 326, 296, 198, 165, 142, 34, 33        \\
2003 & 62             & 540, 452, 439, 363, 282, 228, 134, 84, 51, 39         \\
2004 & 65             & 878, 584, 474, 309, 249, 243, 166, 121, 26, 19        \\
2005 & 102            & 1179, 571, 523, 350, 268, 121, 121, 104, 55, 44       \\
2006 & 98             & 1098, 565, 518, 493, 462, 323, 256, 195, 124, 67      \\
2007 & 106            & 1193, 965, 713, 538, 369, 221, 169, 109, 101, 91      \\
2008 & 152            & 1918, 1183, 856, 758, 639, 470, 370, 285, 166, 152    \\
2009 & 164            & 1835, 1305, 1295, 1130, 815, 669, 380, 361, 273, 268  \\ \hline
\end{tabular}
\end{table}

Once the algorithm completed the processing the evolutionary networks and returned the clusters, the next step was to evaluate these clusters against the ancestors of the actual debutant topics. The evaluator then compared the clusters and the debutants in an \textit{all-vs-all} fashion, creating a matrix of similarity. Then, from this similarity matrix it becomes possible to define the minimum value of similarity above which they are considered similar. 

\begin{figure}[ht]
\centering
  \includegraphics[width=\linewidth]{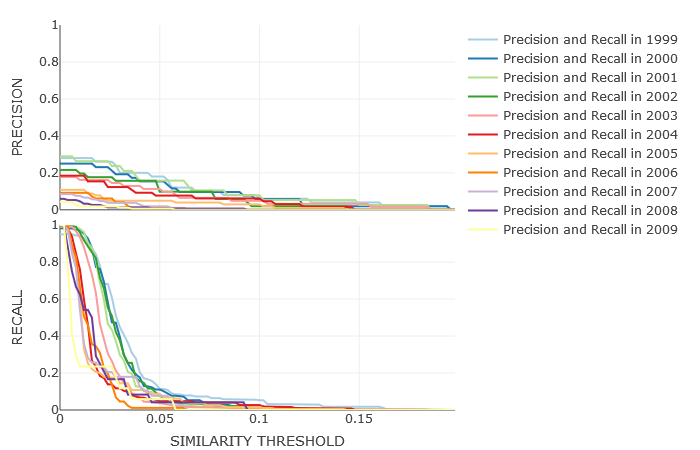}
  \caption{Results in terms of precision and recall obtained from evaluating the Fast Greedy algorithm against the gold standard.}
  \label{fig:results-evaluation-fg}
\end{figure}

As already mentioned, using a low similarity value (near 0), it means that a cluster and a debutant topic will be considered similar even with few mutual topics. A very high similarity value (near 1) means that a cluster must contain almost all and exclusively the ancestors of a debutant topic to be considered a match. However, to choose a suitable similarity threshold, we need to consider the minimum value in which the match is considered meaningful and at the same time it maximises precision and recall. In Fig \ref{fig:results-evaluation-fg}, we report, for each year, all the values of precision and recall depending on the different values of similarity. The lines of precision and recall for a particular year are identified by the same colour. As it can be seen from the figure, regardless of the value of similarity, the precision is always low while the recall is very high for similarity values near 0 but then it quickly drops when we attempt to increase it. According to these results we can understand that because of low precision only a small portion of the returned clusters matched the debutant topics. Increasing the value of similarity worsen the situation since fewer and fewer clusters could match debutant topics. Regarding the recall, we can see that using a low similarity value, almost all the debutant topics were matching the returned clusters, most certainly because only a few mutual topics are needed to consider the two parts similar, and then everything almost matches with everything else. If we increase the similarity value, fewer debutant topics get matched and then the recall drastically decreases. 

An interesting phenomenon highlighted from Fig. \ref{fig:results-evaluation-fg} is that the values of precision and recall depend from the year of analysis. We can see that, although poor, the best performance comes from the evolutionary networks of 1999 to 2001 and then they decrease along time. This can be justified by the fact that there are fewer and fewer debutant topics that can be matched by the retrieved clusters in the last part of the analysed period, as reported in Fig. \ref{fig:results-evaluation-fg}. Thus, even if the results appear to be very poor, they do provide interesting insights.

In conclusion, as the results suggest, the Fast Greedy algorithm is not suitable for the aim of this work as it does not provide good results in terms of precision and recall. 

\subsection{Results with Leading Eigenvector Algorithm}\label{sec:result-le}
The Leading Eigenvector, as the Fast Greedy, is also implemented in the iGraph\footnote{\url{http://igraph.org/r/doc/cluster_leading_eigen.html}} package for R. The algorithm iteratively processes the 11 evolutionary networks, returning, for each year, a set of clusters. Table \ref{tab:leadingeigendata} reports the number of communities the algorithm is able to locate within the different networks, as well as the size of the first ten communities. From the table, we can observe that also this algorithm returns communities whose size resembles a long-tailed distribution. In particular, for years like 1999, 2000, 2006, 2007, 2008 and 2009 the algorithm returns one or two large communities and some other very small ones.
\begin{table}[ht]
\centering
\caption{Number of communities and the size of the first ten communities using the Leading Eigenvector algorithm over the different evolutionary networks}
\label{tab:leadingeigendata}
\begin{tabular}{l|c|l}
\hline
\textbf{Year} & \textbf{\# communities} & \textbf{Dimension of the first 10 communities }        \\ \hline
1999 & 14		      & 2187, 3, 2, 2, 2, 2, 2, 2, 2, 2               \\
2000 & 14             & 2069, 169, 3, 2, 2, 2, 2, 2, 2, 2             \\
2001 & 11             & 1279, 421, 237, 207, 190, 72, 2, 2, 2, 2      \\
2002 & 36             & 567, 553, 369, 347, 279, 207, 100, 49, 6, 6   \\
2003 & 15             & 1661, 765, 306, 44, 36, 4, 3, 2, 2, 2         \\
2004 & 25             & 908, 514, 406, 387, 380, 324, 244, 26, 21, 14 \\
2005 & 22             & 1337, 609, 564, 470, 351, 254, 106, 3, 2, 2   \\
2006 & 22             & 4286, 177, 2, 2, 2, 2, 2, 2, 2, 2             \\
2007 & 17             & 5183, 4, 2, 2, 2, 2, 2, 2, 2, 2               \\
2008 & 12             & 7866, 3, 2, 2, 2, 2, 2, 2, 2, 2               \\
2009 & 24             & 7684, 1448, 691, 2, 2, 2, 2, 2, 2, 2          \\ \hline
\end{tabular}
\end{table}

\begin{figure}[ht]
\centering
  \includegraphics[width=\linewidth]{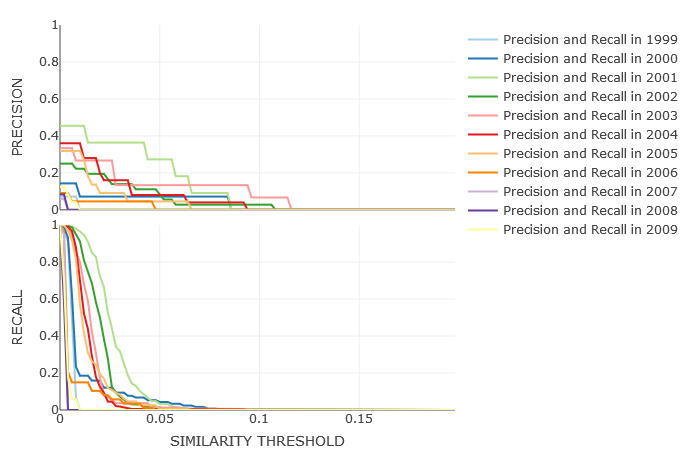}
  \caption{Results in terms of precision and recall obtained from evaluating the Leading Eigenvector algorithm against the gold standard.}
  \label{fig:results-evaluation-le}
\end{figure}

We ran the evaluator over the obtained clusters and Fig. \ref{fig:results-evaluation-le} reports the values of precision and recall depending on the different values of similarity. As we can see, also this approach does not provide valuable results as the precision is below 0.4 regardless of the similarity value and the recall decreases as soon as the value raises. These values of precision and recall are comparable to the results obtained using the Fast Greedy algorithm, mostly because both these algorithms form few giant communities and many other small ones. This phenomenon mainly happens in networks with high number of nodes and edges, because Fast Greedy and the Leading Eigenvector respectively tend to merge and split the communities, trying to optimise a quality function called modularity\index{modularity}. In optimising the modularity function they force small communities into larger ones, offering a misleading characterisation of the underlying community structure \citep{lancichinetti2011}.  
This is a well-known problem in the literature, which pushed the community of scholars to move towards alternative solutions that could work on real-world complex networks \citep{holmstrom2009, botta2016}.

In conclusion, even if modularity offers a first-principle understanding of community structure, an algorithm based on its maximisation cannot provide a good characterisation. It is also worth to remember that both Fast Greedy and Leading Eigenvector belong to the crisp family of community detection algorithm. Although they helped to move the first steps towards the effective detection of emerging trends, these algorithms cannot characterise the community structure of our networks, given that a topic can foster the emergence of more than one other topic. In the next sections we will show the results we obtained using some fuzzy community detection algorithms.

\subsection{Results with Fuzzy C-Means}\label{sec:result-fcm}

Fuzzy C-Means is one of the most famous and widely adopted algorithms for clustering data and, as the name suggests, it belongs to the fuzzy family. An implementation of this algorithm is available in the e1071\footnote{https://www.rdocumentation.org/packages/e1071/versions/1.6-8/topics/cmeans} package for R. The required arguments include the input matrix, the number of clusters and the degree of fuzzification. The input matrix contains information about the evolutionary networks, from which the algorithm will find the clusters. It is structured as matrix, where rows correspond to observations (i.e., topics in the evolutionary network) and columns to variables (i.e., topics to which it is connected). The degree of fuzzification or fuzziness coefficient can be any real number greater than 1\footnote{With m=1, the membership of a node (topic) to a cluster will converge to either 0 or 1, which denotes a crisp partitioning.} and will be set to 2, as commonly done when the domain is not well known \citep{hathaway2001}.
Lastly, the number of clusters is the number of clusters we expected as output from the algorithm. This value is not known a priori but the state of the art offers several methods that can help to determine it, such as the elbow method. Since each evolutionary network can contain a different number of communities, these networks need to be individually analysed to determine their number of clusters.

The elbow criterion is a visual method and consists in running the clustering algorithm with a constantly increasing number of clusters, as showed in Code \ref{alg:fuzzycmeans}. For each run, we calculate the sum of squared errors of prediction (SSE)\footnote{The sum of squared errors of prediction measures quantitatively the discrepancy between the actual data and the predicted model.} and plot it against the number of clusters. If the line curves like an arm, the parameter is set to the minimum number of clusters in which the SSE starts to asymptotically decrease towards 0. This is typically located at the elbow of the curve. However, this method is computationally expensive since one needs to actually run the C-means algorithm several times trying different number of clusters. According to the number of debutant topics per year (see Fig. \ref{fig:emerging-topics}), it is reasonable to have a range of clusters that goes from 5 to 200. However, since each evolutionary network is a very large data sample and the execution of one single instance of C-Means is quite expensive, at each run we increased the number of cluster by 5.
\newline
\newline
\begin{algorithm}[H]
\label{alg:fuzzycmeans}
\setstretch{1}
\SetKwInOut{Input}{Input}\SetKwInOut{Output}{Output}
\SetKw{KwBy}{by}
\Input{List of Evolutionary Networks $evolutionary.networks$, list of years $years$}
\Output{Number of clusters $num.clusters$ along time}
\BlankLine
\ForEach{year of years}{
	ev.network$\leftarrow$ GetEvolutionaryNetworkOfYear(evolutionary.networks, year)\;
	mat$\leftarrow$AsAdjacencyMatrix(ev.network)\tcp{Graph into adjacency matrix}
	i $\leftarrow$ 1\;
	\For{$cl\leftarrow 5$ \KwTo $200$  \KwBy $5$}{
		fclust.obj$\leftarrow$ cmeans(mat,clusters=cl,m=2)\tcp{m = degree of fuzzif.}
		\tcc{$fclust.obj$ contains the memberships, the centroids and the value of the objective function (SSE)}
		sse[i]$\leftarrow$ GetWithinError(fclust.obj)\;	
		x[i]$\leftarrow$ cl\;
		i $\leftarrow$ i + 1\;
	}
	plot(x, sse)\;
	\tcc{The position of the ``elbow'' in the plot determines the number of clusters in that particular year}
}

\caption{Counting the number of clusters within the evolutionary networks using the Elbow Method with Fuzzy C-Means.}
\end{algorithm}
\vspace*{0.7cm}

\begin{figure}[ht]
\centering
  \includegraphics[width=\linewidth]{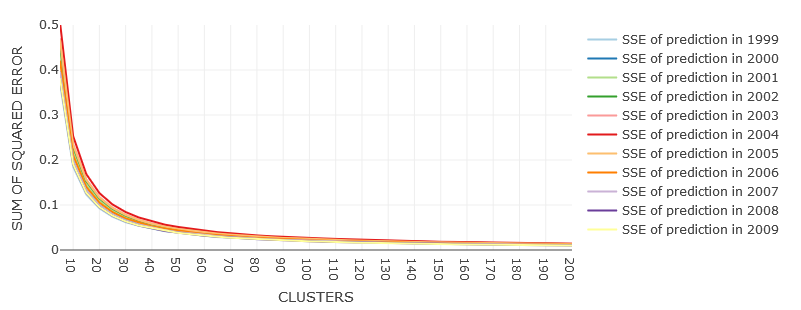}
  \caption{Sum of squared error in prediction (SSE) obtained from evaluating the Fuzzy C-Means with different number clusters (from 5 to 200) over the different evolutionary networks. We can see that for all the evolutionary networks the elbow is located at 25.}
  \label{fig:elbow-results}
\end{figure}

We run the method showed in Code \ref{alg:fuzzycmeans} and we reported in Fig. \ref{fig:elbow-results} all the different SSEs, based on the year and the number of clusters in which the Fuzzy C-Means algorithm has been trained.
From the figure we can observe a common pattern: with 5 clusters the SSE for each evolutionary network is around 0.45 and it exponentially decreases towards 0.01 for 200 clusters. In addition, we can observe that for all the different evolutionary networks, the SSE curves start to bend around 25. Hence, we ran the Fuzzy C-Means over the different evolutionary networks using 25 as number of clusters.

As output, the algorithm returns a membership matrix where rows correspond to the different topics and 25 columns corresponding to the clusters. This matrix reports the likelihood of each topic to belong to a particular cluster. 

To retrieve the final cluster set, for a given cluster, we selected all topics having likelihood (of belonging to it) above a particular threshold (0.04 or \textit{chance level} 1/25). Instead, if for a particular cluster, all the likelihoods were below or equal to such threshold we selected the first 25 topics.

Table \ref{tab:fuzzycmeansmethod} reports the number of communities (as already mentioned, set to 25), as well as the size of the first ten communities. We can observe that also the Fuzzy C-Means fails to characterise the different clusters within the different networks, because for each year it provides one large community and some little ones. Also, the fact that all these little communities have 25 topics, means that some of the topics found in a cluster, are likely to be there by chance. 

\begin{figure}[ht]
\centering
  \includegraphics[width=\linewidth]{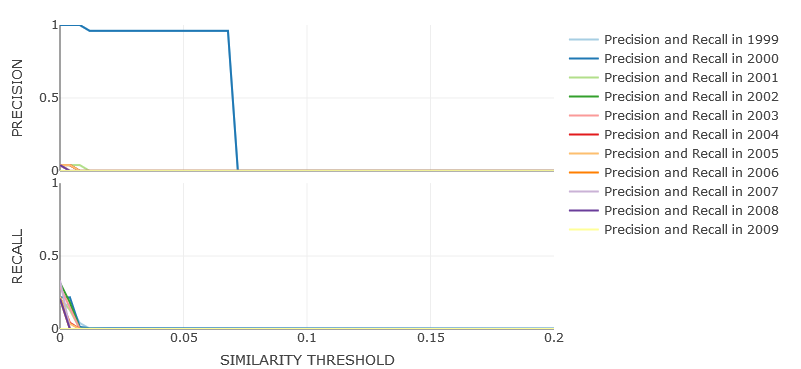}
  \caption{Results in terms of precision and recall obtained from evaluating the Fuzzy C-Means against the gold standard.}
  \label{fig:results-evaluation-fcm}
\end{figure}

Using these clusters, we ran the evaluator and Fig. \ref{fig:results-evaluation-fcm} reports the values of precision and recall depending on the different values of similarity. Overall, we can observe very low values of precision and recall, dropping to zero when the similarity threshold reaches around 0.01. We can find an exception in the year 2000, in which the precision stays around 0.96 for initial values of similarity thresholds and then it collapse when it reaches 0.0638. This phenomenon is due to the fact that all the 24 small communities were containing three similar topics (``telecommunication links'', ``telecommunication networks'', ``channel capacity'') that found a match with the ancestors of \textit{Network Coding} boosting the precision value.  When the similarity threshold increases over 0.0638, the precision decreases because we cannot find matches that exceed such threshold.


\begin{table}[ht!]
\centering
\caption{Number of communities and the size of the first ten communities using the Fuzzy C-Means algorithm over the different evolutionary networks.}
\label{tab:fuzzycmeansmethod}
\begin{tabular}{l|c|l}
\hline
\textbf{Year} & \textbf{\# communities} & \textbf{Dimension of the first 10 communities} \\ \hline
1999 & 25		      & 681, 25, 25, 25, 25, 25, 25, 25, 25, 25 \\
2000 & 25		      & 728, 25, 25, 25, 25, 25, 25, 25, 25, 25 \\
2001 & 25		      & 885, 25, 25, 25, 25, 25, 25, 25, 25, 25 \\
2002 & 25		      & 922, 25, 25, 25, 25, 25, 25, 25, 25, 25 \\
2003 & 25		      & 1050, 25, 25, 25, 25, 25, 25, 25, 25, 25 \\
2004 & 25		      & 1381, 25, 25, 25, 25, 25, 25, 25, 25, 25 \\
2005 & 25		      & 1507, 25, 25, 25, 25, 25, 25, 25, 25, 25 \\
2006 & 25		      & 1791, 25, 25, 25, 25, 25, 25, 25, 25, 25 \\
2007 & 25		      & 2082, 25, 25, 25, 25, 25, 25, 25, 25, 25 \\
2008 & 25		      & 3443, 25, 25, 25, 25, 25, 25, 25, 25, 25 \\
2009 & 25		      & 3853, 25, 25, 25, 25, 25, 25, 25, 25, 25 \\ \hline
\end{tabular}
\end{table}

\subsection{Results with Clique Percolation Method}\label{sec:result-cpm}
The Clique Percolation Method, is another algorithm that falls in the family of fuzzy community detection algorithms. Among all the packages that allow to perform network analysis in \textit{R} there was not an implementation of the Clique Percolation Method. After developing a version\footnote{This version is available in my personal Github repository: \url{https://github.com/angelosalatino/CliquePercolationMethod-R}.} based on the work of \cite{palla2005}, we ran the algorithm over the 11 evolutionary networks, which returned the communities summarised in Table \ref{tab:cliquepercolationmethod}. From the same table, we can observe that also this algorithm returns communities whose size resembles a long-tailed distribution. In particular, we can see that, for each year, there is one big blob of topics, which in most of the cases is over $1\,000$ topics and in some cases over $4\,000$ topics. All other communities have at most 11 topics. 

\begin{table}[ht]
\centering
\caption{Number of communities and the size of the first ten communities using the Clique Percolation Method algorithm over the different evolutionary networks.}
\label{tab:cliquepercolationmethod}
\begin{tabular}{l|c|l}
\hline
\textbf{Year} & \textbf{\# communities} & \textbf{Dimension of the first 10 communities} \\ \hline
1999 & 68		      & 966, 11, 11, 9, 7, 6, 5, 5, 4, 4      \\
2000 & 54             & 1124, 7, 6, 5, 5, 4, 4, 4, 4, 4       \\
2001 & 31             & 1302, 4, 4, 4, 4, 4, 4, 4, 3, 3       \\
2002 & 26             & 1454, 5, 4, 4, 4, 4, 4, 3, 3, 3       \\
2003 & 40             & 1554, 6, 6, 6, 5, 5, 4, 4, 4, 4       \\
2004 & 22             & 1783, 7, 6, 4, 4, 3, 3, 3, 3, 3       \\
2005 & 48             & 1793, 5, 5, 4, 4, 4, 4, 4, 4, 4       \\
2006 & 67             & 2042, 8, 5, 5, 5, 5, 5, 5, 4, 4       \\
2007 & 83             & 2541, 5, 5, 5, 4, 4, 4, 4, 4, 4       \\
2008 & 74             & 4492, 7, 5, 4, 4, 4, 4, 4, 4, 4       \\
2009 & 88             & 6031, 6, 5, 5, 5, 5, 5, 4, 4, 4       \\ \hline
\end{tabular}
\end{table}

\begin{figure}[ht]
\centering
  \includegraphics[width=\linewidth]{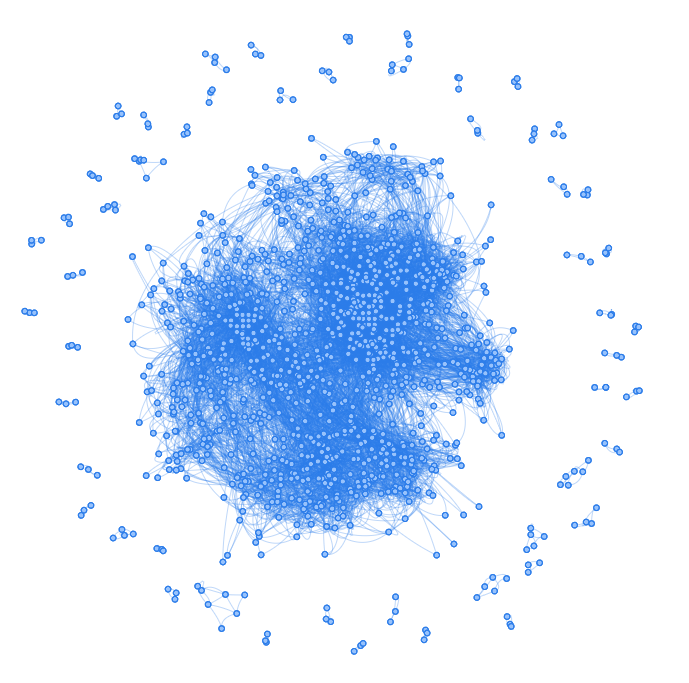}
  \caption{Evolutionary graph of the year 2000 split into communities using the Clique Percolation Method. The largest community counts 1124 topics and 10393 edges.}
  \label{fig:cliquegraph2000}
\end{figure}

Observing in more detail the large communities that the algorithm returned over time, it is possible to see that all the topics belong to the \textit{Computer Science} field. This means that the \textit{Computer Science} field is so tight on a clique level that the algorithm is unable to split it in further subnetworks. We can better appreciate this phenomenon through Fig. \ref{fig:cliquegraph2000}. This figure shows the evolutionary network of year 2000 divided in 54 communities or connected components as per the definition of community for the CPM. From the figure, we can see that there is one very large community and many other smaller ones. The largest community, which contains all the subareas of \textit{Computer Science}, is composed by $1\,124$ topics and $10\,939$ edges. The average degree of the nodes is 19.46, meaning that on average each topic is connected with at least 19 other topics. In addition, the subnetwork is composed by $27\,064$ 3-cliques with an average of 72 cliques per topic, and each couple of cliques shares two topics. Also, the average path length (also known as \textit{degree of separation}) of the network is 3.04, which means that on average two topics are separated by only 2 topics. All these characteristics suggest that this particular area of the evolutionary graph contains a high density of ties, to the extent that the Clique Percolation Method is unable to split it in further subnetworks. In Table \ref{tab:statofevolnetwork}, we report some other statistics that point out the characteristics of both the evolutionary network of year 2000 and its main community of \textit{Computer Science}, also suggesting their dense nature. In particular, while the main community of  \textit{Computer Science} contains half of the nodes, it also contains 82\% of the total number of links. Also, the diameter between the network and the community drastically reduces from 20.67 to 8.95, making it quite tight-knit. 

\begin{table}[!ht]
\centering
\caption{Some statistics for the evolutionary network of year 2000 and its largest community of \textit{Computer Science}.}
\label{tab:statofevolnetwork}
{\renewcommand{\arraystretch}{1.2}%
\begin{tabular}{L{4cm}|L{3cm}|L{3cm}}
\hline
\textbf{Network parameter}     & \textbf{Evolutionary Network (of 2000)} & \textbf{Computer Science community}  \\ \hline
\textit{Nodes}                 & 2263                          & 1124     (49.6 \%)              \\
\textit{Edges}                 & 13327                         & 10939     (82 \%)              \\
\textit{Avg degree}            & 11.77                         & 19.46                  \\
\textit{Avg clustering coeff.} & 0.163                         & 0.198             \\
\textit{Max degree}            & 184                           & 160              \\
\textit{Diameter}              & 20.67                         & 8.95          \\ \hline
\end{tabular}}
\end{table}

\begin{figure}[ht]
\centering
  \includegraphics[width=\linewidth]{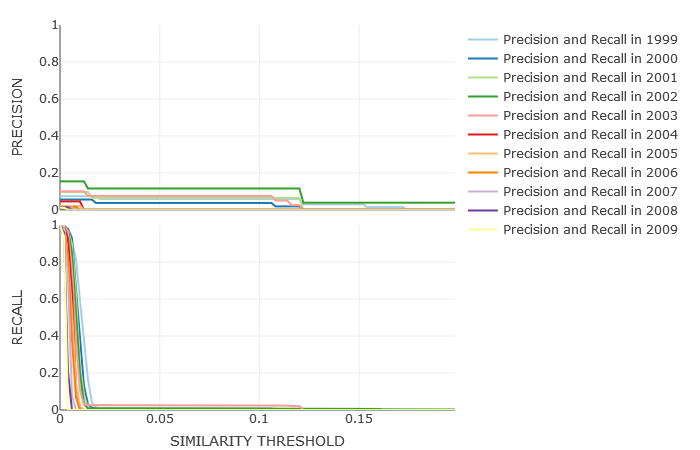}
  \caption{Results in terms of precision and recall obtained from evaluating the Clique Percolation Method against the gold standard.}
  \label{fig:results-evaluation-cpm}
\end{figure}

For the sake of completeness, the returned clusters have been evaluated against the debutant topics whose results are reported in Fig \ref{fig:results-evaluation-cpm}. From the figure, we can observe that both precision and recall exhibit the same patterns of the previously tested approaches. 
In brief, the Clique Percolation Method appears to not be suitable for clustering evolutionary networks. As showed, it tends to split these networks in a set of one very large community and several other small communities. This is mostly due to the fact that the evolutionary network appears to be very dense in specific area and then all the topics within this area are gathered together. Indeed, this phenomenon was already observed running the Fast Greedy and Leading Eigenvector methods, which while trying to optimise the modularity function, merge smaller communities into bigger ones.

Another limitation of the Clique Percolation Method is that it does not take advantage of the weights of the networks as the other algorithms do. 

In conclusion, the Clique Percolation Method as well as the Fast Greedy, Leading Eigenvector and Fuzzy C-Means methods, are not providing high values of precision and recall at high values of similarity. However, it is worth mentioning that these algorithms helped us to understand the structure and peculiarities of the evolutionary network: it is very dense and some algorithms do not take advantage of the link weights. 


\subsection{Results with Advanced Clique Percolation Method}\label{sec:result-acpm}

\begin{figure}[ht]
\centering
  \includegraphics[width=\linewidth]{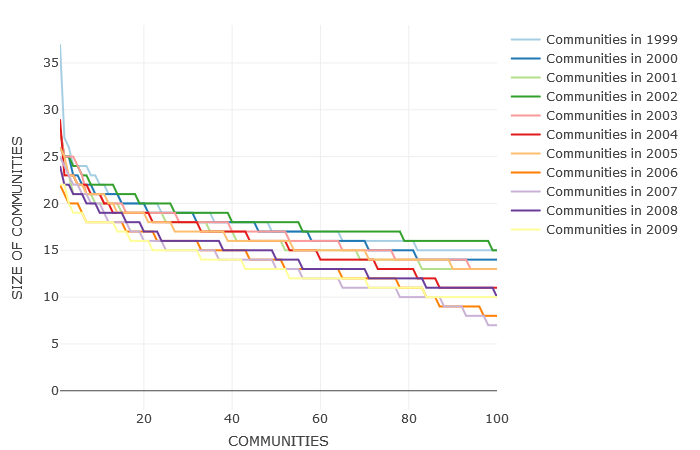}
  \caption{Distribution of the community sizes obtained from the Advanced Clique Percolation Method in the different years.}
  \label{fig:size-clusters-acpm}
\end{figure}

The Advanced Clique Percolation Method (ACPM) is a newly developed community detection algorithm and, as the name suggests, it is inspired by the Clique Percolation Method. In particular, this new version takes into account the weights of the network and redefines the concept of community, as explained in Section \ref{sec:acpm}.
In Table \ref{tab:acpm} are reported the number of communities this algorithm returned each year, as well as the size of the first ten communities sorted in descending order. As we can see, the algorithm is returning communities with a more reasonable size (around 20 topics) compared to the communities retrieved with the Clique Percolation Method, Fast Greedy and so on. 
Figure \ref{fig:size-clusters-acpm} reports more extensively the sizes of the first 100 communities, in which we can observe how the distributions are less long-tailed and more uniform.

\begin{table}[ht]
\centering
\caption{Number of communities and the size of the first ten communities using the Advanced Clique Percolation Method algorithm over the different evolutionary networks.}
\label{tab:acpm}
\begin{tabular}{l|c|l}
\hline
\textbf{Year} & \textbf{\# communities} & \textbf{Dimension of the first 10 communities}  \\ \hline
1999 & 238            & 37, 27, 26, 24, 24, 24, 24, 23, 23, 22 \\
2000 & 201            & 26, 25, 25, 23, 23, 22, 22, 22, 21, 21 \\
2001 & 174            & 25, 24, 23, 21, 21, 21, 21, 21, 20, 20 \\
2002 & 228            & 28, 25, 25, 24, 24, 23, 23, 22, 22, 22 \\
2003 & 159            & 28, 25, 25, 25, 24, 23, 21, 21, 21, 21 \\
2004 & 130            & 29, 23, 23, 23, 22, 22, 22, 21, 21, 21 \\
2005 & 168            & 26, 25, 23, 23, 22, 22, 21, 21, 21, 21 \\
2006 & 110            & 22, 21, 20, 20, 20, 19, 18, 18, 18, 18 \\
2007 & 103            & 25, 24, 23, 22, 22, 21, 21, 20, 20, 20 \\
2008 & 167            & 24, 22, 22, 21, 21, 21, 20, 20, 20, 19 \\
2009 & 177            & 22, 22, 20, 19, 19, 19, 18, 18, 18, 18 \\ \hline
\end{tabular}
\end{table}

\begin{figure}[!htbp]
\centering
  \includegraphics[width=\linewidth]{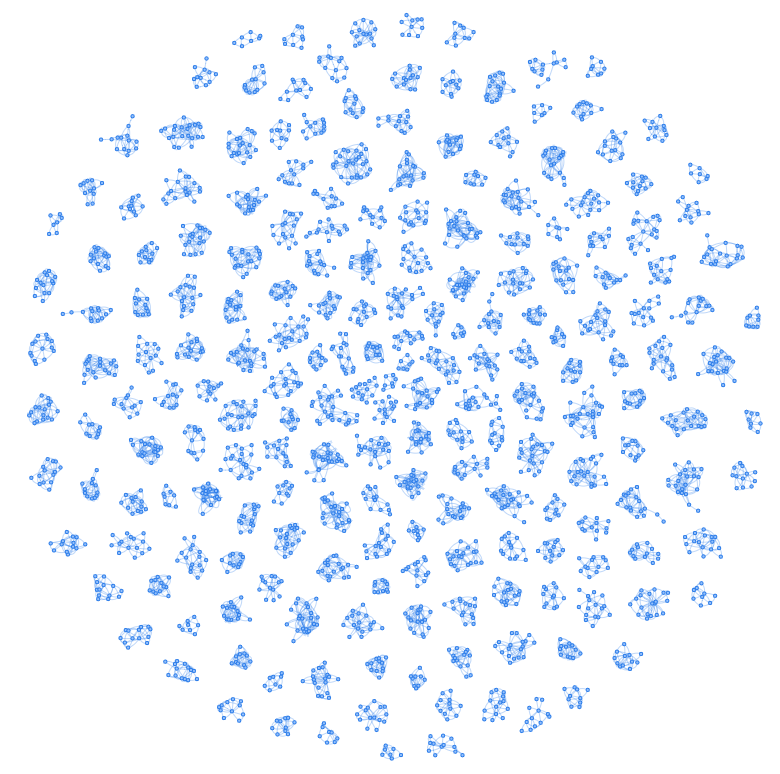}
  \caption{Evolutionary graph of the year 2000 split into communities using the Advanced Clique Percolation Method.}
  \label{fig:size-clusters-acpm-graph}
\end{figure}

These new distributions of size of communities suggests that the algorithm managed to split the area of \textit{Computer Science} in further sub areas. Figure \ref{fig:size-clusters-acpm-graph} shows the evolutionary network of the year 2000, which is split into 201 communities. Differently form Fig. \ref{fig:cliquegraph2000}, we can see that all the topics that where gathered within the large bubble of \textit{Computer Science} are now spread out into several communities. However, since the Advanced Clique Percolation Method falls into the fuzzy community detection algorithms, the same topic can appear into more than one community. This means that it can actively hold several partnerships and eventually lead to the emergence of more than one new area.

\begin{figure}[ht]
\centering
  \includegraphics[width=\linewidth]{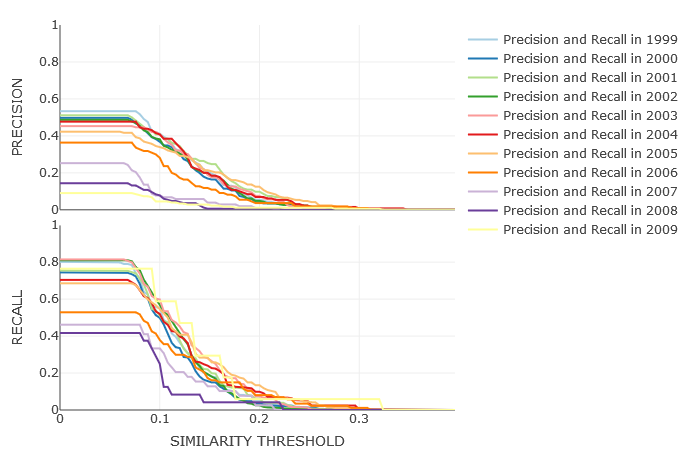}
  \caption{Results in terms of precision and recall obtained from evaluating the Advanced Clique Percolation Method against the gold standard.}
  \label{fig:finale-evaluation-no-filter}
\end{figure}

All the communities that the algorithm returned each year were evaluated against the gold standard, as already performed in the previous tests. The results of the evaluation are reported in Fig \ref{fig:finale-evaluation-no-filter}. Comparing them to the results obtained with the Clique Percolation Method and by extension also with the other approaches, we can observe that for ACPM the overall values of precision increase and also they stay up until almost 0.1 of similarity value. The values of recall instead are fairly good and are no longer affected by this quick fall.
However, although we see a general improvement in both precision and recall we cannot be completely satisfied by these results. The values for precision are still very low and the algorithm needs to be further improved in order to return better results.

To further improve the algorithm we analysed the communities more qualitatively. We realised that in some communities, most of the topics were not sub areas of \textit{Computer Science}. This impacts precision (see Eq. \ref{eq:precision2}), because these communities containing \textit{non-Computer Science} topics where unsuccessfully compared against the ancestors of debutant topics with the \textit{Computer Science} area. To cope with this problem, we filtered the communities semantically.

\newpage
\section{Applying semantic filters}\label{sec:semanticfilters}
In order to improve the results and thus remove the communities where most of the topics do not belong to \textit{Computer Science}, we developed a filter. This filter selects communities for which least 30\% of their topics are found in the Computer Science Ontology. This filtering process is part of the \textit{post-processing} phase, highlighted in the workflow in Fig. \ref{fig:workflow-augur}, at page \pageref{fig:workflow-augur}. As a result, we remove all the communities that can never match with the debutant topics and eventually improve the values of precision.
Indeed, thanks to this filter, the results obtained from all the approaches show an improvement in terms of precision and recall. Table \ref{tab:semantic-filters} reports the number of communities returned by all the analysed approaches, both before and after applying the semantic filter. On average (along the different years), when applying this filter to Fast Greedy, we reduce the number of clusters by 37\%. When filtering the clusters of Leading Eigenvector, they are reduced by 41\%, and by 52\%, 43\% and 37\% for the Fuzzy C-Means, Clique Percolation Method and its Advanced version, respectively.

\begin{table}[!ht]
\centering
\caption{Listing the number of communities obtained using the Fast Greedy (FG), Leading Eigenvector (LE), Fuzzy C-Means (FCM), Clique Percolation Method (CPM) and Advanced Clique Percolation Method (ACPM), before (\# comm.) and after filtering (\# comm. filt.). Below the table is reported the average percentage difference between the initial value of community and the final number of communities after filtering.}
\label{tab:semantic-filters}
{\renewcommand{\arraystretch}{1.4}%
\begin{tabular}{L{1cm}|L{1cm}L{1cm}|L{1cm}L{1cm}|L{1cm}L{1cm}|L{1cm}L{1cm}|L{1cm}L{1cm}}
\cline{2-11}
 & \multicolumn{2}{c|}{FG} & \multicolumn{2}{c|}{LE} & \multicolumn{2}{c|}{FCM} & \multicolumn{2}{c|}{CPM} & \multicolumn{2}{c}{ACPM} \\ \hline
\multicolumn{1}{L{1cm}|}{Years} & \multicolumn{1}{L{1cm}|}{\# comm.} & \# comm. filt. & \multicolumn{1}{L{1cm}|}{\# comm.} & \# comm. filt. & \multicolumn{1}{L{1cm}|}{\# comm.} & \# comm. filt. & \multicolumn{1}{L{1cm}|}{\# comm.} & \# comm. filt. & \multicolumn{1}{L{1cm}|}{\# comm.} & \multicolumn{1}{L{1cm}}{\# comm. filt.} \\ \hline
1999 & 50 & 26 & 14 & 5 & 25 & 25 & 68 & 32 & 238 & 121 \\
2000 & 52 & 29 & 14 & 7 & 25 & 25 & 54 & 22 & 201 & 103 \\
2001 & 38 & 23 & 11 & 9 & 25 & 25 & 31 & 12 & 174 & 98 \\
2002 & 51 & 28 & 36 & 18 & 25 & 1 & 26 & 14 & 228 & 118 \\
2003 & 62 & 44 & 15 & 10 & 25 & 1 & 40 & 25 & 159 & 83 \\
2004 & 65 & 45 & 25 & 17 & 25 & 1 & 22 & 17 & 130 & 70 \\
2005 & 102 & 73 & 22 & 16 & 25 & 1 & 48 & 33 & 168 & 91 \\
2006 & 98 & 68 & 22 & 15 & 25 & 25 & 67 & 45 & 110 & 81 \\
2007 & 106 & 76 & 17 & 10 & 25 & 25 & 83 & 39 & 103 & 94 \\
2008 & 152 & 101 & 12 & 6 & 25 & 1 & 74 & 42 & 167 & 129 \\
2009 & 164 & 96 & 24 & 11 & 25 & 1 & 88 & 58 & 177 & 146 \\ \hline
\%diff &  & 37\% &  & 41\% &  & 52\% &  & 43\% &  & 37\% \\
\end{tabular}}
\end{table}

\begin{table}[]
\centering
\caption{Figures comparison for each approach before and after applying the filter.}
\label{tab:comparison-before-after-filter}
{\renewcommand{\arraystretch}{1.4}%
\begin{tabular}{l|l|l}
\hline
\textbf{Method} & \textbf{Before filtering} & \textbf{After filtering} \\ \hline
FG & Figure \ref{fig:results-evaluation-fg} at p. \pageref{fig:results-evaluation-fg}   & Figure \ref{fig:finale-evaluation-with-filter-fg} at p. \pageref{fig:finale-evaluation-with-filter-fg}   \\
LE & Figure \ref{fig:results-evaluation-le} at p. \pageref{fig:results-evaluation-le}   & Figure \ref{fig:finale-evaluation-with-filter-le} at p. \pageref{fig:finale-evaluation-with-filter-le}   \\
FCM & Figure \ref{fig:results-evaluation-fcm} at p. \pageref{fig:results-evaluation-fcm}   & Figure \ref{fig:finale-evaluation-with-filter-fcm} at p. \pageref{fig:finale-evaluation-with-filter-fcm}   \\
CPM & Figure \ref{fig:results-evaluation-cpm} at p. \pageref{fig:results-evaluation-cpm}  & Figure \ref{fig:finale-evaluation-with-filter-cpm} at p. \pageref{fig:finale-evaluation-with-filter-cpm}   \\
ACPM & Figure \ref{fig:finale-evaluation-no-filter} at p. \pageref{fig:finale-evaluation-no-filter} & Figure \ref{fig:finale-evaluation-with-filter} at p. \pageref{fig:finale-evaluation-with-filter}   \\ \hline
\end{tabular}}
\end{table}

\begin{figure}[!ht]
\centering
  \includegraphics[width=\linewidth]{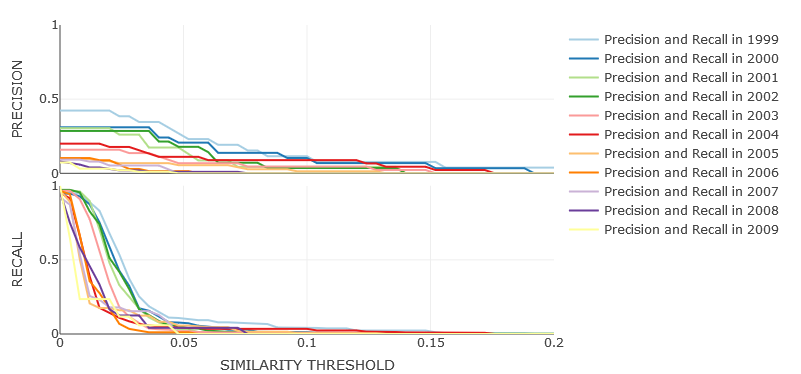}
  \caption{Results in terms of precision and recall obtained from evaluating the Fast Greedy against the gold standard after filtering the communities semantically.}
  \label{fig:finale-evaluation-with-filter-fg}
\end{figure}

\begin{figure}[!ht]
\centering
  \includegraphics[width=\linewidth]{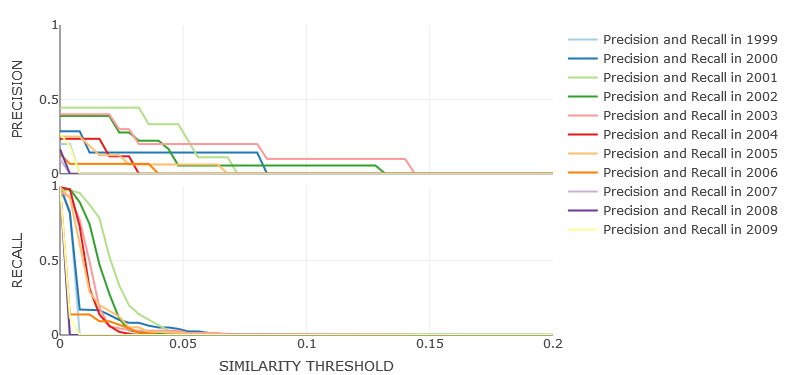}
  \caption{Results in terms of precision and recall obtained from evaluating the Leading Eigenvector against the gold standard after filtering the communities semantically.}
  \label{fig:finale-evaluation-with-filter-le}
\end{figure}

\begin{figure}[!ht]
\centering
  \includegraphics[width=\linewidth]{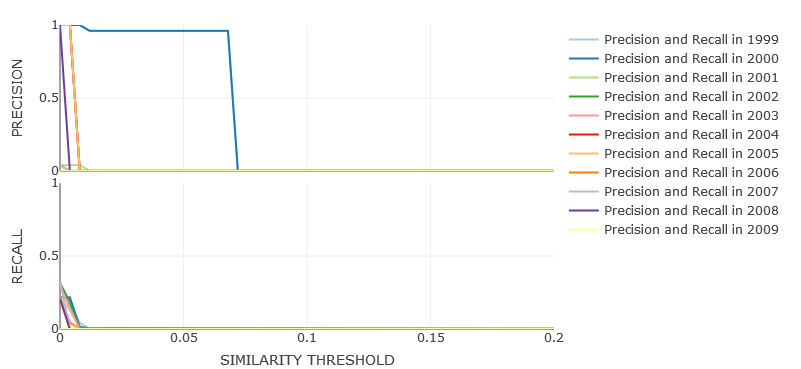}
  \caption{Results in terms of precision and recall obtained from evaluating the Fuzzy C-Means against the gold standard after filtering the communities semantically.}
  \label{fig:finale-evaluation-with-filter-fcm}
\end{figure}

\begin{figure}[!ht]
\centering
  \includegraphics[width=\linewidth]{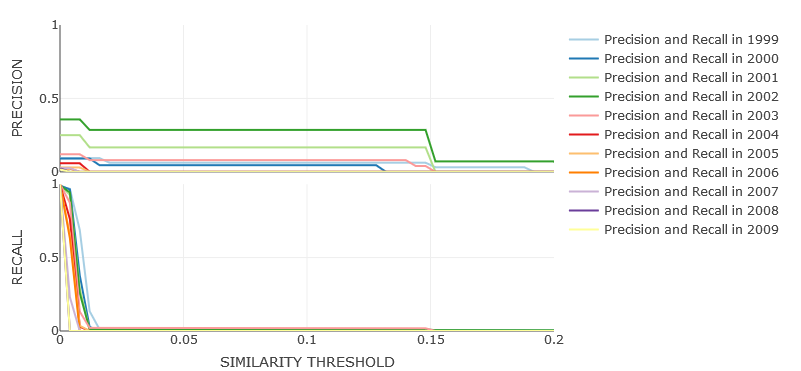}
  \caption{Results in terms of precision and recall obtained from evaluating the Clique Percolation Method against the gold standard after filtering the communities semantically.}
  \label{fig:finale-evaluation-with-filter-cpm}
\end{figure}

\begin{figure}[!ht]
\centering
  \includegraphics[width=\linewidth]{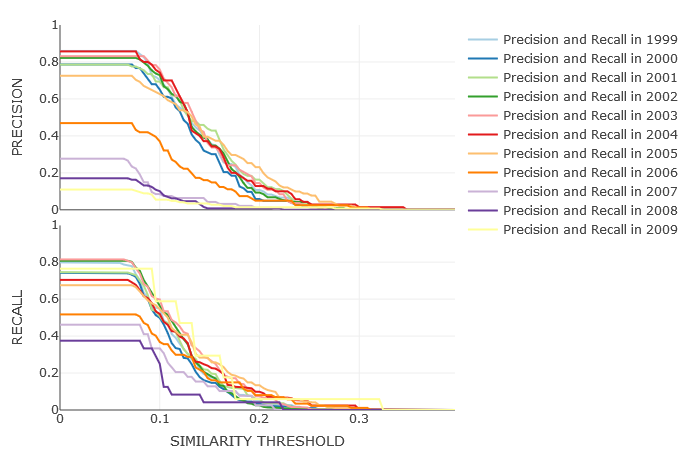}
  \caption{Results in terms of precision and recall obtained from evaluating the Advanced Clique Percolation Method against the gold standard after filtering the communities semantically.}
  \label{fig:finale-evaluation-with-filter}
\end{figure}

All the communities that succeeded the filtering phase were then evaluated against the gold standard. 
In Fig. \ref{fig:finale-evaluation-with-filter-fg}, \ref{fig:finale-evaluation-with-filter-le}, \ref{fig:finale-evaluation-with-filter-fcm}, \ref{fig:finale-evaluation-with-filter-cpm}, \ref{fig:finale-evaluation-with-filter} are reported the new values of precision and recall with different values of similarity, obtained each year from all the tested approaches. Table \ref{tab:comparison-before-after-filter}, for the sake of comparison, provides an index of the figures reporting the results for the various methods, before and after applying the filter. Indeed, comparing these results with the ones obtained without filtering, we can observe an improvement in precision while the values of recall are equivalent. 
Looking at the definitions of precision and recall, described in Section \ref{sec:performance}, these results suggest that we managed to keep all the communities that were matching with the debutant topics and we also managed to remove all those communities that were having a negative impact on the values of precision.

In conclusion, applying a semantic filter we observed a substantial improvement in the results, because it allows us to remove some outlier communities. Nonetheless we would like to investigate whether it is possible to further improve these results and in the next section we will show how these results have been further improved and how semantics keeps playing an important role in these enhancements. At this stage an improvement of the results means to let both precision and recall be high also for higher values of similarity. 
\cleardoublepage
\newpage
\newpage
\newpage
\section{Semantic enhancement}\label{sec:semanticenhancement}

We already observed in several cases that the adoption of semantics, e.g., taking advantage of the Computer Science Ontology, can help to improve the results. The ontology provided all the topics with \textit{same-as} relationships, so that we could produce a better match between clusters and debutant topics (see Eq. \ref{eq:jaccard1}) and it also helped to filter the communities. With the aim of understanding whether the current results, showed in Fig. \ref{fig:finale-evaluation-with-filter-fg}, \ref{fig:finale-evaluation-with-filter-le}, \ref{fig:finale-evaluation-with-filter-fcm}, \ref{fig:finale-evaluation-with-filter-cpm}, can be further improved, both clusters and the ancestors of the debutant topics have been enhanced using the \textit{skos:broaderGeneric} relationship within the Computer Science Ontology. This enhancement consists of merging clusters and debutant topics with their \textit{super-areas}. 
For instance, enhancing a set $Z_x$, which can be either a cluster or the ancestors of a debutant topic, consists of selecting all the super-areas of its topics and including them into the original set, as showed in Eq. \ref{eq:enhancements}. 

\begin{equation}\label{eq:enhancements}
\begin{gathered}
  {Z_x}\,\,:\,given\,set\,(either\,clusters\,or\,ancestors) \hfill \\
  topics({Z_x})\,\,:\, returns\,topics\,in\,that\,set \hfill \\
  SuperAreas({t_x})\,\,:\, returns\,the\,SuperArea\,of\,a\,topic \hfill \\
  enhanced({Z_x}) = topics({Z_x}) \cup SuperAreas(topics({Z_x})) \hfill \\
\end{gathered} 
\end{equation}


Given the possibility of enhancing both clusters and ancestors with their super-areas, we tested all the possible strategies:
\begin{enumerate}[label={Approach \arabic*.},leftmargin=3cm]
\item Enhancement of clusters with super-areas against only the ancestors of debutant topics, labelled also as \textbf{(C $\cup$ Sup) vs. D}
\item Enhancement of debutant topics with super-areas against topics in the cluster, \textbf{C vs. (D $\cup$ Sup)}
\item Enhancement of clusters with super-areas against enhancement of debutant topics with super-areas, \textbf{(C $\cup$ Sup) vs. (D $\cup$ Sup)}

\end{enumerate}

After enhancing both sets of clusters and debutant topics, we ran the different evaluations according to the previous three configurations. To this end, the similarity measure, defined in Eq. \ref{eq:jaccard1} at page \pageref{eq:jaccard1}, needs to be revisited so that it can consider these new enhanced sets. 
Equation \ref{eq:new-jaccard-index} defines the new similarity measure, which combines clusters and debutant topics with their respective enhancements. It computes the Jaccard similary between the two enhanced sets: $({C_i} \cup S{A_i} \cup E{C_i})$ and $({D_k} \cup E{D_k})$. The first enhanced set includes the topics of the original cluster $C_i$, their \textit{same-as} topics $S{A_i}$ and their super-areas in $E{C_i}$. Instead, the second enhanced set includes the ancestors of the debutant topics $D_k$ and their super-areas $E{D_k}$. This new similarity measure is more comprehensive as it can suit all three approaches. 

In particular, when we test the first approach, we do not enhance the ancestors of debutant topic, therefore $E{D_k}\, = \,\emptyset $. Whereas, for the second approach $E{C_i}\, = \,\emptyset $, as we do not enhance the clusters. In addition, we can also observe that when there are no enhancements, i.e., $E{C_i}\, = \,E{D_k}\, = \,\emptyset$, Eq. \ref{eq:new-jaccard-index} is similar to Eq. \ref{eq:jaccard1}. We did not include the \textit{same-as} topics of the cluster $S{A_i}$ in the denominator for the same reasons ponted out when describing Eq. \ref{eq:jaccard1}. 


\begin{equation}\label{eq:new-jaccard-index}
\begin{gathered}
  {C_i} \to i \mhyphen th\,Cluster \hfill \\
  {D_k} \to k \mhyphen th\,Debutant\,topic \hfill \\
  S{A_i} \to \,sameAs\,of\,i \mhyphen th\,Cluster \hfill \\
  E{C_i} \to \,enhancement\,of\,i \mhyphen th\,Cluster \hfill \\
  E{D_k} \to enhancement\,of\,k \mhyphen th\,Debutant\,topic \hfill \\
J\left( {{C_i},{D_k},S{A_i},E{C_i},E{D_k}} \right){\kern 1pt}  = {\kern 1pt} \frac{{\left| {({C_i} \cup S{A_i} \cup E{C_i}) \cap ({D_k} \cup E{D_k})} \right|}}{{\left| {{C_i} \cup E{C_i} \cup {D_k} \cup E{D_k}} \right|}}
 \hfill \\ 
\end{gathered} 
\end{equation}

\begin{figure}[!ht]
\centering
  \includegraphics[width=\linewidth]{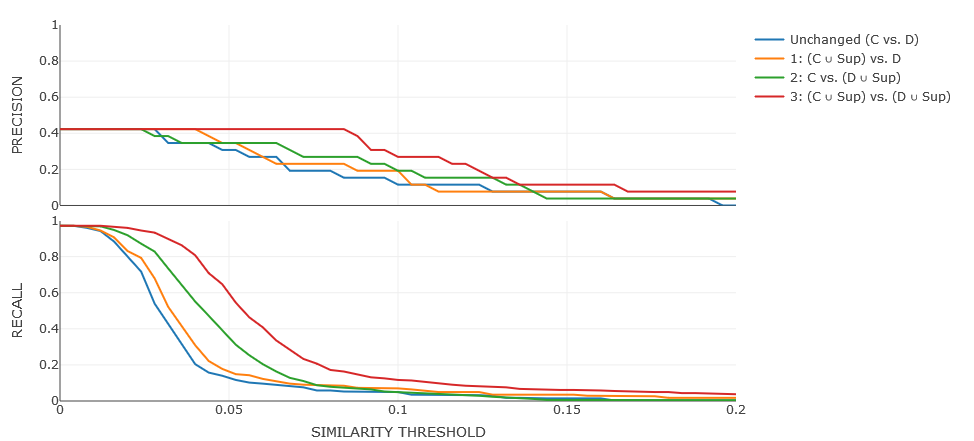}
  \caption{Results in terms of precision and recall obtained from evaluating the Fast Greedy in year 1999 with different solution of semantic enhancement.}
  \label{fig:strategy1999-fg}
\end{figure}
\begin{figure}[!ht]
\centering
  \includegraphics[width=\linewidth]{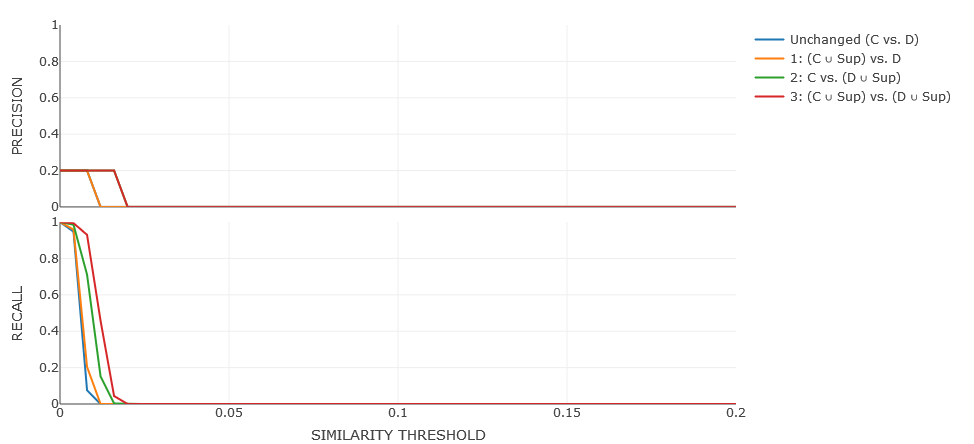}
  \caption{Results in terms of precision and recall obtained from evaluating the Leading Eigenvector in year 1999 with different solution of semantic enhancement.}
  \label{fig:strategy1999-le}
\end{figure}
\begin{figure}[!ht]
\centering
  \includegraphics[width=\linewidth]{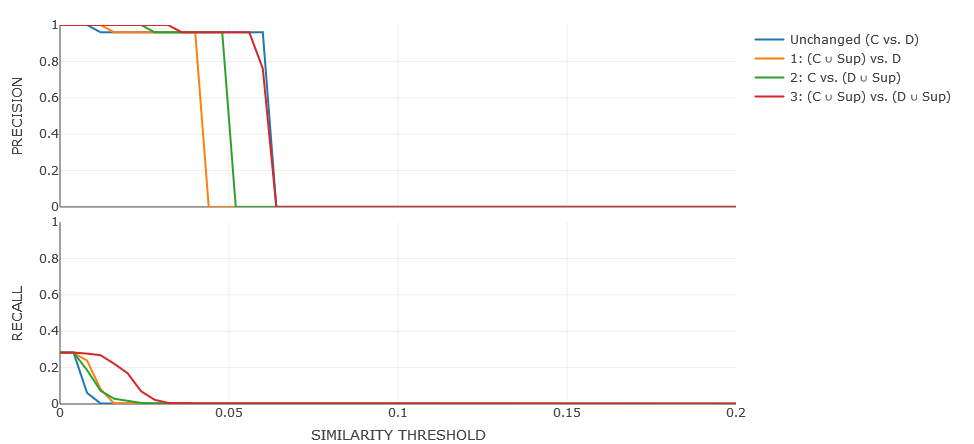}
  \caption{Results in terms of precision and recall obtained from evaluating the Fuzzy C-Means in year 1999 with different solution of semantic enhancement.}
  \label{fig:strategy1999-fcm}
\end{figure}
\begin{figure}[!ht]
\centering
  \includegraphics[width=\linewidth]{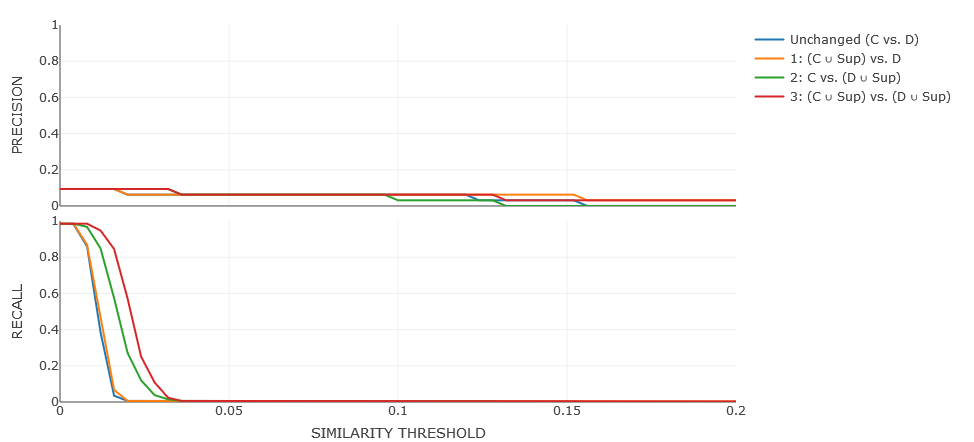}
  \caption{Results in terms of precision and recall obtained from evaluating the Clique Percolation Method in year 1999 with different solution of semantic enhancement.}
  \label{fig:strategy1999-cpm}
\end{figure}
\begin{figure}[!ht]
\centering
  \includegraphics[width=\linewidth]{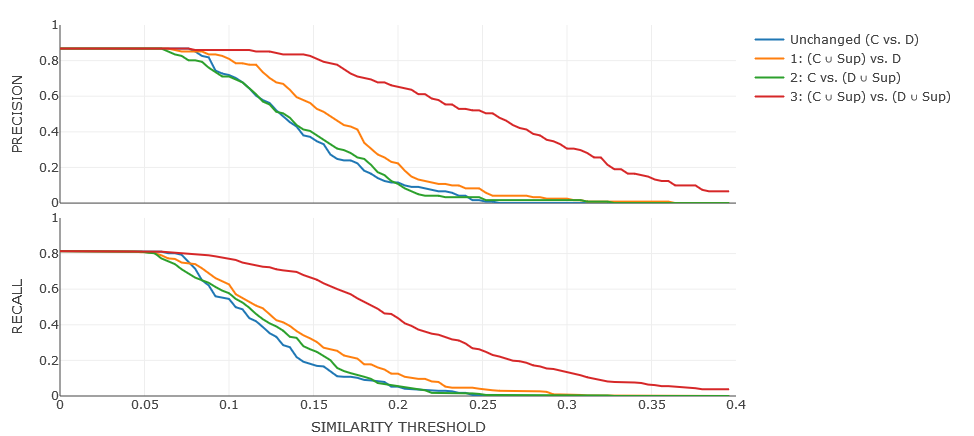}
  \caption{Results in terms of precision and recall obtained from evaluating the Advanced Clique Percolation Method in year 1999 with different solution of semantic enhancement.}
  \label{fig:strategy1999}
\end{figure}

We compared the results obtained from the three different strategies, including also the results obtained when applying semantic filtering (Section \ref{sec:semanticfilters}). In Fig. \ref{fig:strategy1999-fg}, \ref{fig:strategy1999-le}, \ref{fig:strategy1999-fcm}, \ref{fig:strategy1999-cpm} and \ref{fig:strategy1999} are reported the results in terms of precision and recall for the year 1999 for the five community detection algorithms. Each subplot reports four lines, three of different strategies and one, coloured in blue, that reports the results of the unchanged clusters and debutant topics. The results for Advanced Clique Percolation Method are reported in Fig. \ref{fig:finale-evaluation-with-filter}, for the other four algorithms in Fig. \ref{fig:finale-evaluation-with-filter-fg}, \ref{fig:finale-evaluation-with-filter-le}, \ref{fig:finale-evaluation-with-filter-fcm} and \ref{fig:finale-evaluation-with-filter-cpm}.

From all these graph charts we can acknowledge that the adoption of the Computer Science Ontology to enhance the cluster and the debutant topics improved the results. Focusing on the results of the Advanced Clique Percolation Method in Fig. \ref{fig:strategy1999}, choosing a similarity value of 0.1, for the \textit{unchanged} approach (blue line), the precision is around 0.7 and the recall 0.5. Whereas, at the same similarity value, the new strategies have higher values of precision with a peak around 0.85, as well as high values of recall that reaches 0.77 for the third strategy (red line). A similar behaviour occurs for the results of other algorithms, such as the ones for Fast Greedy in Fig. \ref{fig:strategy1999-fg}.

In general, we can see that all these approaches 
provide improved results that also appear to be far from the ones obtained with the unchanged approach (blue line). However, approach 3 (red line), which compares enhanced clusters and enhance debutant topics, provides even better results. Indeed, for all the tested algorithms, we can see that, for equal values of similarity, the third strategy has always higher values of precision and recall. 

\begin{figure}[ht]
\centering
  \includegraphics[width=\linewidth]{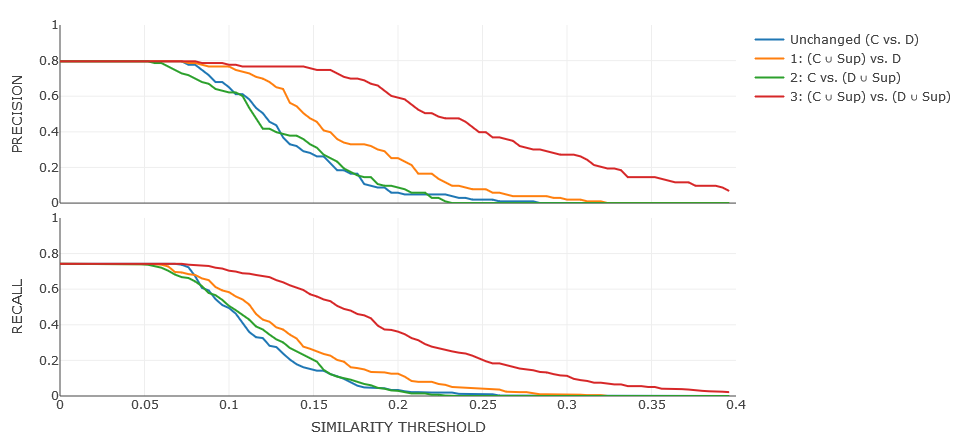}
  \caption{Results in terms of precision and recall obtained from evaluating the Advanced Clique Percolation Method in year 2000 with different solution of semantic enhancement.}
  \label{fig:strategy2000}
\end{figure}

Performing this evaluation in the years from 2000 to 2009, we observed a similar behaviour for all the five community detection algorithms. For instance, in Fig. \ref{fig:strategy2000} are reported the values of precision and recall for year 2000 for the Advanced Clique Percolation Method.

As we can see, the different approaches perform in a similar way and approach 3 outperforms the others. In a similar way, the approach 3 outperforms the other approaches in all the remaining years (2001-2009) and it also does it for the other four community detection algorithms. 

\begin{figure}[ht]
\centering
  \includegraphics[width=\linewidth]{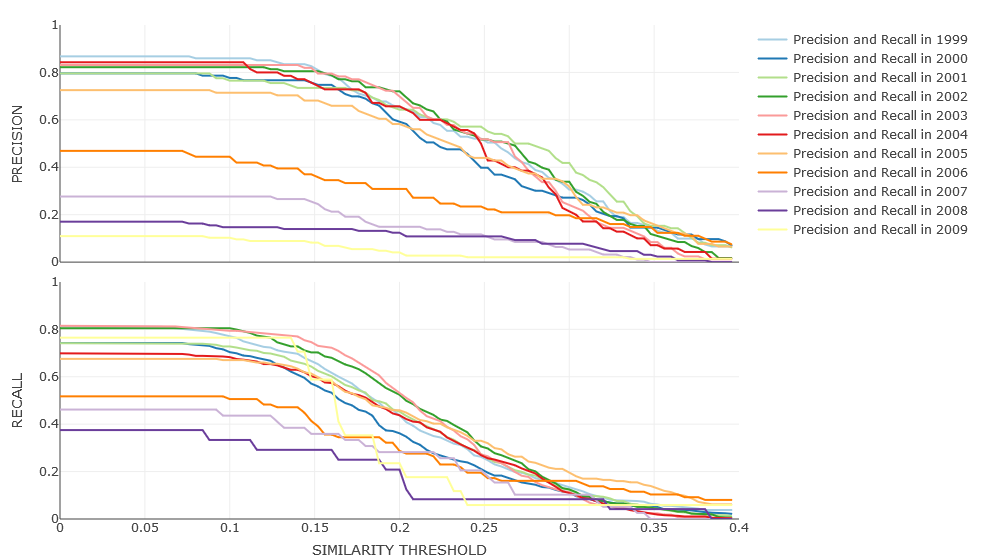}
  \caption{Results in terms of precision and recall obtained from evaluating the Advanced Clique Percolation Method after enhancing both clusters and debutant topics with their super-areas.}
  \label{fig:finale-all}
\end{figure}

Since the third approach is performing better than others, in Fig. \ref{fig:finale-all} are reported the values of precision and recall of its application on the Advanced Clique Percolation Method during the period of analysis (1999-2009). 

From the results, we can observe that from year 1999 to 2005 the values of precision and recall are quite high. Indeed, for a similarity value of 0.2, for those years, the values of precision go from 0.6 to 0.72, while the values of recall go from 0.36 to 0.55. However, those results start to gradually decrease for the years 2006 to 2009. We can see that for year 2006, for a similarity value of 0.2, the precision is 0.3 while the recall is 0.28. For the years from 2007 to 2009 the different values of precision are below 0.2 and values of recall are between 0.2 and 0.3.

\begin{figure}[ht]
\centering
  \includegraphics[width=\linewidth]{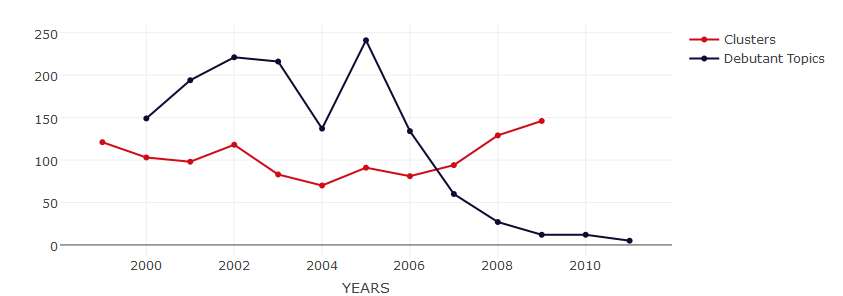}
  \caption{Comparison between the amount of debutant topics and cluster, retrieved each year.}
  \label{fig:debutant-vs-clusters}
\end{figure}

This decrease in performance can be justified by the fact that while the number of clusters returned from the evolutionary networks keeps growing over time, the number of debutant topics is decreasing, as showed in Fig. \ref{fig:debutant-vs-clusters}.

\begin{figure}[ht]
\centering
  \includegraphics[width=\linewidth]{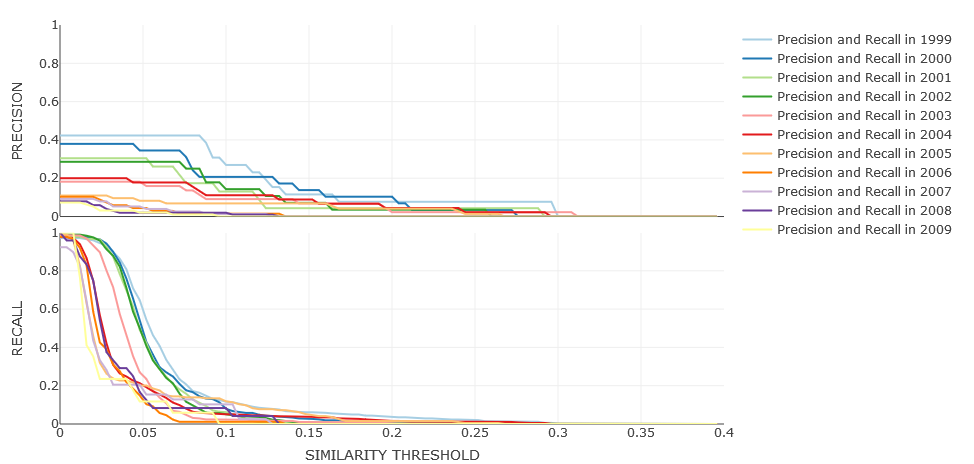}
  \caption{Results in terms of precision and recall obtained from evaluating the Fast Greedy after enhancing both clusters and debutant topics with their super-areas.}
  \label{fig:finale-all-fg}
\end{figure}

\begin{figure}[ht]
\centering
  \includegraphics[width=\linewidth]{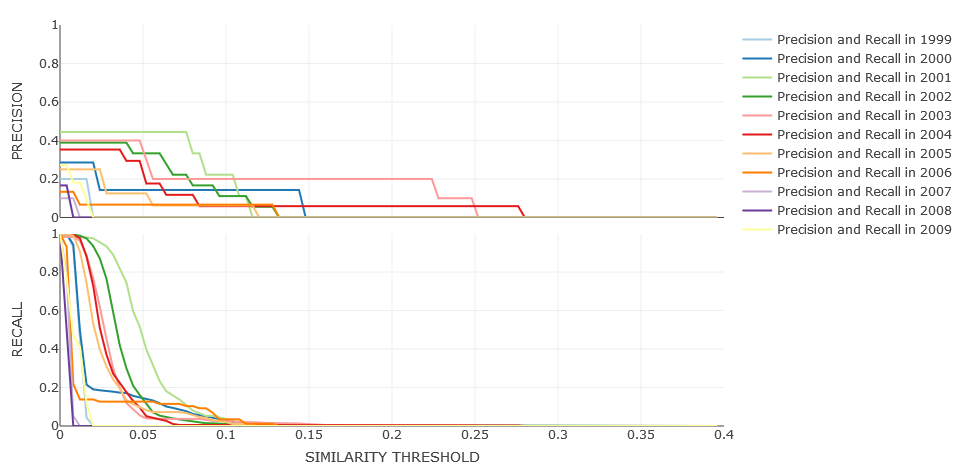}
  \caption{Results in terms of precision and recall obtained from evaluating the Leading Eigenvector after enhancing both clusters and debutant topics with their super-areas.}
  \label{fig:finale-all-le}
\end{figure}

\begin{figure}[ht]
\centering
  \includegraphics[width=\linewidth]{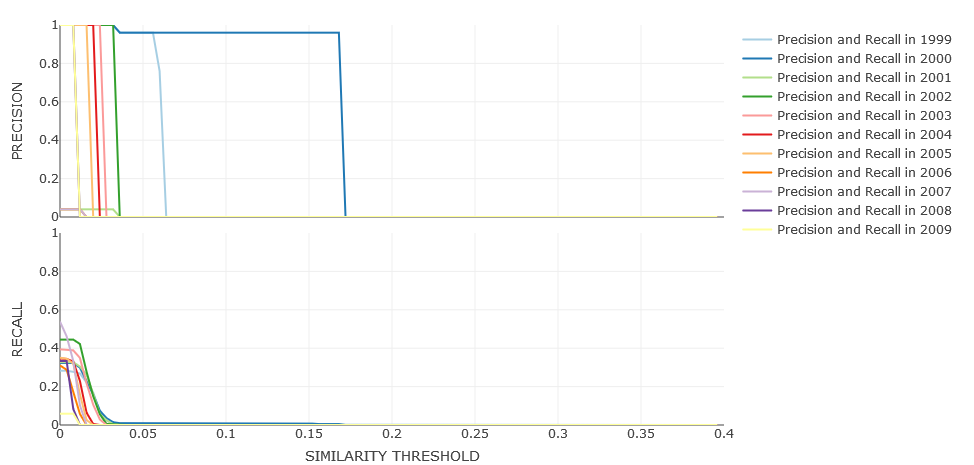}
  \caption{Results in terms of precision and recall obtained from evaluating the Fuzzy C-Means after enhancing both clusters and debutant topics with their super-areas.}
  \label{fig:finale-all-fcm}
\end{figure}

\begin{figure}[ht]
\centering
  \includegraphics[width=\linewidth]{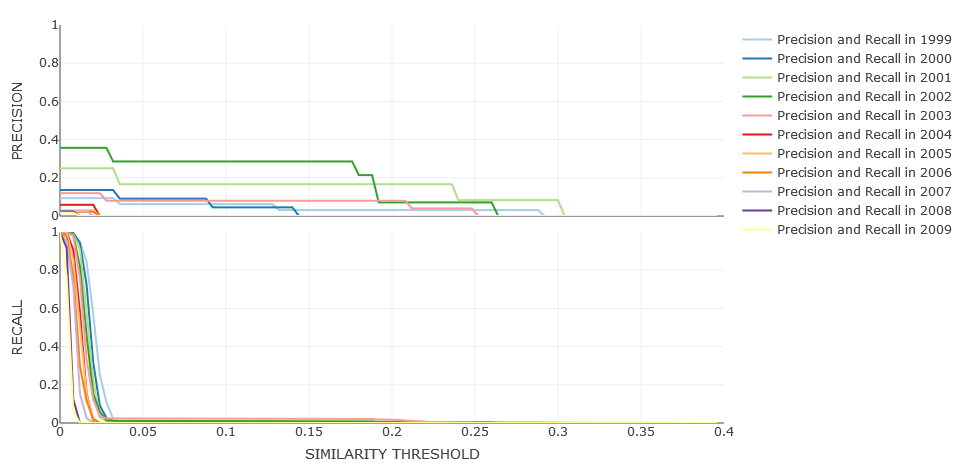}
  \caption{Results in terms of precision and recall obtained from evaluating the Clique Percolation Method after enhancing both clusters and debutant topics with their super-areas.}
  \label{fig:finale-all-cpm}
\end{figure}

In Fig. \ref{fig:finale-all-fg}, \ref{fig:finale-all-le}, \ref{fig:finale-all-fcm}, \ref{fig:finale-all-cpm}, are reported the new values of precision and recall with different values of similarity, when using the third approach respectively Fast Greedy, Leading Eigenvector, Fuzzy C-Means and Clique Percolation Method. 
If we compare these results with those obtained when applying semantic filtering only (see Section \ref{sec:semanticfilters}) we can observe, for all approaches, an increase of performance in both precision and recall.
In Table \ref{tab:complete-comparison-original-semantic} are reported the different indexes of figures showing the results across different sections: only clustering (Section \ref{sec:labelling-the-clusters}), applying semantic filtering (see Section \ref{sec:semanticfilters}) and this semantic enhancement.

\begin{table}[!ht]
\centering
\caption{Figures comparison for each approach, when using only the clustering techniques, when applying semantic filtering and when enhancing semantically the set of clusters and debutants.}
\label{tab:complete-comparison-original-semantic}
{\renewcommand{\arraystretch}{1.4}%
\begin{tabular}{l|l|l|l}
\hline
\textbf{Method} & \textbf{Only clustering} & \textbf{Semantic filtering} & \textbf{Semantic Enhancement} \\ \hline
FG & Figure \ref{fig:results-evaluation-fg} at p. \pageref{fig:results-evaluation-fg}   & Figure \ref{fig:finale-evaluation-with-filter-fg} at p. \pageref{fig:finale-evaluation-with-filter-fg} & Figure \ref{fig:finale-all-fg} at p. \pageref{fig:finale-all-fg} \\
LE & Figure \ref{fig:results-evaluation-le} at p. \pageref{fig:results-evaluation-le} & Figure \ref{fig:finale-evaluation-with-filter-le} at p. \pageref{fig:finale-evaluation-with-filter-le} & Figure \ref{fig:finale-all-le} at p. \pageref{fig:finale-all-le} \\
FCM & Figure \ref{fig:results-evaluation-fcm} at p. \pageref{fig:results-evaluation-fcm} & Figure \ref{fig:finale-evaluation-with-filter-fcm} at p. \pageref{fig:finale-evaluation-with-filter-fcm} & Figure \ref{fig:finale-all-fcm} at p. \pageref{fig:finale-all-fcm} \\
CPM & Figure \ref{fig:results-evaluation-cpm} at p. \pageref{fig:results-evaluation-cpm} & Figure \ref{fig:finale-evaluation-with-filter-cpm} at p. \pageref{fig:finale-evaluation-with-filter-cpm} & Figure \ref{fig:finale-all-cpm} at p. \pageref{fig:finale-all-cpm} \\
ACPM & Figure \ref{fig:finale-evaluation-no-filter} at p. \pageref{fig:finale-evaluation-no-filter} & Figure \ref{fig:finale-evaluation-with-filter} at p. \pageref{fig:finale-evaluation-with-filter} & Figure \ref{fig:finale-all} at p. \pageref{fig:finale-all} \\ \hline
\end{tabular}}
\end{table}

In brief, by observing the results, we can see that applying semantic filters and enhancing both the topics in the communities and the ancestors of debutant topics boost the performance of each approach, with the Advanced Clique Percolation Method outperforming the other four tested algorithm.

Table \ref{tab:final-comparison-results0.1}, \ref{tab:final-comparison-results0.15} and \ref{tab:final-comparison-results0.2} provide another perspective on the obtained results. They show the results with a fixed similarity threshold, Table \ref{tab:final-comparison-results0.1} for similarity threshold equal to 0.1, Table \ref{tab:final-comparison-results0.15} for 0.15, and Table \ref{tab:final-comparison-results0.2} for 0.2. As we can see the ACPM algorithm, for a given similarity threshold outperforms the other four algorithms. 
A noticeable outlier in Table \ref{tab:final-comparison-results0.1} and \ref{tab:final-comparison-results0.15} is in the results of Fuzzy C-Means (FCM) in the year 2000, where we can find 0.96 of precision. This aberration is due to three topics that were constantly encountered, which were matching the ancestor of a debutant topic, as explained in Section \ref{sec:result-fcm}.


\begin{table}[!ht]
\centering
\caption{Values of Precision and Recall for the five approaches along time, at similarity value of 0.1. In bold the best results.}
\label{tab:final-comparison-results0.1}
{\renewcommand{\arraystretch}{1.2}%
\begin{tabular}{L{1cm}|L{1cm}L{1cm}|L{1cm}L{1cm}|L{1cm}L{1cm}|L{1cm}L{1cm}|L{1cm}L{1cm}}
\cline{2-11}
 & \multicolumn{2}{c|}{FG} & \multicolumn{2}{c|}{LE} & \multicolumn{2}{c|}{FCM} & \multicolumn{2}{c|}{CPM} & \multicolumn{2}{c}{\textbf{ACPM}} \\ \hline
\multicolumn{1}{L{1cm}|}{Years} & \multicolumn{1}{L{1cm}|}{Pr} & Re & \multicolumn{1}{L{1cm}|}{Pr} & Re & \multicolumn{1}{L{1cm}|}{Pr} & Re & \multicolumn{1}{L{1cm}|}{Pr} & Re & \multicolumn{1}{L{1cm}|}{\textbf{Pr}} & \textbf{Re}  \\ \hline
1999  & .27 & .11 & .00 & .00  & .00 & .00 & .06 & .01 & \textbf{.86} & \textbf{.76} \\
2000  & .21 & .07 & .14 & .02  & .96 & .01 & .05 & .00 & \textbf{.78} & \textbf{.70} \\
2001  & .13 & .04 & .11 & .01  & .00 & .00 & .17 & .00 & \textbf{.77} & \textbf{.72} \\
2002  & .14 & .04 & .11 & .01  & .00 & .00 & .29 & .01 & \textbf{.82} & \textbf{.80} \\
2003  & .09 & .02 & .20 & .02  & .00 & .00 & .08 & .02 & \textbf{.83} & \textbf{.79} \\
2004  & .11 & .05 & .06 & .00  & .00 & .00 & .00 & .00 & \textbf{.84} & \textbf{.68} \\
2005  & .07 & .11 & .06 & .01  & .00 & .00 & .00 & .00 & \textbf{.71} & \textbf{.66} \\
2006  & .01 & .01 & .07 & .01  & .00 & .00 & .00 & .00 & \textbf{.43} & \textbf{.51} \\
2007  & .01 & .08 & .00 & .00  & .00 & .00 & .00 & .00 & \textbf{.28} & \textbf{.44} \\
2008  & .01 & .04 & .00 & .00  & .00 & .00 & .00 & .00 & \textbf{.15} & \textbf{.33} \\
2009  & .00 & .00 & .00 & .00  & .00 & .00 & .00 & .00 & \textbf{.09} & \textbf{.76}\\ \hline
\end{tabular}}
\end{table}

\begin{table}[!ht]
\centering
\caption{Values of Precision and Recall for the five approaches along time, at similarity value of 0.15. In bold the best results.}
\label{tab:final-comparison-results0.15}
{\renewcommand{\arraystretch}{1.2}%
\begin{tabular}{L{1cm}|L{1cm}L{1cm}|L{1cm}L{1cm}|L{1cm}L{1cm}|L{1cm}L{1cm}|L{1cm}L{1cm}}
\cline{2-11}
 & \multicolumn{2}{c|}{FG} & \multicolumn{2}{c|}{LE} & \multicolumn{2}{c|}{FCM} & \multicolumn{2}{c|}{CPM} & \multicolumn{2}{c}{\textbf{ACPM}} \\ \hline
\multicolumn{1}{L{1cm}|}{Years} & \multicolumn{1}{L{1cm}|}{Pr} & Re & \multicolumn{1}{L{1cm}|}{Pr} & Re & \multicolumn{1}{L{1cm}|}{Pr} & Re & \multicolumn{1}{L{1cm}|}{Pr} & Re & \multicolumn{1}{L{1cm}|}{\textbf{Pr}} & \textbf{Re}  \\ \hline
1999  & .12 & .06 & .00 & .00  & .00 & .00 & .03 & .00 & \textbf{.81} & \textbf{.65} \\
2000  & .14 & .02 & .00 & .00  & .96 & .01 & .00 & .00 & \textbf{.75} & \textbf{.55} \\
2001  & .04 & .00 & .00 & .00  & .00 & .00 & .17 & .00 & \textbf{.73} & \textbf{.63} \\
2002  & .07 & .01 & .00 & .00  & .00 & .00 & .29 & .01 & \textbf{.81} & \textbf{.70} \\
2003  & .07 & .01 & .20 & .01  & .00 & .00 & .08 & .02 & \textbf{.80} & \textbf{.73} \\
2004  & .07 & .03 & .06 & .00  & .00 & .00 & .00 & .00 & \textbf{.74} & \textbf{.58} \\
2005  & .05 & .04 & .00 & .00  & .00 & .00 & .00 & .00 & \textbf{.68} & \textbf{.59} \\
2006  & .00 & .00 & .00 & .00  & .00 & .00 & .00 & .00 & \textbf{.35} & \textbf{.38} \\
2007  & .00 & .00 & .00 & .00  & .00 & .00 & .00 & .00 & \textbf{.24} & \textbf{.36} \\
2008  & .00 & .00 & .00 & .00  & .00 & .00 & .00 & .00 & \textbf{.14} & \textbf{.25} \\
2009  & .00 & .00 & .00 & .00  & .00 & .00 & .00 & .00 & \textbf{.07} & \textbf{.59} \\ \hline
\end{tabular}}
\end{table}

\begin{table}[!ht]
\centering
\caption{Values of Precision and Recall for the five approaches along time, at similarity value of 0.2. In bold the best results.}
\label{tab:final-comparison-results0.2}
{\renewcommand{\arraystretch}{1.2}%
\begin{tabular}{L{1cm}|L{1cm}L{1cm}|L{1cm}L{1cm}|L{1cm}L{1cm}|L{1cm}L{1cm}|L{1cm}L{1cm}}
\cline{2-11}
 & \multicolumn{2}{c|}{FG} & \multicolumn{2}{c|}{LE} & \multicolumn{2}{c|}{FCM} & \multicolumn{2}{c|}{CPM} & \multicolumn{2}{c}{\textbf{ACPM}} \\ \hline
\multicolumn{1}{L{1cm}|}{Years} & \multicolumn{1}{L{1cm}|}{Pr} & Re & \multicolumn{1}{L{1cm}|}{Pr} & Re & \multicolumn{1}{L{1cm}|}{Pr} & Re & \multicolumn{1}{L{1cm}|}{Pr} & Re & \multicolumn{1}{L{1cm}|}{\textbf{Pr}} & \textbf{Re}  \\ \hline
1999  & .08 & .03 & .00 & .00  & .00 & .00 & .03 & .00 & \textbf{.64} & \textbf{.41} \\
2000  & .03 & .00 & .00 & .00  & .00 & .00 & .00 & .00 & \textbf{.57} & \textbf{.33} \\
2001  & .04 & .00 & .00 & .00  & .00 & .00 & .17 & .00 & \textbf{.63} & \textbf{.44} \\
2002  & .04 & .00 & .00 & .00  & .00 & .00 & .07 & .01 & \textbf{.70} & \textbf{.50} \\
2003  & .02 & .00 & .20 & .01  & .00 & .00 & .08 & .02 & \textbf{.70} & \textbf{.51} \\
2004  & .04 & .02 & .06 & .00  & .00 & .00 & .00 & .00 & \textbf{.66} & \textbf{.42} \\
2005  & .04 & .02 & .00 & .00  & .00 & .00 & .00 & .00 & \textbf{.56} & \textbf{.43} \\
2006  & .00 & .00 & .00 & .00  & .00 & .00 & .00 & .00 & \textbf{.30} & \textbf{.28} \\
2007  & .00 & .00 & .00 & .00  & .00 & .00 & .00 & .00 & \textbf{.15} & \textbf{.28} \\
2008  & .00 & .00 & .00 & .00  & .00 & .00 & .00 & .00 & \textbf{.12} & \textbf{.17} \\
2009  & .00 & .00 & .00 & .00  & .00 & .00 & .00 & .00 & \textbf{.03} & \textbf{.24} \\ \hline
\end{tabular}}
\end{table}

\cleardoublepage
\newpage
\newpage
\newpage
\section{Discussion}\label{sec:discussion}

Throughout this chapter, we analysed the different steps that we have undertaken to assess the validity of Augur and the other clustering algorithms. We started with a number of community detection algorithms that were available in the state of the art. We saw that they do not provide good results but, at the same time, they allowed us to understand the specialities of the evolutionary networks we were dealing with. Hence, we realised that we needed a new approach, which could take into account all the characteristics of the networks, e.g., weights, and could also cope with the fact the networks appear to be very dense in certain areas. The Advanced Clique Percolation Method provided good results, and we succeeded in improving them further by means of semantic filtering and semantic enhancement techniques. These allowed us to increase precision and recall at higher values of similarity. 

Nonetheless, there are still some issues that need to be further investigated. Specifically, the sections that follow seek to explore the following questions:
\begin{itemize}
\item What is the appropriate value of similarity for a match?
\item Why precision and recall decrease in the last years of the analysed decade?
\item What else can be done to improve the results?
\end{itemize}
We will discuss these aspects to better assess the validity of Augur.

\subsection{When does a cluster match a debutant topic?}\label{sec:discussion1}
The aim of the evaluation is to determine the effectiveness of Augur in returning a set of clusters containing topics that are increasing their degree of collaboration over time, which will eventually lead to the emergence of a new research area.
In brief, to evaluate such effectiveness, we performed the following steps:
\begin{itemize}
\item run the five community detection algorithms on the evolutionary networks form 1999 to 2009, according to Table \ref{tab:evaluation-schema};
\item for each approach, compare the clusters returned each year with the ancestors of areas that actually emerged in the following two years (see Table \ref{tab:match-evg-deb} at page \pageref{tab:match-evg-deb}) for different similarity values.
\end{itemize} 


Throughout this chapter, the results in terms of precision and recall of each approach, have been presented as a function of the similarity value. In particular, observing the results we obtained for the last approach (see Fig. \ref{fig:finale-all}), a similarity value that offers a good balance in performance and is also valid across the years is being considered 0.2. Given this value, precision stays high (0.6-0.71) for the years from 1999 to 2005, while recall is in the range 0.43 to 0.63.  

Table \ref{tab:comparison-semantic-search} reports two comparisons between communities and ancestors of debutant topics that are considered similar because their similarity is above 0.2. In both cases the clusters match with the \textit{Semantic Search} topic. This phenomenon, in which two or more clusters can match the ancestors of a specific debutant topic, is recurrent in the results and will be further investigated in the next section, while here we will focus on the chosen similarity value.

From the table, we can see that 14 out of 19 topics within the first cluster (marked in bold) match ancestors (also marked in bold) of the topic \textit{Semantic Search}, which debuted in 2003. The similarity value between the community and the debutant topic is 0.425. 
A topic marked in bold and appearing in only one set (either cluster or ancestors fo debutant topic), matched thanks to the semantic enhancement.

For the second community, instead, 16 out of 19 topics in the cluster matched the ancestors of \textit{Semantic Search} and their similarity stands at 0.413. According to these results it seems that a cluster covers particularly well the debutant topics with a similarity above 0.4. So far we observed similarities between clusters and debutant topics that double the chosen similarity value of 0.2. \textbf{So, what can be said when there is a match with similarity of 0.2?}

\begin{table}[!htbp]
\centering
\caption{Matchings between cluster and debutant topics with similarity measure over 0.2. In this case the two clusters match the same debutant topic: \textit{Semantic Search}}
\label{tab:comparison-semantic-search}
\begin{tabular}{|p{5cm}|p{6cm}|p{3cm}|}
\hline
\textbf{Cluster                                                                                                                                                                                                                                                                                                                                                                      } & \textbf{Ancestors of debutant topic                                                                                                                                                                                                                                                                                                                                                                                                                                                      } & \textbf{Debutant topic                                                                             } \\ \hline
\textbf{world wide web}, \textbf{query languages}, \textbf{metadata}, \textbf{content based retrieval}, \textbf{information retrieval}, \textbf{search engines}, \textbf{xml}, \textbf{information systems}, \textbf{information retrieval systems}, \textbf{servers}, \textbf{digital libraries}, \textbf{indexing (of information)}, \textbf{text processing}, \textbf{electronic commerce}, multi agent systems, intelligent agents, information management, web browsers, classification (of information) & \textbf{world wide web}, \textbf{query languages}, \textbf{metadata}, \textbf{content based retrieval}, \textbf{information retrieval}, \textbf{search engines}, \textbf{xml}, \textbf{information systems}, \textbf{information retrieval systems}, \textbf{search engine}, \textbf{hypertext systems}, \textbf{query processing}, \textbf{natural language processing systems}, safety devices, semantics, mathematical expressions, enabling technologies, websites, human computer interaction, social networking (online), current system, ontology, tools, internet, telecommunication equipment & Semantic Search (debuting in 2003 and according to Eq. \ref{eq:new-jaccard-index} the value of similarity is 0.425) \\ \hline
\textbf{world wide web}, \textbf{query languages}, \textbf{metadata}, \textbf{content based retrieval}, \textbf{information retrieval}, \textbf{search engines}, \textbf{xml}, \textbf{information retrieval systems}, \textbf{servers}, \textbf{client server computer systems}, \textbf{digital libraries}, \textbf{indexing (of information)}, \textbf{html}, \textbf{text processing}, \textbf{network protocols}, \textbf{electronic commerce}, java programming language, information management, web browsers            & \textbf{world wide web}, \textbf{query languages}, \textbf{metadata}, \textbf{content based retrieval}, \textbf{information retrieval}, \textbf{search engines}, \textbf{xml}, \textbf{information retrieval systems}, \textbf{search engine}, \textbf{hypertext systems}, \textbf{query processing}, \textbf{natural language processing systems}, safety devices, semantics, mathematical expressions, enabling technologies, websites, human computer interaction, social networking (online), information systems, current system, ontology, tools, internet, telecommunication equipment & Semantic Search (value of similarity is 0.413)                                              \\ \hline
\end{tabular}
\end{table}

\begin{figure}[ht]
\centering
  \includegraphics[width=\linewidth]{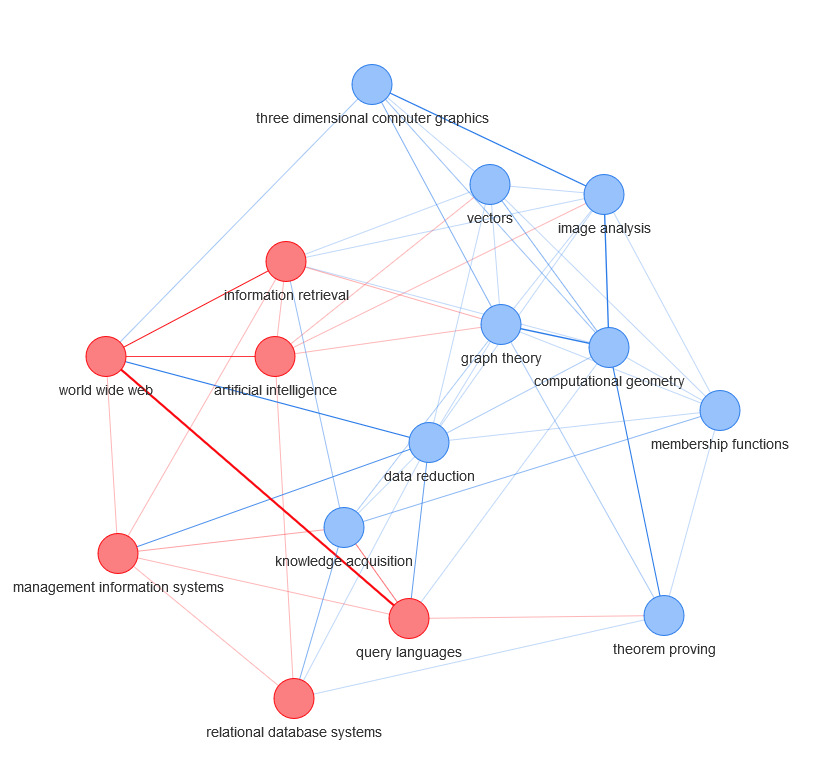}
  \caption{Community extracted from the evolutionary network of year 1999 that matches with \textit{Resource Description Framework} debuting in 2000. Red nodes are the matched topics, blue nodes are the unmatched topics.}
  \label{fig:rdf-graph}
\end{figure}

In Table \ref{tab:match0_2}, we report two comparisons between clusters and debutant topics with values of similarity just above 0.2. As we can see, in the first comparison 6 of the 15 topics in the cluster match the ancestors of \textit{Resource Description Framework} with a similarity measure of 0.21. However, looking at the list of topics within the cluster, among those that did not match there are topics that could be considered very far from the debutant topic. In particular, \textit{Three Dimensional Computer Graphics}, \textit{Theorem Proving}, and \textit{Image analysis} are unrelated to the first topics that match \textit{Resource Description Framework}. Looking at the network in Fig. \ref{fig:rdf-graph}, we can actually observe the existence of two different sections that are grouped together, the topics that matched (and coloured in red) and the ones that did not match (coloured in blue). We can observe that since there are many links that cross the two groups, the algorithm is unable to split this cluster in the two parts. Nonetheless, the results are still valuable.
With a similar value of similarity, the second cluster matches 5 of 9 topics with the ancestors of debutant topics. These results suggest that with only 5 topics, the match between the two sets is still significant. A future improvement of Augur would be the refinement of the Advanced Clique Percolation Method so that it can further split the network in Fig. \ref{fig:rdf-graph} in further subnetworks. 

\begin{table}[!htbp]
\centering
\caption{Matchings between cluster and debutant topics with similarity measure just above 0.2.}
\label{tab:match0_2}
\begin{tabular}{|p{5cm}|p{6cm}|p{3cm}|}
\hline
\textbf{Cluster                                                                                                                                                                                                                                                                                                                                                                      } & \textbf{Ancestors of debutant topic                                                                                                                                                                                                                                                                                                                                                                                                                                                      } & \textbf{Debutant topic                                                                             } \\ \hline
\textbf{query languages}, \textbf{information retrieval}, \textbf{artificial intelligence}, \textbf{world wide web}, \textbf{management information systems}, \textbf{relational database systems}, computational geometry, image analysis, data reduction, knowledge acquisition, three dimensional computer graphics, membership functions, graph theory, theorem proving, vectors & \textbf{query languages}, \textbf{information retrieval}, \textbf{artificial intelligence}, \textbf{world wide web}, \textbf{management information systems}, \textbf{natural language processing systems}, \textbf{multi agent systems}, \textbf{expert systems}, \textbf{xml}, \textbf{search engines}, semantics, metadata, intelligent agents, interoperability, html, algorithms, data communication systems, security of data, schemas, linguistics, resource allocation, constraint theory, software agents, network protocols, websites                                                               & Resource Description Framework (debuting in 2000\footnotemark{} and according to Eq. \ref{eq:new-jaccard-index} the value of similarity is 0.21) \\ \hline
\textbf{database systems}, \textbf{data mining}, \textbf{data structures}, \textbf{world wide web}, \textbf{knowledge engineering}, learning systems, genetic algorithms, multilayer neural networks, decision theory                                                                                                                                                       & \textbf{database systems}, \textbf{data mining}, \textbf{data structures}, \textbf{world wide web}, \textbf{knowledge engineering}, \textbf{information analysis}, \textbf{semantic web}, semantics, ontology, information systems, knowledge management, knowledge representation, management information systems, mathematical models, information technology, computer aided software engineering, metadata, natural language processing systems, taxonomies, knowledge based systems, software engineering, information retrieval, research, data processing, formal languages & Ontology Engineering (debuting in 2002 and according to Eq. \ref{eq:new-jaccard-index} the value of similarity is 0.21)                                                        \\ \hline
\end{tabular}
\end{table}

\addtocounter{footnote}{-1} 
\stepcounter{footnote}\footnotetext{According to Wikipedia (\url{https://en.wikipedia.org/wiki/Resource_Description_Framework}) the topic \textit{Resource Description Framework} debuted in 1997. However, in our dataset, the first paper tagged with this keyword appears in the year 2000.}

Table \ref{tab:comparison-semantic-search}, reports a case in which two (or in some other cases more) communities match the same debutant topic. Analysing qualitatively the matches between clusters and debutant topics, we realised that, as two or more communities can match the same debutant topic, in a similar way one community can match more than one debutant topic, making the relationship between the clusters and debutant topics \textit{many-to-many}.
In brief, this qualitative analysis over the matches between clusters and ancestors of debutant topics allows us to observe that with a similarity of 0.2, we have significant matches (see Table \ref{tab:match0_2}).
However, these results pose a new question on why there is this \textit{many-to-many} relationship between clusters and debutant topics, which will be addressed in the next section.

\subsection{Why clusters can match more than one debutant topic and vice versa?}\label{sec:discussion-cluster-2-more-deb}
The relationship between clusters and debutant topics is \textit{many-to-many}. Indeed, our evaluation metrics (see Section \ref{sec:performance}) were designed to capture such phenomenon.  An example in which two communities match the same debutant topic, \textit{Semantic Search}, is reported in the previous section (Table \ref{tab:comparison-semantic-search}). An example, instead, in which the same community match two debutant topics, like \textit{Lexical Resources} and \textit{Ontology Learning}, is reported in Table \ref{tab:sameclustermatch}. 

The reason why two communities can match the same debutant topic is because even if the communities are different, they can share a good portion of topics, like a signature. Observing the two communities in Table \ref{tab:comparison-semantic-search}, the topics that match with the ancestor of \textit{Semantic Search} (marked in bold) are very alike. Indeed, using Eq. \ref{eq:new-jaccard-index}, the similarity between those two clusters is 0.733.

If we consider how those communities have been extracted (see Section \ref{sec:acpm}), we can see that the reason why there are very similar clusters, like the ones in the example, is because two local maxima in the evolutionary network are very close to each other. This is reasonable since the \textit{Computer Science} environment is very well-knit and highly dynamic. Two topics not so far apart can be the centre of two different movements, but because they are not so far apart, when the algorithm extracts one's neighbourhood, the other maximum with part of its community gets dragged in. This is logical as we are looking for overlapping communities, since the assumption is that one topic can foster the emergence of more than one new research area. However, this could be seen as a resolution issue when extracting the neighbourhoods of local maxima. If the resolution is very low, the cluster might not include some important topics and eventually fail to match any of the ancestors of debutant topics, leading to low precision. On the contrary, if the resolution is very large the cluster will include many topics and then it will match almost every debutant topic. 

\begin{figure}[ht]
\centering
  \includegraphics[width=\linewidth]{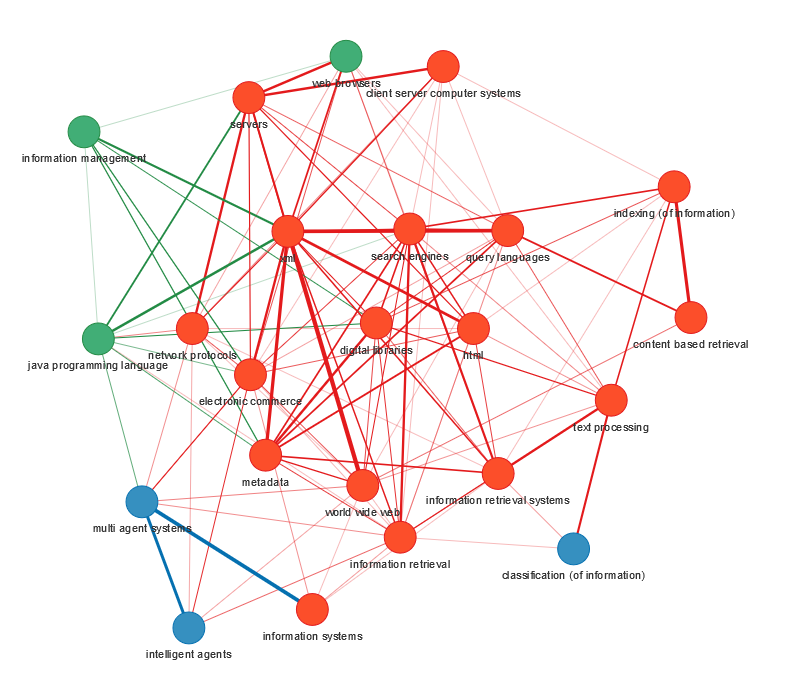}
  \caption{Community extracted from the evolutionary network of year 2003 that matches with \textit{Lexical Resources} and \textit{Ontology Learning}, respectively debuting in 2005 and 2004. Red nodes are the matched topics, while blue and green nodes are the topics that did not match with the ancestors of the two debutant topics.}
  \label{fig:semantic-search-graph}
\end{figure}

Figure \ref{fig:semantic-search-graph} depicts the network which combines the two communities. The nodes in red are those which match the ancestors of the debutant topics, while the nodes in green and blue are those belonging to the respective communities that did not match. As we can see, the two communities share many topics and very likely the two local maxima that generated them were not so far from each other.

Let’s explore now the issue of \textbf{why one or more debutant topics can be matched by the same cluster?} As we can see in the first row of Table \ref{tab:sameclustermatch}, 15 of 18 topics (marked in bold) of a community extracted from the evolutionary network of the year 2003 matched the ancestors of \textit{Lexical Resources}\footnote{\url{https://en.wikipedia.org/wiki/Lexical_resource}} with a similarity measure of 0.393. In the second row, instead, 13 of the 18 topics in the same cluster matched the ancestors of \textit{Ontology Learning} with 0.348 as similarity measure. The two topics debuted respectively in the years 2005 and 2004, and while it may seem obvious that the community is pushing the emergence of new research areas in two different years, there are other communities that can match more than two topics in the same year.

\begin{table}[!htbp]
\centering
\caption{Example of matching between one clusters and multiple debutant topics.}
\label{tab:sameclustermatch}
\begin{tabular}{|p{5cm}|p{6cm}|p{3cm}|}
\hline
\textbf{Cluster                                                                                                                                                                                                                                                                                                                                                                      } & \textbf{Ancestors of debutant topic                                                                                                                                                                                                                                                                                                                                                                                                                                                      } & \textbf{Debutant topic                                                                             } \\ \hline
\textbf{world wide web}, \textbf{query languages}, \textbf{natural language processing systems}, \textbf{database systems}, \textbf{artificial intelligence}, \textbf{information retrieval systems}, \textbf{information retrieval}, \textbf{syntactics}, \textbf{computer programming languages}, \textbf{metadata}, \textbf{xml}, \textbf{multi agent systems}, \textbf{servers}, \textbf{search engines}, \textbf{information management}, electronic commerce, software agents, intelligent agents & \textbf{world wide web}, \textbf{query languages}, \textbf{natural language processing systems}, \textbf{database systems}, \textbf{artificial intelligence}, \textbf{information retrieval systems}, \textbf{information retrieval}, \textbf{syntactics}, \textbf{semantic web}, \textbf{natural language processing}, \textbf{text processing}, \textbf{wordnet}, \textbf{computational linguistics}, semantics, ontology, linguistics, information theory, international conferences, word processing, terminology, natural languages, lexical semantics, algorithms, thesauri, semantic relations         & Lexical Resources (debuting in 2005 and according to Eq. \ref{eq:new-jaccard-index} the value of similarity is 0.393) \\ \hline
\textbf{world wide web}, \textbf{query languages}, \textbf{natural language processing systems}, \textbf{artificial intelligence}, \textbf{information retrieval}, \textbf{information management}, \textbf{search engines}, \textbf{computer programming languages}, \textbf{metadata}, \textbf{syntactics}, \textbf{xml}, \textbf{servers}, \textbf{multi agent systems}, database systems, electronic commerce, information retrieval systems, software agents, intelligent agents & \textbf{world wide web}, \textbf{query languages}, \textbf{natural language processing systems}, \textbf{artificial intelligence}, \textbf{information retrieval}, \textbf{information management}, \textbf{search engines}, \textbf{web services}, \textbf{semantic web}, \textbf{natural language processing}, \textbf{computational linguistics}, \textbf{learning systems}, ontology, semantics, taxonomies, education, information theory, knowledge acquisition, learning algorithms, data mining, linguistics, international conferences, ontology engineering, domain ontologies, technology & Ontology Learning (debuting in 2004 and according to Eq. \ref{eq:new-jaccard-index} the value of similarity is 0.348) \\ \hline
\end{tabular}
\end{table}

This can be justified by the fact that two debutant topics can share some parts of their ``DNA'' as they can have similar ancestors. As a result the community that contains this shared part will match both debutant topics. Indeed, \textit{Ontology Learning} and \textit{Lexical Resources} do share many ancestors, and according to Eq. \ref{eq:new-jaccard-index}, their similarity, in terms of shared ancestors, is 0.515. This means that the two debutant topics are more similar between themselves than each of them is to the cluster.

\begin{figure}[ht]
\centering
  \includegraphics[width=340px]{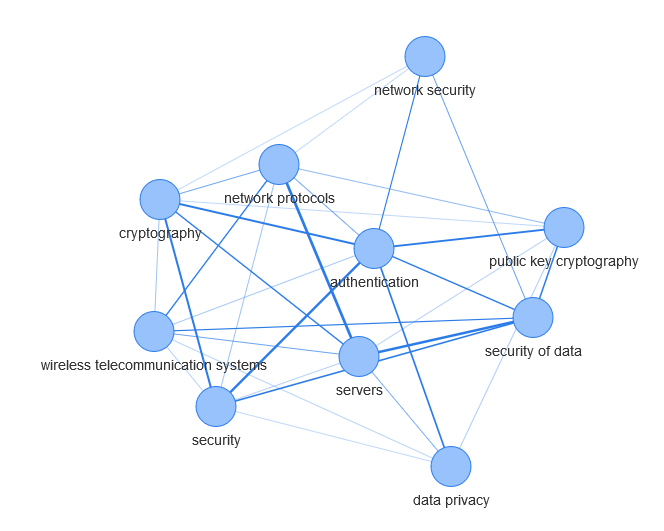}
  \caption{Community of topics extracted in the evolutionary network of year 2004 that matched with more 40 debutant topics in the following two years.}
  \label{fig:graph-security}
\end{figure}

Further, from this analysis it emerged that a cluster extracted from the evolutionary network of year 2004 matched over 40 debutant topics within the years 2005 and 2006. This cluster, whose network is depicted in Fig. \ref{fig:graph-security}, contains 10 topics: \textit{Network Protocols}, \textit{Security}, \textit{Security Of Data}, \textit{Cryptography}, \textit{Authentication}, \textit{Wireless Telecommunication Systems}, \textit{Servers}, \textit{Public Key Cryptography}, \textit{Data Privacy}, \textit{Network Security}. The debutant topics that have been matched with this community are all related to the \textit{Security} field, and are:

\textit{Secure Data} (with similarity = 0.452),\textit{ Authenticated Key Exchange} (0.449), \textit{Symmetric Keys} (0.449), \textit{Man In The Middle Attacks} (0.441), \textit{Bilinear Pairing} (0.433), \textit{Keys (for Locks)} (0.433), \textit{Identity Based Cryptography} (0.425), \textit{Key Pre-Distribution} (0.424), \textit{Key Establishments} (0.424), \textit{Bilinear Map} (0.417), \textit{Anonymous Authentication} (0.417), \textit{Improved Scheme} (0.400), \textit{Security Challenges} (0.391), \textit{Security Frameworks} (0.391), \textit{Sensitive Datas} (0.382), \textit{Various Attacks} (0.382), \textit{Heterogeneous Sensor Networks} (0.364), \textit{Forgery Attacks} (0.351), \textit{Unforgeability} (0.351), \textit{Rfid Systems} (0.333), \textit{Homomorphic-Encryptions} (0.333), \textit{Encrypted Data} (0.324), \textit{Phishing} (0.324), \textit{Secret Information} (0.324), \textit{Diffie-Hellman Assumption} (0.315), \textit{Malicious Activities} (0.296), \textit{Privacy Issue} (0.275), \textit{Phishing Attacks} (0.275), \textit{Side-Channel Analysis} (0.270), \textit{Sybil Attack} (0.265), \textit{Rfid Applications} (0.265), \textit{Cyber-Attacks} (0.260), \textit{Biometric Data} (0.260), \textit{Trojans} (0.260), \textit{Public-Key Encryption Scheme} (0.256), \textit{Malicious Nodes} (0.254), \textit{Malwares} (0.250), \textit{Private Data} (0.242), \textit{Hash Value} (0.239), \textit{Malicious Software} (0.237), \textit{Privacy Preservation} (0.229), \textit{Secret Messages} (0.229), \textit{Sql Injection} (0.216), \textit{K-Anonymity} (0.206), \textit{Network Attack} (0.206), \textit{Privacy Requirements} (0.206).
In this case, we can observe a common movement in an area that can spawn multiple related topics.

\subsection{To what extent the number of ancestors can influence the performance of Augur?}
When extracting the ancestors of each debutant topic, in Section \ref{sec:extraction-related-topics}, we selected the first 25 meaningful topics.
In this analysis we want to understand whether the number of ancestors for a debutant topic can influence the results. To this end, the debutant topics have been recomputed and enhanced with a different number of ancestors. The selection of 25 ancestors may already seem excessive and further increasing this number would mean increasing the chances of drawing some topics that not should necessarily be considered ancestors (see Section \ref{sec:discussion-first-study}). Therefore, we reduced the number of ancestors to 20 and 15. 

\begin{figure}[!htbp]
\centering
  \includegraphics[width=\linewidth]{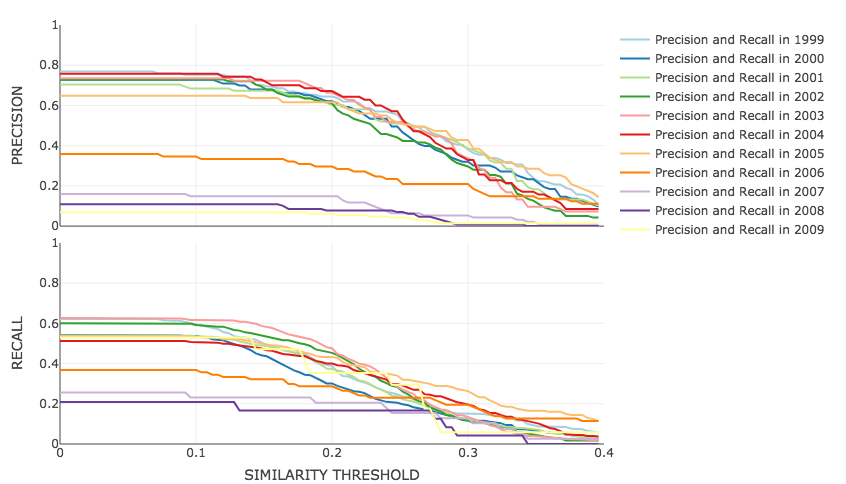}
  \caption{Results in terms of precision and recall obtained from evaluating the Advanced Clique Percolation Method against the gold standard with 15 ancestors. Both clusters and debutant topics have been enhanced with their super-areas.}
  \label{fig:finale-all-15}
\end{figure}

\begin{figure}[!htbp]
\centering
  \includegraphics[width=\linewidth]{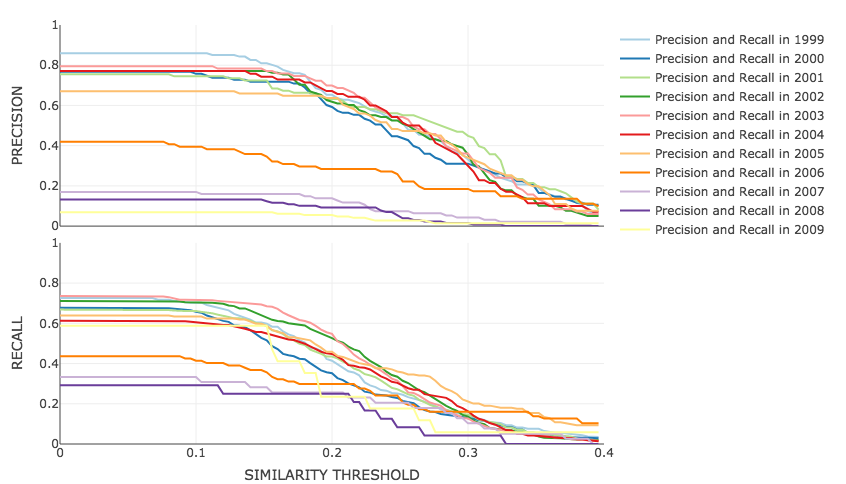}
  \caption{Results in terms of precision and recall obtained from evaluating the Advanced Clique Percolation Method against the gold standard with 20 ancestors. Both clusters and debutant topics have been enhanced with their super-areas.}
  \label{fig:finale-all-20}
\end{figure}

We performed this analysis only with the clusters of the Advanced Clique Percolation Method, as it outperformed the other four community detection algorithms. These clusters have been evaluated against these two newly computed sets of debutant topics and the results are reported in Fig. \ref{fig:finale-all-15} and Fig. \ref{fig:finale-all-20}. We can see that reducing the number of ancestors slightly affects the value of precision and recall, but they are still good results. This suggests that using 25 ancestors for each debutant topic is a good solution.

\subsection{What else can be said regarding the decrease of performance in the last period of analysis?}\label{sec:discussion-decreasing-perf}
\begin{figure}[ht]
\centering
  \includegraphics[width=\linewidth]{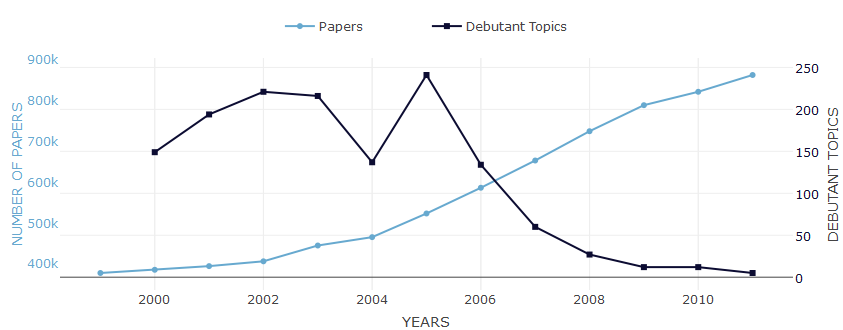}
  \caption{Comparison between the constant increase of published papers and the decrease of emerging topics each year.}
  \label{fig:papersvsdebutant}
\end{figure}

Throughout this chapter, we have seen in many occasions that the performances were decreasing in the last period of analysis. In particular, both precision and recall were affected because fewer and fewer clusters could be effectively matched with the ancestors of debutant topics. The main reason is certainly the lack of data, as showed in Fig. \ref{fig:emerging-topics} on page \pageref{fig:emerging-topics}. During the period of analysis, we have a constantly increasing number of clusters while the number of emerging areas substantially decreases after 2006. In the year 2010 only 12 topics emerge and even fewer in the following years. The reasons of this decrease in the number of debutant topics are uncertain. One might argue that maybe the corpus does not contain enough data that cover those years, but actually, as showed in Fig. \ref{fig:papersvsdebutant}, the number of papers keeps increasing year by year, while debutant topics decrease. Although it is hard to believe that, in reality, the number of debutant topics decreases over time, in particular with the observed rate, \textbf{what else can be said for the clusters that did not find a match because of the lack of emerging topics?
}

Comparing the set of clusters against the set of debutant topics, leads to four different cases, as reported in Section \ref{sec:performance}.

Here we want to focus our attention to the case in which clusters do not match debutant topics: the \textit{false positives}. 

A cluster embodies certain dynamics, topics that are tightening their relationships and their collaboration is becoming more intense. The debutant topic is the result of this close relationship and serves as a label for this intense multidisciplinary activity. However, we can argue that is not always guaranteed that an increased collaboration will always lead to the emergence of a new area. It might be the case that two or more topics are increasing their collaboration towards a particular common aim that eventually will not turn into a new area. Nonetheless, this collaboration can still be meaningful and relevant for the stakeholders. Hence, it might be the case that \textit{even if the cluster does not match any debutant topic, it can still provide important information}.

To this end, we performed a more qualitative investigation over these specific communities to asses if Augur was detecting community of topics that would keep growing and influence the research landscape. For each cluster (at year \textit{t}), we extracted the portion of network containing the topics that generated the cluster (from the evolutionary network at year \textit{t}) and also the same portion of network in the following year (from the evolutionary network at year \textit{t+1}), to understand how these communities developed in time. This analysis allowed me to classify those clusters in four further categories:
\begin{enumerate}[label={Category \arabic*.},leftmargin=2.25cm]
\item \textbf{Strengthening Ties}: The links that those two networks share are increasing their intensity in year \textit{t+1}, meaning that the topics of this cluster are still increasing their pace of collaboration over time
\item \textbf{Network Expansion}: The network at year \textit{t+1} introduces new links between the topics, compared to the network at year \textit{t}, meaning that the network is growing in terms of its relationships
\item \textbf{Strengthening Ties and Network Expansion}: The network \textit{t+1} exhibits a growth in terms of new links and also the old links are increasing in intensity
\item \textbf{No development}: The network at year \textit{t+1} is not showing any development compared to the network at year \textit{t}.
\end{enumerate}

\begin{figure}[h]
\begin{subfigure}{\columnwidth}
\centering
\includegraphics[clip,width=\columnwidth]{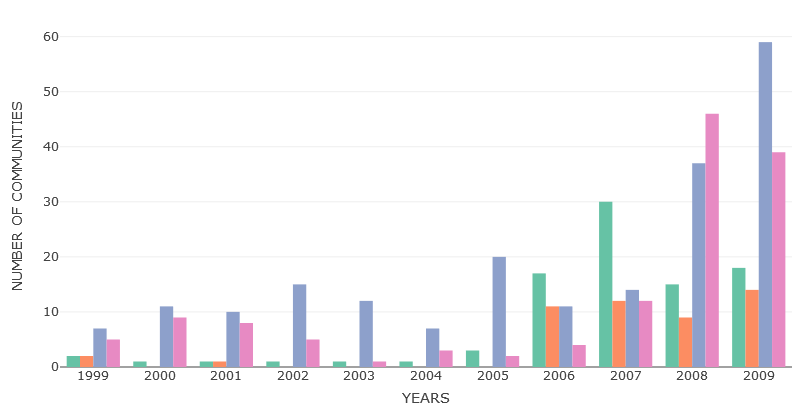}
\caption*{}
\end{subfigure}%

\begin{subfigure}{\columnwidth}
\centering
\includegraphics[clip,width=\columnwidth]{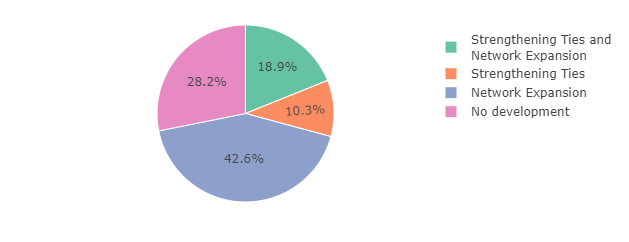}
\caption*{}
\end{subfigure}%
\caption{Differences in how the unmatched clusters, obtained in a certain year, develop in the following year. A cluster can strengthen its ties, expand its network, both expand its network and strengthen its ties, or not developing.}
\label{fig:unmatched-cluster-analysis}
\end{figure}

In Fig. \ref{fig:unmatched-cluster-analysis}, we reported a bar chart that shows how the unmatched cluster can be classified according to the previous categories, yearly. In general, we can see that almost 72\% of the unmatched clusters are amplifying and/or expanding their relationships. The rest remains undeveloped. In accordance to Fig. \ref{fig:finale-all}, in the year 2006 when precision starts to fall, there are more clusters that are unmatched. However, even if these clusters cannot be labelled with emerging topics, we can see that for most of them, their inner topics are still involved in this process of development. Indeed, for the year 2006, there are 39 active clusters against 4. In year 2007, 56 unmatched clusters will further develop in the next year against 12. In 2008 and 2009 the clusters that remain undeveloped increase but their counterpart is still high. Indeed, in year 2008 there are 61 developing clusters against 46 ``static'' ones and in 2009 there are 91 developing clusters against 39.
In conclusion, even if there is not enough data in the last period of analysis, in terms of emerging topics, to support the evaluation in matching clusters, we can still observe that most of the unmatched clusters continue to develop with their topics either amplifying their relationships, or embracing new ones, or both.

%
%
%
%
%

\begin{figure}[h!]
\begin{subfigure}{\columnwidth}
\centering
\includegraphics[clip,width=\columnwidth]{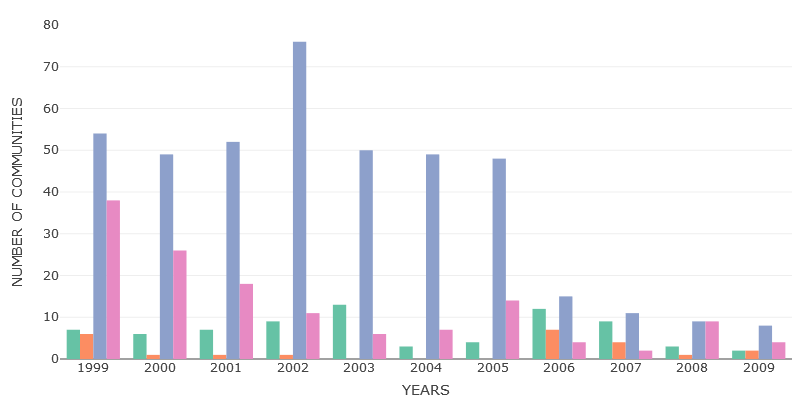}
\caption*{}
\end{subfigure}%

\begin{subfigure}{\columnwidth}
\centering
\includegraphics[clip,width=\columnwidth]{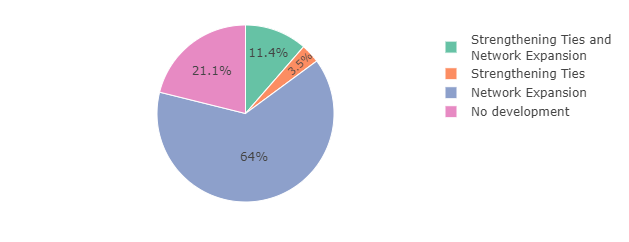}
\caption*{}
\end{subfigure}%
\caption{Differences in how the matched clusters, obtained in a certain year, develop in the following year. A cluster can strengthen its ties, expand its network, both expand its network and strengthen its ties, or not developing.}
\label{fig:matched-cluster-analysis}
\end{figure}

For the sake of curiosity, we conducted a similar analysis over the matched clusters. All the clusters have been analysed and categorised according to the four groups previously mentioned. From Fig. \ref{fig:matched-cluster-analysis}, we can observe that 79\% of clusters continue their development in the following year, while 21\% of them remain undeveloped. From the bar chart, we can observe that, over time, the most prominent form of development is that the inner topics create new relationships: 42.6\% (violet bar), while 19\% of clusters are both strengthening their relationships and expanding the network (green bar). Instead, just 3.5\% of clusters are strengthening their ties only (orange bar). This can be justified by the fact that, as a new area emerges, the topics in the cluster reduce their collaboration as they now might focus on the new area. Nonetheless, new collaborations between these topics can still be established.
From the bar chart, we can also observe the presence of some clusters that did not show any development in the following year (pink bar). For this particular subset of clusters, it might be the case that the new topic is drastically changing the dynamics of the clusters. Since a new topic just stemmed out, most of the topics in the cluster can start to collaborate more with the newborn rather than continuing to collaborate with their peers. In other words, the emerging topic is draining the degree of their collaboration. 

\clearpage

\subsection{Why are there still unmatched debutant topics?}\label{sec:discussion-unmatched-debutants}

From the results reported in Fig. \ref{fig:finale-all}, we observed that the values of recall fluctuate between 0.3 and 0.8 and they also tend to decrease in the last period of analysis. This means that some of the emerging topics do not match the returned clusters. 
Table \ref{tab:matched-unmatched-clusters} reports, for each year, the number of debutant topics that received a match and also those which did not. Understanding the reasons why there are still unmatched debutant topics allows us to understand how we can further improve Augur and perhaps can tell us more about the dynamic of debutant topics.

\begin{table}[ht]
\centering
\caption{Number of matched and unmatched debutant topics in the years from 2000 to 2011}
\label{tab:matched-unmatched-clusters}
\begin{tabular}{l|C{3cm}|C{3cm}}
\hline
\textbf{Year} & \textbf{Matched Debutant topics} & \textbf{Non-matched debutant topics} \\ \hline
2000 & 122                     & 27                          \\
2001 & 165                     & 29                          \\ 
2002 & 184                     & 37                          \\ 
2003 & 179                     & 37                          \\ 
2004 & 123                     & 14                          \\ 
2005 & 206                     & 35                          \\ 
2006 & 110                     & 24                          \\ 
2007 & 44                      & 16                          \\ 
2008 & 16                      & 11                          \\ 
2009 & 5                       & 7                           \\ 
2010 & 9                       & 3                           \\ 
2011 & 4                       & 1                           \\ \hline
\end{tabular}
\end{table}

The reasons why a debutant topic is left unmatched can be twofold. The first possibility is that for such debutant topic we were unable to find its right ancestors. The second possibility is instead that Augur fails to locate the clusters that could eventually match the ancestors of the debutant topics.

To verify the first hypothesis, we manually checked the ancestors of some unmatched debutant topics and see whether there were unrelated topics. Here, we observed that the ancestors appear to be accurate and related to the debutant topic in question. For example, the ancestors of the unmatched topic \textit{Wearable Devices}, which appear to be reasonable, are \textit{Wearable Computers}, \textit{Mobile Computing}, \textit{Sensors}, \textit{Wearable Computing}, \textit{Pervasive Computing}, \textit{Human Engineering}, \textit{Ubiquitous Computing}, \textit{Algorithms}, \textit{Kinematics}, \textit{Forecasting}, \textit{Patient Monitoring}, \textit{Data Acquisition}, \textit{Data Processing}, \textit{Computer Operating Systems}, \textit{Tracking (position)}, \textit{Biomedical Equipment}, \textit{Social Interactions}, \textit{Sensor}, \textit{Accelerometer}, \textit{Potential Applications}, \textit{Wireless Telecommunication Systems}, \textit{Textiles}. Given such evidence we can reject the first hypothesis.

To verify the second hypothesis, 
we computed the overall pace of collaboration of each network of ancestors, related to both matched and unmatched debutant topics.

\begin{figure}[ht]
\centering
  \includegraphics[width=0.9\linewidth]{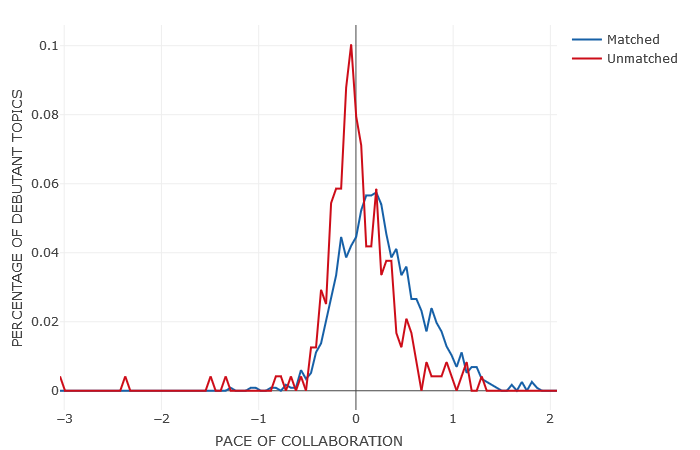}
  \caption{Distributions of pace of collaboration valued from the networks of ancestors of both matched and unmatched debutant topics.}
  \label{fig:pace-of-collaboration-debutants}
\end{figure}

In particular, for each debutant topic, we selected the portion of network containing their ancestors in the year priors to their debut. Then we computed the overall pace of collaboration. Since each link of the network expresses the pace of collaboration between the two nodes (see Section \ref{sec:evolutionary-network}), computing the overall pace of collaboration for this portion of the network involves computing the average of the weights of all the links. Finally, we collected and then compared the pace of collaboration for the two groups, matched and unmatched debutant topics. In Fig. \ref{fig:pace-of-collaboration-debutants} are reported the two distributions of pace of collaboration obtained from the networks of both matched and unmatched debutant topics. Specifically, the line charts report the frequency (in percentage) of the pace of collaboration computed from the networks, related to the two groups. In general, we can see that more than 50\% of the unmatched debutant topics (red line) have pace of collaboration smaller than 0. By contrast, more than 75\% of the matched debutant topics (blue line) have pace of collaboration greater than 0. The mean value of pace of collaboration obtained from networks associated to unmatched topics is 0.02, while the mean value of pace of collaboration of the other group is 0.29.

We analysed the pace of collaboration because Augur returns the communities that exhibit high values of pace of collaboration. If a portion of network, similar to the one containing the ancestors of a debutant topic, has a low value of pace of collaboration, it might be concealed by areas that exhibit higher values. Considering that this analysis allowed us to acknowledge that the pace of collaboration associated to the unmatched debutant topics is in general lower than the pace of collaboration associated to the other group, we can conclude that the reason why there are still unmatched debutant topics is because Augur is not returning all the needed clusters. This can be due to the fact that their ancestors have a low pace of collaboration or because some topics may indeed come into existence more abruptly as new technologies or emerge through new discoveries.

\subsection{Can authors support the identification of emerging areas?}\label{sec:discussion-authors}
The evaluation so far consisted of verifying which of the returned clusters can match one of the emerging topics. This match was performed between the topics of each cluster and the ancestors of the debutant topics. 

However, another source of information that could be employed is the \textit{authors}. Indeed, Augur crawls the different evolutionary networks and, for each of them, returns a set of clusters, containing topics that exhibit an increase in the pace of collaboration, as well as a set of authors that are considered active (see Section \ref{sec:influential-authors2}). The latter information can be used during the evaluation as a way to confirm the matches between the clusters and the debutant topics. 
The idea is that, given a cluster, the active authors are the most active authors in that particular set of topics. They are the leaders of this dense collaboration between the different topics and potentially they will also be the first people to publish in this brand-new area fostered by this multidisciplinarity. 

Based on this assumption, enriching the gold standard with authors who were the first to publish in a debutant topic, might help to further evaluate the clusters obtained from the different evolutionary networks. Practically, for a given debutant topic, the algorithm will select the authors that have published the highest number of publications, in the first five years of its life.

In brief, the algorithm for extracting the main authors firstly selects all the authors that published in that topic, within its first five years, and then it ranks them based on the number of publications and selects the first 100, as showed in Code \ref{alg:extracting-authors}. However, the threshold of 100 authors can be reconsidered during the analysis.
\newline
\newline
\begin{algorithm}[H]
\label{alg:extracting-authors}
\setstretch{1}
\SetKwInOut{Input}{Input}\SetKwInOut{Output}{Output}

\Input{List of Debutant Topics $deb.topics$}
\Output{List of Authors $authors$}
\BlankLine
\ForEach{topic in deb.topics}{
	year$\leftarrow$ GetYearOfSoftDebut(topic)\;
	related.authors$\leftarrow$ SelectAuthorsInTopics(topic, from=year, to=year+5)\;
	\ForEach{author in related.author}{
	papers.author[author]$\leftarrow$ NumPapersPublishedInTopic(topic, author, from=year, to=year+5)\;
}
temp.authors$\leftarrow$ sort(related.author, values=papers.author, decreasing=TRUE)\;
authors[topic]$\leftarrow$temp.authors[1:100]\tcc*[r]{keep only the first 100}
}
return(authors)\;
\caption{Extracting main authors of debutant topic.}
\end{algorithm}
\vspace*{0.7cm}

We performed a quantitative analysis, in which we employed the authors to confirm the matches between clusters and ancestors.
In particular, for each match between clusters and debutant topics, the algorithm evaluated the similarity, using the Jaccard Index, between active authors (described in Section \ref{sec:influential-authors2}) and the authors associated to the debutant topics. Observing the results, we realised that these results could not bring any contribution as the similarity values were below 0.01. This is justified by the fact that the sets of authors can be very large and if there are only few mutual authors, then the similarity between the two sets can be very low. 

The reason for this problem is because we are comparing the similarities obtained from two different kinds of information that are carried by topics and authors. Topics are more coarse-grained, as they can be seen as categories for a collection of papers written on such topics. On the contrary, authors can be seen as single data points and are more fine-grained.
Hence, the question of whether authors can help the identifications of emerging areas, cannot be answered without further work.

\subsection{How does the final result look like?}\label{sec:discussion-final-result}
The final result is the information that the user receives from Augur, which aims to alert him or her to the emergence of a new area and provide sufficient detail to allow him/her to understand the key elements of the topic dynamic.

\begin{figure}[!htbp]
\centering
  \includegraphics[width=\linewidth]{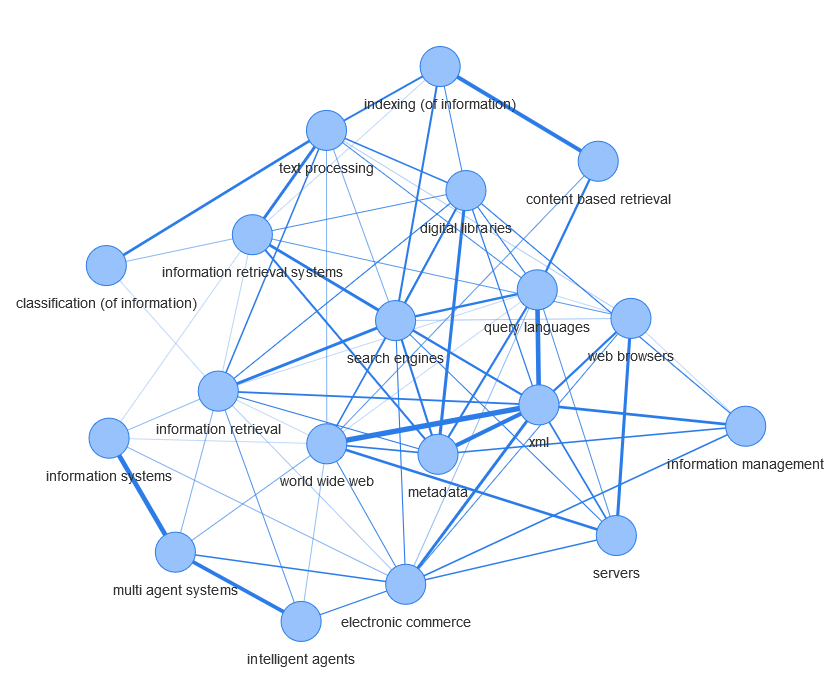}
  \caption{Network representation of the same cluster in Table \ref{tab:cluster-output} extracted from the evolutionary network of year 2002.}
  \label{fig:graph-semantic-search}
\end{figure}

Here follows an example of a cluster produced by Augur from the topic networks in the period 1998-2002. The cluster contains topics, such as, \textit{World Wide Web}, \textit{Query Languages}, \textit{Metadata}, \textit{Content Base Retrieval}, and \textit{Search Engines}, which exhibit a strong increment in their pace of collaboration in the period under analysis and match the ancestors of \textit{Semantic Search}, a topic that debuted in 2003. Therefore, we considered this cluster as correctly predicting \textit{Semantic Search} (with a similarity of 0.38). \textit{Semantic Search} aims to improve search accuracy by understanding the contextual meaning of the query terms and combines research in semantic technologies and information retrieval. In Table \ref{tab:cluster-output}, we report in bold the direct ancestors of \textit{Semantic Search}, but, even among the other ones, we find many topics conducive to \textit{Semantic Search} or that produced technologies adopted by this field, such as \textit{Text Processing}, \textit{Electronic Commerce}, \textit{Digital Libraries}, and \textit{Web Browser}. In Fig. \ref{fig:graph-semantic-search} we show its network representation.

This is an exemplary case of the dynamics exploited by Augur, in which some topics, previously less connected, started to collaborate and moulded a novel research area that inherited their domains (e.g., \textit{Information Retrieval}, \textit{Digital Libraries}), formats (e.g., \textit{Xml}), software (e.g., \textit{Search Engines}), and applications (e.g., \textit{Content-based Retrieval}). As part of the making sense process, in Table \ref{tab:iaip-output} we also show the top 10 authors (top) and the top 5 papers (bottom) relevant to this cluster.

\begin{table}[!htbp]
\centering
\caption{Example of output produced by Augur. The cluster associated with the emergence of the \textit{Semantic Search} topic (in bold the topics that match its ancestors).}
\label{tab:cluster-output}
\begin{tabular}{p{10cm}}
\textbf{Topics within the cluster} \\ \hline
\textbf{world wide web, query languages, metadata, content-based retrieval, information retrieval, search engines, xml, information systems, information retrieval systems}, multi agent systems, intelligent agents, servers, digital libraries, electronic commerce, text processing, information management, indexing, web browsers, classification
\end{tabular}
\end{table}

\clearpage
\newpage
\begin{table}[t]
\centering
\caption{Example of output produced by Augur. On top, there are the top 10 active authors. At the bottom, the top 5 papers.}
\label{tab:iaip-output}
\begin{tabular}{p{4.5cm}|p{9.5cm}}
\textbf{Active Authors} & \textbf{Relevant Papers} \\ \hline
\begin{itemize}[noitemsep,topsep=0pt,leftmargin=*]
\item W. Bruce Croft,  
\item Dieter Fensel, 
\item Dan Suciu, 
\item William W. Cohen, 
\item Berthier Ribeiro-Neto, 
\item Clement T. Yu, 
\item James Allan, 
\item Justin Zobel, 
\item Dragomir R. Radev, 
\item Victor Vianu
\end{itemize} 
 & 
\begin{itemize}[noitemsep,topsep=0pt, leftmargin=*]
\item A Sheth et al. ``\textit{Managing semantic content for the Web}'' (2002)
\item RWP Luk et al. ``\textit{A survey in indexing and searching XML documents}'' (2002)
\item J Kahan et al. ``\textit{Annotea: An open RDF infrastructure for shared Web annotations}'' (2002)
\item R Manmatha et al. ``\textit{Modeling score distributions for combining the outputs of search engines}'' (2001)
\item S Dagtas et al. ``\textit{Models for motion-based video indexing and retrieval}'' (2000)
\end{itemize}
\end{tabular}
\end{table}

\section{Final remarks}

In this chapter, we evaluated Augur on a gold standard of 1,408 debutant topics, which emerged in the 2000-2011 timeframe, and we compared the results of the Advanced Clique Percolation Method against four alternative approaches: Fast Greedy, Leiden Eigenvector, Fuzzy C-Means, and Clique Percolation Method.
In particular, we initially designed an evaluation framework, defining how we compared the clusters of topics obtained from each evolutionary network, against the topics debuting in the following two years. Then, we designed our gold standard that included the selection of 1408 debutant topics, and the extraction of their foster ancestors. At this stage, we firstly evaluated the results obtained by the five clustering algorithms on the gold standard using the similarity index showed in Eq. {\ref{eq:jaccard1}}. The ACPM provided interesting results, however there still was room for improvement. To this end, initially we applied a semantic filter to further clean the results obtained by the different algorithms, and then we applied a more comprehensive similarity index (see Eq. {\ref{eq:new-jaccard-index}}). This new index take fully advantage of the Computer Science Ontology by enhancing both clusters and ancestors of debutant topics with their broader topics.
The final results show that our approach outperforms state of the art solutions and is able to successfully identify clusters that will lead to the emergence of new topics in the two following years.

While these results are satisfactory, our analysis presents some limitations that can be object of future work. In the first instance, the gold standard does not cover well the years after 2007. Indeed, as discussed in Section {\ref{sec:discussion-decreasing-perf}} the number of debutant topics gradually decreases after 2007. We thus intend to consider more up-to-date scholarly datasets and produce a more comprehensive version of the gold standard that could be adopted by the scholarly community to further study this task. 

Another limitation of Augur is that, in its current version, it only focuses on the pace of collaboration between topics and network density. We can argue that relying on only two indicators may not be enough to fully understand and detect the complex dynamics behind the creation of a topic. One way to further improve the results, could be to add other relevant entities, such as authors, venues and so on. 
This may enable the possibility to investigate other kinds of dynamics that could be associated with the emergence of new research areas, such as the patterns of collaboration between prominent authors, the dynamics of citations networks, or changes in the topic distributions of high-tier scientific venues.

\chapter{Conclusion\label{ch:conclusion}}

The main goal of this dissertation was to develop an approach for detecting emerging topics at the embryonic stage, i.e., when they have not yet been labelled or associated with a considerable number of publications. In a first study, we focused on understanding the dynamics that can drive the emergence of a new research topic (Chapter  \ref{ch:firststudy}). The results showed a strong correlation between the pace of collaboration in a topic network and the emergence of new research topics a few years later. We then created Augur, a system for detecting areas of the network of topics characterised by these dynamics (Chapter \ref{ch:secondstudy}). Augur uses the Advanced Clique Percolation Method (ACPM) for analysing the topics involved in these dynamics and clustering them. The output is a number of clusters of topics, each of which has a strong probability to be associated with an emerging new topic. These are then further characterised by providing a list of significant authors and publications. We evaluated Augur and ACPM versus four alternative community detection algorithms on a gold standard of 1,408 debutant topics in the 2000-2011 timeframe. The results show that Augur outperforms state-of-the-art solutions and is able to successfully identify a high percentage of clusters that will produce new topics in the two following years.

\section{Overview of contributions and major findings}
In Chapter \ref{ch:intro}, we introduced six research questions that drove this doctoral work. We will now analyse the answers of these questions and restate our contributions.

\vspace{6mm} 
\textbf{Research Question 1:} Is it possible to precisely define the notion of established topic? (Section \ref{sec:rq1})

To answer this question, we performed an analysis of the body of research in order to explore some ideas in relation to the problem. In Section \ref{sec:topiclifecycle}, we analysed the stages that a research topic goes through. In Chapter \ref{ch:firststudy}, we attempted to determine how a topic can be considered established, from a practical point of view. Even though in principle we could use a more exhaustive list of metrics to define when a topic is considered established, we focused on the age of the topic, the number of relevant papers that it are published each year, and the fact that it is covered in CSO.

This understanding proved to be very valuable, as in our first study, we were then able to compare the dynamics associated with established topics against the dynamics involving recently emerged ones.

\vspace{6mm} 
\textbf{Research Question 2:} How early in the topic lifecycle is it possible to identify an emerging topic? (Section \ref{sec:rq2})

In Section \ref{sec:topiclifecycle}, we analysed the different stages which a new research topic goes through before becoming established on the basis of our first hypothesis, which states ``\textit{before emerging, new topics face an embryonic stage in which researcher are already involved in several activities}''. In the literature, we found enough evidence that would suggest the existence of an embryonic stage. In this particular stage, a topic has not yet been explicitly labelled and recognised by a research community, but it is already taking shape, as evidenced by the fact that researchers from a variety of fields are forming new collaborations, producing new work, and starting to define the challenges and the paradigms associated with the emerging new area.

In Chapter \ref{ch:firststudy}, we performed an analysis in which we observed the dynamics leading to the emergence of new research topics.

In this analysis, we found a correlation between the collaboration pace of existent topics and  the emergence of new emerging topics in the near future (p-value $<$ 0.0001). In addition, because at the time of performing the analysis the emerging topics did not exist yet, we were able then to analyse their emergence within their embryonic stage.

In brief, based on the evidence collected in this work, we can say that it is possible to do better than the current approaches available in the literature, which detect new topics from the early stage onwards, by forecasting their emergence already at the embryonic stage.

\vspace{6mm} 
\textbf{Research Question 3:} What are the indicators that can be exploited to predict the emergence of new topics? (Section \ref{sec:rq3})

This question has been driven by our second hypothesis: “\textit{the emergence of a new research topic is anticipated by specific dynamics between pre-existing topics.}”

In Chapter \ref{ch:firststudy}, we compared the sections of the co-occurrence networks where new topics are about to emerge with a control group of subgraphs associated with established topics. These graphs were analysed by using two novel approaches that integrate both statistics and semantics. The results provide evidence that the emergence of a novel research topic can be anticipated by a significant increase in the pace of collaboration between relevant research areas, which can be seen as the ``ancestors'' of the new topic. In particular, we identified two indicators that are correlated with the emergence of new research topics:
\begin{itemize}
\item pace of collaboration, Eq. \ref{eq:pocdef}, p-value = $4.64 \times 10^{-45}$ 
\item network density, Eq. \ref{eq:growthindex}, p-value = $6.43 \times 10^{-16}$
\end{itemize}

In other words, the development of new topics seems to be encouraged by the cross-fertilisation of established research areas. Of course there may well exist other patterns, which could help to forecast the emergence of a new research area; a hypothesis that we plan to explore in future work.


\vspace{6mm} 
\textbf{Research Question 4:} Is it possible to develop an effective computational method that can support this prediction task? (Section \ref{sec:rq4})
We hypothesised that it is possible to develop an automatic approach for detecting new emerging topics at their embryonic stage by analysing the dynamics of existing topics (i.e., observing their patterns of collaboration).

To test this hypothesis, we performed a second study which led to the development of Augur. As described in Chapter \ref{ch:secondstudy}, we aimed at effectively detecting the emergence of new research areas by analysing topic networks and identifying clusters associated with an overall increase of the pace of collaboration between research areas. 

The evaluation discussed in Chapter \ref{ch:finalevaluation} showed that Augur is able to detect with high precision the topics that will emerge a few years later. In particular, from Table \ref{tab:final-comparison-results0.15}, at page \pageref{tab:final-comparison-results0.15}, we can see how in the years from 1999 to 2005, ACPM yields values of precision around 0.8 and values of recall around 0.7, outperforming the other four algorithms. Indeed, other algorithms such as FG, LE, and FCM do not exceed 0.15 in precision and 0.5 in recall. Only the Clique Percolation Method is able to partially exceed these values with precision nearly 0.3, but a low recall.

These results allow us to affirm:
\begin{itemize}
\item that the ACPM algorithm, designed to identify clusters from evolutionary networks, outperforms other state-of-the-art algorithms (i.e., FG, LE, FCM and CPM) for forecasting the emergence of new research topics;
\item that is possible to develop a computational method for predicting the emergence of new topics at their early stage and forecast their emergence.
\end{itemize}

\vspace{6mm} 
\textbf{Research Question 5:} Are there commonalities between our approach to predicting the emergence of new topics and epistemological theories of research dynamics? (Section \ref{sec:rq5})


Thomas \cite{kuhn1970}\index{Kuhn Thomas} stated that science proceeds within certain paradigms or, in other words, sets of concepts or thought patterns. When some paradigms cannot cope with certain problems, a paradigm shift can lead to the emergence of a new scientific discipline. Other theories mention that different communities of researchers can be engaged in an exchange of tools, methodologies and theories, leading then to the emergence of a new research area \citep{surowiecki2005,becher2001}. All these theories support the idea that a new topic comes into existence even before it is explicitly labelled and recognised by a research community. In this phase, the topic is already taking shape, as researchers from a variety of fields are forming new collaborations and producing new work, starting to define the challenges and the paradigms associated with the emerging new area.

This doctoral work has many commonalities with what historians of Science and philosophers have been arguing in the last 60 years. Indeed, in designing and developing an approach for the detection of new emerging topics, we have been significantly influenced by these theories. In our first study, we observed that within the embryonic stage of topics, their related topics (i.e., their ancestors) are intensifying their activity of collaboration (p-value = $4.64 \times 10^{-45}$) as well as establish new collaborations (p-value = $6.43 \times 10^{-16}$). These findings suggest that an interdisciplinary environment is the driving force for the creation of novel research topics.

These findings bring new empirical evidence to the relevant theories in \textit{Philosophy of Science}, highlighting the important role of multidisciplinarity in the creation of new research topics.

\vspace{6mm} 
\textbf{Research Question 6:} What evaluation mechanisms are appropriate for this task? (Section \ref{sec:rq6})

This question focused on how to make sure that the induced clusters (i.e., groups of topics) could be validated and evaluated as valid fertile areas that are fostering the emergence of a new ones. In order to answer this question, we faced several challenges, including developing a gold standard and defining the right evaluation measures. In particular, for the gold standard, only knowing the topics that emerged in a particular period, does not provide enough information to evaluate the results of Augur, or at least to do so automatically. Therefore, we associated these emerging topics with their ancestors that eventually would provide the match for the topics found in clusters. 

The second challenging task was to define the evaluation measures to assess the validity of Augur. In Section \ref{sec:discussion-cluster-2-more-deb} we discussed that a cluster can potentially match multiple debutant topics, and vice versa (see also Fig. \ref{fig:matching-clusters-debutants} at page \pageref{fig:matching-clusters-debutants}). This \textit{many-to-many} relationship prevents us to adopt the standard values of precision and recall. For this reason, in Section \ref{sec:performance}, we redefined precision (Eq. \ref{eq:precision}) as the fraction of clusters that have been successfully matched with debutant topics, and recall (Eq. \ref{eq:recall}) as the fraction of topics that have been successfully matched by the clusters.

Another aspect that we investigated is the semantic similarity between the clusters and the ancestors of debutant topics. To investigate similarity between these two sets we relied on the Jaccard index. This index matches the samples within two sets in a syntactic way and thus it discards any semantics. Hence, in Eq. \ref{eq:jaccard1} and Eq. \ref{eq:new-jaccard-index} we proposed an enhanced version of the Jaccard index which considers also the semantic component. In the first instance, we enhanced the set of clusters with the relevant same-as relations. In this way, an ancestor can match either a topic in the cluster or other topics in a same-as relation with it. To this end, Eq. \ref{eq:new-jaccard-index}, we developed a more semantic index. In particular, both clusters and ancestors of debutant topics are enhanced with their super-areas, with the assumption that if the two sets share many of their super-areas it is very likely that they belong to the same sample.

Using the semantic enhanced index of Eq. \ref{eq:new-jaccard-index}, we were able to improve the matching process between sets of clusters generated by Augur and ancestors of debutant topics, so that the ACPM was able to reach excellent results, up to a precision of 0.8 and recall of 0.76 in the same period. 


\section{Limitations of the study and future directions}
In this section, we will discuss some limitations of this doctoral work and discuss future work.

\subsection{Gold standard}
One of the main issues we faced when evaluating Augur, was that towards the end of the analysed period the number of debutant topics in our dataset decreases drastically (see Figure \ref{fig:emerging-topics} at page \pageref{fig:emerging-topics}). For the sake of completeness, we also included in the analysis the topics emerged in the years after 2006, however this issue prevents us from fully trusting the results of the evaluation for the years in that period. As future work, we plan to acquire more up-to-date scholarly datasets and to produce a more comprehensive version of the gold standard.

\subsection{Evolutionary network}
The evolutionary network is a key component of Augur's framework. It structures the data so that they can be easily processed in the following step of the process. In particular, we create semantic enhanced topic networks from publication metadata, and then we convert them into evolutionary networks, which track the pace of collaboration between research topics over a period of five years. All the assumptions that we made when designing these evolutionary networks, such as number of years and how to measure the pace of collaboration, derive from the results of the first study (Chapter \ref{ch:firststudy}). In future work, we intend to explore other configurations with the aim of further improving the performances of Augur. In particular, we plan to investigate the following questions: what is the number of consecutive topic networks that can be merged to have a good estimate of the pace of collaboration between topics? What is the best function that allows to approximate the pace of collaboration between research topics? How does this approximation function change according to the amount of network to process?

\subsection{Clustering Algorithm}
The second phase of Augur is the identification of areas exhibiting an intense activity of collaboration, which eventually leads to the emergence of new research topics. To this end, we developed the Advanced Clique Percolation Method with the aim of detecting clusters of topics that exhibit a significant increase in collaboration pace. We designed ACPM to address the inability of the standard Clique Percolation Method to deal with very dense networks. Since the retrieved clusters necessitate post-processing (by going through the third stage), as a future work we should also aim at improving this clustering phase. This implies either improving ACPM or seeking new solutions for clustering the evolutionary networks. The field of \textit{Network Science} is constantly growing with new techniques to analyse large and dense networks, which may be relevant here. Furthermore, we can explore other community detection algorithms and eventually extend them, as we did for CPM, to try and improve the performance of Augur. 

\subsection{Post processing and sense making}
In the third and final stage of Augur, we filter the set of clusters and we enhance them with information regarding active authors and relevant papers. This information supports users in making sense of the results. Work to improve this stage  will focus on understanding qualitatively the kind of information that can better support the user in making sense of the results. By definition, the set of authors returned by Augur comprise researchers are engaged in the formation of the new topic, whereas the resulting papers are important research outputs that are laying the basis of this new area. 
We believe that it would be interesting to explore whether other information can be provided to the user, to assist him or her in making sense of the dynamics associated with the emergence of a new area.

\subsection{Trying different windows of years}
In Chapter \ref{ch:finalevaluation}, we designed the evaluation of Augur, aiming at investigating its ability to forecast the emergence of new research topics in the two years following the year of analysis. As showed in Table \ref{tab:match-evg-deb} (at page \pageref{tab:match-evg-deb}), an analysis performed in the year 2000, would allow us to detect topics emerged in the years 2001 and 2002. This choice was determined by the results obtained in the first study. For a particular emerging topic, we analysed the relationship between its ancestors in the five years prior its emergence. In Augur, we also used five sequential topic networks for detecting emerging topics in the following year. However, we estimated that with such amount of consecutive topic networks there would be enough information to forecast topics also in the following second year. In future work we plan to explore how far in the future is possible to have accurate forecasting and to what extent the performance of Augur changes over a larger timeframe.

\subsection{Testing new version of CSO}
In June 2018, our research team released a new version of CSO \citep{salatino2018b}. This new version was produced by running Klink-2 on a up-to-date corpus of publications, and it contains about 26K topics and 226K relationships (see Section \ref{sec:cso}). Compared to the previous version, this new version is more fine-grained and comprehensive in terms of relationships. 

In the studies presented in this dissertation, including the development of Augur, we heavily relied on the Computer Science Ontology in several stages, such as 
\begin{enumerate*}[label=(\roman*)]
\item creation of the semantic enhanced topic networks, 
\item creating the evolutionary networks, 
\item filtering clusters, and 
\item semantic enhancement in the evaluation. 
\end{enumerate*}

In the future we plan to evaluate whether the use of a more comprehensive and robust ontology allows us to improve the performance of Augur.

In particular, in Chapter \ref{ch:finalevaluation}, we used Eq. \ref{eq:new-jaccard-index} to measure the semantic similarity between a given cluster and ancestors of debutant topics. It will be interesting to observe how this measure changes when adopting the new version of the ontology.

\subsection{Introducing further dynamics}
Another limitation of Augur is that in this current version, it only focuses on the pace of collaboration between topics. It is possible that using only this indicator might not be enough to fully understand and detect the complex dynamics behind the creation of a topic. Augur uses information coming from authors and papers to support the user in making sense of the given results, however this information is not actually used in the process associated with the early detection of emerging topics. 

We believe that taking in consideration other entities, like authors and venues, may further improve Augur performances. This will enable the possibility to investigate other kinds of dynamics that could be associated with the emergence of new research areas, such as the patterns of collaboration between prominent authors, the dynamics of citations networks, or the change in the topic distributions of high-tier scientific venues.

The introduction of these new dynamics would require performing new studies like the ones presented in this dissertation. There should be a first study, similar to the one proposed in Chapter \ref{ch:firststudy}, to understand how a particular dynamic correlates with the emergence of new research area. Then, there should be a second study to analyse how to explore this new dynamic to effectively detect the emergence of new research topics. Eventually, there should be a third study to understand how to integrate the new findings within Augur.

\subsection{Scope}

At the moment, Augur works within the domain of \textit{Computer Science}. This limitation is due to two main reasons. Firstly, because of the limited availability of scholarly metadata at our disposal. The Rexplore dataset is not as comprehensive for other fields as it is for \textit{Computer Science}. The second reason is that the ontology (CSO) we adopted to define our topics also covers only the field of \textit{Computer Science}. In order to broaden the scope of Augur we need both to seek new scholarly datasets and topic ontologies for other disciplines. This is partially achievable in some fields like Medicine and Physics. 
For instance, in the field of Medicine, we can use Medline\footnote{Medline/PubMed data \url{https://www.ncbi.nlm.nih.gov/pubmed/}} as scholarly dataset and the Medical Subject Heading\footnote{Medical Subject Heading \url{https://meshb.nlm.nih.gov/}} (MeSH) for defining topics. Indeed, we are currently planning to adopt Augur to work in this domain, since both resources are available.
In the field of Physics, instead, we can use the Physics Subject Headings\footnote{Physics Subject Headings \url{https://physh.aps.org/}} (PhySH) as vocabulary for subject areas and arXiv\footnote{arXiv Physics \url{https://arxiv.org/archive/physics}} for the corpus. In this case, however, there are other challenges to face, because for instance the arXiv corpus does not store metadata and therefore they must be parsed from the \textit{pdf} files.

Since the design of Augur was based on robust epistemological theories, we are confident that it will achieve satisfactory results also in other scientific fields. However, given that \textit{Computer Science} is a highly dynamic area compared to other scientific fields, we expect that some parameters of the system will need to be attuned.

In brief, here we list a few necessary requirements in order to adopt Augur in other fields of Science:
\begin{itemize}
\item \textbf{Ontology or research topics}, this is important as it will be used as glossary of topics and therefore support the definition of research topic in the investigated field;
\item \textbf{Corpus of scientific articles}, covering the new field, upon which performing the analysis. As in Rexplore, for each paper, the corpus should contain information about either keywords or topics and the year of publication. This information is required for creating the semantic enhanced topic network from which building the several evolutionary networks, and extracting the clusters of collaborating topics. If topics are not available, there is a need for a further step which extracts topics from papers using their metadata, such as title and abstract or full-content;
\item \textbf{Tuning of parameters}, since the parameters of Augur are domain dependent, and need to be adjusted to accommodate the explored field. These parameters include the number of links that define the cluster, the similarity measure used to merge similar clusters, and others showed in Chapter \mbox{\ref{ch:secondstudy}}. 

Another factor that should be taken into account when exploring the dynamics in other fields is their publishing behaviour. Each field possesses its own publishing preferences. For instance, journal or conference articles are poor vehicles of communication in the field of \textit{Social Science} as they are not suitable to discuss complex issues. Hence preferring books. However, many other fields investigating narrow area of study, with discrete and separable problems, e.g., \textit{Physics}, \textit{Medicine}, \textit{Engineering}, and others, predominantly use articles. We do not have enough evidence to claim how the behaviour of Augur could be affected by this factor.
\end{itemize}

\subsection{Scalability}

It is also worth discussing a possible scalability issue that could arise while processing different topic networks. 

As reported in Table {\ref{tab:statofevolnetwork}} the evolutionary network of the year 2000 consists of 2263 nodes and 13327 links, and both these numbers tended to grow over time. Indeed, the evolutionary network of the year 2009 consists of 9865 nodes and 99660 links. However, the computational time to process each one of these networks did not exceed 3-4 hours. Working in other domains of science can lead to process larger networks affecting the scalability. For instance, speaking with colleagues at Chan Zuckerberg Initiative, in which they investigate trends in the field of \textit{Medicine}, they deal with keyword networks containing more than 15K unique keywords, which poses real challenges in processing them. 

We are currently evaluating Augur on a more recent data using Microsoft Academic Search with number of nodes and link about three times larger than our previous evolutionary networks. Processing such large evolutionary network takes around 24 hours.

We are thus confident that scalability should not be an issue when applying Augur on other research areas. 

\subsection{Human-based evaluation}

In Chapter {\ref{ch:finalevaluation}} we evaluated Augur on a gold standard of historical data, and in order to do so, we first we designed an evaluation framework, which defines when a returned cluster matches the ancestors of a debutant topic. Then, we extracted the gold standard.
These two steps allowed us to evaluate Augur both in an automatic and a quantitative way. However, it may be possible to evaluate Augur more qualitatively with the support of domain experts. 

A domain expert can analyse the interaction of some topics in a certain year and suggest, based on his or her judgement, that such interaction might lead to the emergence of new topics. However, to pursue such evaluation we would need several domain experts with different areas of expertise analysing the different clusters returned by Augur. This was not attainable within the timescale provided for this research. Indeed, this would have required a set of time consuming activities such as: (i) finding available domain experts for each relevant topic, (ii) understanding their domain of expertise, (iii) selecting candidate clusters that could fit his or her expertise, (iv) engaging with the experts during the process of evaluation, and (v) making sure the experts complete the evaluation with high quality.

In the future, we plan to run a user study in which we will collect the opinion of several domain experts, including senior editors at Springer Nature, regarding the results obtained by Augur. 

\section{Concluding remarks}
In this dissertation we presented the work of three years of research on the early detection of new research topics. This problem has been analysed from different angles, putting us in front of several challenges, that made this doctoral work particularly interesting and rewarding.

We firstly analysed the literature and formulated the set of hypotheses discussed in Section \ref{sec:hypotheses}. We then conducted an empirical study which aimed at uncovering some of the patterns that can lead to the emergence of new research topics. As discussed in Chapter \ref{ch:firststudy}, we initially analysed the relationship between an emerging topic and the collaboration between already existing topics, and how the development of the latter can influence the former. This understanding allowed us to design Augur, an approach to detecting research topics in their embryonic stage, as described in Chapter \ref{ch:secondstudy} and evaluated in Chapter \ref{ch:finalevaluation}. Augur yielded excellent results and outperformed alternative approaches. 

This doctoral work opens up several interesting research directions and we look forward to further developing the techniques presented here.


\cleardoublepage
\ifdefined\phantomsection
  \phantomsection  
\else
\fi
\addcontentsline{toc}{chapter}{Bibliography}

\bibliography{references/thesis}
\clearpage
\addcontentsline{toc}{chapter}{Index}
\printindex

\end{document}